\documentclass[a4paper,11pt]{book}
\usepackage[english]{babel}
\usepackage{amsmath,amssymb}
\usepackage{graphicx}
\usepackage{subfig}
\usepackage{appendix}
\usepackage{hyperref}
\usepackage{tikz}
\usepackage{epigraph}
\usepackage{adjustbox}
\usepackage{booktabs}
\usepackage[uline]{hhtensor}
\usepackage[inline]{asymptote}
\usepackage{multibib}
\newcites{publist}{Publication list}
\usepackage{cmbright}

\newcommand\PP{\mathbf{P}}
\newcommand\pp{\mathbf{p}}
\newcommand\pphat{\mathbold{\hat{p}}}
\newcommand\ppbar{\mathbf{\bar{p}}}
\newcommand\LL{\mathbf{L}}
\newcommand\SSS{\mathbf{S}}
\newcommand\FF{\mathbf{F}}

\newcommand\GG{\mathbf{G}}
\newcommand\GGm{\mathbf{G_-}}
\newcommand\GGp{\mathbf{G_+}}
\newcommand\Gsix{\begin{bmatrix}\GGp\\\GGm\end{bmatrix}}

\newcommand\Curlsix{\begin{bmatrix}\nabla\times&0\\0&-\nabla\times\end{bmatrix}}
\newcommand\CurlsixEH{\begin{bmatrix}0&\nabla\times\\-\nabla\times&0\end{bmatrix}}

\newcommand\RSold{\frac{1}{\sqrt{2}}\begin{bmatrix}I&iZ\\I&-iZ\end{bmatrix}}
\newcommand\HH{\mathbf{H}}
\newcommand\DD{\mathbf{D}}
\newcommand\EE{\mathbf{E}}
\newcommand\BB{\mathbf{B}}
\newcommand\JJ{\mathbf{J}}
\newcommand\jj{\mathbf{j}}
\newcommand\Hel{\mathbf{J}\cdot\mathbf{P}/|\mathbf{P}|}
\newcommand\Helfrac{\frac{\mathbf{J}\cdot\mathbf{P}}{|\mathbf{P}|}}
\newcommand\Hsix{\begin{bmatrix}i\partial_t&0\\0&i\partial_t\end{bmatrix}}
\newcommand\HsixEH{\begin{bmatrix}\partial_t&0\\0&\partial_t\end{bmatrix}}
\newcommand\rr{\mathbf{r}}
\newcommand\xhat{\mathbf{\hat{x}}}
\newcommand\lhat{\mathbf{\hat{l}}}
\newcommand\yhat{\mathbf{\hat{y}}}
\newcommand\zhat{\mathbf{\hat{z}}}
\newcommand\what{\mathbf{\hat{w}}}
\newcommand\rhat{\mathbf{\hat{r}}}
\newcommand\uhat{\mathbf{\hat{u}}}
\newcommand\shat{\mathbf{\hat{s}}}

\newcommand\phat{\mathbf{\hat{p}}}
\newcommand\nhat{\mathbf{\hat{n}}}
\newcommand\Mr{\widehat{\mathbf{M}}_\nu(\mathbf{r})}
\newcommand\Nr{\widehat{\mathbf{N}}_\nu(\mathbf{r})}
\newcommand\Mreta{\mathbf{M}_\nu}
\newcommand\Nreta{\mathbf{N}_\nu}
\newcommand\plane{\exp\left(i\pp\cdot\rr\right)}
\newcommand\planewt{\exp\left(i\left(\pp\cdot\rr-\omega t\right)\right)}
\newcommand\pwplus{\mathbf{\hat{e}}_+}
\newcommand\pwminus{\mathbf{\hat{e}}_-}
\newcommand\Longint{\int_0^{\infty} dk\int_0^{\pi}d\theta\sin\theta\int_{-\pi}^{\pi}d\phi}
\newcommand\Longintchange{\int_0^{\infty} dk\int_0^{\pi}d\theta\sin\theta\int_{-\pi+\beta}^{\pi+\beta}d\phi}

\newcommand\krho{p_{\rho}}
\newcommand\qrho{q_{\rho}}
\newcommand\kz{p_z}

\newcommand\Cnkrho{\mathbf{C}_{n\krho}}
\newcommand\Dnkrho{\mathbf{D}_{n\krho}}

\newcommand\pwket{|\pp\ \lambda\rangle}

\newcommand\ed{\mathbf{p}}
\newcommand\md{\mathbf{m}}
\newcommand\MdE{{\matr{\alpha}}_{\ed\mathbf{E}}}
\newcommand\MdH{{\matr{\alpha}}_{\ed\mathbf{H}}}
\newcommand\MmE{{\matr{\alpha}}_{\md\mathbf{E}}}
\newcommand\MmH{{\matr{\alpha}}_{\md\mathbf{H}}}
\newcommand\rrp{\mathbf{r_0}}
\newcommand\Eed{\mathbf{E_{\ed}}}
\newcommand\Emd{\mathbf{E_{\md}}}
\newcommand\Hed{\mathbf{H_{\ed}}}
\newcommand\Hmd{\mathbf{H_{\md}}}
\newcommand\edbar{\mathbf{\bar{p}}}
\newcommand\Eedbar{\mathbf{E_{\edbar}}}
\newcommand\Hedbar{\mathbf{H_{\edbar}}}
\newcommand\duetoref[1]{\stackrel{(\ref{#1})}{=}}
\newcommand\impliesdueto[1]{\stackrel{#1}{\implies}}

\newcommand\ket[1]{|#1\rangle}

\newcommand{\Ef}{{\mathcal{E}}}
\newcommand{\Hf}{{\mathcal{H}}}

\newcommand\Shat{\mathbf{\hat{S}}}
\newcommand\Lhat{\mathbf{\hat{L}}}
\newcommand\pwsthetaphi[2]{\widehat{\textbf{s}}_{#1,#2}}
\newcommand\pwpthetaphi[2]{\widehat{\textbf{p}}_{#1,#2}}
\newcommand\phik{\phi_{\pp}}
\newcommand\thetak{\theta_{\pp}}
\newcommand\smallkrho{\frac{\krho}{k}\rightarrow 0}
\newcommand\smallqrho{\frac{q_\rho}{k}\rightarrow 0}

\newcommand\Field{|\omega\ \lambda\rangle}
\newcommand\Gen{\left(-i\beta_x\Lambda P_x/|\PP|\right)}
\newcommand\Gentwo{\left(-i\beta_x P_x/|\PP|\right)}
\newcommand\Genthree{\left(-i(\lambda\beta_x/\omega) P_x\right)}

\newcommand{\shortminus}{\scalebox{0.75}[1.0]{\( - \)}}

\newcommand\zbra{\langle\bar{\lambda}\ p\zhat|}
\newcommand\mzbra{\langle\bar{\lambda}\ \shortminus p\zhat|}
\newcommand\pmz{|\pm p\zhat \ \lambda\rangle}
\newcommand\pmzbra{\langle\lambda\ \pm p\zhat|}
\newcommand\z{|p\zhat \ \lambda\rangle}
\newcommand\mz{|\shortminus p\zhat \ \lambda\rangle}

\newcommand\vhat{\mathbf{\hat{v}}}
\newcommand\aaa{a_{\pp}^{\ppbar}}
\newcommand\ddd{d_{\pp}^{\ppbar}}
\providecommand{\e}[1]{\ensuremath{\times 10^{#1}}}
\newcommand\dtheta{\chi_{\pp\ppbar}}

\newcommand\F{\mathcal{F}}
\newcommand\G{\mathcal{G}}

\makeatletter{}\usetikzlibrary{arrows}
\newcommand{\rc}[4]{
\draw [thick,->,>=triangle 45,shift={(#1,#2)},rotate=#3,scale=#4] (0,0) -- (2,0);
\draw [thick,->,>=stealth,shift={(#1,#2)},rotate=#3,scale=#4](1.55,0.1) arc (20:340:0.10 and 0.25);
}
\newcommand{\lc}[4]{
\draw [thick,->,>=triangle 45,shift={(#1,#2)},rotate=#3,scale=#4] (0,0) -- (2,0);
\draw [thick,<-,>=stealth,shift={(#1,#2)},rotate=#3,scale=#4](1.55,0.1) arc (20:340:0.10 and 0.25);
}

\newlength{\drop}
\newcommand*{\titleDS}{\begingroup\drop=0.1\textheight
\flushleft
{\Huge\bfseries Helicity and duality symmetry}\vspace*{0.35cm}
{\Huge\bfseries in light matter interactions:}\vspace*{0.35cm}
{\Huge\bfseries Theory and applications}\par
\vspace{13cm}
\flushright
{\bf \Large Ivan Fernandez i Corbaton}\\
\vspace{2cm}
{Thesis accepted by Macquarie University\\for the degree of\\
	{\large Doctor of Philosophy}\\
Department of Physics and Astronomy\\}\par

\endgroup}

\usepackage[pages=some]{background}
\backgroundsetup{
scale=2,
angle=0,
opacity=1,  contents={\includegraphics[height=\paperheight]{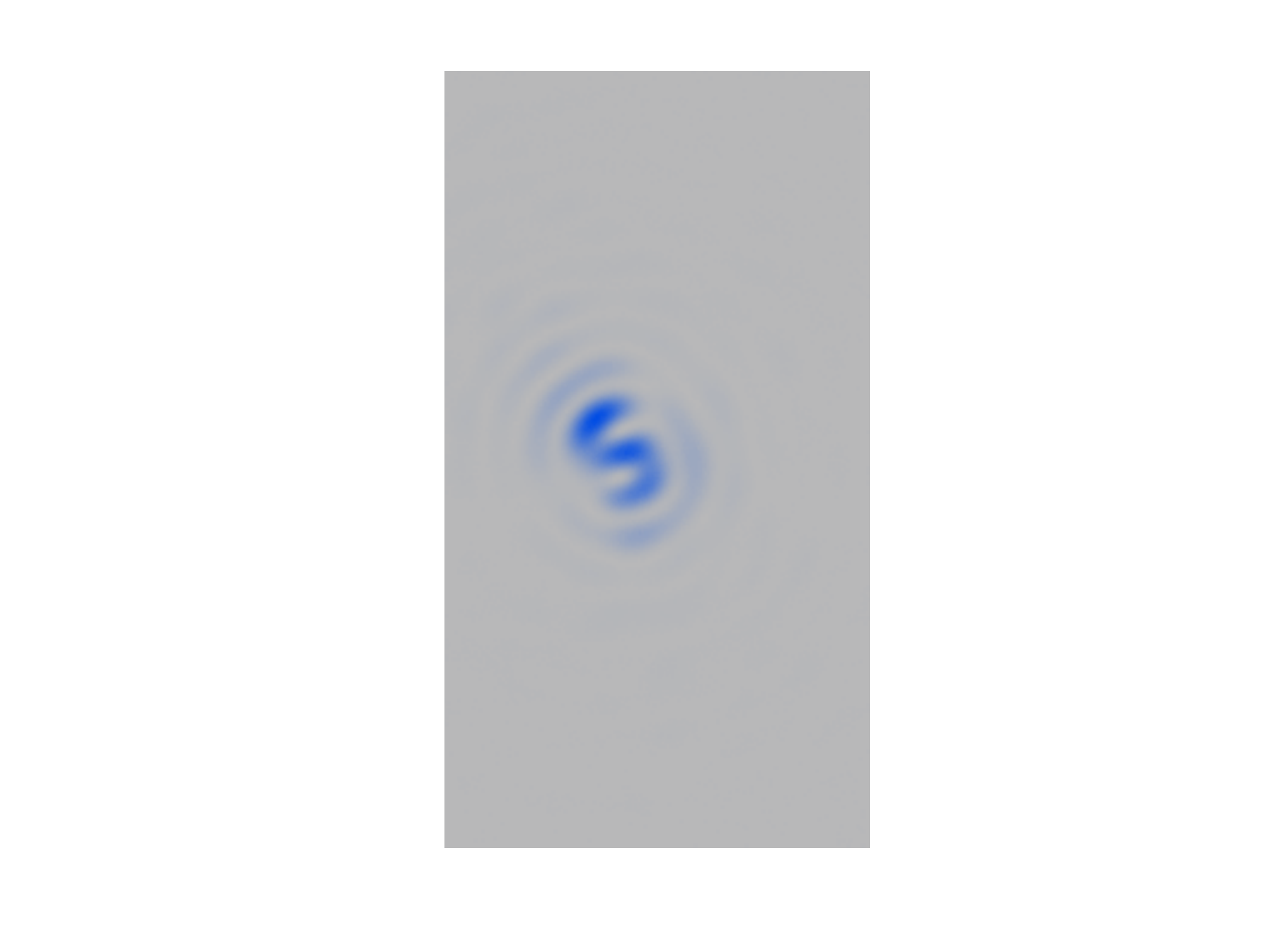}}
}
\begin{document}
\thispagestyle{empty}
\BgThispage
\titleDS

\newpage\thispagestyle{empty}
\hbox{}
\frontmatter
 
\makeatletter{}\chapter*{}
Except where acknowledged in the customary manner, the material presented in this thesis is original, to the best of my knowledge, and has not been submitted in whole or part for a degree in any university.
		\vspace{0.5cm}
\hrule
		\vspace{0.5cm}
{\begin{flushright} Ivan Fernandez i Corbaton \end{flushright}
\chapter{Acknowledgments}
First and foremost I want to thank my wife Magda for agreeing to embark in an adventure that was bound to change both our lives. It indeed has. Her support, in many different ways, has been crucial.

I also want to thank my mother, Neus, for her amazing encouragement and support during all the years that I have been studying, working, and again studying in different countries; some of them quite far away from our home town of Almacelles (Catalonia). My brother Darius, and my extended family have also always been supportive of my choices. I particularly want to mention Julia, my goddaughter, to whom I owe a few steak dinners. A debt which I fully intend to settle.

I have the fortune to have a few lifelong friends. I would trust and help them with anything. Every time that I see them I can feel my bond with them growing stronger, overcoming the effects of large spatio-temporal distances.

I have made new friends in Macquarie University, to whom I wish the best for the future. I will make special mention of Xavier Zambrana-Puyalto, Nora Tischler and Mauro Cirio. I always enjoy talking with them about physics, conversations that have no doubt contributed to my thesis. I also very much enjoy our conversations about (so many!) other matters.

I also would like to thank my friend and long time colleague Srikant Jayaraman. I learned a lot from Srikant during the years that we worked together in Qualcomm. Also, Srikant was the one that put the idea of a PhD in my head.

Last, but not least by any measure, I thank my advisor A. Prof. Gabriel Molina-Terriza. His guidance, insights, flexibility and wide field of view have made this thesis possible. I want to particularly thank him for taking a risk with me. He took an electrical engineer as his student and approved of him venturing in uncharted territory quite early on. Judging from my short experience, Gabriel seems to me the prototype of an all around physicist, blurring the divide between experimentalists and theorists. I have greatly benefited from his world class knowledge of both disciplines.  

\chapter{Agra\"iments}
Primer de tot, vull agrair-li a la meva dona Magda que acced\'is a comen\c{c}ar una aventura que havia de canviar les nostres vides. Com efectivament ha passat. El seu suport, de molts tipus, ha estat crucial. 

Tamb\'e vull donar-li les gr\`acies a la meva mare Neus pel seu suport i encoratjament durant tots els anys en que he estat estudiant, treballant i estudiant altre cop a diferents pa\"isos; alguns d'ells molt lluny del nostre poble d'Almacelles. El meu germ\`a Dario i la resta de la meva fam\'ilia tamb\'e han estat sempre comprensius envers les meves anades i vingudes. Vull mencionar particularment la meva fillola J\'ulia, a qui dec alguns sopars. Un deute el qual tinc la intenci\'o de saldar.

Tinc la sort de tenir alguns amics de per vida, als quals confiaria qualsevol cosa. Cada cop que els veig sento que la meva connexi\'o amb ells es fa m\'es forta, superant els efectes deguts a grans dist\`ancies espai-temporals.

He fet alguns amics nous a Macquarie University, als quals desitjo el millor per al futur. Mencionar\'e en particular en Xavier Zambrana-Puyalto, la Nora Tischler i en Mauro Cirio. Sempre disfruto parlant amb ells de f\'isica, unes conversacions que han contribu\"it a la meva tesi. Tamb\'e disfruto de les conversacions amb ells sobre (tants!) altres temes.

Tamb\'e vull donar-li les gr\`acies al meu amic, i company de feina durant molt de temps, Srikant Jayaraman. Vaig apendre molt d'en Srikant durant els anys en que v\`arem treballar junts a Qualcomm. A m\'es a m\'es, ell fou el que em va ficar al cap la idea de fer un doctorat.

Per \'ultim, per\`o no pas pel que fa a la import\`ancia, vull donar les gr\`acies al meu director de tesi, el professor associat Gabriel Molina-Terriza. Les seves indicacions, visi\'o i flexibilitat han fet possible aquesta tesi. Volia agra\"ir-li especialment haver pres riscos amb mi. Va acceptar com a estudiant un enginyer en telecomunicacions i va deixar que s'endins\'es en territori desconegut molt r\`apidament. Desde la meva curta experi\`encia en f\'isica, en Gabriel em sembla el prototip del f\'isic complet, que esborrona la l\'inia que divideix els experimentalistes dels te\`orics. M'he beneficiat enormement del seu coneixement, de nivell mudial, de les dues disciplines.

\chapter{Publication list}
\begingroup
\renewcommand{\chapter}[2]{}%
\newcommand{\etalchar}[1]{$^{#1}$}

\endgroup

\chapter{Abstract}
The understanding of the interaction between electromagnetic radiation and matter has played a crucial role in our technological development. Solar cells, the internet, cell phones, GPS and X-rays are examples of it. In all likelihood this role will continue as we strive to build better solar cells, millimeter sized laboratories and more sensitive medical imaging systems, among other things. Many of these new applications are stretching the capabilities of the tools that we use for studying and engineering the interaction of electromagnetic radiation and matter. This is particularly true at the meso-, nano- and micro-scales. 

My thesis is an attempt to build a new tool for studying, understanding and engineering the interaction of electromagnetic radiation with material systems. The strategy that I have followed is to approach interaction problems from the point of view of symmetries and conservation laws. The main novelty is the systematic use of the electromagnetic duality symmetry and its conserved quantity, the electromagnetic helicity. Their use allows to treat the electromagnetic polarization degrees of freedom in a straightforward way and makes the framework useful in practice.

Since the tool is based on symmetries, the results obtained with it are very general. In particular, they are often independent of the electromagnetic size of the scatterers. On the other hand, they are often mostly qualitative. When additional quantitative results are required, more work needs to be done after the symmetry analysis. Nevertheless, one then faces the task armed with a fundamental understanding of the problem.

In my thesis, I first develop the theoretical basis and tools for the use of helicity and duality in the study, understanding and engineering of interactions between electromagnetic radiation and material systems. Then, within the general framework of symmetries and conservation laws, I apply the theoretical results to several different problems: Optical activity, zero backscattering, metamaterials for transformation optics and nanophotonics phenomena involving the electromagnetic angular momentum. I will show that the tool provides new insights and design guidelines in all these cases. 

\chapter{Preface}\label{chap1}
I have tried to write my thesis so that it may be useful for as many people as possible. In doing so, I have included material that will appear unnecessary to readers that are familiar with the study of symmetry transformations and Hilbert spaces. Such material is contained mostly in the background chapter (Chap. \ref{chap2}) and the first section of the theory chapter (Sec. \ref{secc3:gf}). I hope that those parts will allow the readers that are not familiar with their content to better judge whether the core parts of my thesis, namely the rest of the theory chapter and the application chapters, are of any use to them. 

Later in this preface I give a short account of how my research changed from its initial direction to the development of a tool for the study of the interaction between electromagnetic radiation and matter. Chapter \ref{chap2} is the background and introduction chapter. Chapter \ref{chap3} contains the theoretical basis and tools for the use of helicity and duality in the study, understanding and engineering of the interaction between electromagnetic radiation and material systems. Subsequent chapters contain applications of the theoretical framework that have lead to new insights in phenomena related to angular momentum (Chap. \ref{chap4}), zero backscattering (Chap. \ref{chap5}), molecular optical activity (Chap. \ref{chap6}) and metamaterials for transformation optics (Chaps. \ref{chap7}). Chapter \ref{chap8} contains a summary of the main contributions to the field contained in this thesis, conclusions and outlook. The most relevant publications are attached at the end.

I sincerely hope that you find some of the contents of my thesis useful for your work.

\makeatletter{}\section*{Why develop a new tool?}
My research was not initially aimed at the development of a new framework for the study of light matter interactions. The following is an account of how the subject of my thesis changed. I include it here because the process may interest some readers. In short, what happened was that the analysis of some numerical and experimental results revealed inconsistencies in the state-of-the art theoretical explanation of those results. Those inconsistencies were completely solved using the point of view of symmetries and conservation laws and, in particular, the electromagnetic duality symmetry and helicity conservation. Motivated by this initial success, my advisor and I agreed to redirect my research. 

The initial title of my thesis was ``Exploring the limits of spatially entangled photons''. I was supposed to study the properties of photons entangled in their momentum degrees of freedom. To start my project, my advisor A. Prof. Gabriel Molina-Terriza suggested to get hold of the radiation diagram of a nanohole in a metallic film. The idea was to explore the interaction of momentum entangled photons interacting with subwavelength size structures. The classical radiation diagram of the nanohole was therefore a necessary first step. The high degree of symmetry of the system suggested the existence of an analytical expression for its radiation diagram. Such expression does not currently exist. Even though the literature about interaction of light with nanoapertures is vast (\cite{Ebbesen1998,Sonnichsen2000,Martin-Moreno2001,Genet2007} and references therein), there is no exact analytical solution for the radiation diagram of a nanohole in a metallic film. Had there been one, my thesis would be very different.

The next best thing was a numerical approach. None of the existing techniques (notably \cite{Paulus2001,Bonod2005}) seemed to provide what I was after, i.e., the plane wave decomposition of the field scattered by the nanohole upon excitation by an arbitrary plane wave. I decided to try to develop a method and succeeded in devising a suitable semi-analytical approach. The technique allows to obtain the plane wave decomposition of the field scattered by objects embedded in a planar multilayer structure under general illumination \cite{FerCor2011}. 
After I wrote and tested the code to implement the technique, my advisor suggested to illuminate the hole with modes of different polarization and spatial phase dependence. These are modes whose expression in the collimated limit\footnote{The optical axis being the $z$ axis.} is dominated by a term to $\exp(il\theta)\lhat$ or $\exp(il\theta)\rhat$, where $l$ is an integer, $\theta=\arctan(y/x)$ and $\lhat$ and $\rhat$ are the left and right circular polarization vectors, respectively. An $\exp(il\theta)$ phase dependence implies the existence of a phase singularity with zero intensity and topological charge $l$ in the center $(x=0,y=0)$. These modes are the well known ``doughnut'' beams.

With additional simulation code, I modeled the focusing of the input mode, its interaction with the nanohole (using its radiation diagram) and the action of a collection objective on the transmitted light. Figure \ref{figc1:sim} shows exemplary results of the amplitude and phase of the two circular polarizations at the output. The results show polarization conversion\footnote{The percentage of conversion is not relevant for this discussion.}. Crucially, an invariant quantity is found by assigning the value +1 to $\lhat$ and $-1$ to $\rhat$, and summing it to the azimuthal phase number $l$. This sum is preserved: Its value is the same for the input and the two output polarizations.

\begin{figure}[h!]

\begin{center}
	 \begin{tabular}{cc|cc}
	\subfloat[$(-1,\lhat)\rightarrow(-1,\lhat)$]{\includegraphics[width=0.23\textwidth]{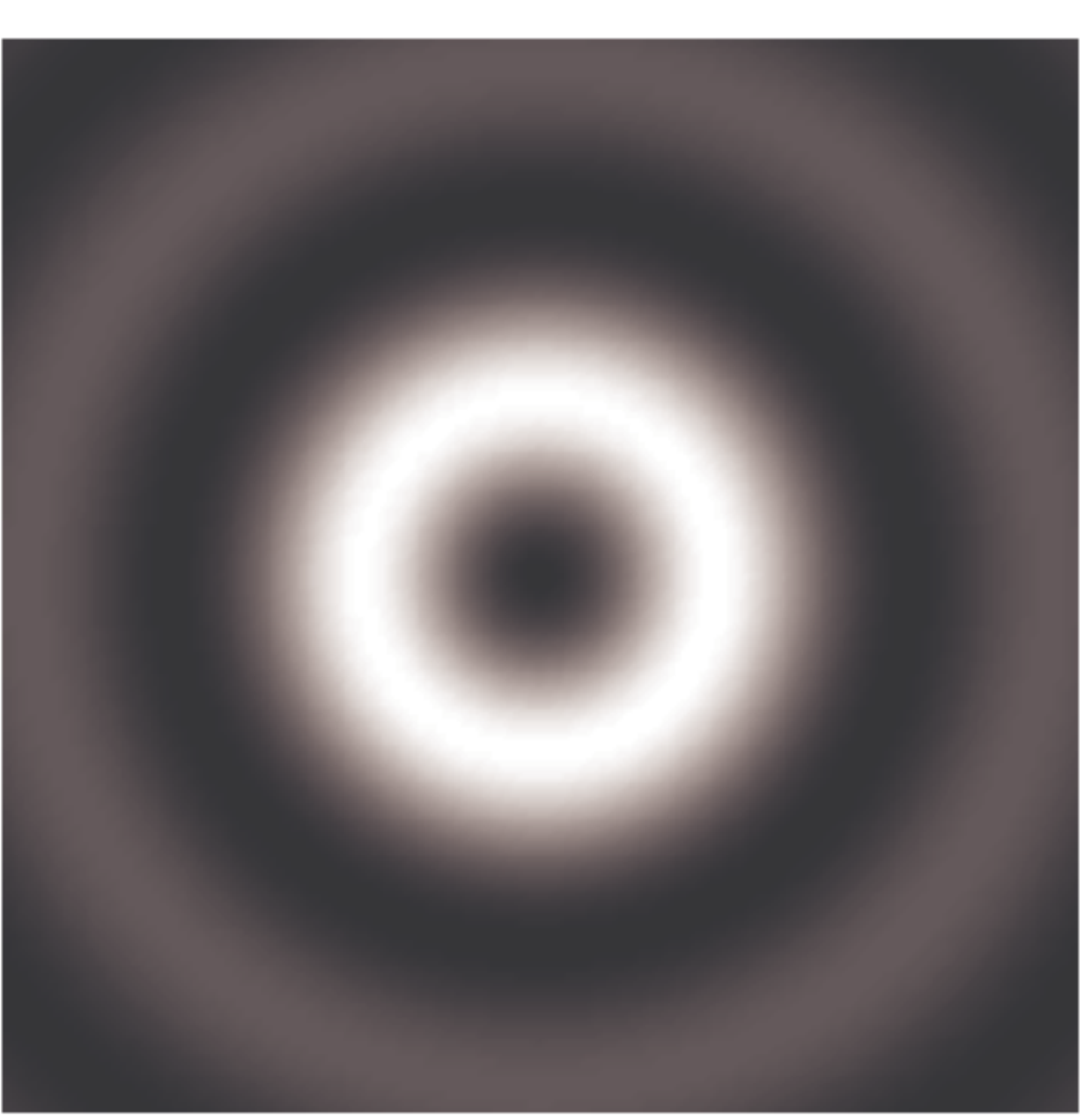}}
	&
	\subfloat[$(-1,\lhat)\rightarrow(1,\rhat)$]{\includegraphics[width=0.23\textwidth]{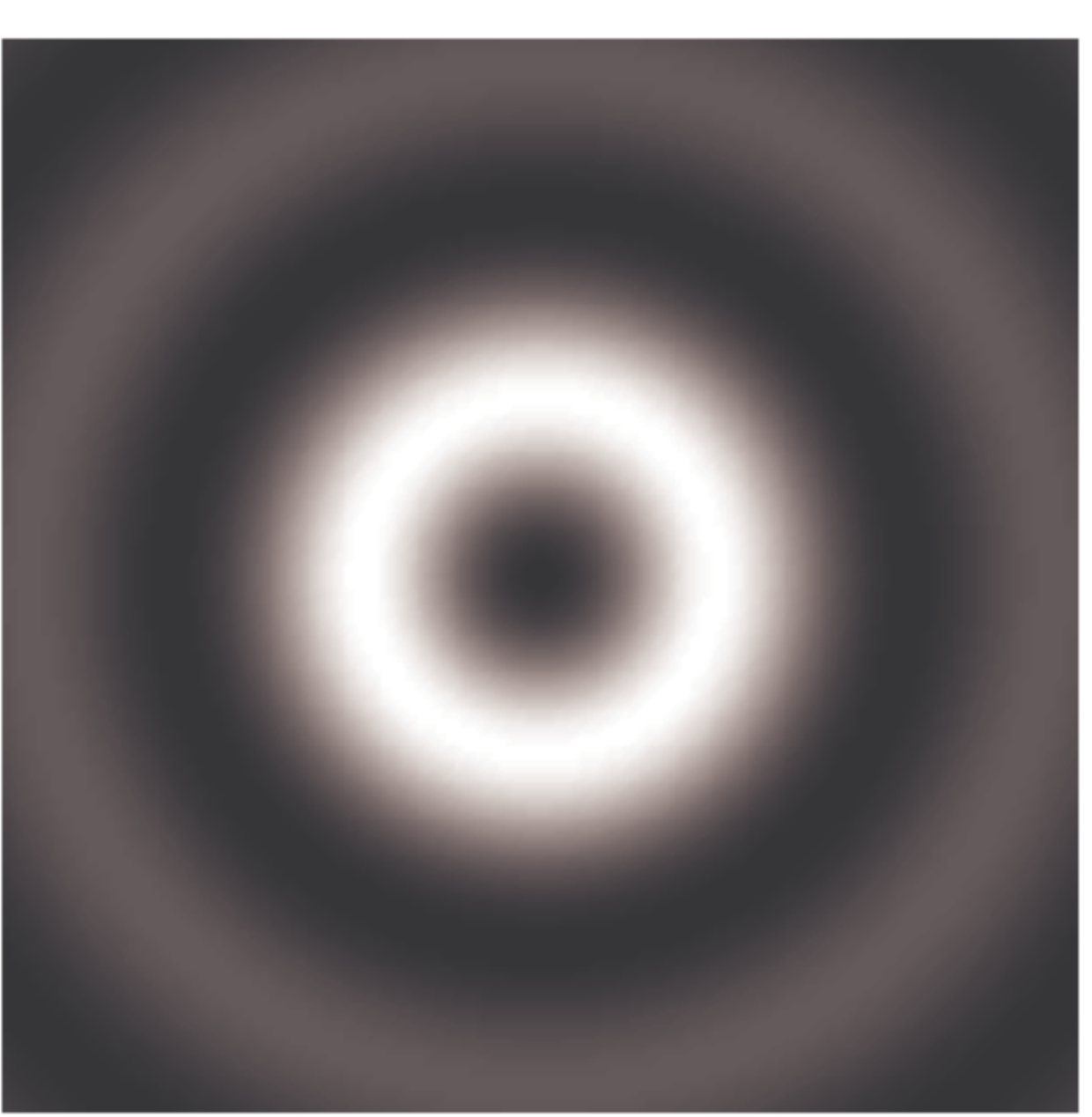}}
	&
	\subfloat[$(0,\rhat)\rightarrow(-2,\lhat)$]{\includegraphics[width=0.23\textwidth]{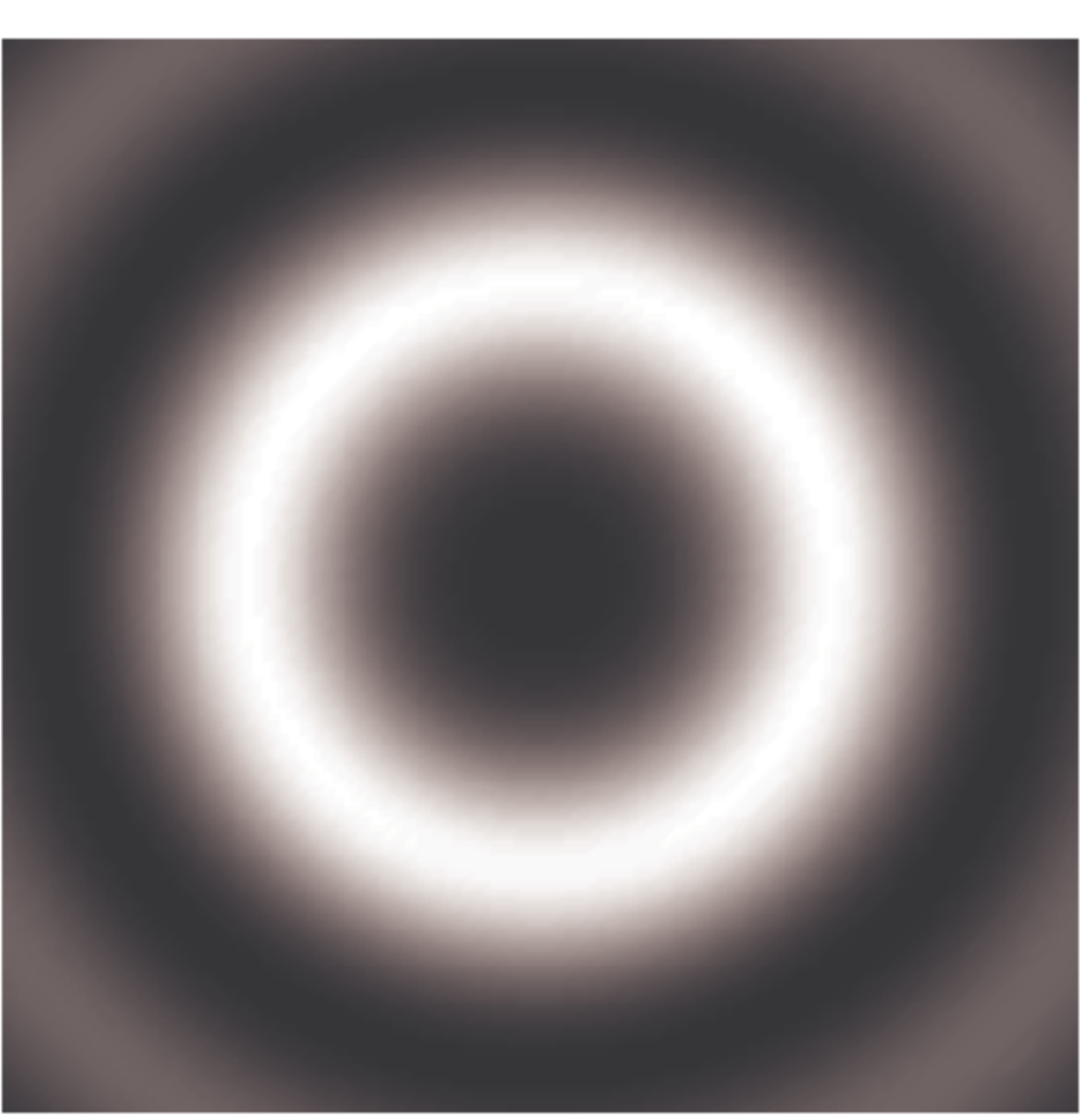}}
	&
	\subfloat[$(0,\rhat)\rightarrow(0,\rhat)$]{\includegraphics[width=0.23\textwidth]{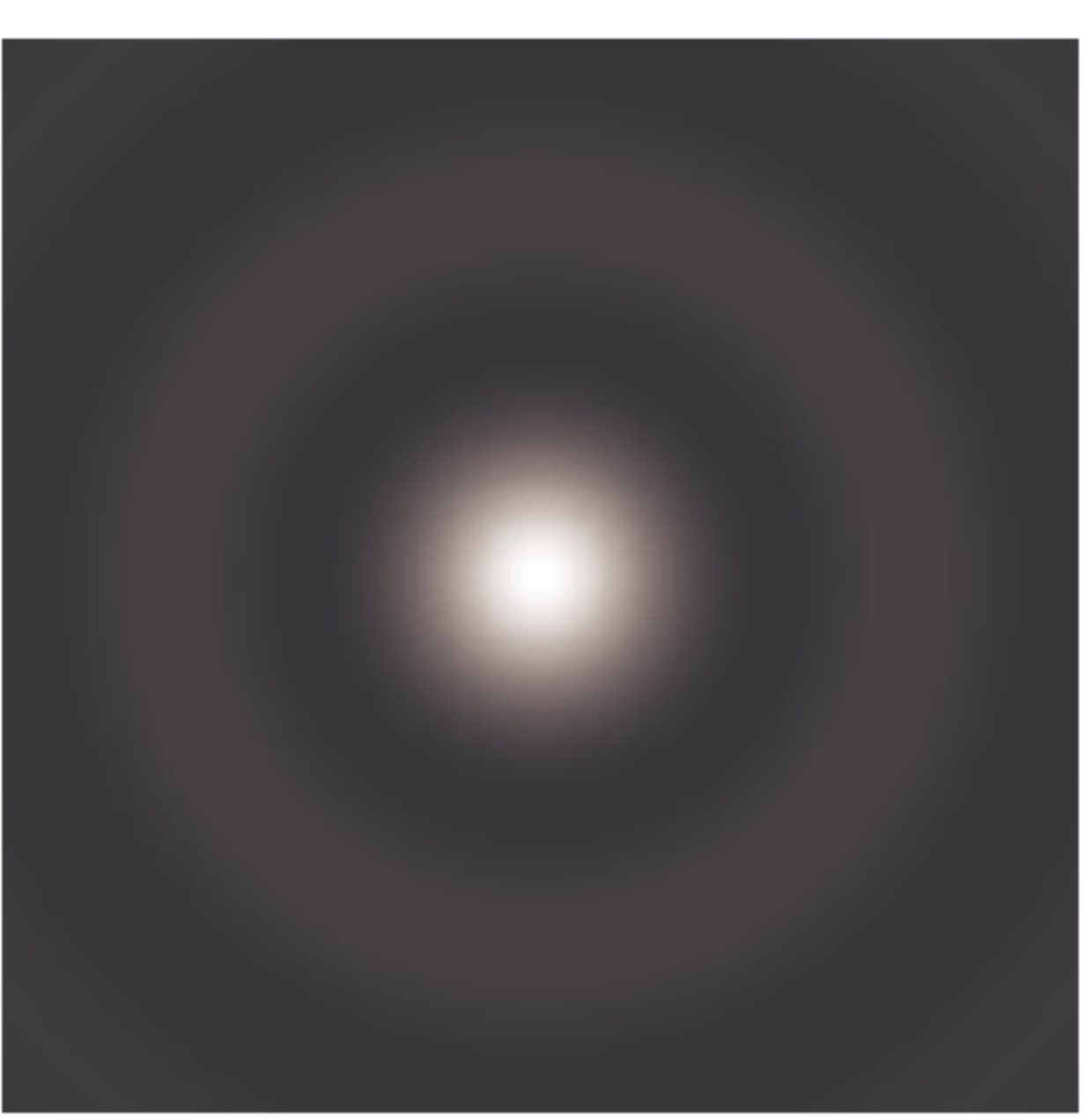}}
\\
	\subfloat[$(-1,\lhat)\rightarrow(-1,\lhat)$]{\includegraphics[width=0.23\textwidth]{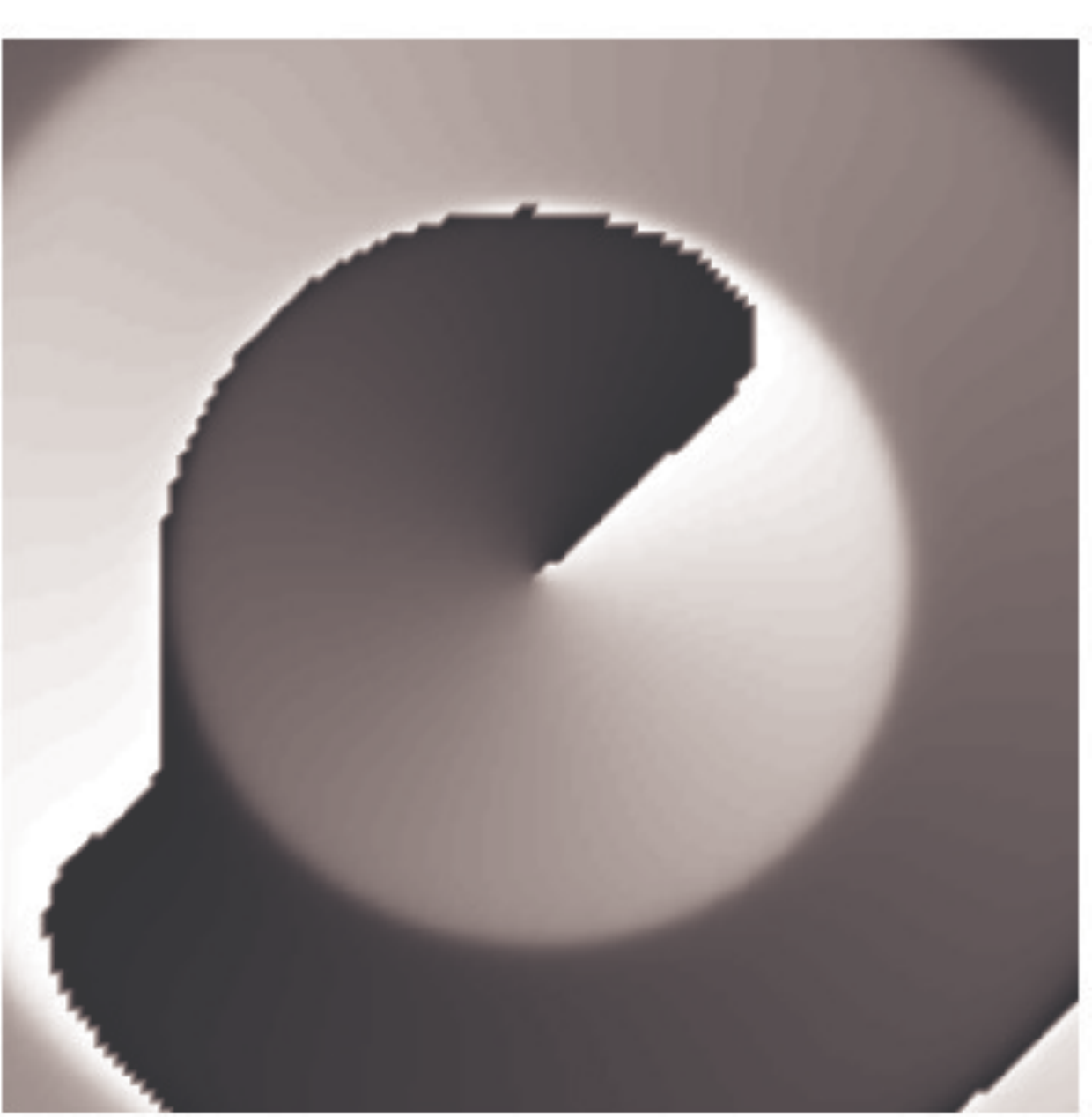}}
	&
	\subfloat[$(-1,\lhat)\rightarrow(1,\rhat)$]{\includegraphics[width=0.23\textwidth]{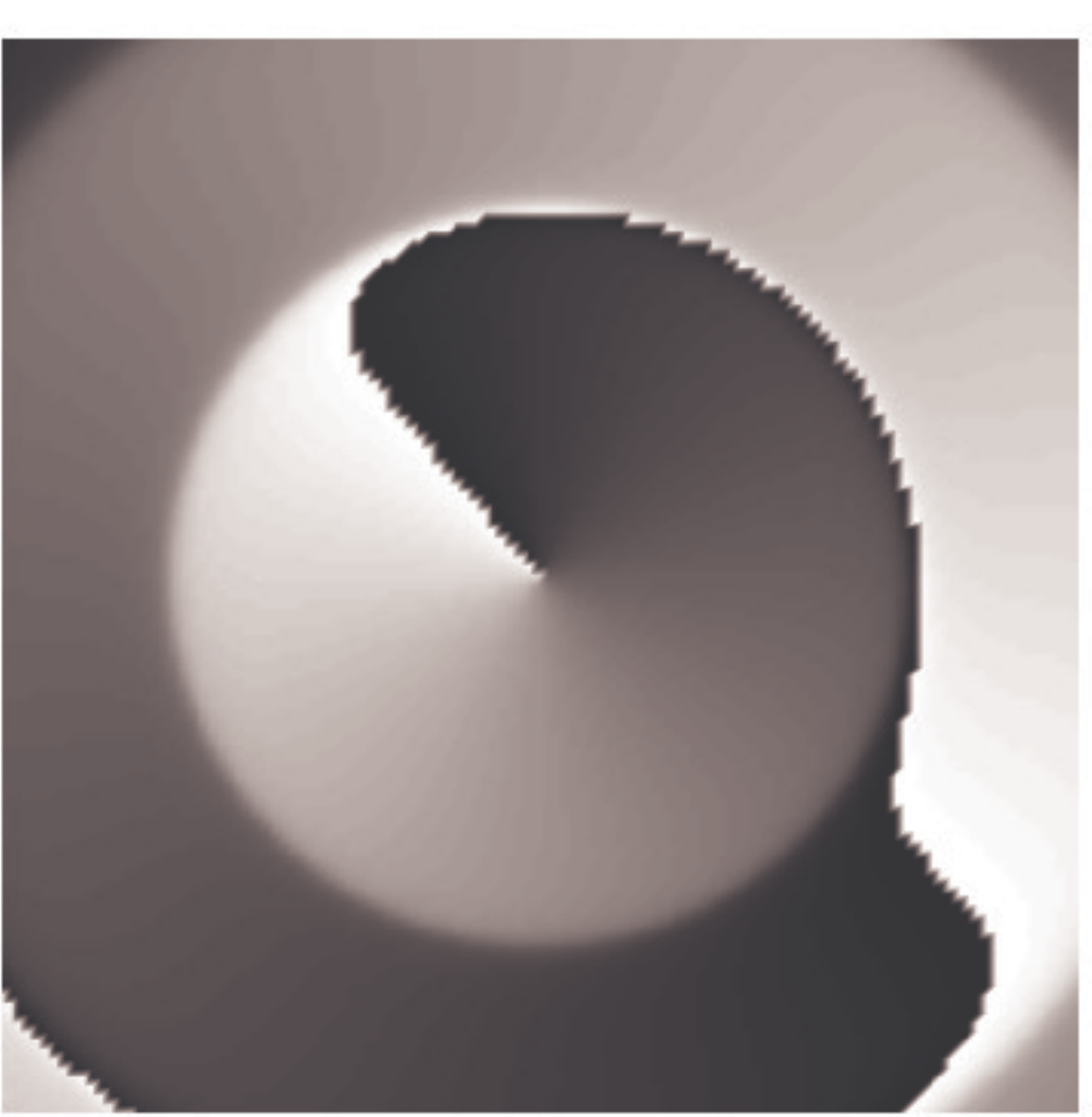}}
	&
	\subfloat[$(0,\rhat)\rightarrow(-2,\lhat)$]{\includegraphics[width=0.23\textwidth]{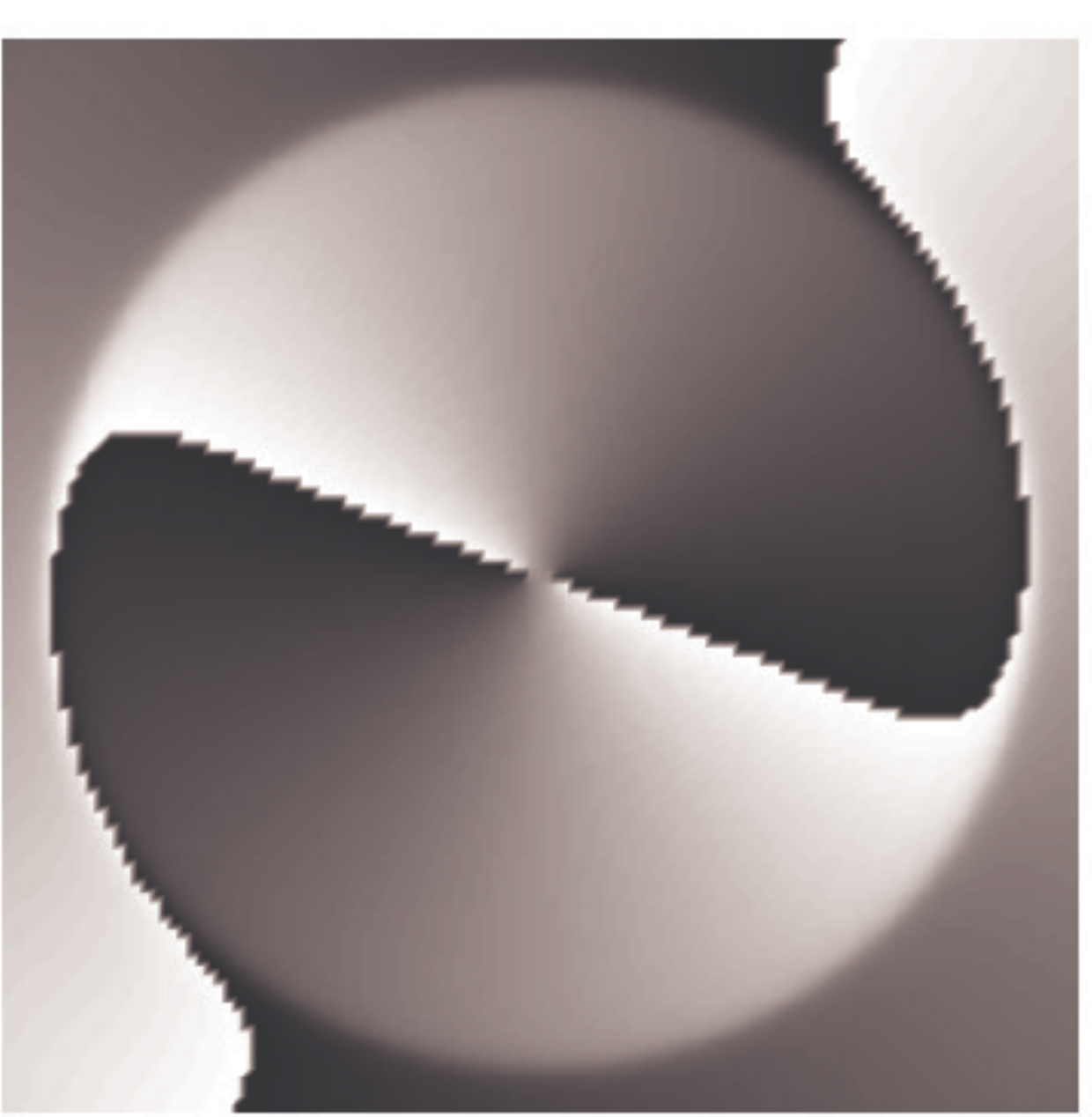}}
	&
	\subfloat[$(0,\rhat)\rightarrow(0,\rhat)$]{\includegraphics[width=0.23\textwidth]{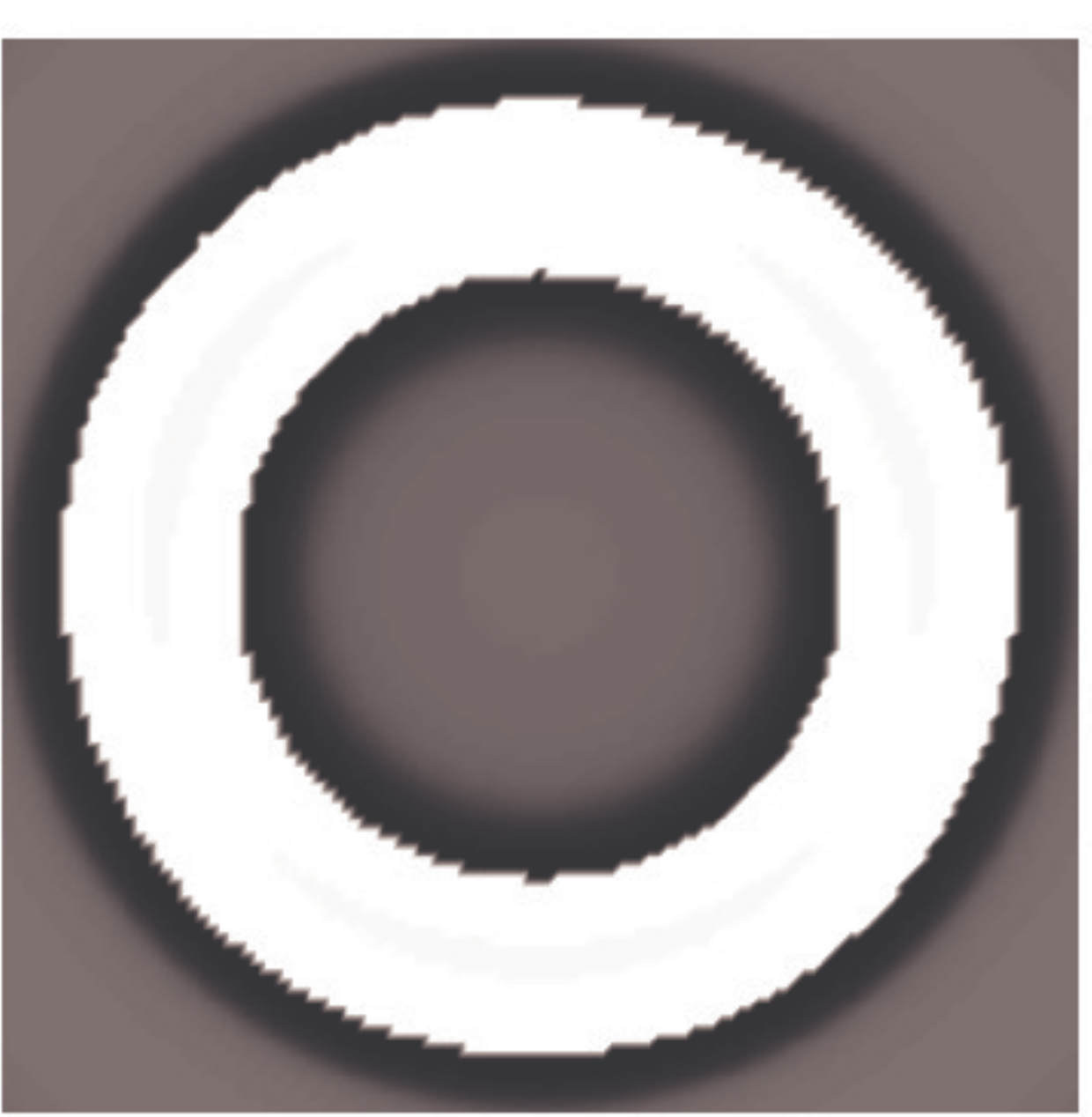}}
\\
\midrule
\subfloat[$(1,\lhat)\rightarrow(1,\lhat)$]{\includegraphics[width=0.23\textwidth]{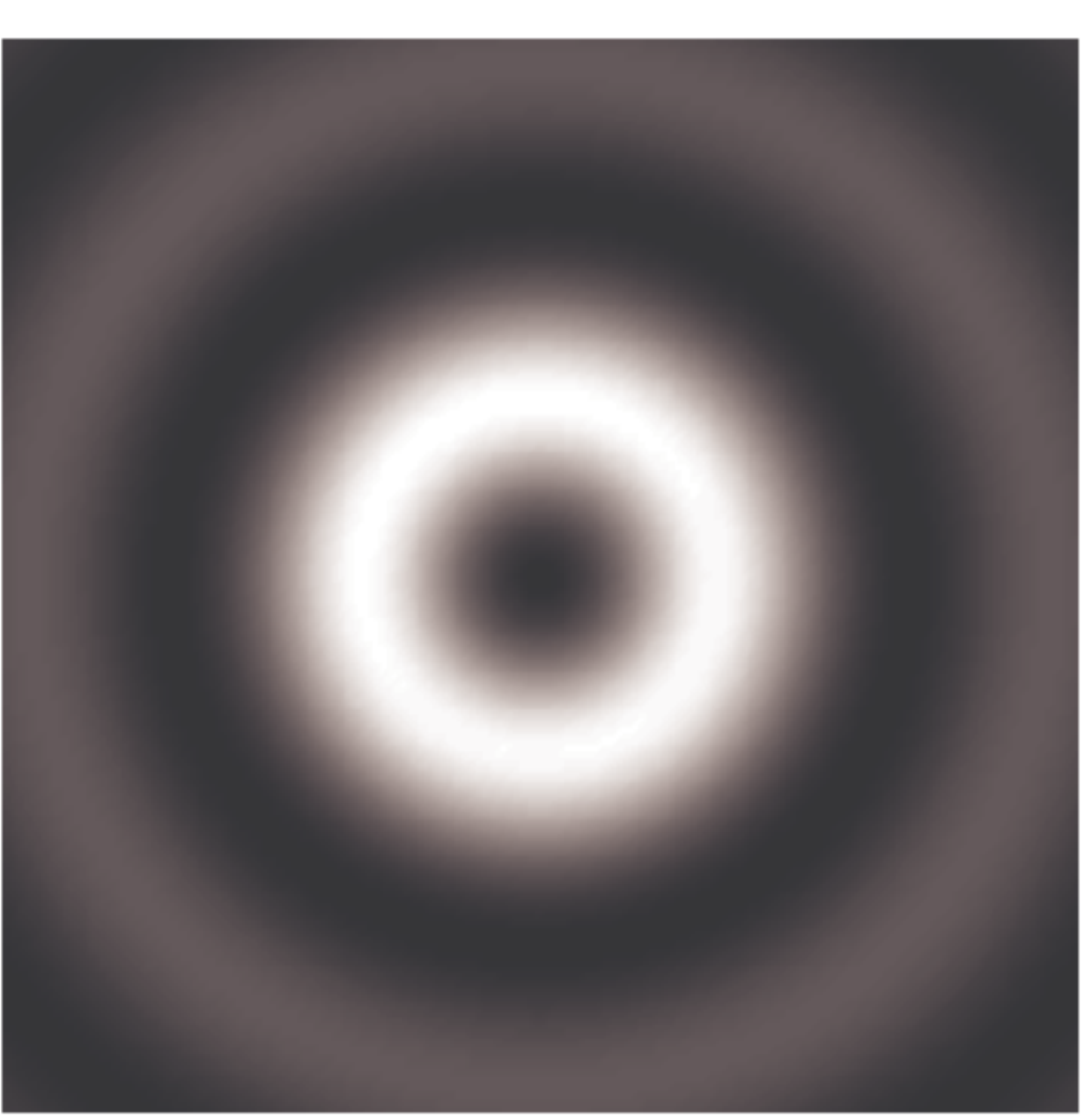}}
&	
\subfloat[$(1,\lhat)\rightarrow(3,\rhat)$]{\includegraphics[width=0.23\textwidth]{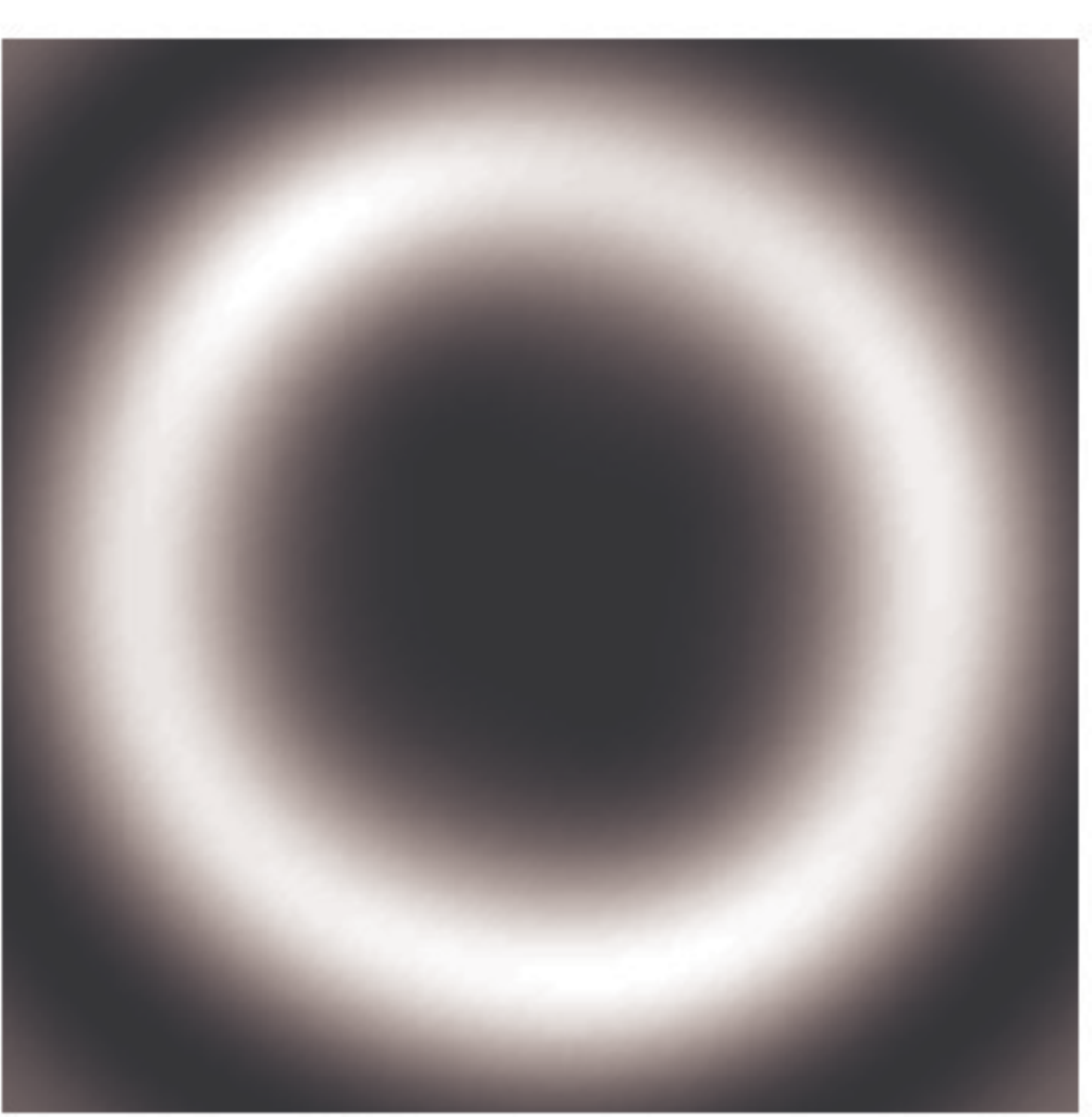}}
	&
	\subfloat[$(2,\rhat)\rightarrow(0,\lhat)$]{\includegraphics[width=0.23\textwidth]{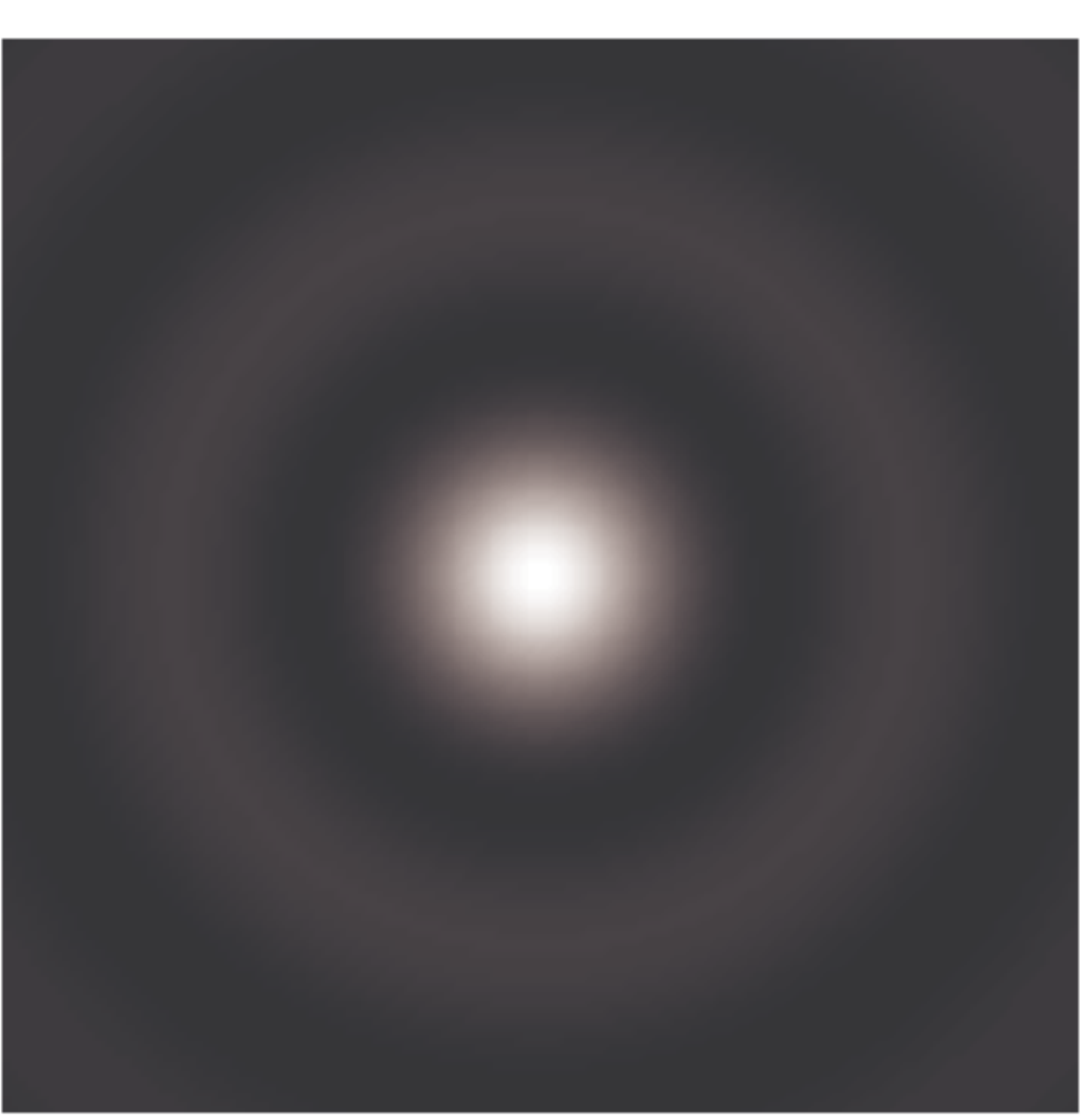}}
	&
	\subfloat[$(2,\rhat)\rightarrow(2,\rhat)$]{\includegraphics[width=0.23\textwidth]{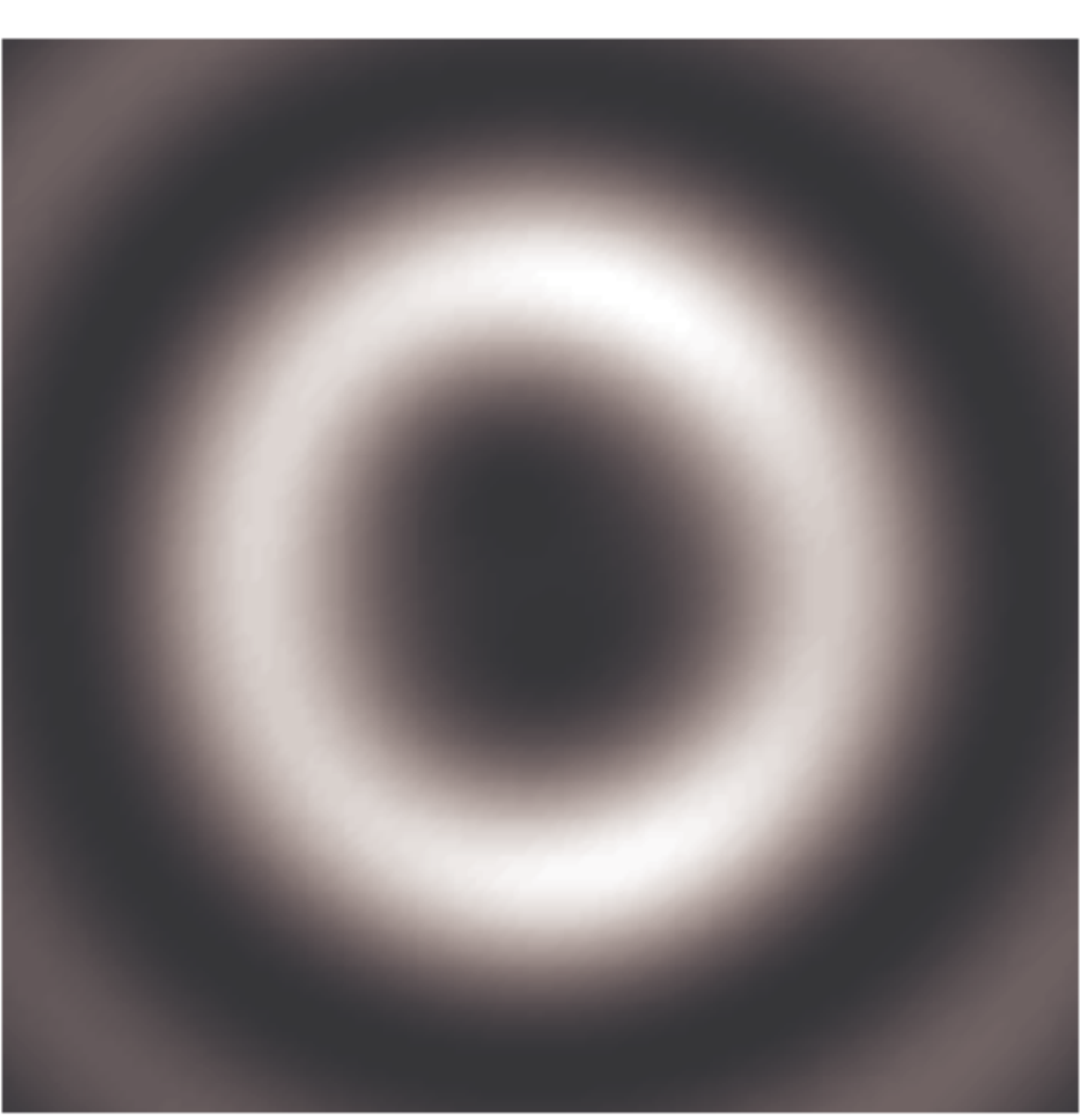}}
\\
	\subfloat[$(1,\lhat)\rightarrow(1,\lhat)$]{\includegraphics[width=0.23\textwidth]{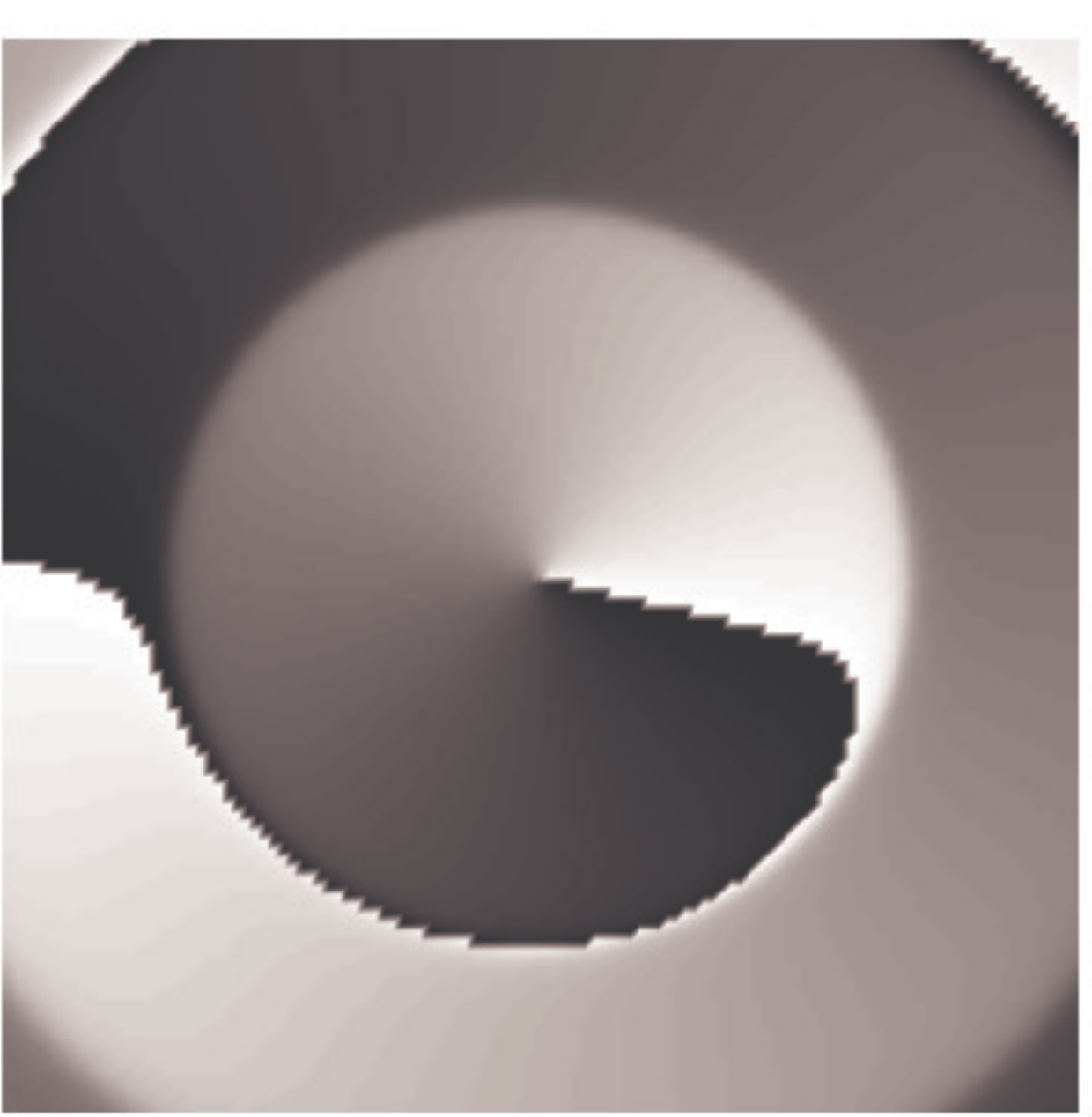}}
	&
	\subfloat[$(1,\lhat)\rightarrow(3,\rhat)$]{\includegraphics[width=0.23\textwidth]{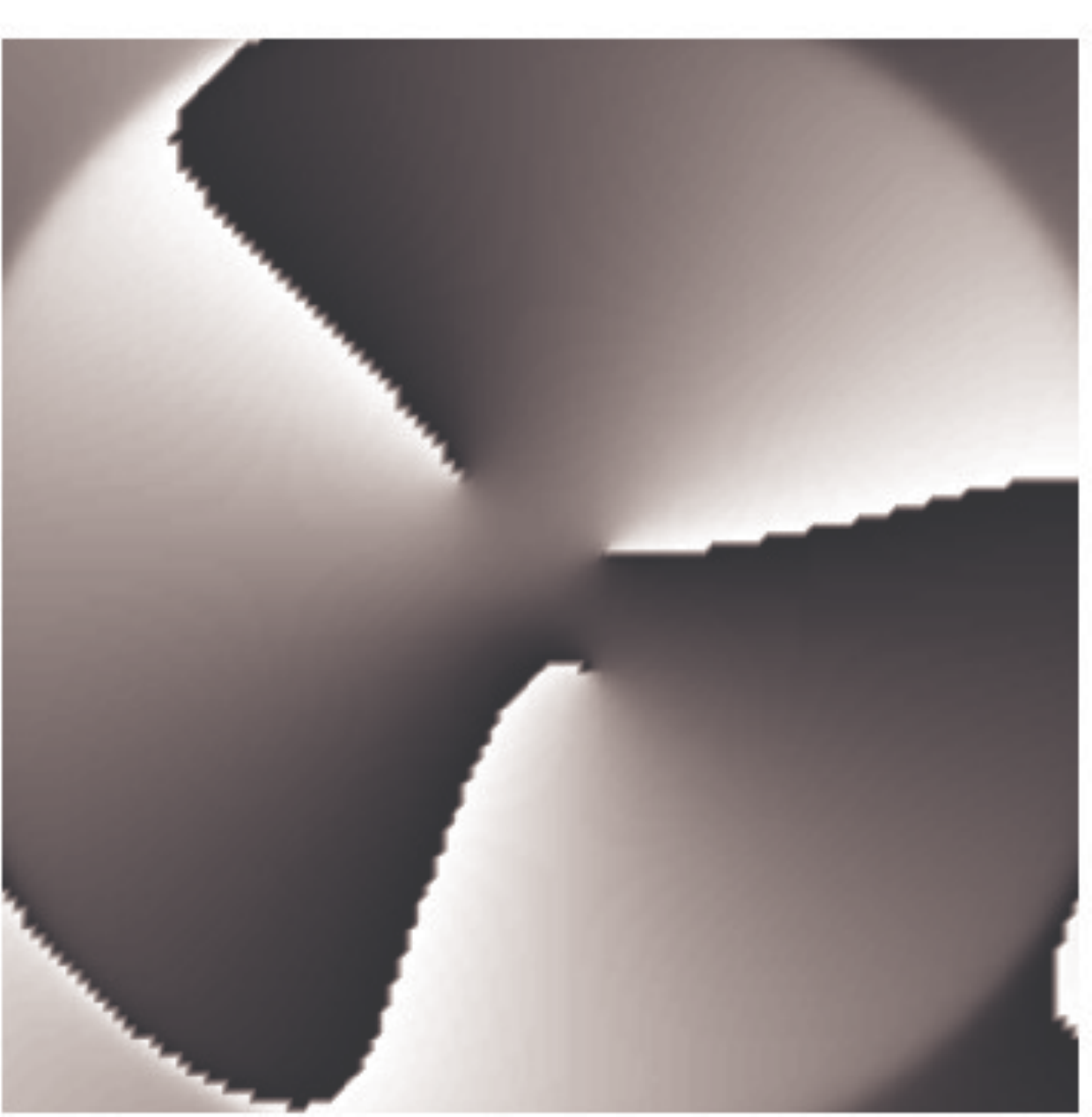}}
	&
	\subfloat[$(2,\rhat)\rightarrow(0,\lhat)$]{\includegraphics[width=0.23\textwidth]{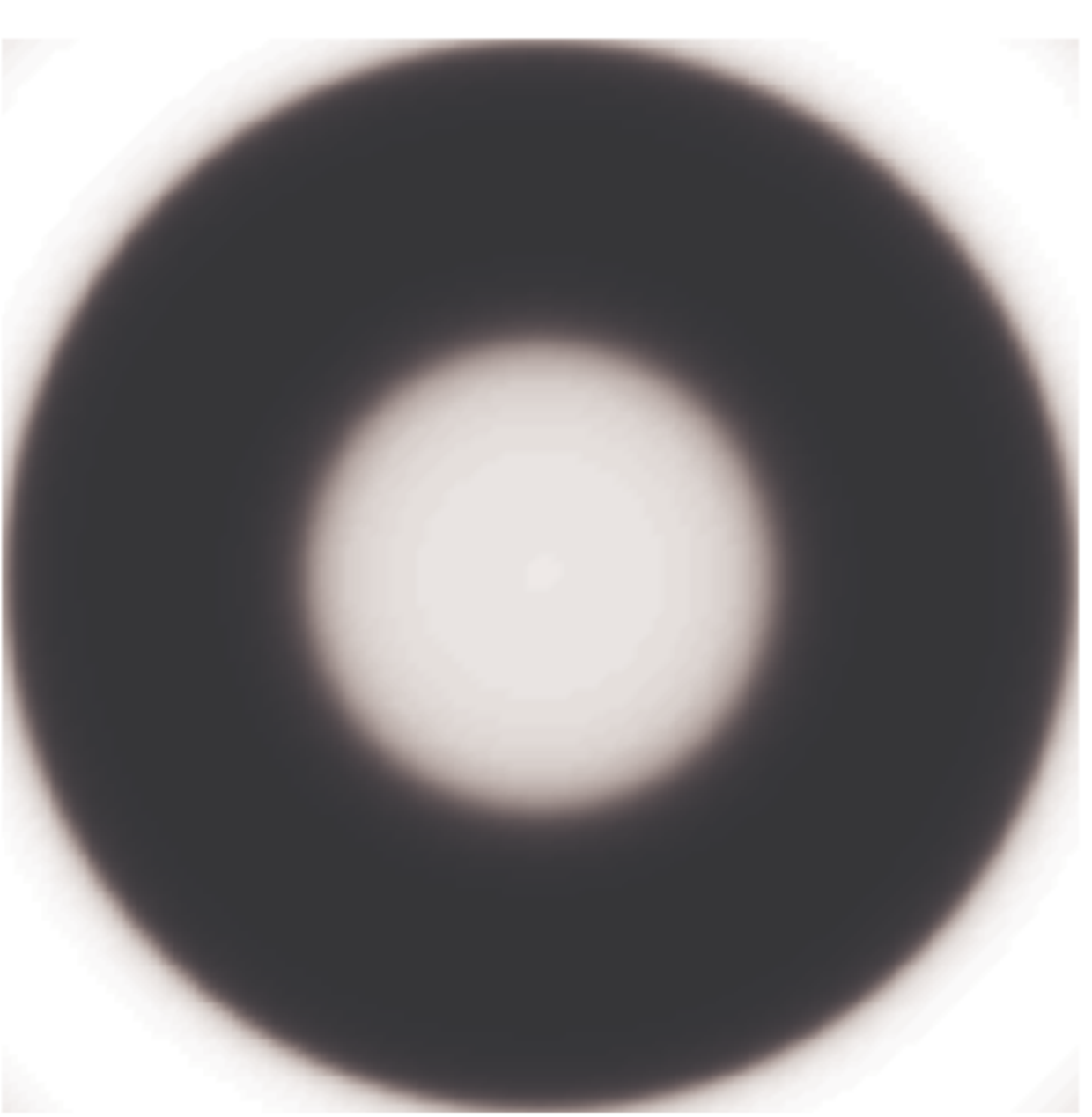}}
	&
	\subfloat[$(2,\rhat)\rightarrow(2,\rhat)$]{\includegraphics[width=0.23\textwidth]{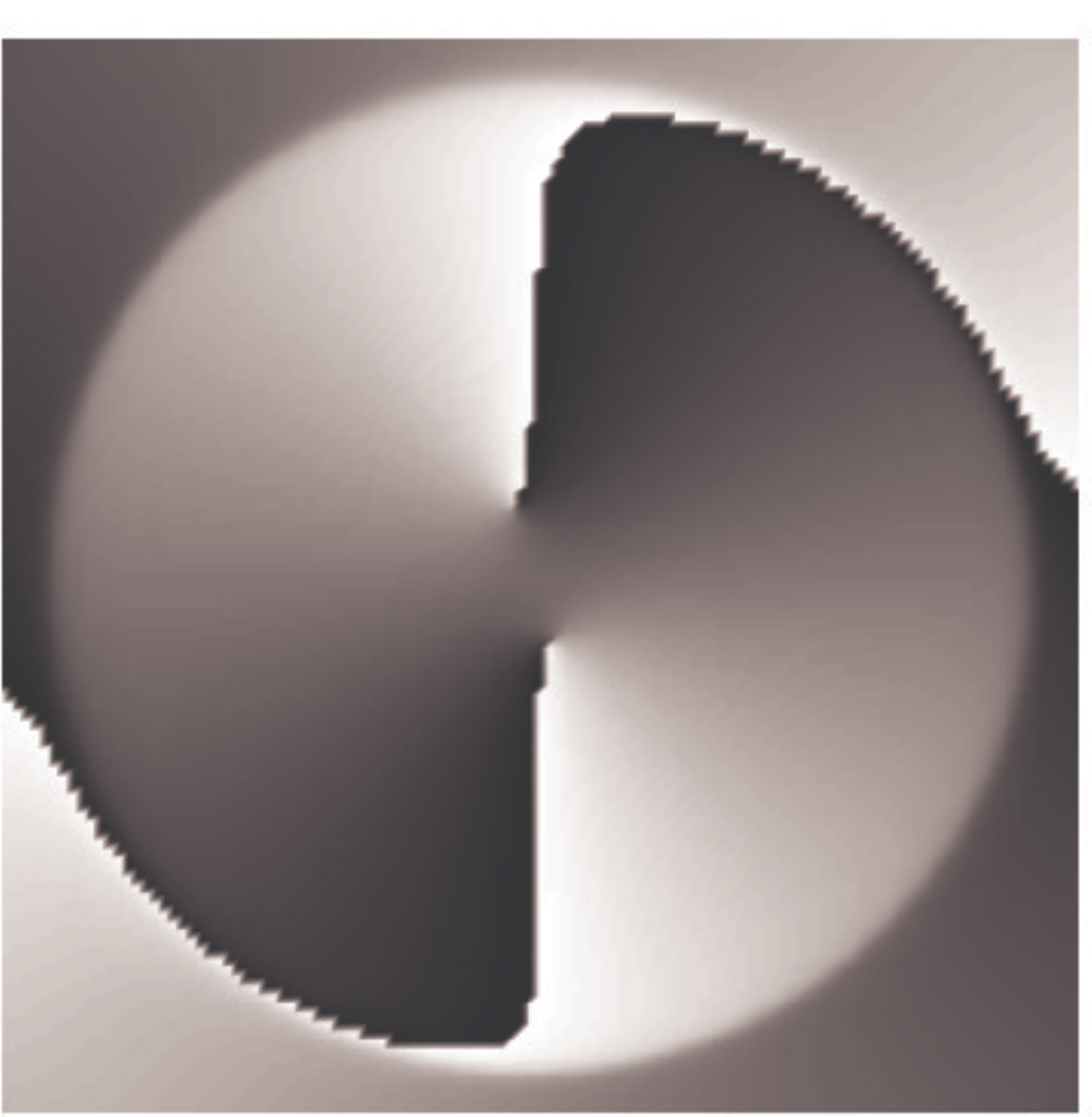}}
\end{tabular}
\end{center}
	\caption[Polarization components after nanohole scattering.]{\label{figc1:sim} Numerical results. Amplitude and phase plots of the output circular polarizations resulting from input modes with different azimuthal phase dependence and polarization. The collimated input mode is focused onto the nanohole and the transmitted light collected and collimated by a second objective. See the setup in Fig. \ref{figc1:setup}-(a). In the notation $(l,\vhat)\rightarrow(\bar{l},\what)$, $(l,\vhat)$ represents the input azimuthal number $l$ and circular polarization $\vhat$, and $(\bar{l},\what)$ the output ones. The output azimuthal number $\bar{l}$ can be visually inferred by the number of times that the phase wraps around in the plots and the sense towards which it increases. The first and third rows are amplitude plots with arbitrary units. The second and fourth rows are phase plots where the gray scale goes from white to dark as the phase goes from $0$ to $2\pi$. Figures (a)(b)(e)(f) correspond to a $(-1,\lhat)$ input, (c)(d)(g)(h) to a $(0,\rhat)$ input, (i)(j)(m)(n) to a $(1,\lhat)$ input and (k)(l)(o)(p) to a $(2,\rhat)$ input. The quantity $s(\vhat)+l$, where $s(\lhat)=-1$ and $s(\rhat)=-1$, is preserved in all cases: It has the same value for the input and the two outputs of different circular polarization.} 
\end{figure}

It was then time to try to observe this preservation effect in the laboratory. My colleague Xavier Zambrana-Puyalto and I worked in the experimental setup for a few weeks. It was the first serious contact with an optics laboratory for both of us. Figure \ref{figc1:setup} is the schematic representation of the setup and Figs. \ref{figc1:picture1} and \ref{figc1:pictures} show pictures of parts of the actual setup.

\begin{figure}[h]
	\begin{minipage}{0.80\linewidth}
\centering
\subfloat[]{\begin{tikzpicture}[thick,scale=1.41,every node/.style={scale=0.8}]
		\makeatletter{}\draw (-0.25,0.35) -- (0,0.35);\draw (-0.5,0) node {LASER};
\draw (-0.25,-0.35) -- (0,-0.35);
\draw (0,-0.5) rectangle (0.25,0.5);\draw (0.125,0.75) node {LP};
\draw (0.25,0.35) -- (0.5,0.35);
\draw (0.25,-0.35) -- (0.5,-0.35);
\draw (0.5,-0.5) rectangle (0.75,0.5);\draw (0.675,0.75) node {QWP};
\draw (0.75,0.35) -- (1.25,0.35);
\draw (0.75,-0.35) -- (1.25,-0.35);
\draw[very thick,<->,>=stealth] (1.25,-0.5) -- (1.25,0.5);\draw (1.25,0.75) node {Lens};
\draw[->] (1.25,-0.35) -- (1.75,-0.15);
\draw[->] (1.25,0.35) -- (1.75,0.15);
\draw[shift={(0.5,0)},scale=0.25]  (5,0.5) rectangle (5.5,4);
\draw[shift={(0.5,0)},scale=0.25] (5,-0.5) rectangle (5.5,-4);
\draw[shift={(0.5,0)},scale=0.25] (5.5,-4) rectangle (7,4);
\draw[->] (2.25,0.15) -- (2.5,0.35);
\draw[->] (2.25,-0.15) -- (2.5,-0.35);
\draw[very thick,<->,>=stealth] (2.5,-0.5) -- (2.5,0.5);\draw (2.5,0.75) node {Lens};
\draw (2.5,0.35) -- (3,0.35);
\draw (2.5,-0.35) -- (3,-0.35);
\draw (3,-0.5) rectangle (3.25,0.5);\draw (3.125,0.75) node {QWP};
\draw (3.25,-0.35) -- (3.5,-0.35);
\draw (3.25,0.35) -- (3.5,0.35);
\draw (3.5,-0.5) rectangle (3.75,0.5);\draw (3.675,0.75) node {LP};
\draw (3.75,0.35) -- (4.35,0.35);
\draw (3.75,-0.35) -- (4.35,-0.35);
\draw (4.35,-0.5) rectangle (4.75,0.5);
\draw (4.75,-1.75) rectangle (5.75,1.75);\draw (5.25,0) node {CCD};
 
	\end{tikzpicture}}
\end{minipage}
	\begin{minipage}{0.15\linewidth}
		\subfloat[]{\includegraphics[scale=0.15]{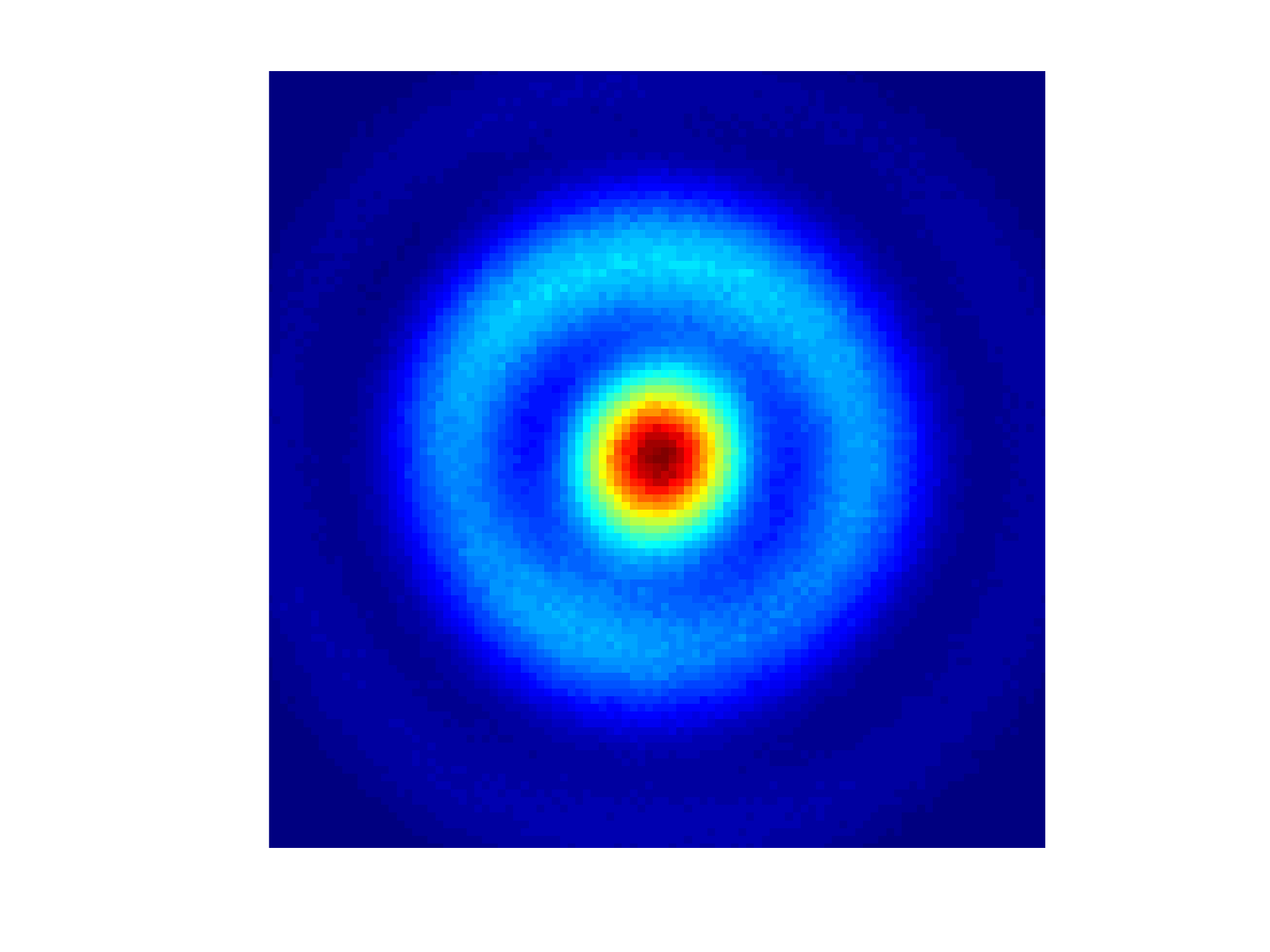}}\\
		\subfloat[]{\includegraphics[scale=0.15]{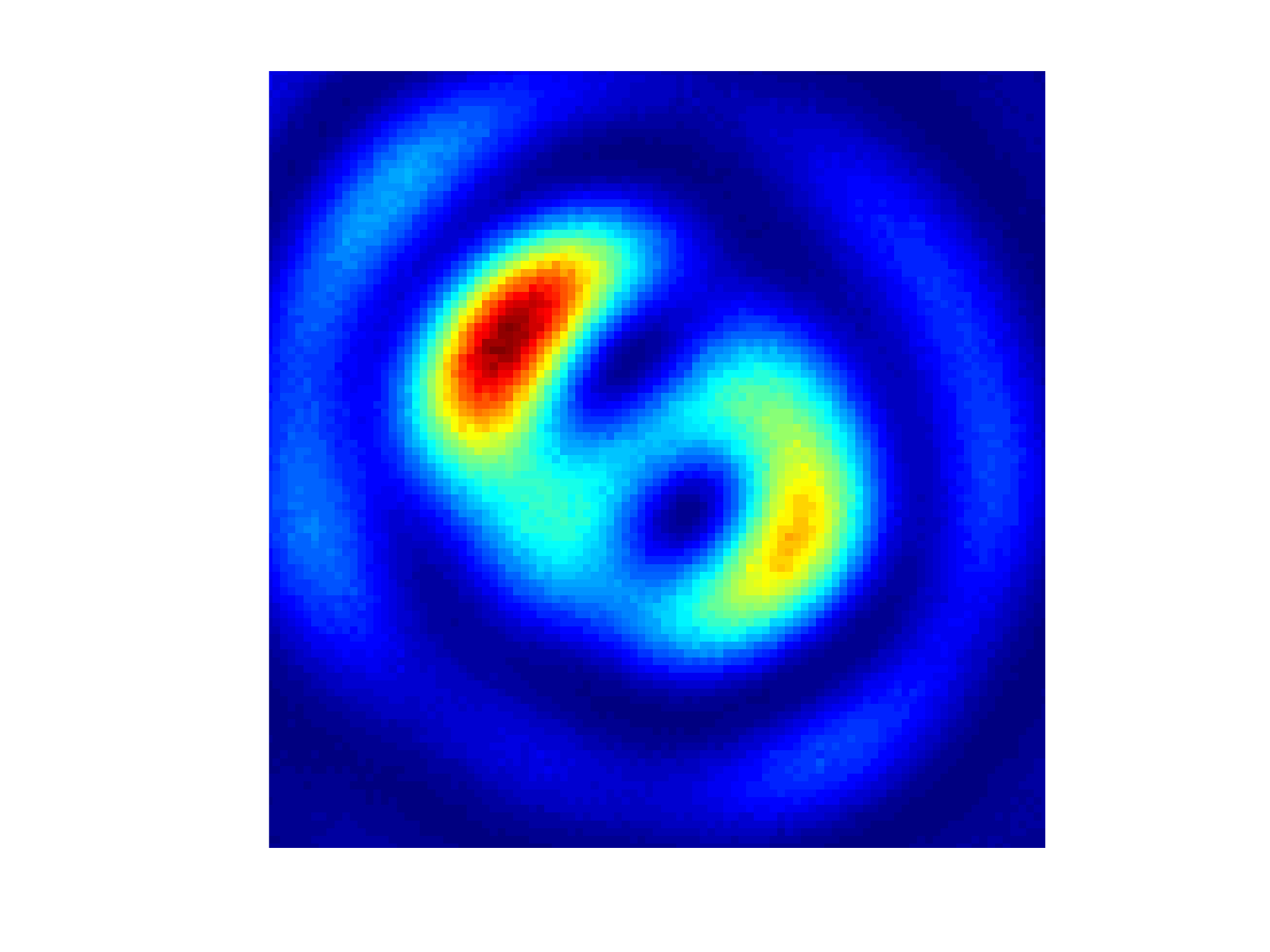}}
	\end{minipage}
	\caption[Setup and results for helicity changes in nanohole scattering.]{\label{figc1:setup} (a) Schematic representation of the experimental setup. The output of a He-Ne laser (with wavelength equal to 632.8 nm) is passed through a set of Linear Polarizer (LP) and Quarter Wave Plate (QWP) and focused with a microscope objective of Numerical Aperture (NA) of 0.5 onto the sample. The sample is a cylindrical hole of 400nm of diameter on a 200 nm thick gold layer on top of a 1 mm glass substrate. The transmitted light is collected and collimated with another microscope objective of the same NA, analyzed with another set of QWP and LP, and imaged with a Charged Coupled Device (CCD) camera. (b) CCD image when the axis of the second LP is set to select the input polarization. (c) CCD image when the axis of the second LP is set to select the polarization orthogonal to the input one.}
\end{figure}

\begin{figure}[h!]
	\begin{center}
		\includegraphics[width=\textwidth]{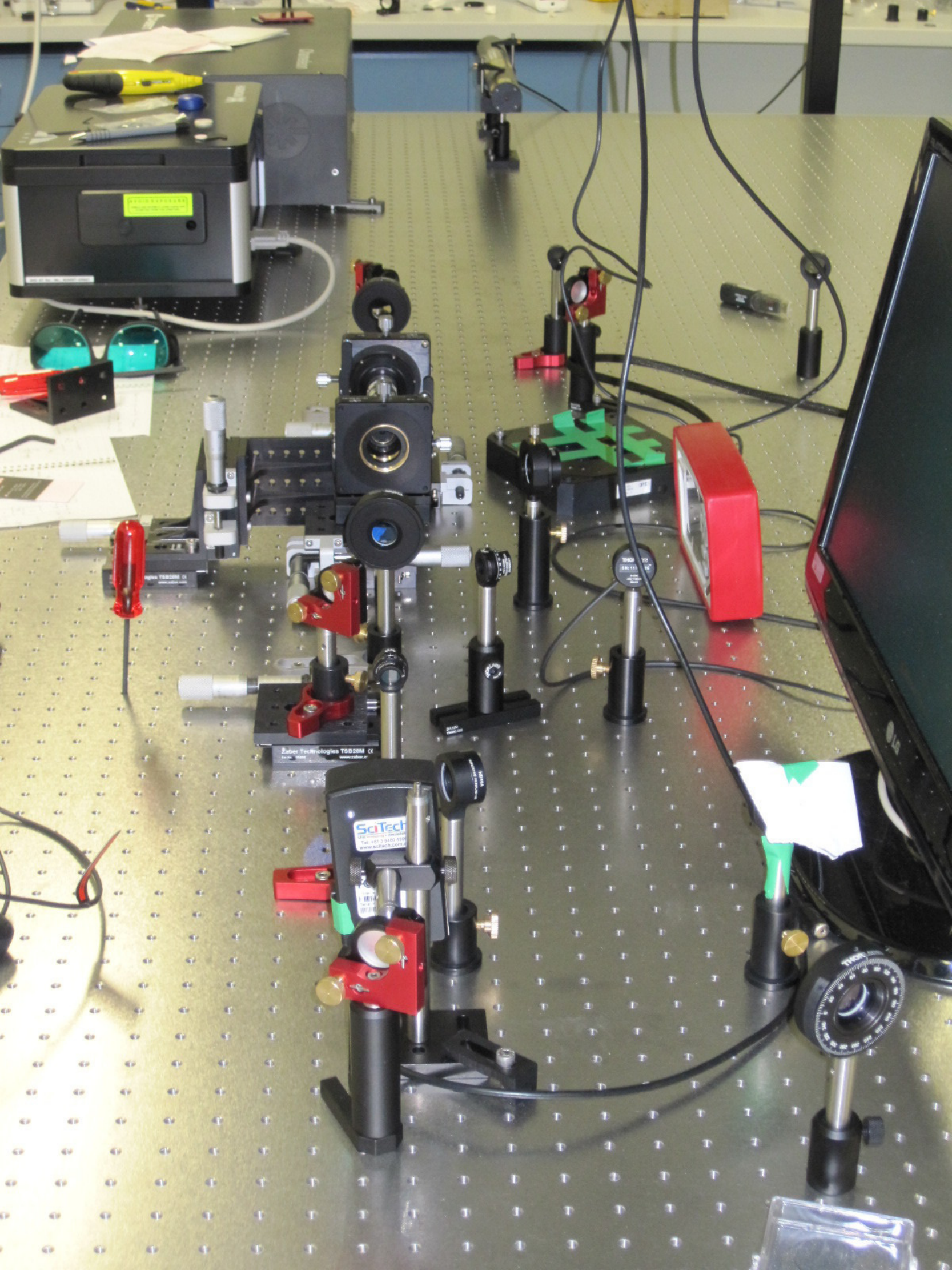}
	\end{center}
	\caption[Picture of the laboratory setup]{\label{figc1:picture1} Picture of the setup without the sample. The laser is at the far end of the table (cylinder next to the right of the big boxes) and the camera at the near end (it has a white ``SciTech'' label on its back). The nanopositioning stage containing the sample is laying flat to the right of the setup, distinguishable by four strips of green sticky tape (see Fig. \ref{figc1:pictures}-(b)). Please refer to the caption of Fig. \ref{figc1:setup} for a detailed explanation of the setup.}
\end{figure}
\begin{figure}[h!]
	\begin{center}
		\subfloat[]{\includegraphics[width=0.75\textwidth]{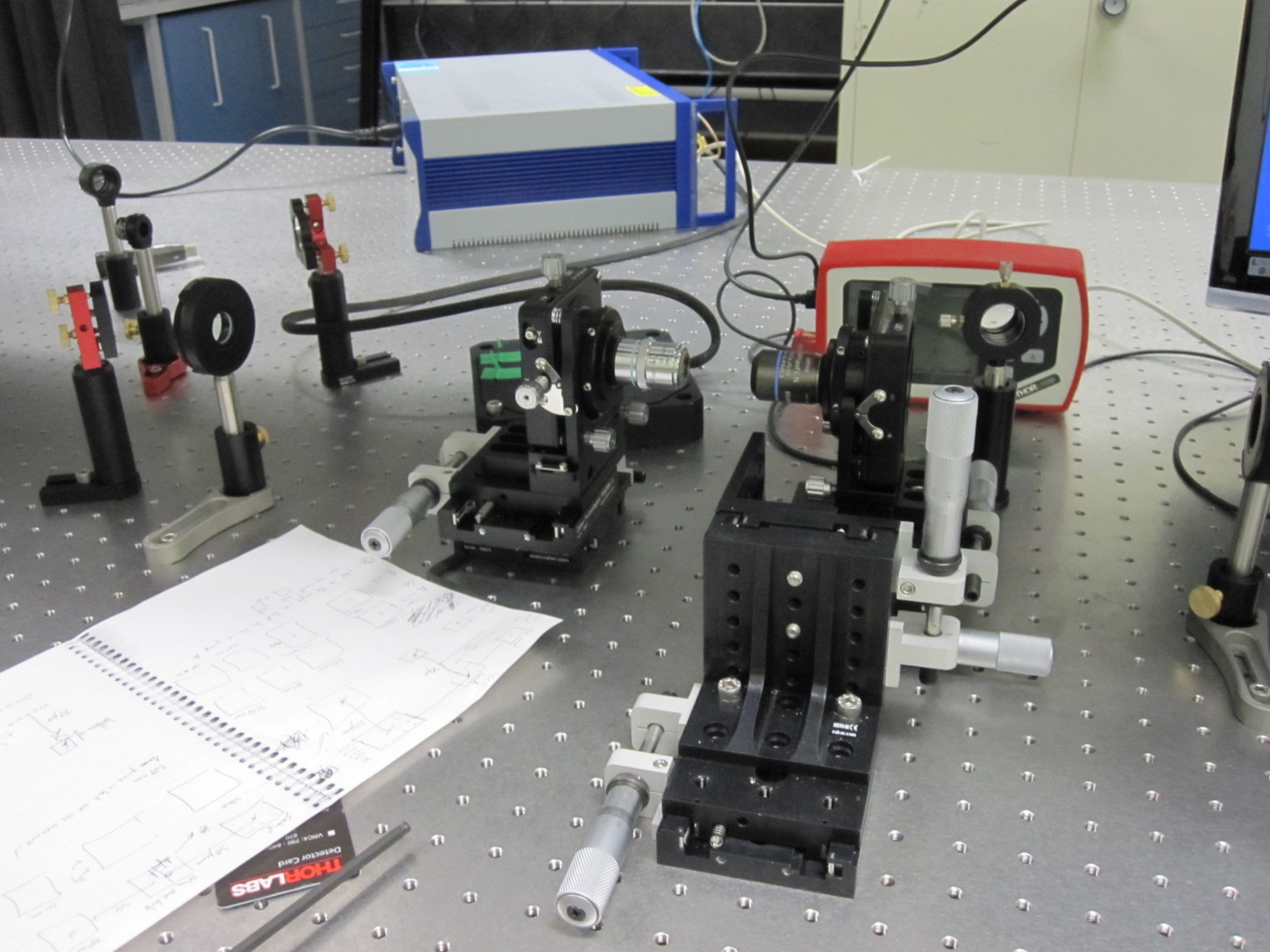}}
		\\
		\subfloat[]{\includegraphics[width=0.75\textwidth]{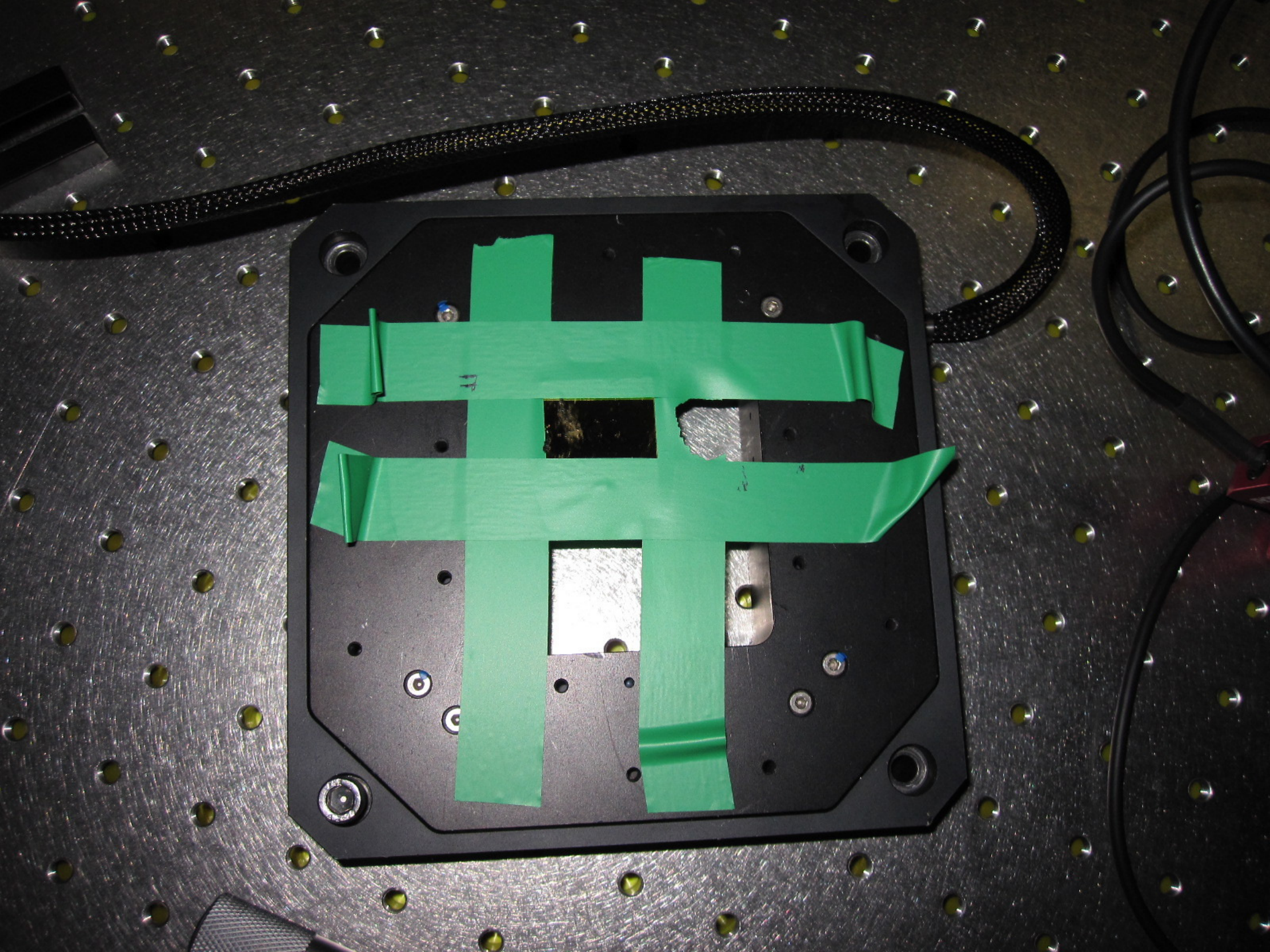}}
	\end{center}
	\caption[Detail of the central part of the experiment.]{\label{figc1:pictures} (a) Detail of the central part of the experiment with the two microscope objectives facing each other. (b) The sample fixed to a nanopositioning stage. The nanopositioning stage containing the sample was mounted on the stage structure visible at the forefront of (a). The sample is a 200nm thick gold film on top of a 1 mm glass substrate. Arrays of different size holes had been milled in the gold film. The holes in the array were sufficiently spaced so that the interaction with the focused beam would occur with a single hole.}
\end{figure}

The aim of the experiment was to observe the signature of a phase singularity of charge 2 in the CCD camera when the analyzing polarizer was set to select the polarization opposite to the one carried by the input Gaussian beam. This corresponds to the simulated cases in Fig. \ref{figc1:sim}(c)(d)(g)(h). After a few discouraging days, it turned out that the lack of results was due to a faulty servo in the nanopositioning stage (Fig. \ref{figc1:pictures}-(b)). The sample was moving too much. Turning it off immediately produced an image with two intensity nulls, like the one in Fig. \ref{figc1:setup}-(c). This is the intensity signature of a charge 2 phase singularity after splitting into two charge 1 singularities because of noise, something that higher order singularities tend to do \cite{Ricci2012}. It was an exciting moment.

Simulation and experiment were in agreement. What about the theory? The literature did offer an explanation of the results based on the separation of the electromagnetic angular momentum $J_z$ into the spin $S_z$ and orbital $L_z$ angular momenta \cite{Gorodetski2009,Vuong2010,Bliokh2011}. In such explanation, $S_z$ is associated with circular polarization and takes values 1 for $\lhat$ and -1 for $\rhat$. $L_z$ is associated with the azimuthal phase dependence and takes the value $l$. The explanation sustains that, while the total angular momentum $J_z$ has to be preserved due to cylindrical symmetry, there is a transfer between spin and orbital angular momentum in the interaction. For example, a $(0,\rhat)$ input beam originates two outputs, one with the same $(0,\rhat)$ values and another with $(-2,\lhat)$. In the latter beam, the value of $L_z$ has to decrease by 2 units in order to compensate the increase of $S_z$ by 2 units. All the examples in Fig. \ref{figc1:sim} and the experimental results fit this explanation. Nevertheless, there are several problems with it. 

First of all, there was no single explanation for the actual cause of spin to orbital angular momentum transfer. It seemed to happen in the interaction of focused beams with nanoapertures \cite{Gorodetski2009,Vuong2010} and with semiconductor microcavities \cite{Manni2011}, during the focusing itself \cite{Zhao2007}, and also for a collimated beam in inhomogeneous and anisotropic media \cite{Marrucci2006}. The question of {\em why} it happened does not have a single answer in this framework.

Then, there is authoritative literature against the separate consideration of $L_z$ and $S_z$. For example, on page 50 of Cohen-Tannoudji et. al's {\em Photons and atoms} \cite{Cohen1997} we read: ``Let us show that $\LL$ and $\SSS$ are not separately physically observable as $\JJ$ is.'' Also, in Sec. 16 of the fourth volume of the Landau and Lifshiftz course of theoretical physics {\em Quantum Electrodynamics} \cite{Berestetskii1982} it says that: ``In the relativistic theory the orbital angular momentum $\LL$ and the spin $\SSS$ of a moving particle are not separately conserved. Only the total angular momentum is. The component of the spin in any fixed direction (taken as the $z$-axis) is therefore not conserved and cannot be used to enumerate the polarization (spin) states of the moving particle.''

This, to me, was enough evidence to discard the $S_z/L_z$ explanation for the experimental and numerical results. The results clearly showed that something was changing. A portion of the beam changed into a very different kind of beam. What was then changing? The answer is also in \cite{Berestetskii1982}, only one paragraph below the one I have copied above: 

''The component of the spin in the direction of the momentum {\em is} conserved, however: since $\LL=\rr\times\PP$ the product $\SSS\cdot\PP/|\PP|$ is equal to the conserved product $\JJ\cdot\PP/|\PP|$. This quantity is called {\em helicity}; [...]. Its eigenvalues will be denoted by $\lambda$ ($\lambda=-s,\ldots,+s$) and states of a particle having definite values of $\lambda$ will be called {\em helicity states}.''

Helicity is therefore an observable quantity. Could it be that it was the one changing? The answer is yes. An analysis of the experiment by means of {\em helicity states} showed that the results were consistent with helicity changes in the interaction with the nanohole. It also explained other instances of ``spin to orbital angular momentum transfer'' (see Chap. \ref{chap4}). 
After understanding {\em what} was happening, the question was {\em why} was helicity changing? The answer is: Because electromagnetic duality symmetry was broken by the sample. The results could be explained by quite simple considerations of symmetries and conserved quantities in the system \cite{FerCor2012b}. Helicity is expected to change in such a setup in the same way that a non cylindrically symmetric target is expected to change angular momentum: The scattering breaks the symmetry associated with the corresponding conservation law.

This first example where the symmetry approach using helicity and duality allowed to gain new insight was an encouraging sign: Maybe it would also be useful in other problems. That was the point where my thesis changed definitively. It turned into the development of a framework based on symmetries and conservation laws for the study of interactions of electromagnetic radiation with matter. Helicity and duality ended up playing a crucial role in it.

The framework is proposed in \cite{FerCor2012b}, where it is used to clarify the ``spin to orbital angular momentum transfer'' explanation.  The discussion about helicity preservation and duality symmetry in the presence of matter is in \cite{FerCor2012p} and also in \cite{FerCor2013}. The framework has indeed produced results in different areas: Optical activity \cite{FerCor2012c}, metamaterials \cite{FerCor2013}, zero backscattering \cite{FerCor2013c} and nanophotonics \cite{Zambrana2013}.

\listoffigures
\listoftables
\tableofcontents
\mainmatter
\makeatletter{}\chapter{Background}
\label{chap2}
\epigraph{La pregunta ha de ser clara, i ha de permetre una resposta expl\'icita.\\The question has to be clear, and has to allow an explicit answer.}{Carme Forcadell (catalan political activist)}

We have cell phones, GPS, high speed internet, X-rays for medical diagnose, radiotherapy, the microwave oven, 3D movies, solar cells ... . We have been able to develop these technologies because we understand fairly well how electromagnetic radiation interacts with matter. 

We want more efficient solar cells, nanomachines able to seek and destroy cancer cells from inside our bodies, millimeter sized laboratories that use only a nanoliter of blood, and the means to manipulate electromagnetic radiation at will. These are some of the current research areas that push our understanding of light matter interactions.

My thesis is an attempt to build a tool for studying, understanding and engineering light matter interactions. The basis of the tool is a concept that is central in physics: Symmetry. In the first part of this chapter I will go over the concepts of symmetry, invariance and conservation laws. I will make no attempt to be either exhaustive or formally rigorous. My only aim is to provide a simple introduction to the ideas, language and notation that I will be using in the next chapters. These ideas are, roughly, that:
\begin{itemize}
	\item We can apply transformations to a physical system.
	\item If a transformation leaves a system unchanged, it implies that a particular property of external objects interacting with the system does not change either. It is preserved by the system.
\end{itemize}
In the second part of the chapter, I will introduce the two main characters of this thesis and provide some background information on them. They are a transformation for electromagnetic fields called {\bf electromagnetic duality} and the property that is preserved by systems that are unchanged by the duality transformation: {\bf Helicity}. Both of them together allow to study light matter interactions by means of symmetries and conserved quantities in a relatively simple way.

\section{Symmetry, invariance and conservation laws}\label{secc2:sym}
\begin{figure}[h]
\makeatletter{}\def\ASYprefix{}
\newbox\ASYbox
\newdimen\ASYdimen
\long\def\ASYbase#1#2{\leavevmode\setbox\ASYbox=\hbox{#1}\ASYdimen=\ht\ASYbox\setbox\ASYbox=\hbox{#2}\lower\ASYdimen\box\ASYbox}
\long\def\ASYaligned(#1,#2)(#3,#4)#5#6#7{\leavevmode\setbox\ASYbox=\hbox{#7}\setbox\ASYbox\hbox{\ASYdimen=\ht\ASYbox\advance\ASYdimen by\dp\ASYbox\kern#3\wd\ASYbox\raise#4\ASYdimen\box\ASYbox}\put(#1,#2){#5\wd\ASYbox 0pt\dp\ASYbox 0pt\ht\ASYbox 0pt\box\ASYbox#6}}\long\def\ASYalignT(#1,#2)(#3,#4)#5#6{\ASYaligned(#1,#2)(#3,#4){
}{
}{#6}}
\long\def\ASYalign(#1,#2)(#3,#4)#5{\ASYaligned(#1,#2)(#3,#4){}{}{#5}}
\def\ASYraw#1{
currentpoint currentpoint translate matrix currentmatrix
100 12 div -100 12 div scale
#1
setmatrix neg exch neg exch translate}
 
\makeatletter{}\setlength{\unitlength}{1pt}
\makeatletter\let\ASYencoding\f@encoding\let\ASYfamily\f@family\let\ASYseries\f@series\let\ASYshape\f@shape\makeatother{\catcode`"=12\includegraphics{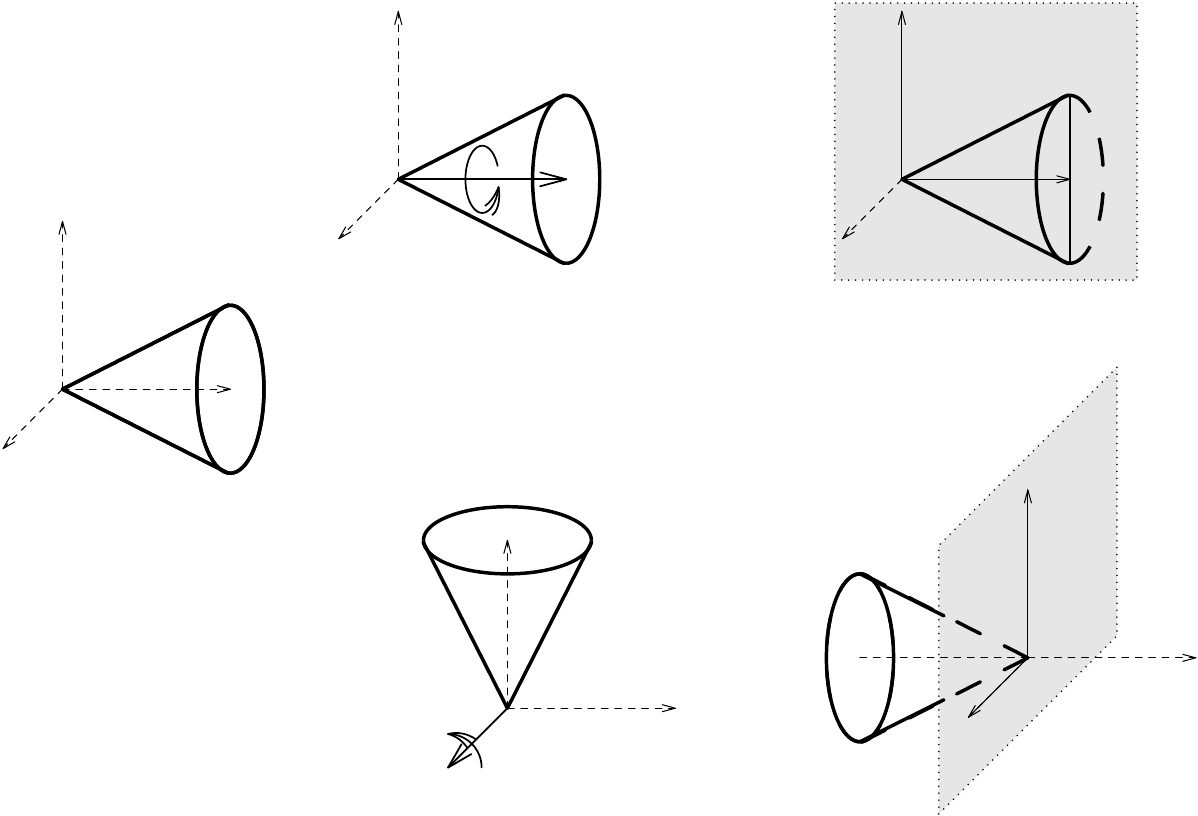}}\definecolor{ASYcolor}{gray}{0.000000}\color{ASYcolor}
\fontsize{12.000000}{14.400000}\selectfont
\usefont{\ASYencoding}{\ASYfamily}{\ASYseries}{\ASYshape}\ASYalignT(-324.133124,147.326394)(0.000000,-0.500000){0.800000 -0.000000 -0.000000 0.800000}{$\hat{x}$}
\definecolor{ASYcolor}{gray}{0.000000}\color{ASYcolor}
\fontsize{12.000000}{14.400000}\selectfont
\ASYalignT(-303.465890,119.459161)(-0.500000,-1.000000){0.800000 -0.000000 -0.000000 0.800000}{$\hat{z}$}
\definecolor{ASYcolor}{gray}{0.000000}\color{ASYcolor}
\fontsize{12.000000}{14.400000}\selectfont
\ASYalignT(-338.858471,117.024982)(-1.000000,0.218749){0.800000 -0.000000 -0.000000 0.800000}{$\hat{y}$}
\definecolor{ASYcolor}{gray}{0.000000}\color{ASYcolor}
\fontsize{12.000000}{14.400000}\selectfont
\ASYalign(-301.039167,164.313458)(-0.500000,-0.250000){0)}
\definecolor{ASYcolor}{gray}{0.000000}\color{ASYcolor}
\fontsize{12.000000}{14.400000}\selectfont
\ASYalign(-196.345138,180.127245)(-0.500000,-1.000000){$\theta$}
\definecolor{ASYcolor}{gray}{0.000000}\color{ASYcolor}
\fontsize{12.000000}{14.400000}\selectfont
\ASYalign(-165.142658,205.567755)(-0.500000,-0.250000){a)}
\definecolor{ASYcolor}{gray}{0.000000}\color{ASYcolor}
\fontsize{12.000000}{14.400000}\selectfont
\ASYalign(-70.500447,205.567755)(-0.500000,-0.250000){b)}
\definecolor{ASYcolor}{gray}{0.000000}\color{ASYcolor}
\fontsize{12.000000}{14.400000}\selectfont
\ASYalignT(-219.876311,23.391040)(-1.000000,-0.163770){1.400000 -0.000000 -0.000000 1.400000}{$\frac{\pi}{2}$}
\definecolor{ASYcolor}{gray}{0.000000}\color{ASYcolor}
\fontsize{12.000000}{14.400000}\selectfont
\ASYalign(-167.569381,76.951417)(-0.500000,-0.250000){c)}
\definecolor{ASYcolor}{gray}{0.000000}\color{ASYcolor}
\fontsize{12.000000}{14.400000}\selectfont
\ASYalign(-87.487510,79.378140)(-0.500000,-0.250000){d)}
 
\caption[Different transformations applied to a cone.]{\label{figc2:conetx}Different transformations applied to the cone in (0). The two in the upper row (a,b) leave the cone unchanged. The two in the lower row (c,d) have a noticeable effect on the cone.}
\end{figure}

Look at the cone in Fig. \ref{figc2:conetx}-(0). The other four sub-figures show the cone after a transformation has been applied to it.
\begin{itemize}
\item (a): Rotation along the $z$ axis by an arbitrary angle.  
\item (b): Mirror reflection across the $xz$ plane.
\item (c): Rotation of 90 degrees ($\pi$ radians) along the $y$ axis.
\item (d): Mirror reflection across the $xy$ plane.
\end{itemize}
We immediately see that the first two transformations have no effect on the cone: After the transformation, the cone looks exactly as it did before in Fig. \ref{figc2:conetx}-(0). The second two transformations have a distinguishable effect on the cone. After the transformation, the cone looks different than before it. We say that the cone is invariant under the first two transformations and is not invariant under the second two. We also say that the cone has rotational symmetry along the $z$ axis and mirror symmetry across the $xz$ plane and that it breaks rotational symmetry along the $y$ axis and mirror symmetry across the $xy$ plane. 

Similarly simple ideas about transformations and symmetries are nowadays regarded as the most fundamental basis of our description of Nature \cite{Gross1995}. From the standard model of particle physics, through the study of atoms, molecules and crystals to the movement of the stars: Symmetry is the concept that allows us to study and attempt to understand these systems.

According to David Gross \cite{Gross1995} ``Einstein's great advance in 1905 was to put symmetry first, to regard the symmetry principle as the primary feature of nature that constrains the allowable dynamical laws''. Einstein brought the concepts of symmetry and invariance to the forefront of theoretical physics. First with his special theory of relativity \cite{Einstein1905}, which is based on the invariant validity of physical law upon changes of inertial reference frame. Later, with his general theory of relativity: The idea that the laws of physics are invariant under changes of space-time coordinates resulted in our best model of space and time. Many spectacular predictions originated from it and have been experimentally verified. For example that the measure of time is affected by the presence of massive objects or that the light of the stars bends around the sun, to name just two. Understanding the first one allows satellite navigation systems like GPS to be as precise as they are. 

Just a bit later in the century, in 1939, Wigner proposed to define an elementary particle as an ``object'' with some properties, like mass, that are invariant under a particular set of transformations \cite{Wigner1939}. His idea is the theoretical cornerstone of the current standard model of elementary particles, and the extensions under consideration like supersymmetry, string theory or M-theory. Wigner did not stop there, and used the concept of symmetry to formalize our modern understanding of atoms \cite{Wigner1959}. 

Many areas have followed suite in employing the concept of symmetry and the mathematical branch which allows its formalization: Group theory \cite{Tung1985}. This has brought uncountable advances. For example, we understand the Higgs boson, the spin of the electron, the atomic Zeeman and Stark effects, the rules for exciting a molecule with light and the behavior of matter waves in crystals thanks mainly to the study of symmetry.

\begin{figure}[h]
\makeatletter{}\def\ASYprefix{}
\newbox\ASYbox
\newdimen\ASYdimen
\long\def\ASYbase#1#2{\leavevmode\setbox\ASYbox=\hbox{#1}\ASYdimen=\ht\ASYbox\setbox\ASYbox=\hbox{#2}\lower\ASYdimen\box\ASYbox}
\long\def\ASYaligned(#1,#2)(#3,#4)#5#6#7{\leavevmode\setbox\ASYbox=\hbox{#7}\setbox\ASYbox\hbox{\ASYdimen=\ht\ASYbox\advance\ASYdimen by\dp\ASYbox\kern#3\wd\ASYbox\raise#4\ASYdimen\box\ASYbox}\put(#1,#2){#5\wd\ASYbox 0pt\dp\ASYbox 0pt\ht\ASYbox 0pt\box\ASYbox#6}}\long\def\ASYalignT(#1,#2)(#3,#4)#5#6{\ASYaligned(#1,#2)(#3,#4){
}{
}{#6}}
\long\def\ASYalign(#1,#2)(#3,#4)#5{\ASYaligned(#1,#2)(#3,#4){}{}{#5}}
\def\ASYraw#1{
currentpoint currentpoint translate matrix currentmatrix
100 12 div -100 12 div scale
#1
setmatrix neg exch neg exch translate}
 
\makeatletter{}\setlength{\unitlength}{1pt}
\makeatletter\let\ASYencoding\f@encoding\let\ASYfamily\f@family\let\ASYseries\f@series\let\ASYshape\f@shape\makeatother{\catcode`"=12\includegraphics{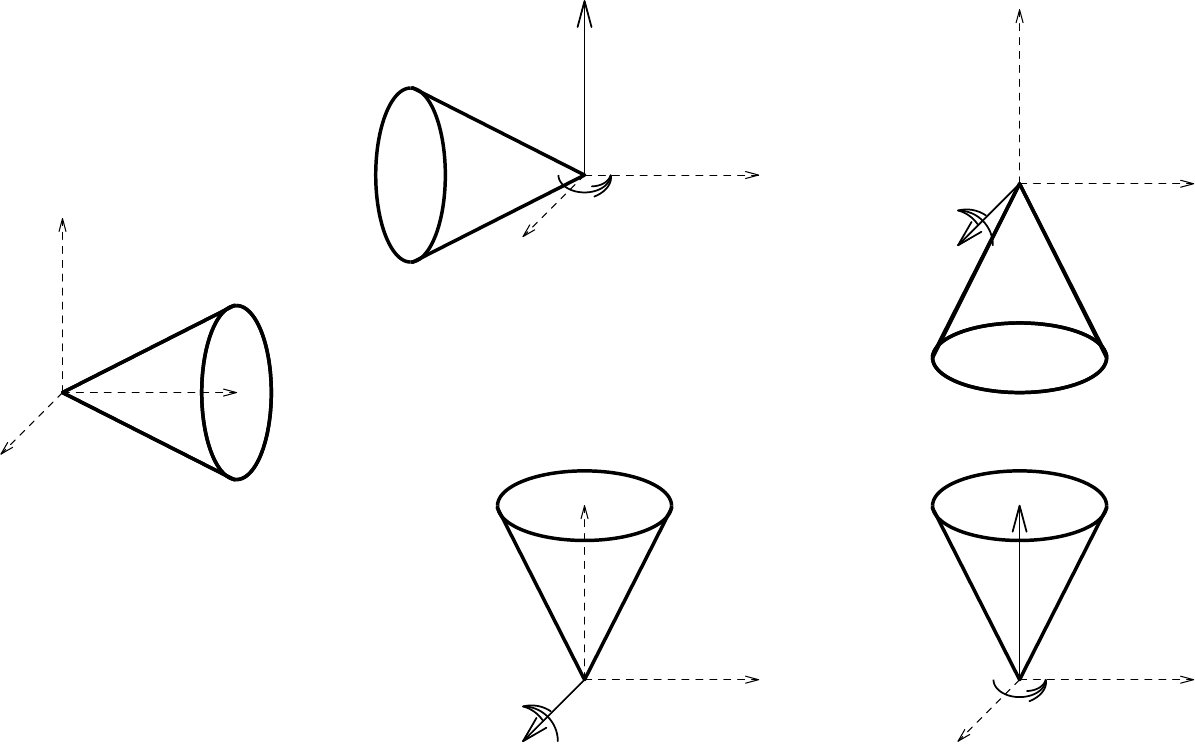}}\definecolor{ASYcolor}{gray}{0.000000}\color{ASYcolor}
\fontsize{12.000000}{14.400000}\selectfont
\usefont{\ASYencoding}{\ASYfamily}{\ASYseries}{\ASYshape}\ASYalignT(-323.507188,126.596669)(0.000000,-0.500000){0.800000 -0.000000 -0.000000 0.800000}{$\hat{x}$}
\definecolor{ASYcolor}{gray}{0.000000}\color{ASYcolor}
\fontsize{12.000000}{14.400000}\selectfont
\ASYalignT(-301.954749,97.844229)(-0.500000,-1.000000){0.800000 -0.000000 -0.000000 0.800000}{$\hat{z}$}
\definecolor{ASYcolor}{gray}{0.000000}\color{ASYcolor}
\fontsize{12.000000}{14.400000}\selectfont
\ASYalignT(-338.545503,95.097083)(-1.000000,0.218749){0.800000 -0.000000 -0.000000 0.800000}{$\hat{y}$}
\definecolor{ASYcolor}{gray}{0.000000}\color{ASYcolor}
\fontsize{12.000000}{14.400000}\selectfont
\ASYalign(-299.439505,144.203377)(-0.500000,-0.250000){0)}
\definecolor{ASYcolor}{gray}{0.000000}\color{ASYcolor}
\fontsize{12.000000}{14.400000}\selectfont
\ASYalign(-176.192549,155.694841)(-0.500000,-1.000000){$\pi$}
\definecolor{ASYcolor}{gray}{0.000000}\color{ASYcolor}
\fontsize{12.000000}{14.400000}\selectfont
\ASYalign(-146.009621,202.053989)(-0.500000,-0.250000){a1)}
\definecolor{ASYcolor}{gray}{0.000000}\color{ASYcolor}
\fontsize{12.000000}{14.400000}\selectfont
\ASYalignT(-71.815810,154.085600)(-1.000000,-0.163770){1.400000 -0.000000 -0.000000 1.400000}{$\frac{\pi}{2}$}
\definecolor{ASYcolor}{gray}{0.000000}\color{ASYcolor}
\fontsize{12.000000}{14.400000}\selectfont
\ASYalign(-25.277909,204.569233)(-0.500000,-0.250000){a2)}
\definecolor{ASYcolor}{gray}{0.000000}\color{ASYcolor}
\fontsize{12.000000}{14.400000}\selectfont
\ASYalignT(-197.578009,10.716692)(-1.000000,-0.163770){1.400000 -0.000000 -0.000000 1.400000}{$\frac{\pi}{2}$}
\definecolor{ASYcolor}{gray}{0.000000}\color{ASYcolor}
\fontsize{12.000000}{14.400000}\selectfont
\ASYalign(-143.494377,56.169837)(-0.500000,-0.250000){b1)}
\definecolor{ASYcolor}{gray}{0.000000}\color{ASYcolor}
\fontsize{12.000000}{14.400000}\selectfont
\ASYalign(-50.430349,9.810689)(-0.500000,-1.000000){$\pi$}
\definecolor{ASYcolor}{gray}{0.000000}\color{ASYcolor}
\fontsize{12.000000}{14.400000}\selectfont
\ASYalign(-17.732177,56.169837)(-0.500000,-0.250000){b2)}
 
\caption[Non commutative transformations.]{\label{figc2:commu}The end result of two successive transformations can depend on the order in which the transformations are applied. In the upper (a1,a2) and lower (b1,b2) row, the same two transformations are applied to the cone in (0) but in different order. The end result is different (a2,b2).}
\end{figure}

Having briefly highlighted the importance of symmetry in modern physics, let us go back to the example of the cone and discuss two important concepts related to transformations: Commutativity and the difference between discrete and continuous transformations. In Fig. \ref{figc2:commu}, two successive transformations are applied to the cone: Rotation by $\pi$ along the $x$ axis ($R_x(\pi)$) and rotation by $\pi/2$ along the $y$ axis ($R_y(\pi/2)$). The difference between the upper and lower rows in the figure is only the order in which the two rotations are applied. Figs. \ref{figc2:commu}-(a2) and (b2) clearly show that the end result is not the same. The order of application matters. If, in order to allow a more formal discussion, we take the convention that $R_y(\pi/2)R_x(\pi)$ means that we first apply $R_x(\pi)$ and then we apply $R_y(\pi/2)$ to the result of the first transformation, what the figure shows is that:
\begin{equation}
\begin{split}
 R_y(\pi/2)R_x(\pi)&\neq R_x(\pi)R_y(\pi/2)\implies \\
& R_y(\pi/2)R_x(\pi)-R_x(\pi)R_y(\pi/2)\equiv[R_y(\pi/2),R_x(\pi)]\neq0,
\end{split}
\end{equation}
where the commutator between two transformations $[A,B]=AB-BA$ is defined. When the commutator is not zero, as for the rotations in the figure, we say that the two transformations do not commute. When the commutator is zero, the order of application of $A$ and $B$ does not matter, we say that the transformations commute. For example, spatial translations in a plane always commute: if in a flat football field you run straight for forty meters and then turn left and run for another ten meters you will meet a team mate of yours having started from the same initial point and first turned left and run for ten meters, and then turn right and run for forty meters. 

Let us go back to Fig. \ref{figc2:conetx} where I applied rotations and mirror reflections to the cone. There is a fundamental difference between these two transformations. A rotation is a continuous transformation. After choosing an axis, we can rotate the cone by infinitely many different angles which we can select from a continuous interval extending from $0$ to $2\pi$. A mirror reflection is a discrete transformation. We either do the reflection or not. In this case, the choice that we have is only binary. 

The continuous nature of rotations allows to subdivide them into an ever growing number of successive rotations by an ever smaller angle. For example:
\begin{equation}
R_z(\alpha)=R_z(\alpha/2)R_z(\alpha/2)=R_z(\alpha/4)R_z(\alpha/4)R_z(\alpha/4)R_z(\alpha/4),
\end{equation}
and so on. When the angle of the constitutive rotations gets infinitely small, there is an infinite number of rotations, and each of them is just an infinitesimal perturbation on the identity transformation $I$ (whose action is to leave any object unchanged):

\begin{equation}
R_z(\alpha)=\lim_{N\rightarrow\infty}\left(I-i\frac{\alpha}{N}J_z\right)^N=\exp\left(-i\alpha J_z\right).
\end{equation}
The second equality follows from the definition of the exponential. The reason why we write the perturbation as $-i\frac{\alpha}{N}J_z$ instead of $\frac{\alpha}{N}\bar{J_z}$ is not really important. What is important is that this object, $J_z$, is the only thing that is needed in order to manufacture a rotation along the $z$ axis by an arbitrary angle. We say that $J_z$ is the generator of rotations along the $z$ axis. We also say that $J_z$ is the $z$ component of angular momentum. 

Every continuous transformation has a generator. For example, linear momentum generates translations in space and the energy operator generates translations in time. The generators are also operators. The consideration of how objects transform under the action of a generator and its corresponding transformation before and after interacting with another system leads me to one of the most profound consequences of invariance: Conservation laws. 

For this task, instead of the cone, I need to consider abstract mathematical objects, which we will call vectors. These vectors ``live'' in some abstract space $\Omega$. The properties of $\Omega$ and the vectors in it are much like the properties of the three dimensional space with coordinates $(x,y,z)$ and the three dimensional vectors which ``live'' in it. Adding two vectors from $\Omega$ results in another vector in $\Omega$, as does multiplying a vector by a number. These are the properties of a linear vector space. We can also transform vectors like I did with the cone. In linear vector spaces the transformations are represented\footnote{Transformations are abstract entities. In order to apply them to concrete mathematical objects, they have to be mapped into appropriate operators for those objects. For example: We can rotate an electron and we can rotate a Higgs boson. The rotation operator for a Dirac four spinor is formally very different from that which rotates a scalar. The representation is different but the essence of the transformation is the same. Unfortunately, the formula {\em operator that represents transformation X} is too long and I will just use {\em transformation X} when I mean the former.} by operators. They take the vectors in $\Omega$ and map them back to possibly different vectors in $\Omega$. For an original vector $|\Psi\rangle$, an operator $S$ and a resulting vector $|\Phi\rangle$ we write:
\begin{equation}
\label{eqc2:interaction}
|\Phi\rangle=S|\Psi\rangle.
\end{equation}
In general, we would end up with a vector $|\Phi\rangle$ with no obvious relationship with $|\Psi\rangle$, but, given an operator that represents a transformation, some of the vectors in $\Omega$ are special with respect to it in that they transform very simply by just acquiring a phase. Take for example $R_z(\alpha)$. There exists a set of vectors $|\Psi_n\rangle$ such that \cite[Chap. 7]{Tung1985}:
\begin{equation}
\label{eqc2:rz}
R_z(\alpha)|\Psi_n\rangle=\exp\left(-i\alpha n\right)|\Psi_n\rangle,
\end{equation}
where $n$ is an integer. If we make the angle infinitesimally small $\alpha\rightarrow d\alpha$, we can figure out how do the $|\Psi_n\rangle$ transform under the action of the generator of rotations $J_z$:
\begin{equation}
\begin{split}
R_z(d\alpha)|\Psi_n\rangle&=\exp\left(-id\alpha J_z\right)|\Psi_n\rangle=\exp\left(-id\alpha n\right)|\Psi_n\rangle\\
&\impliesdueto{d\alpha\rightarrow 0}(I-id\alpha J_z)|\Psi_n\rangle=(1-id\alpha n)|\Psi_n\rangle\\
&\implies J_z|\Psi_n\rangle=n|\Psi_n\rangle.
\end{split}
\end{equation}
When the vector $|\Phi\rangle$ resulting from the action of an operator $X$ on an initial vector $|\Psi\rangle$ is $|\Phi\rangle=c|\Psi\rangle$ for a scalar $c$, we say that $|\Phi\rangle$ is an eigenvector or eigenstate of $X$ with eigenvalue $c$. In the above case, we say that the $|\Psi_n\rangle$ are the eigenvectors of $J_z$ with eigenvalues equal to $n$. We also say that $|\Psi_n\rangle$ has a well defined, sharp, or definite angular momentum equal to $n$. A counter example is, for instance $|\Psi_2\rangle+|\Psi_{-1}\rangle$, which is a linear combination of two eigenvectors with different eigenvalue: Its angular momentum is not well defined. 

This formalism can be used to study interaction problems in physical situations. I will discuss this point at large in Chap. \ref{chap3}. For example, an electron travels towards a sample, interacts with it, and as a result of the interaction some properties of the electron, like its momentum, change. Roughly speaking, in Eq. \ref{eqc2:interaction}, $|\Psi\rangle$ represents the electron before the interaction, $S$ the action of the sample and $|\Phi\rangle$ the electron after the interaction. Assume that $S$ commutes with $R_z(\alpha)$, that is $[S,R_z(\alpha)]=SR_z(\alpha)-R_z(\alpha)S=0$. This implies that the operator $S$ is invariant under rotations along the $z$ axis. It also implies that the sample that $S$ represents exhibits such invariance as well.

To discuss conservation laws, let us examine the properties of $S|\Psi_n\rangle$ after the action of $R_z(\alpha)$. Before the action of $S$, the $|\Psi_n\rangle$ had very simple and definite transformation properties under rotations (\ref{eqc2:rz}). Did the action of $S$ change this? In other words: are the $S|\Psi_n\rangle$ still eigenvectors of $R_z(\alpha)$ with eigenvalue $\exp(-in\alpha)$? They are, because:
\begin{equation}
\label{eqc2:cons}
	R_z(\alpha)S|\Phi_n\rangle=SR_z(\alpha)|\Phi_n\rangle=\exp(-i\alpha n)S|\Phi_n\rangle,
\end{equation}
where the first equality follows from the assumption that $SR_z(\alpha)=R_z(\alpha)S$ and the second one from (\ref{eqc2:rz}). So, after interaction with a system which is invariant under rotations, an initial eigenstate of $R_z(\alpha)$ is still an eigenstate of $R_z(\alpha)$ with the same eigenvalue $\exp(-i\alpha n)$ as before. The same can be said about $J_z$: After the interaction, each eigenstate of $J_z$ is still an eigenstate of $J_z$ with the same eigenvalue that it had before the interaction. This is the expression of a conservation law: We say that a system that commutes with a transformation $T$ is invariant under $T$ and it preserves the eigenstates of the symmetry transformation and its generator. Please refer to \cite[Sec. 4.1]{Sakurai1993} for a rigorous discussion. For discrete transformations, like mirror reflections, a conservation law is expressed as the preservation of the eigenstates can corresponding eigenvalues of the transformation in question. Conversely, when an eigenstate of a transformation interacts with a system $S$ which {\bf is not} invariant under such transformation, the resulting state is, in general, a linear superposition containing not only the original but also all the other eigenstates,
\begin{equation}
\label{eqc2:sup}
	S|\Psi_n\rangle=\sum_m c_m|\Psi_m\rangle.
\end{equation}
The transformation properties of $S|\Psi_n\rangle$ under the original transformation are not simple anymore.

The original result on invariance and conservation laws is due to Emily N\"other \cite{Noether1918}, who in 1918 derived her celebrated theorem. Using the original form of the theorem and its posterior extensions, each invariance of the time evolution equations of a system can be linked to a conservation law. In the previous discussion, I have hidden the evolution of time by using terms like {\em before the action} and {\em after the interaction}. This trick is the basis of the study of collisions: Scattering theory \cite[Chap. 3]{Weinberg1995}, \cite[Chap. XIX]{Messiah1958}. Picture an object traveling on a collision path towards a target. As mentioned before, the action of the target on the object is modeled by a scattering operator $S$ acting on the vector space $\Omega$ where each vector represent a different state of the object. The {\em before the action} initial state $|\Phi\rangle$ is taken to be the state of the object an ``infinite'' time before the collision, and the {\em after the interaction} final state $S|\Phi\rangle$ represents the object an ``infinite'' time after the collision. The idea is that much before or much after the collision, the object can be described ignoring the presence of the target. From a practical point of view, this is a useful simplification and, among many other important applications, is used in spectroscopy or high energy particle accelerators. From a theoretical point of view, it may be argued that the exact details of what happens during the collision of, for example, two protons traveling at 99.99\% of the speed of light are unknown or even outside the domain of applicability of the theories that we have \cite[Chap. 12]{Weinberg1995}. When it suffices for one's purposes, the scattering picture is quite convenient, and much can be learned from the study of the symmetries of $S$. Nevertheless, a statement like {\em the scattering operator $S$ commutes with transformation $T$} is related to N\"other's theorem because the definition of $S$ involves the time evolution operator of the system \cite[Chap. 3.2]{Weinberg1995}. Ultimately, such a statement reflects the invariance of the dynamical equations upon transformation with $T$. 

What does {\em invariance of the dynamical equations} mean? Consider the following time evolution equations for the position of a point-like object A in a two dimensional world interacting with another object B:

\begin{equation}
\label{eqc2:tev}
	\begin{split}
		x(t)&=\frac{d}{dt} y(t),\\
		y(t)&=\frac{d}{dt} x(t).
	\end{split}
\end{equation}
Let me now transform them using the spatial inversion $(x,y)\rightarrow (\tilde{x}=-x,\tilde{y}=-y)$. The transformation of the equations, which you can think of as a change of basis, goes like this:
\begin{equation}
	\begin{split}
		\begin{bmatrix}x(t)\\y(t)\end{bmatrix}&=\begin{bmatrix}0&\frac{d}{dt}\\\frac{d}{dt}&0\end{bmatrix}\begin{bmatrix}x(t)\\y(t)\end{bmatrix}\longrightarrow\\
	\begin{bmatrix}-1&0\\0&-1\end{bmatrix}\begin{bmatrix}x(t)\\y(t)\end{bmatrix}&=\begin{bmatrix}-1&0\\0&-1\end{bmatrix}\begin{bmatrix}0&\frac{d}{dt}\\\frac{d}{dt}&0\end{bmatrix}\begin{bmatrix}-1&0\\0&-1\end{bmatrix}^{-1}\begin{bmatrix}-1&0\\0&-1\end{bmatrix}\begin{bmatrix}x(t)\\y(t)\end{bmatrix}.\\
		\begin{bmatrix}\tilde{x}(t)\\\tilde{y}(t)\end{bmatrix}&=\begin{bmatrix}0&\frac{d}{dt}\\\frac{d}{dt}&0\end{bmatrix}\begin{bmatrix}\tilde{x}(t)\\\tilde{y}(t)\end{bmatrix}.
	\end{split}
\end{equation}
The form of the evolution equations for the $(\tilde{x},\tilde{y})$ variables is exactly the same as that of the evolution equations for the $(x,y)$ variables. This is what {\em invariance of the dynamical equations} means. This does not happen for every transformation. Let me now take a rotation $(x,y)\rightarrow (\tilde{x}=x\cos\theta-y\sin\theta,\tilde{y}=y\cos\theta+x\sin\theta)$, repeat the above steps and reach:
\begin{equation}
	\label{eqc2:rot}
	\begin{bmatrix}\tilde{x}(t)\\\tilde{y}(t)\end{bmatrix}=\begin{bmatrix}-2\sin\theta\cos\theta\frac{d}{dt}&(\cos^{2}\theta-\sin^{2}\theta)\frac{d}{dt}\\(\cos^{2}\theta-\sin^{2}\theta)\frac{d}{dt}&2\sin\theta\cos\theta\frac{d}{dt}\end{bmatrix}\begin{bmatrix}\tilde{x}(t)\\\tilde{y}(t)\end{bmatrix},
\end{equation}
to see a counter example. 

These straightforward arguments allow us to categorically affirm that object B is not rotationally invariant but has spatial inversion symmetry ... and we do not even know what object B looks like !

Besides being the tell tale sign of symmetry, this form invariance of the equations has a very practical application. If a system of equations is invariant under a given transformation, and you know one solution of the system, you can produce a new solution by transforming the one you know with the said transformation. A little reflection on the fact that the equations look the same for both the original and the transformed variables should convince you of the general validity of the previous statement. You can see explicitly what this means by assuming that you have a couple of functions $(x(t),y(t))$ which solve Eq. (\ref{eqc2:tev}) and verifying that $(-x(t),-y(t))$ is also a solution of (\ref{eqc2:tev}). You just doubled the number of solutions at your disposal. Try to produce a new solution by rotating the initial one. It does not work for an arbitrary angle, reflecting what we learned from (\ref{eqc2:rot}). It works for $\theta=\pi$. The reason is that, in this case, a rotation by $\pi$ results in $(x(t),y(t))\rightarrow (-x(t),-y(t))$.

In summary, the existence or breaking (lack of) of a symmetry allows us to know whether or not a system preserves the eigenstates of the transformation in question. For continuous transformations it implies the preservation of the eigenstates and corresponding eigenvaluesand corresponding eigenvalues of its generator as well. We can also use the knowledge that a symmetry exist to produce new solutions of a problem from known ones. Symmetry, invariance and conservation laws have many more theoretical and practical implications and applications. The excellent text by Wu-Ki Tung \cite{Tung1985} has allowed me to scratch the surface of this beautiful subject.

In my thesis, symmetries and conservation laws are mostly used as a tool to study, predict and understand the results of light matter interactions. Sometimes, careful considerations of the symmetries and conserved quantities is all that is needed in order to pinpoint the fundamental reason of some ``mysterious'' effect (Sec. \ref{secc5:kerker}), make new experimental predictions about light scattering (Chaps. \ref{chap5}, \ref{chap6}) or to be able to isolate the actual reasons for observations that lacked a consistent explanation until then (Chap. \ref{chap4}). In other cases, it gives a very solid theoretical stepping stone for the posterior investigation of the details of the problem with analytical, numerical or experimental techniques (Chaps. \ref{chap4}, \ref{chap7}) . Also, analyzing the symmetries of the equations can provide valuable design guidelines and constraints (Chap. \ref{chap7}). 

When faced with a new problem, Wigner taught us that it pays off to consider its symmetries first: I hope that, after reading the applications chapters, you will agree that this is indeed a fruitful approach to the study of light matter interactions.

Let me now introduce a continuous transformation for electromagnetic fields, duality, and its generator, helicity. These two are the main characters in my thesis. Thanks to them, the study of light matter interactions using symmetries and conservation laws can be made relatively simple.
\section{Duality and its generator: Helicity}
\label{secc2:dualhel}
The transformations that I have considered in the previous section are {\em geometrical}: they rotate, invert and translate an object in three dimensional space. These transformations are easy to understand because we ourselves inhabit the same space in which they act. When we also include the time axis, new transformations are possible in this fourth dimensional space that did not exist in three dimensions. The action of some of them is also easy to grasp. For example, we can imagine the effects of time going backwards quite easily. Others pose a much bigger challenge to our imagination: Can you imagine the effect of a rotation along an axis which, instead of being perpendicular to a spatial plane, is perpendicular to the plane formed by the spatial $x$ axis and the time $t$ axis? This is not easy to do. And then, there are even more abstract kinds of transformations. They do not act on space or time coordinates but in the extra dimensions that the experimental observations of the last century have forced us to include in the models. For example, transformations of the spin of the electron, ``rotations'' that turn quarks of one kind onto quarks of another kind or the exchange of the labels of the two photons in a two photon state. These transformations, their generators (for the continuous ones), and the abstract spaces in which they act are crucial in physics. Electromagnetic duality is one of these more ``strange'' transformations which does not act on space or time but on a different, more abstract, dimension of electromagnetism. 

Here is the action of duality on the electric and magnetic fields in free space \cite{Zwanziger1968}, \cite[Eq. 6.151]{Jackson1998}:
\begin{equation}
\label{eqc2:gendual}
\begin{split}
\EE(\rr,t)&\rightarrow \EE_\theta(\rr,t)=\EE(\rr,t)\cos\theta  - Z_0\HH(\rr,t)\sin\theta , \\
Z_0\HH(\rr,t)&\rightarrow Z_0\HH_\theta(\rr,t)=\EE(\rr,t)\sin\theta + Z_0\HH(\rr,t)\cos\theta ,
\end{split}
\end{equation}
where $\theta$ is a real angle, $Z_0=\sqrt{\mu_0/\epsilon_0}$ and $(\epsilon_0,\mu_0)$ are the vacuum permittivity and permeability constants. Duality first appeared in print more than a century ago \cite{Heaviside1894}, which probably makes it the first of this kind of abstract transformations to be considered. From the beginning, it was seen to be a symmetry of the free space Maxwell's equations\footnote{Note that (\ref{eqc2:maxwell}) are dynamical evolution equations for $\EE$ and $\HH$ plus the extra conditions $\nabla\cdot\EE=\nabla\cdot\HH=0$.}: 
\begin{equation}
\label{eqc2:maxwell}
	\begin{split}
		\nabla\cdot\EE=0&,\ \nabla\cdot\HH=0,\\
	\partial_t\EE=\nabla\times\HH&,\ \partial_t\HH=-\nabla\times\EE.\\
	\end{split}
\end{equation}
The form of (\ref{eqc2:maxwell}) is invariant under the transformation (\ref{eqc2:gendual}). As we discussed in the previous section, this form invariance means that if the electromagnetic field  $(\EE(\rr,t),\HH(\rr,t))$ is a solution of the free space Maxwell's equations, then the field $(\EE_\theta(\rr,t),\HH_\theta(\rr,t))$ is also a solution for any value of $\theta$. This mixing of electricity and magnetism is inherent in the duality transformation. In 1968 Zwanziger \cite{Zwanziger1968} studied the duality transformation in the context of a quantum field theory with both electric and magnetic charges. He used a generalization of the microscopic Maxwell's equations which included magnetic ($g$) sources in addition to the common electric ($e$) sources. The equations contain the two types of scalar charge densities $\rho$ and vector current densities $\jj$, related by $\nabla\cdot \jj_{e,g}=-\partial_t \rho_{e,g}$.
\begin{equation}
\label{eqc2:micro}
	\begin{split}
		\nabla\cdot\EE=\rho_e&,\ \nabla\cdot\HH=\rho_g,\\
	\partial_t\EE=\nabla\times\HH-\jj_e&,\ \partial_t\HH=-\nabla\times\EE-\jj_g.\\
	\end{split}
\end{equation}
Zwanziger studied the transformation properties of (\ref{eqc2:micro}) under the simultaneous action of the duality transformation of the fields in (\ref{eqc2:gendual}) and the following similar extra transformation of the sources:
\begin{equation}
\label{eqc2:sourcedual}
\begin{split}
	\begin{bmatrix}\rho_e\\\jj_e\end{bmatrix}\rightarrow \begin{bmatrix}\rho_e\\\jj_e\end{bmatrix}_\theta&=\begin{bmatrix}\rho_e\\\jj_e\end{bmatrix}\cos\theta -\begin{bmatrix}\rho_g\\\jj_g\end{bmatrix}\sin\theta , \\
	   \begin{bmatrix}\rho_g\\\jj_g\end{bmatrix}\rightarrow \begin{bmatrix}\rho_g\\\jj_g\end{bmatrix}_\theta&=\begin{bmatrix}\rho_g\\\jj_g\end{bmatrix}\cos\theta -\begin{bmatrix}\rho_e\\\jj_e\end{bmatrix}\sin\theta.\\
\end{split}
\end{equation}
He found that the whole electrodynamic theory is invariant under simultaneous action of the two transformations. He called it the {\em chiral equivalence theorem}. In the next chapter, I will discuss what it actually means to perform the extra source transformation and what type of questions can and cannot be addressed when using it. As an advance, let me say that this thesis is about duality symmetry without using the extra source transformation. I will show that, under certain conditions, material systems can be dual symmetric {\bf without the extra source transformation}. By a dual symmetric system I mean one whose electromagnetic equations are invariant under the transformation (\ref{eqc2:gendual}) alone.

In the same paper, Zwanziger showed that helicity was the generator of the duality transformation. Just three years earlier, Calkin \cite{Calkin1965} had proved the same result for the source free equations. What Calkin and Zwanziger showed is that helicity is the generator of duality in the same sense that angular momentum is the generator of rotations or linear momentum is the generator of translations. The helicity operator is defined \cite[Sec. 8.4.1]{Tung1985} as the projection of the total angular momentum $\mathbf{J}$ onto the linear momentum direction, i.e. $\Lambda=\Hel$. Curiously, neither Calkin nor Zwanziger used the name ``helicity'' in their seminal papers nor the definition of helicity that I just wrote. On top of this, they worked in different vector spaces and reached two different formal expressions for the generator of duality (\cite[expr. 18]{Calkin1965},\cite[expr. 2.17]{Zwanziger1968}). Possibly because of this heterogeneity of the initial treatment and naming, the connection between helicity and duality has been reported several times. Here are the ones that I am aware of \cite{Deser1976,Drummond1999,Salom2006,Cameron2012}.
\subsection{Using duality}
The exploration of electromagnetic duality has turned out to be a fruitful theoretical endeavor in fundamental physics. 

Zwanziger \cite{Zwanziger1968} and Schwinger \cite{Schwinger1969} used the invariance of electromagnetism under simultaneous application of (\ref{eqc2:gendual}) and (\ref{eqc2:sourcedual}) to refine the famous argument by Dirac on electric charge quantization \cite{Dirac1948}. Dirac showed that the mere existence of a single magnetic monopole (a particle with magnetic charge) implies the quantization of electric charge. Nobody knows why, in isolated particles, electric charge is only observed in quantized multiples of the charge of the electron. This empirical truth is one of the remaining mysteries in physics. Finding a magnetic monopole would solve it. 

In an application to a different field, the magnetic monopole idea and a generalization of electromagnetic duality allowed Montonen and Olive to formulate their celebrated conjectured unification of supersymmetry and the theories of elementary particle interactions \cite{Montonen1977}. 

In a more practical sense, the duality symmetry has been skillfully exploited for the study of objects with zero back scattering in the context of light matter interactions: The authors in \cite{Lindell2009,Karilainen2012} find that when a plane wave impinges on an scatterer with both duality symmetry and $\pi/2$ discrete rotational symmetry along the axis defined by the momentum of the plane wave, there is no energy reflected in the backwards direction. I shall show in Chap. \ref{chap5} that when the other piece of the conservation law, helicity, is taken into account, those results can be fully understood using only symmetries and conservation laws. Additionally, the consideration of helicity will allow me to show that any dual symmetric object with a discrete rotational symmetry $2\pi/n$ for $n\ge 3$ will exhibit zero backscattering.

\subsection{Using helicity}
\label{subsecc2:usinghelicity}
The helicity operator $\Hel$ does not only act in vector spaces representing electromagnetism. Helicity is routinely used in particle and high energy physics. Whether we are talking about photons, electrons, neutrinos, or many of the other inhabitants of the zoo, helicity has a well defined meaning. Its importance is apparent from the get go since helicity is the operator that is typically chosen to represent the {\em internal} degrees of freedom of elementary particles, often called {\em spin or polarization} degrees of freedom. For example, an electron state or a photon state can be specified by fixing its four momentum and helicity \cite[Sec. 10.4]{Tung1985},\cite[Chap. 2]{Weinberg1995}. In the analysis of high energy collisions, helicity allows a unified treatment of massive and massless particles \cite{Jacob1959}. 

For massless particles like the photon, helicity is quite special: It can only take two different values ($\pm 1$ for the photon, $\pm 2$ for the graviton), and is a Poincar\'e invariant of the field. That is, helicity is invariant under space and time translations, spatial rotations and Lorentz boosts. Note that this is not the case for other properties like, for instance, momentum. A rotation changes the momentum vector. It is also different from the helicity of massive particles. The helicity of a massive particle takes $2S+1$ different values, where $S$ is the intrinsic spin of the particle, and it is not a Poincare invariant because a boost mixes the different helicities \cite[Chap. 2]{Weinberg1995},\cite[secs. 10.4.2-10.4.4]{Tung1985}. Take for instance an electron with helicity equal to $1/2$ in some reference frame. Since $\Lambda=\Hel$, I can always ``boost myself'' along the electron's propagation direction until I advance it to a reference frame where $\PP\rightarrow-\PP,\JJ\rightarrow\JJ$, and the sign of helicity will flip and become $-1/2$. This argument does not work for massless particles: A photon, for example, cannot be advanced. 
In some sense, two photons of different helicity can be seen as two different elementary particles. The different helicity value denotes a different representation of the Poincare group, which is the criteria used to define an elementary particle \cite[Chap. 10]{Tung1985}. Why should we then say that the two photons of different helicity are the same particle? Here is why: If the spatial inversion (parity) transformation is added to the Poincare group, the two photons of different helicity merge into one particle with two possible helicities because the parity operator ($\Pi$) interchanges the two helicity eigenstates $|\pm\rangle$:
\begin{equation}
\Pi|+\rangle=|-\rangle,\ \Pi|-\rangle=|+\rangle.
\end{equation}
Even though parity is not a symmetry of Nature \cite[2.6, 2.7]{Feynman2011}, it is considered a symmetry in electromagnetism \cite[6.10]{Jackson1998}. The photon must hence be a single particle with two possible helicities in order to keep up with the action of $\Pi$. Consequently, the electromagnetic field has two possible helicities states. 

All this may sound too abstract and separated from light matter interactions. It is not. Chapters \ref{chap5}, \ref{chap6} and \ref{chap7} contain practical engineering guidelines for building zero backscattering objects, polarization rotation objects and metamaterials for transformation optics. These guidelines are obtained by symmetry considerations involving helicity and duality.

Helicity has been exploited by Prof. Iwo Bialynicki-Birula in his works on theoretical electromagnetism. While showing us how to construct a proper real space wave function for the photon \cite{Birula1994,Birula1996}, he makes use of the division of the field in its two helicity components, studies the condition of helicity preservation in inhomogeneous and isotropic media and realizes that helicity is conserved in a general gravitational field. All these ideas are indispensable in my thesis. I also make extensive use of the Riemann-Silberstein formulation of electromagnetic fields that Prof. I. Bialynicki-Birula \cite{Birula1994,Birula1996} together with Prof. Z. Bialynicka-Birula promote \cite{Birula2013}. I had the pleasure and honor of meeting and interacting with both of them during their month long visit to the group of my advisor at Macquarie University (Sydney).

I want to finish this chapter by stating that duality is almost always broken and helicity is almost never preserved in light matter interactions. You may consequently think that this conservation law is almost surely useless. But, if you keep reading, I can hopefully show you that it is not. For now let me say two things. First, broken symmetries can be as crucial for our understanding as unbroken ones: One example of this is the Higgs mechanism. Knowing that duality is broken and that helicity is expected to change can help you understand something about your observations in the laboratory. And second, very interesting phenomena happen when duality is an actual symmetry of the system and helicity is preserved either naturally or as a result of engineering.
 
\makeatletter{}\chapter[Theory]{Theory of helicity and duality symmetry in light matter interactions}
\label{chap3}
\epigraph{{\em In fact the natural correspondence between the basis vectors of unitary irreducible representations of the Poincar\'e group and quantum mechanical states of elementary physical systems stands out as one of the remarkable monuments to unity between mathematics and physics.}}{Wu Ki Tung, ``Group Theory in Physics''}
In this chapter, I aim to establish the theoretical foundations for the use of helicity and duality in the study of light matter interactions. I also develop the tools that I will use in the application chapters. 

Sections \ref{secc3:hilbert} and \ref{secc3:heltransprop} contain a collection of results on Hilbert spaces and transformation properties which are needed for this chapter. They are taken from \cite[Chaps. V, VII and VIII]{Messiah1958}, \cite[Chaps. 1]{Sakurai1993} and \cite{Tung1985}. These results are presented in a way that I believe most appropriate for my thesis. Section \ref{secc3:hilbert} also contains examples which are useful for understanding the use of symmetries and conservation laws in the study of scattering problems. The rest of the sections contain original material, to the best of my knowledge and except for the cited work. 

\section{General framework}
\label{secc3:gf}
\begin{figure}[h]
	\begin{center}\begin{tikzpicture}[scale=0.8,every node/.style={scale=1.2},>=latex]
		\makeatletter{}\draw (0,0) ellipse (0.25 and 0.75);
\draw[rotate=-90](-0.75,0) parabola bend (0,3.5) (0.75,0);
\draw(1.75,0) node{$\GG_{in}$};
\filldraw[red!20!white, draw=red!50!black] (4.5,0) .. controls (4.75,1) and (5.25,2) .. (5.5,1) 
			  .. controls (5.75,0) and (6.25,0.5) .. (6.5,0)
			  .. controls (7,-1) and (5.75,-2) .. (5.5,-1) 
			  .. controls (4 ,-3 ) and (3.5,1) .. (4.5,0)--cycle;
\draw(5.25,-0.2) node{{\Large$\mathbb{S}$}};

\draw[thick,dashed] (3.5,1) .. controls (4,3) .. (5,2);
\draw[thick,dashed] (5,2) .. controls (6,3) and (9,3) .. (7,1);
\draw[thick,dashed] (7,1) .. controls (10,0) and (9,-3) .. (6,-2);
\draw[thick,dashed] (6,-2) .. controls (4.5,-3) and (4,-3) .. (3.5,-1);
\draw(8,-0.5) node{$\GG_{out}$};
 
	\end{tikzpicture}\end{center}
\caption[Electromagnetic scattering.]{\label{figc3:scatt}The interaction of an incident electromagnetic field $\GG_{in}$ with an object $\mathbb{S}$ produces a scattered field $\GG_{out}$.}
\end{figure}
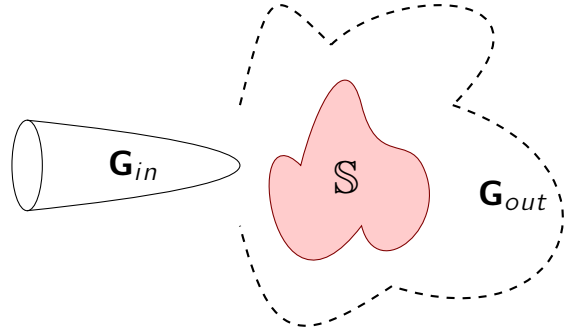
Take a look at Fig. \ref{figc3:scatt}. An incident electromagnetic field $\GG_{in}$ impinges onto a material scatterer $\mathbb{S}$. As a result of the interaction between the field and the material system, a scattered field $\GG_{out}$ is produced. There are many reasons why this problem is interesting. For example, the scattered field contains information about $\mathbb{S}$ which may be used in microscopy to image the system, or in spectroscopy to analyze its components. We may also want to engineer the object $\mathbb{S}$ so that $\GG_{out}$ is some desired function of $\GG_{in}$ like in transformation optics \cite{Leonhardt2009}.

The underlying theme in my thesis is to study this problem by means of symmetries and conservation laws. In order to be able to do this, I will need to introduce some formal machinery, namely that of Hilbert spaces. As I will show you later, both 
$\GG_{in}$ and $\GG_{out}$ can be modeled by vectors in a Hilbert space (represented by $|\Phi\rangle$) and the action of the object $\mathbb{S}$ by means of a linear operator $S$ in such space. This is the main assumption/restriction in my thesis: The action of $\mathbb{S}$ is linear. This means that the operator $S$ meets:
\begin{equation}
	\label{eqc3:linear}
	S\left(c_1|\Phi_1\rangle+c_2|\Phi_2\rangle\right)=c_1S|\Phi_1\rangle+c_2S|\Phi_2\rangle.
\end{equation}
The symmetries of $\mathbb{S}$, which are reflected in the properties of $S$, will play a central role through their corresponding conservation laws. As vectors in a Hilbert space, the fields can be expanded in a basis of the space. Each vector of a basis has four numbers that identify it. These numbers are the eigenvalues of at least four independent commuting operators.  The symmetries of $S$ and their corresponding conservation laws will tell us which of those numbers are going to be maintained between input and output and which ones are allowed to change. Helicity will enter the picture as one of the four numbers, the one that labels the polarization of the field. The polarization is the non-scalar degree of freedom that a scalar field like for instance temperature does not have.
\subsection[Hilbert spaces, transformations and operators]{Hilbert spaces, vector basis, transformations and operators}\label{secc3:hilbert}
A Hilbert space is a linear vector space with an inner product having certain properties \cite[Chap. V \S2]{Messiah1958}. In a linear vector space, the addition of any two vectors $\ket{\Psi}+\ket{\Phi}$ of the space results in a vector that also belongs to the linear vector space. The same is true for the multiplication of a vector by a number $c\ket{\Phi}$. The inner product is a function which takes two vectors $|\Psi\rangle,|\Phi\rangle$ and produces a number $b$:
\begin{equation}
\label{eqc3:inner}
b=\langle \Psi|\Phi\rangle.
\end{equation}
What is $\langle \Psi|$? If $|\Psi\rangle$ belongs to a vector space $\Omega$, $\langle \Psi|$ is a member of the vector space $\Omega^{*}$ dual to $\Omega$. In my thesis, this is an unfortunate naming coincidence. The word ``dual'' here does not have anything to do with the duality transformation. To avoid confusion, and following Sakurai \cite[Chap. 1.2]{Sakurai1993}, I will use the acronym DC for this {\em dual correspondent} between vector spaces. Operators also have DC versions. Tab. \ref{tabc3:DC} contains some of the correspondences between some objects and their DC version. 
\begin{table}[h!]
\begin{center}\begin{tabular}{cc} \toprule
Vector space & DC vector space \\ \midrule
$|\Psi\rangle$ & $\langle \Psi|$\\
$b|\Psi\rangle$ & $b^*\langle \Psi|$\\
$X|\Psi\rangle$ & $\langle \Psi|X^\dagger$\\\bottomrule
\end{tabular}\end{center}
\caption[Dual correspondences of vectors and operators.]{\label{tabc3:DC}DC versions of a vector $|\Psi\rangle$, a vector times a complex number $b$ ($b^*$ is the complex conjugate of $b$), and a vector $|\Psi\rangle$ acted upon by an operator $X$. $X^\dagger$ is called the hermitian adjoint. It is defined as the operator $X^\dagger$ such that $\langle \Psi|X|\Phi\rangle=\langle \Phi|X^\dagger|\Psi\rangle^*$ for all $|\Psi\rangle$ and $|\Phi\rangle$.}
\end{table}

When $\langle \Psi|\Phi\rangle=0$, we say that the two vectors are orthogonal. In a Hilbert space ($\mathbb{H}$) we have the concept of mutually orthogonal basis: A set $\{|\eta\rangle\}$ of orthogonal vectors in $\mathbb{H}$ which can produce any vector in $\mathbb{H}$ by linear combinations, like:
\begin{equation}
	\label{eqc3:expand}
|\Phi\rangle=\sum_\eta c_\eta|\eta\rangle,
\end{equation}
where the numbers $c_\eta$ are $c_\eta=\langle\eta|\Phi\rangle$. The different basis vectors are mutually orthogonal:
\begin{equation}
	\label{eqc3:zero}
	\langle \eta|\bar{\eta}\rangle = 0,\ if\ \eta\neq\bar{\eta}.
\end{equation}
If they are scaled so that $\langle \eta|\eta\rangle = 1\ \forall\ \eta$, we say that they form an orthonormal basis. 

There are many different orthonormal bases that expand (can produce any vector in) a given Hilbert space. As I will show you later, when the vectors of the space represent electromagnetic modes, its basis have one characteristic in common. Each of their vectors is uniquely identified by four numbers. These four numbers are the eigenvalues of four independent commuting operators. The symbolic index $\eta$ contains then four numbers. Some of those numbers may take continuous values and others may take discrete values. The summation over $\eta$ symbol ($\sum_\eta$) in (\ref{eqc3:expand}) is a short hand notation for whatever discrete summations or continuous integrals are needed, for example it may mean
\begin{equation}
	\sum_{\eta}\equiv \int d\eta_1\sum_{\eta_2}\sum_{\eta_3}\int d\eta_4,
\end{equation}
or
\begin{equation}
	\sum_{\eta}\equiv \sum_{\eta_1}\sum_{\eta_2}\sum_{\eta_3} \int d\eta_4,
\end{equation}
etc.

The difference between continuous and discrete indexes is subtle and complex. For starters, vectors with a continuous index have an infinite norm: $\langle \psi | \psi \rangle\rightarrow \infty$. This means that they do not belong to the Hilbert space since one of the requirements for membership is finite norm. Please refer to \cite[Chap. V \S 8; Chap. VII \S4,\S9]{Messiah1958} for the discussion about why it is nonetheless possible to use these outsiders as members of a basis for expanding proper vectors in the Hilbert space. The discussion makes use of the theory of distributions. For practical purposes we may say that they have the same properties (\ref{eqc3:expand})-(\ref{eqc3:zero}) as the basis vectors of finite norm, but, in order to express their orthogonality properties, one has to substitute the Kronecker deltas ($\delta_{nm}=0$ if $n\neq m$, and $\delta_{nn}=1$) by the Dirac delta distributions $\delta(x-x_0)$. I will not discuss this anymore after this paragraph\footnote{If you do not want to read the sections from Messiah's book that I have indicated, but feel a bit uneasy, here is a quick non-rigorous palliative: In an expansion with $\sum_{\eta}$, the infinite norm ``vectors'' will be multiplied by the corresponding infinitesimally small differential $d\eta_i$, and so an infinite sum of infinite norm ``vectors'' can result in a finite norm vector: A proper member of the Hilbert space.}. This serious complication is worth it because of the very simple properties that some of those outsiders have under relevant transformations.  From now on, I will call {\em vectors} both the proper members of the Hilbert space and the infinite norm ones.

Same as vectors, linear operators acting in $\mathbb{H}$ can be expanded using other operators which can be built from the vectors of a basis and their DC versions. Since any vector in $\mathbb{H}$ can be expanded by a chosen basis $|\eta\rangle$, a linear operator $S$ can be expanded as:

\begin{equation}
	\label{eqc3:S}
	S=\sum_\eta\sum_{\bar{\eta}}s_{\eta}^{\bar{\eta}}|\bar{\eta}\rangle\langle\eta|,
\end{equation}
where the $s_{\eta}^{\bar{\eta}}$ are numbers. By appropriately choosing $s_{\eta}^{\bar{\eta}}$ you can produce any linear map from $\mathbb{H}$ to $\mathbb{H}$. 

While I go over all this machinery, I want to give some examples of its use. The machinery is general, but in order not to use abstract examples, I will, from now on, implicitly assume the context of electromagnetic scattering: An input vector $|\Phi\rangle=\sum_\eta c_\eta|\eta\rangle$ produces an output vector $|\bar{\Phi}\rangle$ after the interaction with the object $\mathbb{S}$ which is represented by the linear operator $S$ (see Sec. \ref{secc3:scattop}).
\begin{equation}
	\begin{split}
		|\bar{\Phi}\rangle=S|\Phi\rangle&\stackrel{\ref{eqc3:S}}{=}\sum_\eta\sum_{\bar{\eta}}s_{\eta}^{\bar{\eta}}|\bar{\eta}\rangle\langle\eta|\left(\sum_{\hat{\eta}} c_{\hat{\eta}}|\hat{\eta}\rangle\right)=\\
																&\sum_\eta\sum_{\bar{\eta}}\sum_{\hat{\eta}}s_{\eta}^{\bar{\eta}}c_{\hat{\eta}}|\bar{\eta}\rangle\langle\eta|\hat{\eta}\rangle\stackrel{\ref{eqc3:zero}}{=}\sum_{\bar{\eta}}\left(\sum_\eta s_{\eta}^{\bar{\eta}}c_\eta\right)|\bar{\eta}\rangle,
	\end{split}
\end{equation}
The numbers on top of the equal signs are the references to the information (previous expressions, tables, etc ...) needed to take the corresponding step. I will use this notation quite often. For example, $\duetoref{eqc3:zero}$ above means that the equality follows because the $\eta$ and $\hat{\eta}$ indexes collapse due to Eq. (\ref{eqc3:zero}): $\langle \eta|\hat{\eta}\rangle=0$ unless $\eta=\hat{\eta}$. 

By construction, the set of all the $s_{\eta}^{\bar{\eta}}$ completely define the operator $S$. To recover one of them, say $s_{\eta_0}^{\bar{\eta}_0}$ we do:
\begin{equation}
	s_{\eta_0}^{\bar{\eta}_0}=\langle \bar{\eta}_0|S|\eta_0\rangle.
\end{equation}
The $s_{\eta}^{\bar{\eta}}$ are called the {\em matrix elements} of the operator $S$ between states of the $|\eta\rangle$ basis.

In chapter \ref{chap2}, we saw that symmetry transformations are represented by operators that act on vectors. They also act on other operators. Here is how the operator $S$ is transformed by an operator $T$ representing a given transformation \cite[Chap. 1.1]{Tung1985}:
\begin{equation}
	\label{eqc3:tst}
	\bar{S}=T^{-1}ST.
\end{equation}
$T$ needs to be invertible. This is not a problem since any transformation that leaves a physical system invariant must be represented by a unitary operator \cite[Chap. XV \ \S 1]{Messiah1958}, \cite[Sec. 3.3]{Tung1985}. A unitary operator $T$ meets
\begin{equation}
	\label{eqc3:herm}
T^{-1}T=T^{\dagger}T=I. 
\end{equation}
where $I$ is the identity operator. So, any operator susceptible to represent a symmetry of the system has an inverse, and the inverse is its hermitian adjoint. This is the case for unitary transformations like translations, rotations, parity, duality, time reversal and Lorentz boosts. It is also the case for time inversion, which is antilinear and antiunitary. For operators representing symmetry transformations, I will often use $T^\dagger$ instead of $T^{-1}$.

In both unitary and antiunitary cases, invariance of $S$ under transformation $T$ means then 
\begin{equation}
\label{eqc3:inv}
\bar{S}=T^{-1}ST=T^\dagger ST=S,
\end{equation}
which is also equivalent to saying that $S$ and $T$ commute:
\begin{equation}
T^{-1}ST=S\implies ST=TS\implies [S,T]=0. 
\end{equation}

Equations (\ref{eqc3:tst})-(\ref{eqc3:inv}), together with Tab. \ref{tabc3:DC} and the associativity of the product of operators
\begin{equation}
	\langle \Phi|\left(XYZ|\Phi\rangle\right)=\langle \Phi|X\left(YZ|\Phi\rangle\right)=\langle \Phi|XY\left(Z|\Phi\rangle\right)=\langle \Phi|XYZ\left(|\Phi\rangle\right),
\end{equation}
is all we need to start exploiting the consequences of symmetry on the scattering operator $S$. The methodology is general but, to make the discussion less abstract, I will give examples with specific transformations.

Let us say that the object $\mathbb{S}$ is, like the cone in Fig. \ref{figc2:conetx}, invariant under rotations along the $z$ axis. What this means for its scattering operator is that $S=R_z^{\dagger}(\alpha)SR_z(\alpha)$. Let me we expand the incoming field in Fig. \ref{figc3:scatt} in a basis where one of the four indexes in $\eta$ is the eigenvalue of $J_z$. Without explicitly writing the other three indexes, we have that:
\begin{equation}
	\begin{split}
		s^n_{\bar{n}}&=\langle {\bar{n}} |S|n\rangle\stackrel{\ref{eqc3:inv},\ref{eqc3:herm}}{=}\langle {\bar{n}} |R_z^{\dagger}(\alpha)SR_z(\alpha)|n\rangle\stackrel{}{=}\\
			   &\exp(i\alpha \bar{n})\langle {\bar{n}}|S|n\rangle\exp(-i\alpha n)=s^n_{\bar{n}}\exp(i\alpha({\bar{n}}-n)).
	\end{split}
\end{equation}
 
Since $s^n_{\bar{n}}=s^n_{\bar{n}}\exp(i\alpha({\bar{n}}-n))$ must be true for all $\alpha$, this means that $s^n_{\bar{n}}=0$ unless ${\bar{n}}=n$. This is the manifestation of invariance under symmetry transformations (Sec. \ref{secc2:sym}): Eigenstates of $J_z$ before the interaction are still eigenstates of $J_z$ with the original eigenvalue after the interaction. The number of $s^n_{\bar{n}}$ coefficients needed to describe $S$ is drastically reduced if we choose our basis according to the symmetries of the system. This is going to be a recurring theme.

Imagine now that the scatterer does not have the full cylindrical symmetry but is only invariant under discrete rotations with angles $2\pi/m$ for integer $m\neq 0$. For example, for $m=4$, this is the symmetry of a square prism. Repeating the above steps leads to the conclusion that $s^n_{\bar{n}}=0$ unless ${\bar{n}}-n=qm$ for integer $q$. 

Let us now switch to translations. Consider an infinite slab of material parallel to the $XY$ plane. Because of the invariance of the infinite wall to any transverse translation $(\Delta_x,\Delta_y)$, the matrix elements of $S$ between plane waves with different components of momentum parallel to the wall must vanish.

\begin{equation}
	\label{eqc3:transinv}
	\begin{split}
		s^{p_xp_y}_{\bar{p}_x\bar{p}_y}&=\langle {\bar{p}_x,\bar{p}_y} |S|p_x,p_y\rangle=\langle {\bar{p}_x,\bar{p}_y} |T^{\dagger}_x(\Delta_x)T^{\dagger}_y(\Delta_y)ST_y(\Delta_y)T_x(\Delta_x)|p_x,p_y\rangle=\\
								   &\exp(i(\bar{p}_x\Delta_x+\bar{p}_y\Delta_y))\langle {\bar{p}_x,\bar{p}_y}|S|p_x,p_y\rangle\exp(-i(p_x\Delta_x+p_y\Delta_y))=\\
								   &s^{p_xp_y}_{\bar{p}_x\bar{p}_y}\exp(i(\Delta_x(\bar{p}_x-p_x)+\Delta_y(\bar{p}_y-p_y)))\implies \\ & s^{p_xp_y}_{\bar{p}_x\bar{p}_y}=0 \textrm{ unless } (p_x,p_y)=(\bar{p}_x,\bar{p}_y).\\
	\end{split}
\end{equation}
 In a three dimensional lattice, for instance a cubic lattice of ions, discrete translations can be used in much the same way as we have used discrete rotations above to figure out which are the momenta that are invariant upon scattering through a lattice. In crystallography, these momenta form what is called the {\em reciprocal lattice}.

These simple examples show the importance of taking into account the symmetries of the problem when choosing the working basis in $\mathbb{H}$. 

Enough for now on the effect of symmetries in scattering. Section (\ref{secc3:scattcone}) contains a complete example which better shows the capabilities of the technique. Helicity and duality are needed for it and for all the other applications in the later chapters. We will also need to know how different transformations affect helicity.

\subsection[Helicity transformation properties]{Transformation properties of helicity and other generators}\label{secc3:heltransprop}
In this section, I provide a table with the commutation properties of the helicity operator $\Lambda=\JJ\cdot\PP/|\PP|$ \cite[Chap. 8.4.1]{Tung1985}, some of the other generators and two discrete transformations, spatial inversion (parity, $\Pi:\rr\rightarrow\tilde{\rr}=-\rr$) and time inversion ($Tr:t\rightarrow \tilde{t}=-t$). Time inversion is the only antiunitary transformation that appears in my thesis. 

In the table, I use the anticommutator between two operators
\begin{equation}
	\{A,B\}=AB+BA.
\end{equation}
If $\{A,B\}=0$ we say that $A$ and $B$ anticommute. If$\{A,B\}=0$ and $A$ has an inverse, then $A^{-1}BA=-B$.

From the commutation properties of the generators one can obtain those of the generated transformations following a few rules.
\begin{itemize}
	\item If two generators commute, their generated transformations also commute, and each of the generators commutes with the transformation generated by the other generator. 
	\item If $B$ is a generator and $A$ a unitary discrete transformation
		\begin{itemize}
			\item $[A,B]=0\implies A^{-1}\exp(-i\theta B) A=\exp(-i\theta B)$
			\item $\{A,B\}=0\implies A^{-1}\exp(-i\theta B) A=\exp(i\theta B)$
		\end{itemize}

	\item If $B$ is a generator and $A$ an antiunitary discrete transformation
		\begin{itemize}
			\item $[A,B]=0\implies A^{-1}\exp(-i\theta B) A=\exp(i\theta B)$
			\item $\{A,B\}=0\implies A^{-1}\exp(-i\theta B) A=\exp(-i\theta B)$
		\end{itemize}
\end{itemize}

For example, in the table we see that parity anticommutes with momentum and hence flips its sign. This means that when transforming a translation with the parity operator, the result is a translation in the opposite direction, which makes intuitive sense: $\Pi^{\dagger}T_x(\Delta)\Pi=T_x(-\Delta)$.

Since $[A,A]=0$ and $[A,B]=-[B,A]$, and if $\{A,B\}=0\implies\{B,A\}=0$ many cells in Tab. \ref{tabc3:gentrans} are empty to avoid redundant entries. 
\begin{table}[!h]\small
	\makebox[\textwidth]{		
		\begin{tabular}{ccccccc}
&Helicity&Energy&Momentum&$\begin{array}{l}\text{Angular}\\\text{momentum}\end{array}$&Parity&$\begin{array}{l}\text{Time}\\\text{inversion}\end{array}$\\
	&$\Lambda$&H&$\PP$&$\JJ$&$\Pi$&$Tr$\\\midrule
 $\Lambda$&&$[\Lambda,H]=0$&$[\Lambda,P_n]=0$&$[\Lambda,J_n]=0$&$\{\Lambda,\Pi\}=0$&$[\Lambda,Tr]=0$\\
	   $H$&&&$[H,P_n]=0$&$[H,J_n]=0$&$[H,\Pi]=0$&$[H,Tr]=0$\\
  $\PP$&&&&$[P_m,J_n]=i\epsilon^{mnl}P_l$&$\{P_n,\Pi\}=0$&$\{P_n,Tr\}=0$\\
  $\JJ$&&&&&$[J_n,\Pi]=0$&$\{J_n,Tr\}=0$\\
  $\Pi$&&&&&&$[\Pi,Tr]=0$\\
\end{tabular}
}
\caption[Commutation rules for generators and discrete transformations.]{\label{tabc3:gentrans} Commutation rules for some generators and discrete transformations. The $n$ index takes the values 1,2,3. The $\epsilon^{mnl}$ is the totally antisymmetric tensor with $\epsilon^{123}=1$; $\epsilon^{mnl}=1$ if $mnl$ is an even permutation of $123$ like $312$ or $231$ (there is an even number of position swaps between two elements to get from $123$ to $312$ or $231$), $\epsilon^{mnl}=-1$ if $mnl$ is an odd permutation of $123$ like $213$ or $321$ (there is an odd number of position swaps between two elements to get from $123$ to $213$ or $321$), and $\epsilon^{mnl}=0$ if there is any repeated index, $\epsilon^{112}=0$. The commutation rules in the table are valid in general. For the massless case, helicity commutes with the generators of Lorentz boosts as well. This is not the case for massive particles or fields, like the electron.}
\end{table}

The helicity operator commutes with rotations, translations and time inversion. It anticommutes with parity, which flips the helicity. On the other hand, parity commutes with angular momentum. This difference in the behavior under parity between angular momentum and helicity is worth discussing a bit more. It is one of the differences between turning and twisting.

\subsection{Turning versus twisting}
An ice skater spinning around in a fixed position and a spinning top after a skilled kid pulls out the cord: These systems are turning. What an ant does if it wants to walk along a wine-opener or the movement of a screwdriver when you tighten or loosen a screw. This is twisting. 

We appreciate the difference intuitively. In order for me to turn I only need to rotate ($\JJ$), but, if I want to twist, I need to rotate ($\JJ$) and advance ($\PP$) at the same time. We also see intuitively that there are two possible kinds of twist, left-handed and right-handed. Helicity $\Lambda=\Hel$ {\bf describes the sense of twist}. Its name is quite appropriate in relation with a helix.

The transformation properties of turns and twists are quite different, and correspond to those of $\JJ$ and $\Lambda$. Turns can change upon rotation while twists do not. In the formal language: the components of $\JJ$ do not commute with each other but they all commute with $\Lambda$ (Tab. \ref{tabc3:gentrans}). Imagine that the spinning ice skater is able to do a back flip and start spinning on her head. If you have good spatial intuition you will realize that now she is spinning on the sense opposite to the one before the pirouette. On the other hand, turning a wine-opener on its head does not change its sense of twist. Now, if the ice was clear enough for you to see the ice skater reflected on it while she turns, you would see the ``mirrored'' ice skater turning in the same sense as the real person. Taking your wine-opener-ant system next to a mirror shows that the sense of twisting changes ... {\bf no matter how the mirror and the wine opener are oriented relative to each other}. In the formal language: Any inversion of coordinates flips helicity while it does not necessarily change angular momentum. 

Note that a mirror reflection across a plane perpendicular to axis $\uhat$, $M_\uhat$, can be written as parity times a rotation of 180 degrees along $\uhat$. The order does not matter because rotations and parity commute:
\begin{equation}
	\label{eqc3:Mu}
	M_\uhat=\Pi R_\uhat(\pi)=R_\uhat(\pi)\Pi.
\end{equation}
The transformation properties of $\JJ$ and $\Lambda$ under mirror reflections can now be worked out using (\ref{eqc3:Mu}) and Tab. \ref{tabc3:generators}. Fig. \ref{figc3:twistturn} illustrates the differences between the transformation properties of helicity and those of angular momentum.

\begin{figure}[h!]
	\begin{center}
\makeatletter{}\def\ASYprefix{}
\newbox\ASYbox
\newdimen\ASYdimen
\long\def\ASYbase#1#2{\leavevmode\setbox\ASYbox=\hbox{#1}\ASYdimen=\ht\ASYbox\setbox\ASYbox=\hbox{#2}\lower\ASYdimen\box\ASYbox}
\long\def\ASYaligned(#1,#2)(#3,#4)#5#6#7{\leavevmode\setbox\ASYbox=\hbox{#7}\setbox\ASYbox\hbox{\ASYdimen=\ht\ASYbox\advance\ASYdimen by\dp\ASYbox\kern#3\wd\ASYbox\raise#4\ASYdimen\box\ASYbox}\put(#1,#2){#5\wd\ASYbox 0pt\dp\ASYbox 0pt\ht\ASYbox 0pt\box\ASYbox#6}}\long\def\ASYalignT(#1,#2)(#3,#4)#5#6{\ASYaligned(#1,#2)(#3,#4){
}{
}{#6}}
\long\def\ASYalign(#1,#2)(#3,#4)#5{\ASYaligned(#1,#2)(#3,#4){}{}{#5}}
\def\ASYraw#1{
currentpoint currentpoint translate matrix currentmatrix
100 12 div -100 12 div scale
#1
setmatrix neg exch neg exch translate}
 
\makeatletter{}\setlength{\unitlength}{1pt}
\makeatletter\let\ASYencoding\f@encoding\let\ASYfamily\f@family\let\ASYseries\f@series\let\ASYshape\f@shape\makeatother{\catcode`"=12\includegraphics[width=\linewidth]{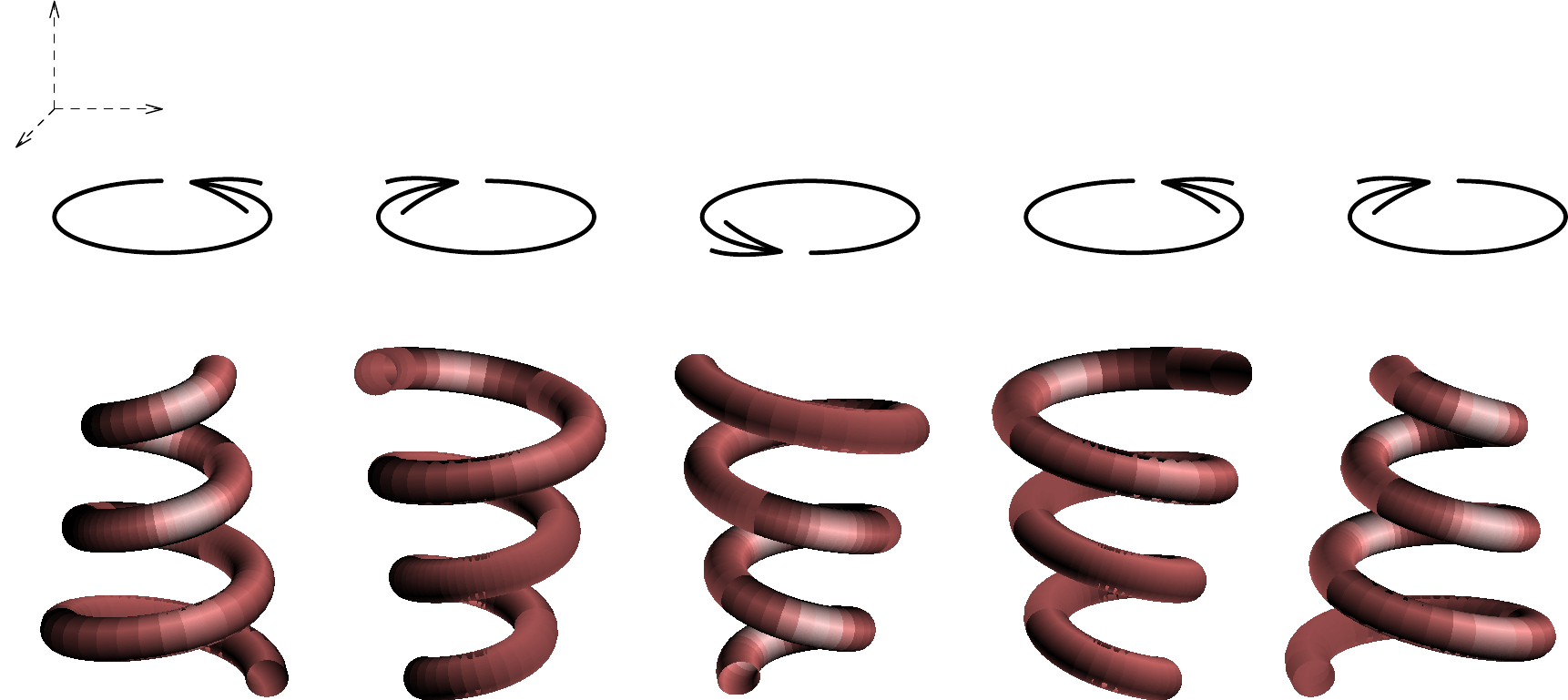}}\definecolor{ASYcolor}{gray}{0.000000}\color{ASYcolor}
\fontsize{12.000000}{14.400000}\selectfont
\usefont{\ASYencoding}{\ASYfamily}{\ASYseries}{\ASYshape}\ASYalignT(-346.115581,149.249868)(0.000000,-0.500000){0.800000 -0.000000 -0.000000 0.800000}{$\hat{x}$}
\definecolor{ASYcolor}{gray}{0.000000}\color{ASYcolor}
\fontsize{12.000000}{14.400000}\selectfont
\ASYalignT(-337.243662,133.177949)(-0.500000,-1.000000){0.800000 -0.000000 -0.000000 0.800000}{$\hat{z}$}
\definecolor{ASYcolor}{gray}{0.000000}\color{ASYcolor}
\fontsize{12.000000}{14.400000}\selectfont
\ASYalignT(-356.670655,134.914045)(-1.000000,0.218749){0.800000 -0.000000 -0.000000 0.800000}{$\hat{y}$}
\definecolor{ASYcolor}{gray}{0.000000}\color{ASYcolor}
\fontsize{12.000000}{14.400000}\selectfont
\ASYalign(-324.771743,149.249868)(-0.500000,-0.250000){0)}
\definecolor{ASYcolor}{gray}{0.000000}\color{ASYcolor}
\fontsize{12.000000}{14.400000}\selectfont
\ASYalign(-249.940231,149.249868)(-0.500000,-0.228418){a) $R_y(\pi)$}
\definecolor{ASYcolor}{gray}{0.000000}\color{ASYcolor}
\fontsize{12.000000}{14.400000}\selectfont
\ASYalign(-175.108718,149.249868)(-0.500000,-0.250000){b) $\Pi$}
\definecolor{ASYcolor}{gray}{0.000000}\color{ASYcolor}
\fontsize{12.000000}{14.400000}\selectfont
\ASYalign(-100.277205,149.249868)(-0.500000,-0.250000){c) $M_{\hat{x}}$}
\definecolor{ASYcolor}{gray}{0.000000}\color{ASYcolor}
\fontsize{12.000000}{14.400000}\selectfont
\ASYalign(-25.445692,149.249868)(-0.500000,-0.250000){d) $M_{\hat{z}}$}
 
\caption[Angular momentum versus helicity: Turning versus twisting.]{\label{figc3:twistturn} Illustration of the transformation properties of turns (top row of figures) and twists (bottom row of figures). The initial turn and twist (column 0)) are transformed by a rotation (a), parity (b), and two different mirror reflections (c,d). Using Tab. \ref{tabc3:generators} and the text in this section, one can check that the transformation properties of the turn match those of $J_x$ and the transformation properties of the twist those of $\Lambda$. Note how the sense of the twist, or sense of screw, is preserved in (a) and changes in (b), (c) and (d). On the other hand, the turn keeps the same sense in (b) and (c) and changes in (a) and (d).}
	\end{center}
\end{figure}

\section{The Hilbert space of transverse Maxwell fields $\mathbb{M}$}\label{secc3:M}
Consider the source free Maxwell equations for an infinite homogeneous and isotropic medium with scalar electric and magnetic constants $\epsilon$ and $\mu$:
\begin{equation}
\label{eqc3:iso}
	\begin{split}
		\nabla\cdot\left(\epsilon\EE\right)=0&,\ \nabla\cdot\left(\mu\HH\right)=0,\\
	\partial_t\EE=\frac{\nabla\times}{\epsilon}\HH&,\partial_t\HH=-\frac{\nabla\times}{\mu}\EE.\\
	\end{split}
\end{equation}
These equations are linear and homogeneous in $(\EE,\HH)$. It follows that adding two of their solutions $(\EE_1,\HH_1)$ and  $(\EE_2,\HH_2)$ produces another valid solution. The same is true for multiplying a solution by a number. The set of all solutions of (\ref{eqc3:iso}) is hence a linear vector space. Together with an inner product between two solutions $(\EE_1,\HH_1)$ and  $(\EE_2,\HH_2)$, which I will discuss shortly, they form a Hilbert space \cite[Chap. 13.3]{Morse1953}, which I will call $\mathbb{M}$.

From a formal point of view, it is preferable to work with vectors and operators in the abstract space as much as possible. I will do so in the application chapters. Nevertheless, sometimes one needs to use a concrete representation. There are several representations of $\mathbb{M}$. For example, a linearly polarized monochromatic plane wave propagating in vacuum along the $z$ axis, like
\begin{equation}
	\label{eqc3:r}
\GG(\rr,t)=\xhat\exp(p_zz-\omega_0 t),
\end{equation}
is nothing but a concrete representation of the abstract plane wave vector $|(p_x=0,p_y=0,p_z=\omega_0/c_0),s\rangle$\footnote{The meaning of the polarization index $s$ will become clear in Sec. \ref{secc3:heli}.}. And so is this one:
\begin{equation}
	\label{eqc3:p}
	\mathcal{G}(\pp)=\delta(\omega-\omega_0)\delta(p_x)\delta(p_y)\delta(p_z-\omega_0/c_0)\xhat,
\end{equation}
where $\delta()$ is the Dirac delta distribution and $c_0=1/\sqrt{\epsilon_0\mu_0}$ is the speed of light in vacuum, which I will set to 1 from now on by setting $\epsilon_0=\mu_0=1$. Eq. (\ref{eqc3:p}) is (proportional to) the Fourier transform of (\ref{eqc3:r}) ($(\rr,t)\rightarrow(\pp,\omega)$). The same information is contained in both expressions. We say that (\ref{eqc3:r}) belongs to the {\em real space} representation of transverse Maxwell fields and that (\ref{eqc3:p}) belongs to the momentum space representation. Equations (\ref{eqc3:iso}) are written in the coordinate representation. Maxwell's equations in the momentum representation can be found, for example, in \cite[I.B.1.2]{Cohen1997} or \cite[Sec. 7.1]{Jackson1998}. 

With respect to the inner product, when Poincar\'e invariance of the resulting scalar is required, there is only one choice. Its expression in momentum space is \cite{Gross1964}:
\begin{equation}
	\label{eqc3:inner}
	\int \frac{d\pp}{|\pp|} \Ef_1(\pp)^\dagger\Ef_2(\pp) + \Hf_1(\pp)^\dagger\Hf_2(\pp)
\end{equation}
where, for $k=1,2$:
\begin{equation}
	\begin{split}
		\Ef_k(\pp)&=\frac{1}{2\pi}\int d\rr \EE_k(\rr,t=0)\exp(-i\pp\cdot\rr),\\
	 \Hf_k(\pp)&=\frac{1}{2\pi}\int d\rr \HH_k(\rr,t=0)\exp(-i\pp\cdot\rr).\
	\end{split}
\end{equation}
In the real space representation, Eq. (\ref{eqc3:inner}) results in 
\begin{equation}
	\label{eqc3:innerreal}
	\frac{2\pi}{(2\pi)^6}\int d\rr\int d\rr' \frac{\epsilon\EE_1(\rr,0)^\dagger\cdot \EE_2(\rr',0)+\mu\HH_1(\rr,0)\cdot \HH_2(\rr',0)}{|\rr-\rr'|^2}.
\end{equation}

As with vectors, operators also have different guises in different representations. Tab. \ref{tabc3:generators} contains the expressions of some of the generators of transformations in $\mathbb{M}$. The first three are well known, the expression for helicity is less well-known. 

\begin{table}[h!]
\begin{center}\begin{tabular}{lll} \toprule
Generator $G$ & Transformation $\exp(-i\theta G)$ & Expression\\ \midrule
Linear momentum $\mathbold{P}$  & Spatial translations&$-i\nabla$\\
Energy $H$ & Time translations&$i\partial_t$\\
Angular momentum $\JJ$ & Rotations &$-i\rr\times\nabla-i\epsilon_{knm}$\\
Helicity $\Lambda$ & Electromagnetic duality &$\frac{\nabla\times}{k}$ (*) \\ \bottomrule
\end{tabular}\end{center}
\caption[Generators in the coordinate representation.]{\label{tabc3:generators}Expressions of some generators of transformations in the coordinate representation of space time varying fields $\GG(\rr,t)$. $\epsilon_{knm}$ is the totally antisymmetric tensor with $\epsilon_{123}=1$. (*) The expression given here for helicity is only valid for monochromatic fields (see Sec. \ref{secc3:heli}).}
\end{table}

\subsection{An abstract derivation of Maxwell's equations}\label{secc3:abstract}
Equations (\ref{eqc3:iso}), which define $\mathbb{M}$, can be reached using only abstract manipulations starting from three assumptions. I will now go through the derivations for three reasons. One, because it highlights the structure underlying Maxwell's equations, which is independent of the representation. Two, because it is interesting to see the abstract form of well known equations of electromagnetism in the coordinate representation. And three, because in this process both helicity and duality appear in a natural manner.

The three assumptions about the members of $\mathbb{M}$ are:
\begin{enumerate}
	\item That they are massless, i.e, they are eigenstates of the mass squared operator $(H^2-c^2\PP^2)$ with eigenvalue zero.
	\item That they have non-scalar degrees of freedom that can be represented by objects which transform as vectors under rotations\footnote{I am imposing that they transform as the spin 1 representation of the spatial rotations group SO(3). Other objects commonly used in physics transform as different representations of SO(3). For example, Pauli spinors, Dirac spinors and gravitons transform as the spin 1/2, the direct sum of two spin 1/2 and the spin 2 representations, respectively \cite[Chap. 7.6]{Tung1985}.}. I will refer to them as vectorial degrees of freedom. 
	\item That their energy is positive, i.e, the eigenvalues of the energy operator are positive.
\end{enumerate}
The third assumption reflects the fact that, in electromagnetism, the information contained in the positive energies is repeated in the negative energies. As Jackson says ``the sign of the frequency has no physical meaning'' \cite[Chap. 14.5]{Jackson1998}. This is related to the fact that {\em the photon does not have charge and is its own antiparticle} \cite[\S 3.1]{Birula1996}. From now on, and unless specifically mentioned, I will always assume positive energies only. Actually, I will assume energies strictly bigger than zero. The zero case corresponds to electro and magneto statics, which I do not treat in my thesis. 

I will now reach Maxwell's equations from these three assumptions. The helicity operator will appear during the process.

Let me start by using the first assumption. Since we are after a massless object, all the vectors $|\Phi\rangle$ in $\mathbb{M}$ must meet
\begin{equation}
	\label{eqc3:massless}
	(H^2-c^2\PP^2)|\Phi\rangle = 0,
\end{equation}
where $\PP^2=P_x^2+P_y^2+P_z^2$ and $c^2=(\epsilon\mu)^{-1}$. Eq. (\ref{eqc3:massless}) says that the four momentum length squared (or mass squared) of $|\Phi\rangle$ is zero. This means that $H^2|\Phi\rangle=c^2\PP^2|\Phi\rangle$ for all $|\Phi\rangle$. The two operators, $H^2$ and $c^2\PP^2$ are therefore equivalent for members of $\mathbb{M}$. 
Equation (\ref{eqc3:massless}) is the abstract form of the wave equation. Using Tab. \ref{tabc3:generators} I can write (\ref{eqc3:massless}) in the coordinate representation:
\begin{equation}
	\label{eqc3:waveeq}
	\left(i\partial_ti\partial_t-c^2(-i\nabla)(-i\nabla)\right)\GG(\rr,t) = 0\implies
	\left(-\partial^2_{t}+c^2\nabla^2\right)\GG(\rr,t)=0.
\end{equation}
This is the wave equation. The only thing one needs to do to get from (\ref{eqc3:waveeq}) to the Helmholtz\footnote{The common way of getting to (\ref{eqc3:helm}) is through Maxwell's equations for a homogeneous and isotropic medium \cite[Chap. 2.6]{Novotny2006}. From (\ref{eqc3:iso}), assume monochromatic fields, isolate the electric field in one of the curl equations and substitute it in the other, use that $\nabla\times\nabla\times=-\nabla^2+\nabla\nabla$ and it is done. It works for both electric and magnetic fields. In my opinion, the abstract procedure (\ref{eqc3:massless})-(\ref{eqc3:waveeq}) allows to better appreciate the link to the massless character of the field.} equation is use fields of the type $\GG(\rr,t)=\widehat{\GG}(\rr)\exp(-i\omega t)$, take the time derivatives, cancel the time dependence from both sides and substitute $\omega^2=c^2k^2$, where $k^2$ is the eigenvalue of $\PP^2$:
\begin{equation}
	\label{eqc3:helm}
\nabla^2\widehat{\GG}(\rr)+k^2\widehat{\GG}(\rr)=0.
\end{equation}
The positive square root of $k^2$ is called the wavenumber.

A field like $\GG(\rr,t)=\widehat{\GG}(\rr)\exp(-i\omega t)$ is an eigenstate of the energy operator $H\equiv i\partial_t$ with eigenvalue $\omega$:
\begin{equation}
	H\GG(\rr,t)=i\partial_t\widehat{\GG}(\rr)\exp(-i\omega t)=\omega\widehat{\GG}(\rr)\exp(-i\omega t)=\omega\GG(\rr,t),
\end{equation}
The $\exp(-i\omega t)$ time dependence is typically called harmonic time dependence, and the fields $\widehat{\GG}(\rr)\exp(-i\omega t)$ are called monochromatic fields. With the convention of using only positive energies, $\omega$ must be bigger than zero.

To continue with the derivation, note that the operator $(H^2-c^2\PP^2)$ does not necessarily have to act on non-scalar objects. There are scalar waves like the sound that meet (\ref{eqc3:waveeq}). According to the second starting assumption, the objects in $\mathbb{M}$ have vectorial (which are non-scalar) degrees of freedom. I will revisit the difference between scalar and non-scalar degrees of freedom in \ref{secc3:heli}. Due to their massless character, their non-scalar (polarization) degrees of freedom are of the transverse kind \cite[Chap. 10.4.4]{Tung1985}. For the case of vectorial fields, Messiah gives us one way to define longitudinal and transverse by means of abstract operators \cite[Chap. XXI, \S 29]{Messiah1958}. For a transverse\footnote{The transversality condition given by Messiah reads, in the coordinate representation: $\nabla\left(\nabla\cdot\right)=\nabla\times\nabla\times-\nabla^2=0$. Under suitable boundary conditions at infinity, it  is equivalent to the null divergence $\nabla\cdot=0$ condition of Maxwell's equations.} field $\left((\SSS\cdot\PP)^2-\PP^2\right)|\Phi\rangle=0$, while for a longitudinal field $\left(\SSS\cdot\PP\right)^2|\Phi\rangle=0$. Instead of $\SSS\cdot\PP$ I will use $\JJ\cdot\PP$, which is equivalent: $\JJ\cdot\PP=(\rr\times\PP+\SSS)\cdot\PP=\SSS\cdot\PP$, because $\rr\times\PP$ is orthogonal to $\PP$. With this choice, I will be using $\JJ$, the generator of rotations, and avoid using $\SSS$, which, in the relativistic theory is not a proper operator, does not generate meaningful transformations and, in the present context, gains meaning only when projected along $\PP$ \cite[\S 16]{Berestetskii1982}, \cite[p.50]{Cohen1997}.

Since we are looking for objects with transverse polarization: 
\begin{equation}
	\label{eqc3:transverse}
\left((\JJ\cdot\PP)^2-\PP^2\right)|\Phi\rangle=0\implies (\JJ\cdot\PP)^2|\Phi\rangle=\PP^2|\Phi\rangle.
\end{equation}
Having excluded the static case ($\omega=0$), $|\PP|^{-1}$ is not singular\footnote{The expansion of the inverse $O^{-1}$ of a hermitian operator $O=\sum_{\eta}o_\eta|\eta\rangle\langle\eta|$ is $O^{-1}=\sum_{\eta}\frac{1}{o_\eta}|\eta\rangle\langle\eta|$, which exists if $O$ does not have null eigenvalues.} and therefore I can write (\ref{eqc3:transverse}) as
\begin{equation}
\label{eqc3:helsq}
	\left(\Helfrac\right)^2|\Phi\rangle=|\Phi\rangle,
\end{equation}
where the helicity operator $\Lambda=\Hel$ appears explicitly. Eq. \ref{eqc3:transverse} says that its square $\Lambda^2$ is the identity for all the vectors in $\mathbb{M}$: This can be taken as the transversality condition that the members in $\mathbb{M}$ must meet. In the language of group theory, $\Lambda^2$ is a Casimir operator. Since $\Lambda^2=I$, $\Lambda$ has eigenvalues equal to $\pm1$. We can use the eigenvalue of $\Lambda$ to distinguish between two different kinds of members of $\mathbb{M}$:
\begin{equation}
	\label{eqc3:helfrac}
	\Helfrac|\Phi_{\pm}\rangle=\pm|\Phi_{\pm}\rangle.
\end{equation}
Therefore, there is a natural symmetry operation in $\mathbb{M}$. The one generated by $\Lambda$:
\begin{equation}
	\label{eqc3:duality}
	D_\theta|\Phi_{\pm}\rangle=\exp\left(-i\theta\Lambda\right)|\Phi_{\pm}\rangle=\exp(\mp i\theta)|\Phi_{\pm}\rangle.
\end{equation}
I will now use the third assumption to obtain the evolution equations for $|\Phi_{\pm}\rangle$. Generally, $H^2=c^2\PP^2\implies H=\pm c|\PP|$, but if $H$ is to have positive eigenvalues only, it must be that 
\begin{equation}
	\label{eqc3:posen}
	H=c|\PP| 
\end{equation}
Then, (\ref{eqc3:helfrac}) can be written as
\begin{equation}
	\label{eqc3:key}
\pm c(\JJ\cdot\PP)|\Phi_{\pm}\rangle=H|\Phi_{\pm}\rangle.
\end{equation}
Since $H$ is the generator of time translations, (\ref{eqc3:key}) are the time evolution equations for each helicity. 

Let me write this in the coordinate representation using that $\JJ\cdot\PP=\SSS\cdot\PP\equiv \nabla \times$ \cite[XIII.93]{Messiah1958}:
\begin{equation}
	\label{eqc3:magic}
\pm c\nabla\times \GG(\rr,t)=i\partial_t \GG(\rr,t).
\end{equation}
Note that $\GG(\rr,t)$ are implictly restricted to positive frequencies only.

The two equations in (\ref{eqc3:magic}) are equivalent to Maxwell's curl equations. To show this, I will use the Riemann-Silberstein representation of electromagnetic fields \cite{Birula1994,Birula1996,Birula2013}, which is obtained by the transformation ($Z=\sqrt{\mu/\epsilon}$)
\begin{equation}
	\label{eqc3:RS}
	\GG_{\pm}=\frac{1}{\sqrt{2}}\left(\EE\pm iZ\HH\right).
\end{equation}
Then, we can starting from Maxwell's curl equations
\begin{equation}
	\begin{bmatrix}\partial_t&0\\0&\partial_t\end{bmatrix}\begin{bmatrix}\EE\\\HH\end{bmatrix}={\begin{bmatrix}0&\frac{\nabla\times}{\epsilon}\\-\frac{\nabla\times}{\mu}&0\end{bmatrix}}\begin{bmatrix}\EE\\\HH\end{bmatrix},
\end{equation}
and transform them with the change in (\ref{eqc3:RS}):
\begin{equation}
	\begin{split}
		&\RSold\begin{bmatrix}\partial_t&0\\0&\partial_t\end{bmatrix}\RSold^{-1}\RSold\begin{bmatrix}\EE\\\HH\end{bmatrix}=\\&\RSold{\begin{bmatrix}0&\frac{\nabla\times}{\epsilon}\\-\frac{\nabla\times}{\mu}&0\end{bmatrix}}\RSold^{-1}\RSold\begin{bmatrix}\EE\\\HH\end{bmatrix},
	\end{split}
\end{equation}
into
\begin{equation}
	\label{eqc3:maxrs}
	\Hsix\Gsix=c\Curlsix\Gsix,
\end{equation}
which is (\ref{eqc3:RS}). Therefore Eq. (\ref{eqc3:key}) are the Maxwell curl equations.

What is the expression of the transformation $D_\theta|\Phi_{\pm}\rangle=\exp(\mp i\theta)|\Phi_{\pm}\rangle$ in the coordinate representation? In the form of (\ref{eqc3:maxrs}), it is just:
\begin{equation}
	\label{eqc3:D}
	\Gsix_\theta=D_\theta\Gsix\duetoref{eqc3:duality}\begin{bmatrix}I\exp(-i\theta)&0\\0&I\exp(i\theta)\end{bmatrix}\Gsix.
\end{equation}
When expressed with $\EE$ and $\HH$, we find the duality transformation \cite[Chap. 6.11]{Jackson1998}:
\begin{equation}
\begin{bmatrix}\EE\\ Z\HH\end{bmatrix}_\theta=\begin{bmatrix}I\cos\theta&-I\sin\theta\\I\sin\theta&I\cos\theta\end{bmatrix}\begin{bmatrix}\EE\\ Z\HH\end{bmatrix}
\end{equation}.

Let me recapitulate. From requiring them to be vectorial positive energy massless objects I have shown that the vectors in $\mathbb{M}$ are those that meet the transversality condition:
\begin{equation}
\left(\Helfrac\right)^2|\Phi_{\pm}\rangle=|\Phi_{\pm}\rangle,
\end{equation}
that they can be classified according to their helicity eigenvalue
\begin{equation}
	\Helfrac|\Phi_{\pm}\rangle=\pm|\Phi_{\pm}\rangle,
\end{equation}
and that their time evolution equations are 
\begin{equation}
	\label{eqc3:curlrs}
\pm c(\JJ\cdot\PP)|\Phi_{\pm}\rangle=H|\Phi_{\pm}\rangle.
\end{equation}
In the coordinate representation, $|\Phi_{\pm}\rangle$ are the Riemann-Silberstein combinations (\ref{eqc3:RS}), and (\ref{eqc3:curlrs}) are the Maxwell's curl equations. Helicity and duality appear naturally in the derivation.

The block diagonal structure of Eq. (\ref{eqc3:maxrs}) is suggestive of the direct sum of two representations \cite[App. II.2]{Tung1985}. This is indeed the case for the proper Lorentz group which is comprised of spatial rotations and boosts. Since helicity commutes with angular momentum (Tab. \ref{tabc3:gentrans}), rotations do not mix $\GG_+$ and $\GG_-$. Also, the already mentioned fact that a boost does not mix the two helicities of the electromagnetic field can be deduced from the way that the electric and magnetic fields transform. Namely, if the boost is in the $\boldsymbol\beta/|\boldsymbol\beta|$ direction and $\gamma=(1-|\boldsymbol\beta|^2)^{-1/2}$, the fields\footnote{For this discussion I will use $\BB$ instead of $\HH$.} transform as \cite[Eq. 11.149]{Jackson1998}:
\begin{equation}
	\begin{split}
\EE'&\rightarrow \gamma\left(\EE+\boldsymbol\beta\times\BB\right)-\frac{\gamma^2}{\gamma+1}\boldsymbol\beta\left(\boldsymbol\beta\cdot\EE\right),\\
\BB'&\rightarrow \gamma\left(\BB-\boldsymbol\beta\times\EE\right)-\frac{\gamma^2}{\gamma+1}\boldsymbol\beta\left(\boldsymbol\beta\cdot\BB\right).
	\end{split}
\end{equation}
In the $\GG_{\pm}$ basis the rules are hence \cite[Sec. 3.2]{Birula1996}:
\begin{equation}
	\begin{split}
	\GG_{+}'&\rightarrow \gamma\left(\GG_+-i\boldsymbol\beta\times\GG_+\right)-\frac{\gamma^2}{\gamma+1}\boldsymbol\beta\left(\boldsymbol\beta\cdot\GG_+\right),\\
 \GG_{-}'&\rightarrow \gamma\left(\GG_-+i\boldsymbol\beta\times\GG_-\right)-\frac{\gamma^2}{\gamma+1}\boldsymbol\beta\left(\boldsymbol\beta\cdot\GG_-\right),\\
	\end{split}
	\end{equation}
which show that a boost does not mix $\GG_+$ and $\GG_-$.

This discussion relates directly to the transformation properties of the second rank antisymmetric tensor $F^{\mu\nu}$ formed with the components of $\EE$ and $\BB$ \cite[Eq. 11.137]{Jackson1998}. $F^{\mu\nu}$ transforms as the (1,0)$\oplus$(0,1) representation of the proper Lorentz group \cite[Chap. 10.5.1]{Tung1985}, but the fields $\EE$ and $\BB$ cannot be the vectorial objects corresponding to the (1,0) and (0,1) components because $\EE$ and $\BB$ get mixed under boosts, negating the direct sum. Clearly, the two components correspond to $\GG_+$ and $\GG_-$.

A word on the Riemann-Silberstein representation is in order. Prof. Iwo Bialynicki-Birula \footnote{You can find his publications in $http://www.cft.edu.pl/~birula/$.} has used the Riemann-Silberstein formalism to construct a bona fide photon wave function in the coordinate representation \cite{Birula1994,Birula1996}. Together with Prof. Zofia Bialynicka-Birula, they have explored the many uses that this formalism has in classical and quantum electromagnetism \cite{Birula2006,Birula2013}. An equation equivalent to (\ref{eqc3:magic}) can be found in \cite[\S 2.2]{Birula1996}. The discussion about the well defined positive and negative helicities of $\GG_{\pm}$ under the assumption of positive energies can be found in \cite[\S 2.1]{Birula1996}.  

I am going to use
\begin{equation}
\label{eqc3:RS3}
\begin{split}
	\GG_{\pm}&=\frac{1}{\sqrt{2}}\left(\pm\EE+iZ\HH\right),\\
\end{split}
\end{equation}
instead of the combinations in (\ref{eqc3:RS}). There is nothing profound about this change. I make it so that the parity operator exchanges the two helicity states without adding a minus sign. If you use the transformation properties of $\EE$ and $\HH$ under parity\footnote{These transformation properties result from the convention that the electric charge does not change sign under parity. This is the convention used by Jackson that I adopt in my thesis.} \cite[Tab. 6.1]{Jackson1998}:
\begin{equation}
\begin{split}
	\EE\stackrel{\Pi}{\rightarrow} -\EE&, \HH\stackrel{\Pi}{\rightarrow} \HH,\\
\end{split}
\end{equation}
you get that,  for the original combinations in (\ref{eqc3:RS}) 
\begin{equation}
	\GGp\stackrel{\Pi}{\rightarrow} -\GGm, \GGm\stackrel{\Pi}{\rightarrow} -\GGp.
\end{equation}
The combinations in (\ref{eqc3:RS3}) get rid of the minus signs.
\begin{equation}
	\GGp\stackrel{\Pi}{\rightarrow} \GGm, \GGm\stackrel{\Pi}{\rightarrow} \GGp,
\end{equation}.

Now that $\mathbb{M}$ is characterized, it is time to construct basis for it.

\subsection{Construction of basis in $\mathbb{M}$}
\label{secc3:heli}
In the coordinate representation, there is an elegant method for finding monochromatic solutions of the Maxwell equations in a source free, isotropic and homogeneous medium. In \cite[Sec. 13.1]{Morse1953}, \cite[Chap. VII]{Stratton1941}, we learn that, under suitable conditions of the chosen spatial coordinate system, for each orthogonal solution of the scalar Helmholtz equation
\begin{equation}
	\label{eqc3:helmsc}
	\nabla^2\psi_\nu+k^2\psi_\nu=0,
\end{equation}
($\nu$ labels different scalar solutions), we can obtain three orthogonal solutions of the vectorial Helmholtz equation  (\ref{eqc3:helm}): One longitudinal and two transverse. In my thesis, I will only consider the transverse degrees of freedom\footnote{The zero mass condition forbids the longitudinal solution for the free field. It can be shown (\cite[I.B.5]{Cohen1997},\cite[Chap. XXI,\S22]{Messiah1958}) that the longitudinal degrees of freedom of the electromagnetic field can always be seen as belonging to the sources.}.

The two transverse solutions are obtained from $\psi_\nu$ by
\begin{equation}
\label{eqc3:MN}
\Mr=\nabla \times (\mathbf{\hat{w}} \psi_\nu) \text{ and } \Nr=\frac{\nabla \times \Mr}{k},
\end{equation}
where $\mathbf{\hat{w}}$ is a fixed unit vector. Vectorial solutions obtained from different scalar solutions are orthogonal. Since transverse solutions of the Helmholtz equation multiplied by $\exp(-ickt)$ are solutions of Maxwell's curl equations and meet the transversality condition, this method allows to build complete vector bases in $\mathbb{M}$. There are six different coordinate systems for which an orthonormal basis for transverse electromagnetic fields can be built in this way \cite[Sec. 13.1]{Morse1953}. Plane waves, multipoles and Bessel beams result from using cartesian, spherical and cylindrical coordinates with $\what$ equal to $\zhat$, $\rr/|\rr|$ and $\zhat$, respectively \cite[Sec. VII]{Stratton1941}. In those three reference systems, the $\Mreta$ and $\Nreta$ modes are commonly referred to as transverse electric (TE or $s$ waves) and transverse magnetic (TM or $p$ waves) modes \footnote{In the case of the multipoles, there is also another popular naming convention. Electric multipoles and magnetic multipoles correspond to TM and TE modes, respectively \cite[expr. 9.116-9.117]{Jackson1998}.}, respectively. 

The scalar solution $\psi_\nu$ determines the scalar properties shared by the two vectorial solutions. For example, the scalar plane wave $\exp(i(\pp\cdot\rr))$ is a simultaneous eigenstate of the three components of $\PP$: $|p_x,p_y,p_z\rangle$. The extra label $(s/p)$ distinguishes between the two orthogonal transverse vectorial solutions, i.e, it represents a non-scalar property:
\begin{equation}
	\label{eqc3:pw}
	|p_x,p_y,p_z,s/p\rangle.
\end{equation}
In exactly the same way, the multipoles and Bessel beams get their first three identification numbers from the scalar $\psi_\nu$ and the fourth one is produced by (\ref{eqc3:MN}). For the multipoles, the first three numbers can be chosen to be eigenvalues of the energy $H$, the square norm of the angular momentum vector operator $J^2=J_x^2+J_y^2+J_z^2$, whose eigenvalues are $j(j+1)$ for integer $j>0$, and the third component of angular momentum $J_z$, whose eigenvalues I denote by the integer $n$,
\begin{equation}
	\label{eqc3:mp}
	|\omega,j(j+1),n,s/p\rangle.
\end{equation}
For Bessel beams the first three numbers can be chosen to be eigenvalues of $H,P_z$ and $J_z$:
\begin{equation}
	\label{eqc3:bba}
	|\omega,p_z,n,s/p\rangle.
\end{equation}

\subsection{Helicity as a polarization index}\label{secc3:tetmhelhel}
Let me turn my attention to the $\frac{\nabla\times}{k}$ operator at the core of the method in (\ref{eqc3:MN}). For monochromatic fields, this is the helicity $\Lambda$ operator:
\begin{equation}
	\label{eqc3:helnabla}
	\Lambda=\Helfrac\equiv\frac{\nabla\times}{k}.
\end{equation}
This can be seen recalling that $\JJ\cdot\PP\equiv\nabla\times$, and realizing that, for monochromatic fields 

\begin{equation}
	|\PP|^{-1}|\Phi\rangle\duetoref{eqc3:posen}cH^{-1}|\Phi\rangle=\frac{c}{\omega}|\Phi\rangle=\frac{1}{k}|\Phi\rangle.
\end{equation}
I can write $\Mr\duetoref{eqc3:helsq}\Lambda^2\Mr\duetoref{eqc3:MN}\Lambda\Nr$. So
\begin{equation}
	\Lambda\Mr=\Nr,\Lambda\Nr=\Mr.
\end{equation}
Therefore, $\Lambda$ changes TE modes into TM modes and vice versa, {\bf without affecting the scalar degrees of freedom}. This means that helicity can be used to label the polarization degrees of freedom. Since $\Mr$ and $\Nr$ are orthogonal, we have that 
\begin{equation}
	\frac{1}{\sqrt{2}}\left(\Mr\pm\Nr\right)
\end{equation}
are two orthogonal modes of well defined helicity equal to $\pm 1$. Therefore besides using the TE/TM character to describe the polarization of vectors ((\ref{eqc3:pw})-(\ref{eqc3:bba})), we can also use helicity:
\begin{equation}
\label{eqc3:helbasis}
	\begin{split}
		|p_x,p_y,p_z,\pm\rangle&=\frac{1}{\sqrt{2}}\left(|p_x,p_y,p_z,s\rangle \pm |p_x,p_y,p_z,p\rangle\right),\\
		|\omega,j(j+1),n,\pm\rangle&=\frac{1}{\sqrt{2}}\left(|\omega,j(j+1),n,s\rangle \pm |\omega,j(j+1),n,p\rangle\right),\\
		|\omega,p_z,n,\pm\rangle&=\frac{1}{\sqrt{2}}\left(|\omega,p_z,n,s\rangle \pm |\omega,p_z,n,p\rangle\right).
	\end{split}
\end{equation}
In Sec. \ref{secc3:tetmhelhel} I will discuss the fundamental difference between TE/TM and helicity as polarization descriptors.
I will write a plane wave of well defined helicity as $\pwket$. Note that plane waves are also eigenstates of the energy because they are eigenstates of $\PP^2$. For the plane waves $p_x^2+p_y^2+p_z^2=|\pp|^2=k^2=\omega^2/c^2$. 

The coordinate representation expressions for plane waves of well defined helicity can be found in Sec. \ref{secc3:shatphat}, those for Bessel beams of well defined helicity in Sec. \ref{secc3:jz}

\subsection{Eigenvectors of helicity in the plane wave basis}
\label{secc3:shatphat}
In the application chapters, I will use the plane wave basis often. It is worth spending some space on it.

If you take the solutions of the scalar Helmholtz equation (\ref{eqc3:helmsc}) to be $\psi_{p_xp_yp_z}=\exp(ip_xx)\exp(ip_yy)\exp(ip_zz)$ and apply the recipe
\begin{equation}
	\label{eqc3:pwhelm1}
	\widehat{\mathbf{M}}_{p_xp_yp_z}(\mathbf{r})=\nabla\times(\zhat\psi_{p_xp_yp_z}),\ \widehat{\mathbf{N}}_{p_xp_yp_z}(\mathbf{r})=\frac{\nabla\times\widehat{\mathbf{M}}_{p_xp_yp_z}(\mathbf{r})}{k}
\end{equation}
for all $(p_x,p_y,p_z)$, you will obtain all the possible plane waves and have constructed a basis of $\mathbb{M}$. The explicit expressions in the cartesian $[\xhat,\yhat,\zhat]$ basis, are:
\begin{equation}
	\label{eqc3:pwhel0}
	\begin{split}
		\widehat{\mathbf{M}}_{p_xp_yp_z}(\mathbf{r})&  = \frac{i}{p_{\rho}} \left( p_y \xhat - p_x\yhat \right) \plane=\\
		&i\left(\sin\phi\xhat-\cos\phi\yhat\right)\plane=\shat \plane\\
\widehat{\mathbf{N}}_{p_xp_yp_z}(\mathbf{r})& =   \left( \frac{ - p_z \left( p_x \xhat + p_y \yhat \right) + p_\rho^2 \zhat} { k p_\rho} \right) \plane=\\
		&\left(-\cos\theta\cos\phi\xhat-\cos\theta\sin\phi\yhat+\sin\theta\zhat\right)\plane=\phat  \plane ,
\end{split}
\end{equation}
where $p_\rho=\sqrt{p_x^2+p_y^2},\ \theta=\arccos\frac{p_z}{k},\ \phi=\arctan\frac{p_y}{p_x}$. 

As discussed before, sum and subtraction of TE and TM modes result in states of well defined helicity:
\begin{equation}
\label{eqc3:pwhel}
\begin{split}
\pwplus\plane&=\frac{1}{\sqrt{2}}\left(\shat +\phat\right)\planewt,\\
\pwminus\plane&=\frac{1}{\sqrt{2}}\left(\shat-\phat\right)\planewt.
\end{split}
\end{equation}
For example, for a plane wave with momentum pointing to the positive $z$ direction, the two states of definite helicity are 
\begin{equation}
	\begin{split}
		|(0,0,p_z),+\rangle&\equiv -\frac{\xhat+i\yhat}{\sqrt{2}}\exp(i(p_zz-\omega t))=-\lhat\exp(i(p_zz-\omega t),\\
	 |(0,0,p_z),-\rangle&\equiv \frac{\xhat-i\yhat}{\sqrt{2}}\exp(i(p_zz-\omega t))=\rhat\exp(i(p_zz-\omega t)).
	\end{split}
\end{equation}

For the plane waves, the helicity eigenstates coincide with the states of definite polarization handedness in the plane perpendicular to their momentum vector.
I can now introduce an operational definition of a beam with well defined helicity when expressed in the plane wave basis.

Any vector in $\mathbb{M}$ can be expanded in the plane wave basis $\pp=(p_x,p_y,p_z)$:
\begin{equation}
	|\Phi\rangle=\int dp_x\int dp_y\int dp_z\alpha^{p_xp_yp_z}_+ |(p_x,p_y,p_z),+\rangle+\alpha^{p_xp_yp_z}_- |(p_x,p_y,p_z),-\rangle
\end{equation}
To obtain an eigenvalue of helicity $\Lambda|\Phi\rangle=|\Phi\rangle$ or $\Lambda|\Phi\rangle=-|\Phi\rangle$, we need either $\alpha^{p_xp_yp_z}_- =0$ or $\alpha^{p_xp_yp_z}_+ =0$ for all $(p_x,p_y,p_z)$, respectively. Fig. \ref{figc3:helicity} illustrates this.

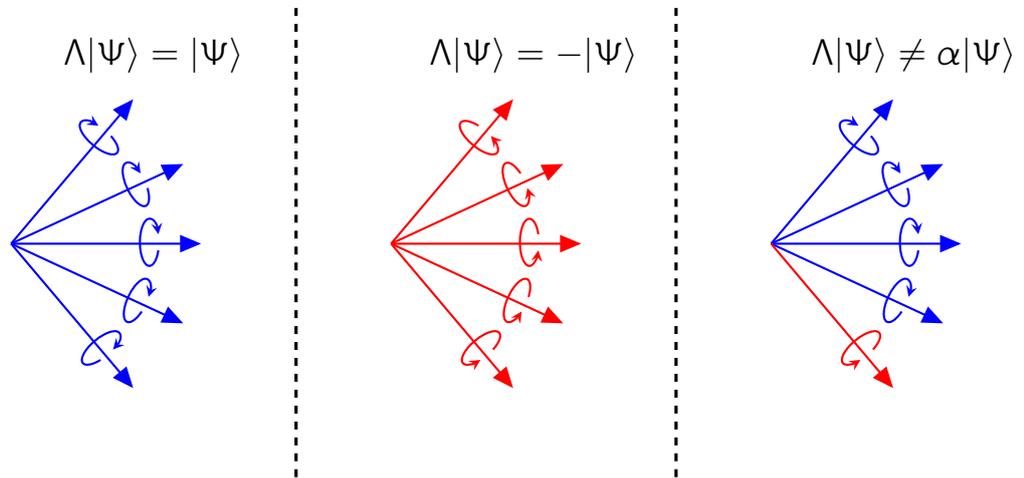
\begin{figure}[h!]
	\begin{center}\begin{tikzpicture}[scale=1.25,every node/.style={scale=1.25},>=latex]
		\makeatletter{}\begin{color}{blue} 
\lc{0}{0}{0}{1}
\lc{0}{0}{25}{1}
\lc{0}{0}{-25}{1}
\lc{0}{0}{50}{1}
\lc{0}{0}{-50}{1}
\end{color}{blue} 
\draw (1.5,2) node {$\Lambda|\Psi\rangle=|\Psi\rangle$};
\draw[very thick,dashed] (3,2.5) -- (3,-2.5);
\begin{color}{red} 
\rc{4}{0}{0}{1}
\rc{4}{0}{25}{1}
\rc{4}{0}{-25}{1}
\rc{4}{0}{50}{1}
\rc{4}{0}{-50}{1}
\end{color}{red} 
\draw (5.5,2) node {$\Lambda|\Psi\rangle=-|\Psi\rangle$};
\draw[very thick,dashed] (7,2.5) -- (7,-2.5);
\begin{color}{blue} 
\lc{8}{0}{0}{1}
\lc{8}{0}{25}{1}
\lc{8}{0}{-25}{1}
\lc{8}{0}{50}{1}
\end{color}{blue} 
\begin{color}{red} 
\rc{8}{0}{-50}{1}
\end{color}
\draw (9.5,2) node {$\Lambda|\Psi\rangle\neq \alpha|\Psi\rangle$};
 
	\end{tikzpicture}\end{center}
\caption[Operational definition of helicity.]{\label{figc3:helicity}A field composed by the superposition of five plane waves has definite helicity equal to one if, with respect to their momentum vectors, all the plane waves are left hand polarized (left part), equal to minus one if they are all right hand polarized (central part), and does not have a definite helicity if all the plane waves do not have the same polarization handedness (right part).}
\end{figure}

Helicity can be seen as polarization handedness in momentum space. It is important not to confuse this interpretation with the polarization in real space which we typically think about in the laboratory. In general, a vector of well defined helicity has the three real space polarization components. For example the left and central panels of Fig. \ref{figc3:helicity}, or a multipole or a Bessel beam of well defined helicity. In a few cases, for example for a single isolated plane wave, the correspondence is simple, but those are only exceptions to the general situation. Also, the property that is connected to the duality symmetry is helicity, not real space polarization. Helicity is a fundamental property in electromagnetism. The same cannot be said about polarization in the real space.

\subsubsection{Transformation properties of plane waves of well defined helicity}
In this subsection, I provide a table with the transformation properties of plane waves of well defined helicity under translations, rotations, parity and time inversion. I will refer to this table often in the application chapters. Except for translations, which are trivial because plane waves are eigenstates of both space and time translations, the phase relationships between the original and the transformed plane wave states must be derived carefully. The phase relations in Tab. \ref{tabc3:pwheltrans} are valid for plane wave vectors in $\mathbb{M}$. Their derivation and more general expressions that are valid for massive fields can be found in \cite{Tung1985}: Chaps. 10, 11 and 12, for rotations, spatial inversion, and time inversion, respectively. 
\begin{table}[h!]
\makebox[\textwidth]{
\begin{tabular}{ll}
Transformation&Action\\\midrule
	Spatial translation&$T_\uhat(\Delta)\pwket=\exp\left(-i\Delta\uhat\cdot\PP\right)\pwket=\exp(-i\Delta\uhat\cdot\pp)\pwket$\\
	   Time translation&$U(\tau)\pwket=\exp\left(-i\tau H\right)\pwket=\exp(-i\tau c |\pp|)\pwket$\\
	Rotation &$R_\uhat(\alpha)\pwket=\exp(-i\lambda \psi)|R_\uhat(\alpha) \pp\ \lambda\rangle$\\
	Parity&$\Pi\pwket=|-\pp\ -\lambda\rangle$\\
	Time inversion&$Tr\pwket=-|-\pp\ \lambda\rangle$\\
  \\
\end{tabular}}

\caption[Transformation properties of plane wave vectors in $\mathbb{M}$.]{\label{tabc3:pwheltrans} Transformation properties of plane wave vectors in $\mathbb{M}$. $\uhat$ is a general unit vector. $|R_\uhat(\alpha)\pp\ \lambda\rangle$ is a plane wave with momentum equal to the vector that results from rotating the original momentum vector $\pp$ with the matrix $R_\uhat(\alpha)$. The phase $\exp(-i\lambda \psi)$ acquired in a rotation depends on $\pp$, $\alpha$ and $\uhat$ \cite[expr. 9.7-12,10.4-25]{Tung1985}. I will mostly use the case: $R_z(\phi)R_y(\theta)|k\zhat \ \lambda\rangle$, for which $\psi=0$.}
\end{table}

\section{Polarization and $\pm \EE + iZ\HH$}
Helicity can be used to describe the polarization of the field. The connection between polarization and the vectors of well defined helicity $|\Phi_{\pm}\rangle\equiv\pm \EE +iZ\HH$ has important practical implications. Assuming positive energies and that the electric field is a circularly polarized plane wave, Maxwell's equations can be used to compute the magnetic field and then compute $\pm \EE + iZ\HH$. One of the two components, the one of helicity opposite to the helicity of the initial plane wave, is zero. The other one is non-zero and is still a circularly polarized plane wave with the initial handedness. This happens independently of the momentum of the plane wave. Therefore, not only plane waves, but {\bf any vector of well defined helicity has a zero component in one of the two $\GG_{\pm}$ components of the coordinate representation}. On the other hand, starting with a linearly polarized plane wave as the electric field, produces components in both $\GG_{\pm}$ and both of them will be circularly polarized plane waves with opposite polarization handedness.

The separation achieved by $\pm \EE + iZ\HH$ is quite general and can be put to practical use, for example in figuring out which dipolar sources emit radiation of well defined helicity (\ref{secc3:dipolar}). Also, it is useful in the analysis of the helicity of any field. For example, say that you use COMSOL or a similar program to compute the near field of a complicated structure. COMSOL will produce the electric and magnetic complex fields in the region that you specify. Now, what can you do if you want to analyze the helicity content of such (possibly very) complicated field?: Using $\pm \EE + iZ\HH$ is likely going to be the best option. Doing a Fourier transform of the obtained fields to get an expansion in plane waves of well defined helicity is certainly much more complicated. My colleague Nora Tischler exploited the fundamental helicity separation in $\pm \EE + iZ\HH$ to figure out the helicity decomposition of a complex near field. I will reproduce the figures that she obtained in section \ref{subsecc3:shape}.

\section{TE / TM and Helicity}\label{secc3:tetmhel}
It looks like we have two choices for representing the polarization of the field or, according to previous discussions, the {\em degree of freedom that scalar fields do not have}: TE/TM or helicity ($\pm$). 

A given operator in $\mathbb{M}$ can be expanded in the two different polarization bases. Denoting now by $\eta$ the other three numbers and by $\uparrow/\downarrow$ the TE/TM character we can write (\ref{eqc3:S}) either:
\begin{equation}
	\makebox[\textwidth][c]{$
S=\int\int d\bar{\eta}d\eta\left(
\alpha_\eta^{\bar{\eta}}|\bar{\eta}\ \uparrow\rangle\langle \uparrow \ \eta|+
\beta_\eta^{\bar{\eta}}|\bar{\eta}\ \uparrow\rangle\langle \downarrow \ \eta|+
\chi_\eta^{\bar{\eta}}|\bar{\eta}\ \downarrow\rangle\langle \uparrow \ \eta|+
\gamma_\eta^{\bar{\eta}}|\bar{\eta}\ \downarrow\rangle\langle \downarrow \ \eta|
\right)$},
\end{equation}

or, using helicity $\pm$
\begin{equation}
	\makebox[\textwidth][c]{$
S=\int\int d\bar{\eta}d\eta\left(
a_\eta^{\bar{\eta}}|\bar{\eta}\ +\rangle\langle  +\ \eta|+
b_\eta^{\bar{\eta}}|\bar{\eta}\ +\rangle\langle  -\ \eta|+
c_\eta^{\bar{\eta}}|\bar{\eta}\ -\rangle\langle  +\ \eta|+
d_\eta^{\bar{\eta}}|\bar{\eta}\ -\rangle\langle  -\ \eta|
\right).$}
\end{equation}

For each pair $(\bar{\eta},\eta)$ there is a 2$\times$2 matrix which contains the information about polarization changes:
\begin{equation}
	S_\eta^{\bar{\eta}}(\pm)=
	\begin{bmatrix}a_\eta^{\bar{\eta}}&b_\eta^{\bar{\eta}}\\c_\eta^{\bar{\eta}}&d_\eta^{\bar{\eta}}\end{bmatrix},\
	S_\eta^{\bar{\eta}}(\uparrow\downarrow)=\begin{bmatrix}\alpha_\eta^{\bar{\eta}}&\beta_\eta^{\bar{\eta}}\\\chi_\eta^{\bar{\eta}}&\gamma_\eta^{\bar{\eta}}\end{bmatrix}.\
\end{equation}

The relationship between the two bases is easy enough (\ref{eqc3:helbasis}) and the matrix to change between them is
\begin{equation}
	\label{eqc3:tetmhel}
	\frac{1}{\sqrt{2}}\begin{bmatrix} 1&1\\1&-1\end{bmatrix}.
\end{equation}

\subsection{Helicity preservation in the TE/TM basis}
\label{secc3:helprestetm}
Many systems have been traditionally analyzed in the TE/TM basis. Examples of those are: The Fresnel formulas for planar multilayer systems, the Mie scattering coefficients for spheres and the analysis of propagating modes of waveguides. The question that I now address is, what are the conditions for helicity preservation when the scattering operator $S$ is expressed in the TE/TM basis? 

For a helicity preserving $S$, all of the $S_\eta^{\bar{\eta}}(\pm)$ are diagonal. In the $\uparrow/\downarrow$ basis, this means that:
\begin{equation}
\label{eqc3:prestetm}
	\begin{split}
		&\frac{1}{\sqrt{2}}\begin{bmatrix} 1&1\\1&-1\end{bmatrix}^{-1}\begin{bmatrix}a_\eta^{\bar{\eta}}&0\\0&d_\eta^{\bar{\eta}}\end{bmatrix}\frac{1}{\sqrt{2}}\begin{bmatrix} 1&1\\1&-1\end{bmatrix}=\\
		&S_\eta^{\bar{\eta}}(\uparrow\downarrow)=\frac{1}{2}\begin{bmatrix}a_\eta^{\bar{\eta}}+d_\eta^{\bar{\eta}}&a_\eta^{\bar{\eta}}-d_\eta^{\bar{\eta}}\\a_\eta^{\bar{\eta}}-d_\eta^{\bar{\eta}}&a_\eta^{\bar{\eta}}+d_\eta^{\bar{\eta}}\end{bmatrix}=\begin{bmatrix}\alpha_\eta^{\bar{\eta}}&\beta_\eta^{\bar{\eta}}\\\beta_\eta^{\bar{\eta}}&\alpha_\eta^{\bar{\eta}}\end{bmatrix}
	\end{split}
\end{equation}
So, helicity is preserved when the scatterer ``treats'' the TE and TM polarizations equally: Maintains them and flips them with the same coefficients. This result will become handy when analyzing the effect that a microscope objective has on helicity (Chap. \ref{chap4}). 

\subsection{The symmetries connected to the TE/TM character}\label{subsecc3:tetmsym}
Duality symmetry implies the preservation of helicity. What is the TE/TM number preserved by? In other words, what is the symmetry linked to TE/TM? I now explore this question. I will now show that the TE/TM number is related to spatial inversion symmetries but cannot be linked to a unique symmetry transformation. This is an important difference between TE/TM and helicity.

To start, let me consider only the polarization degree of freedom. The TE/TM vectors are linear combinations of the eigenvectors of helicity:
\begin{equation}
	|\uparrow\rangle =\frac{1}{\sqrt{2}}\left(|+\rangle+| -\rangle\right),\ |\downarrow\rangle =\frac{1}{\sqrt{2}}\left(|+\rangle-| -\rangle\right).
\end{equation}
Since parity flips helicity (see Tab. \ref{tabc3:gentrans}), $\Pi|\pm\rangle=|\mp\rangle$, we would have that the $|\uparrow\downarrow\rangle$ are the eigenstates of parity
\begin{equation}
	\Pi|\uparrow\rangle=|\uparrow\rangle,\ \Pi|\downarrow\rangle=-|\downarrow\rangle,
\end{equation}
and that parity is the symmetry linked to TE/TM. At this point, though, I have not considered the scalar degrees of freedom. When they are included, the story is not so simple anymore.

Obviously, we cannot ignore the action of $\Pi$ in the other three numbers describing the scalar properties. For example, parity changes momentum: $|\pp\ (s/p)\rangle$ are not eigenstates of parity. On the other hand, the momentum of a plane wave is invariant under any mirror reflection across a plane containing it. Hence, the TE/TM plane waves are eigenstates of one such mirror operation $M_\uhat=\Pi R_\uhat(\pi)$
\begin{equation}
	\begin{split}
		M_\uhat |\pp \uparrow\rangle&=M_\uhat\frac{1}{\sqrt{2}} \left(|\pp \ +\rangle + |\pp \ -\rangle\right)=|\pp \uparrow\rangle,\\
M_\uhat |\pp \downarrow\rangle&=M_\uhat \frac{1}{\sqrt{2}}\left(|\pp \ +\rangle - |\pp \ -\rangle\right)=-|\pp \downarrow\rangle.
	\end{split}
\end{equation}
where $\uhat\cdot\pp=0$ and the facts that helicity is invariant under rotations and flips under parity have been used. The planar multilayer problem is diagonal in the TE/TM plane wave. Here is the symmetry reason for it: Take any plane wave as the input of the scattering problem of a planar multilayer system. There is always a plane that contains the momentum of the input plane wave and is perpendicular to the multilayer. Reflection across this plane is a symmetry of the multilayer, and the TE/TM plane wave is an eigenstate of this reflection. Besides the TE/TM character of the polarization, if $\zhat$ is the stacking direction, the system must also preserve $(p_x,p_y)$ because of transverse translational symmetry. This is exactly what the Fresnel formulas reflect \cite[Chap. 2.8.1]{Novotny2006}.

In the multipole basis, the three indexes referring to scalar properties are invariant under parity because $H$ and $\JJ$ commute with $\Pi$. This shows why the TE/TM multipoles are eigenstates of parity and the Mie problem is completely diagonal in this basis. All four numbers of a TE/TM multipole are preserved upon scattering off a sphere.

In conclusion, the TE/TM description of polarization is related to spatial inversion symmetries but, contrary to the helicity case, cannot be tied to a unique transformation. 

Actually, the transformation built from the TE/TM eigenstates 
\begin{equation}
	\label{eqc3:tetm}
	|\uparrow\rangle\langle\uparrow|-|\downarrow\rangle\langle\downarrow|,
\end{equation}
is not even a symmetry of Maxwell's equations (\ref{eqc3:iso}). The action of (\ref{eqc3:tetm}) is to flip helicity without doing anything else. To see that Maxwell's equations are not invariant under (\ref{eqc3:tetm}), we only need to transform (\ref{eqc3:maxrs}) with the matrix 
\begin{equation}
	\begin{bmatrix}0&I\\I&0\end{bmatrix},
\end{equation}
to see that the form of the equations changes (the signs of the two curl operators change).

One thing is always true, though: sums and subtraction of modes with well defined helicities are the relevant constructions for the analysis of systems with any spatial inversion symmetry; be it parity, mirror reflections or improper axis of rotation.

Therefore, there are important differences between TE/TM and helicity. The TE/TM character does not, by itself, define a symmetry in electromagnetism. To obtain a valid symmetry, the transformation derived from the TE/TM character must be extended to include changes in the scalar degrees of freedom. There is no unique extension, although the different extensions are always some form of spatial inversion. On the other hand, helicity generates the electromagnetic duality symmetry, which acts only on the non-scalar degree of freedom, and the TE/TM eigenstates are produced as sum and subtraction of two modes which differ only in their helicity eigenvalue. In light of this, and of the contents of Sec. \ref{secc3:abstract}, one may say that helicity and duality are the fundamental operator and symmetry transformation of the electromagnetic non-scalar degree of freedom, i.e, the polarization.

\section{General vectors with well defined $J_z$}\label{secc3:jz}

Bessel beams with well defined helicity will have a prominent role in one of the application chapters. In this section, I will derive them by first constructing the most general vector with well defined third component of angular momentum $J_z$, and then particularizing it to obtain Bessel beams. At the end of the section, I will talk about the special case of a single plane wave, and how in this case, helicity and angular momentum are correlated, contrary to the general situation.

I start with a way to build the plane wave basis of well defined helicity alternative to the way in (\ref{eqc3:pwhelm1})-(\ref{eqc3:pwhel}). Any plane wave of momentum $\pp$ and helicity $\lambda$ can be obtained from $|(0,0,|\pp|),\lambda\rangle$ by rotation (\cite[expr. 9.7-12]{Tung1985}, Tab. \ref{tabc3:pwheltrans} and \cite{Gabi2008}):
\begin{equation}
\label{eqc3:rotpw}
	\pwket=R_z(\phi)R_y(\theta)|(0,0,|\pp|=k),\lambda\rangle,
\end{equation}
Upon rotation, the momentum of the initial plane wave changes: Ends up pointing towards that $(\theta,\phi)$ direction. The helicity does not change. Momentum does not commute with rotations, but helicity does (Tab. \ref{tabc3:pwheltrans}). Additionally, these particular rotations (\ref{eqc3:rotpw}) do not imprint any phase term to the resulting plane wave. Therefore, the whole plane wave basis can be built with suitable rotations of $|(0,0,|\pp|=k),\lambda\rangle$ states for $k\in(0,\infty)$.

Consider now the following construction:
\begin{equation}	
\begin{split}
	\label{eqc3:m}
	&|\Phi_n\rangle=\Longint\exp(in\phi)R_z(\phi)R_y(\theta)\\	
	&\left(c_+^{k\theta}|(0,0,k),+\rangle+c_-^{k\theta}|(0,0,k),-\rangle\right).
	\end{split}
\end{equation}

The beam in (\ref{eqc3:m}) is an eigenstate of $J_z$ with eigenvalue $n$. This can be seen by applying a rotation $R_z(\beta)|\Phi_n\rangle$, and verifying that the state transforms into itself times a phase factor $\exp(-in\beta)$ with the following steps: Using that a rotation is a linear operator, that two successive rotations along the same axis are equivalent to a single rotation by the sum of the two angles and the change of the $\phi$ integration variable $\phi\rightarrow \bar{\phi}-\beta$. The important action is in red:
\begin{equation}
\makebox[\textwidth][c]{$
	\label{eqc3:jznolambda}
	\begin{split}
&R_z(\beta)|\Phi_n\rangle=\\	
 &\Longint\exp(in\phi){\color{red}R_z(\beta)}R_z(\phi)R_y(\theta)\left(c_+^{k\theta}|k\zhat,+\rangle+c_-^{k\theta}|k\zhat,-\rangle\right)=\\
 &\Longintchange\exp(in\phi){\color{red}R_z(\beta+\phi)}R_y(\theta)\ \left(c_+^{k\theta}|k\zhat,+\rangle+c_-^{k\theta}|k\zhat,-\rangle\right)=\\
 &\Longintchange\exp(in({\color{red}\bar{\phi}-\beta})){\color{red}R_z(\bar{\phi})}R_y(\theta)\ \left(c_+^{k\theta}|k\zhat,+\rangle+c_-^{k\theta}|k\zhat,-\rangle\right)=\\
 &{\color{red}\exp(-in\beta)}\Longintchange\exp(in\bar{\phi})R_z(\bar{\phi})R_y(\theta)\ \left(c_+^{k\theta}|k\zhat,+\rangle+c_-^{k\theta}|k\zhat,-\rangle\right)\\
 &=\exp(-in\beta)|\Phi_n\rangle
	\end{split}
$}
\end{equation}
To take the last step, I have used the fact that the shift in the integration interval due to the change $\phi\rightarrow \bar{\phi}-\beta$ is irrelevant because the argument inside the integral is $2\pi$-periodic in the integration variable. 

Equation (\ref{eqc3:m}) can be extended to include evanescent modes without affecting the validity of the derivation in (\ref{eqc3:jznolambda}). This is worth explaining in some detail. Evanescent electromagnetic modes exist at the interfaces of different media and exhibit a rapid decay of intensity in the direction perpendicular to the interface. For example, if the eigenvalue of $P_z$ is an imaginary number with positive imaginary part, the $\exp(i p_z z)$ spatial dependence results in exponentially decaying intensity for increasing $z$. In order to obtain an imaginary $p_z$ we need the norm of the transverse momentum to be larger than the wave number: $p_x^2+p_y^2=p_\rho^2>k^2$. Then $p_z/k=+\sqrt{1-p_\rho^2/k^2}$ is an imaginary number while $k^2=p_\rho^2+p_z^2$ is still met. Since $\sin\theta=p_\rho/k$, imaginary $p_z$'s can be achieved in (\ref{eqc3:m}) by changing the integral in $\theta$ to one in $\frac{p_\rho}{k}$ with an infinite upper limit
\begin{equation}
	\int_0^\pi \sin\theta\ d\theta=\int_0^1\frac{p_\rho}{\sqrt{1-p_\rho^2}} \ \frac{dp_\rho}{k}   \rightarrow \int_0^\infty \frac{p_\rho}{\sqrt{1-p_\rho^2}} \ \frac{dp_\rho}{k} ,
\end{equation}
and the rotation $R_y(\theta)$ to $R_y(\arcsin\frac{p_\rho}{k})$. The evanescent components will be those with $\frac{p_\rho}{k}>1$. It is easy to verify that when $\frac{p_\rho}{k}>1$ the first rotation $R_y(\arcsin\frac{p_\rho}{k})$ changes $p\zhat$ into exactly the desired momentum vector $\pp=[p_\rho,0,i\sqrt{p_\rho^2-|\pp|^2}]$.

When the evanescent modes are included, and since the coefficients $c_{\pm}^{k\theta}$ are arbitrary, the construction in (\ref{eqc3:m}) is the most general beam with well defined $J_z$. I will use this construction in Chap. \ref{chap4} to discuss the relationship between angular momentum and polarization. 

\subsection{Bessel beams of well defined helicity}\label{secc3:bbeamshel}
Bessel beams of well defined helicity will play a central role in Chap. \ref{chap4}. The following constructive procedure allows us to appreciate their well defined quantities.

By setting
\begin{equation}
\label{eqc3:bb}
c_\lambda^{k\theta}=\delta(k-k_0)\delta(\theta-\theta_0), c_{-\lambda}^{k\theta}=0,
\end{equation}
in (\ref{eqc3:m}), we obtain modes of well defined helicity $\lambda$, energy $ck_0$, third component of linear momentum $p_z=k\cos\theta_0$ and third component of angular momentum equal to $n$. These are Bessel beams of well defined helicity \cite{Hacyan2006,Matula2013}.

\begin{figure}[h!]
	\begin{center}
\makeatletter{}\def\ASYprefix{}
\newbox\ASYbox
\newdimen\ASYdimen
\long\def\ASYbase#1#2{\leavevmode\setbox\ASYbox=\hbox{#1}\ASYdimen=\ht\ASYbox\setbox\ASYbox=\hbox{#2}\lower\ASYdimen\box\ASYbox}
\long\def\ASYaligned(#1,#2)(#3,#4)#5#6#7{\leavevmode\setbox\ASYbox=\hbox{#7}\setbox\ASYbox\hbox{\ASYdimen=\ht\ASYbox\advance\ASYdimen by\dp\ASYbox\kern#3\wd\ASYbox\raise#4\ASYdimen\box\ASYbox}\put(#1,#2){#5\wd\ASYbox 0pt\dp\ASYbox 0pt\ht\ASYbox 0pt\box\ASYbox#6}}\long\def\ASYalignT(#1,#2)(#3,#4)#5#6{\ASYaligned(#1,#2)(#3,#4){
}{
}{#6}}
\long\def\ASYalign(#1,#2)(#3,#4)#5{\ASYaligned(#1,#2)(#3,#4){}{}{#5}}
\def\ASYraw#1{
currentpoint currentpoint translate matrix currentmatrix
100 12 div -100 12 div scale
#1
setmatrix neg exch neg exch translate}
 
\makeatletter{}\setlength{\unitlength}{1pt}
\makeatletter\let\ASYencoding\f@encoding\let\ASYfamily\f@family\let\ASYseries\f@series\let\ASYshape\f@shape\makeatother{\catcode`"=12\includegraphics{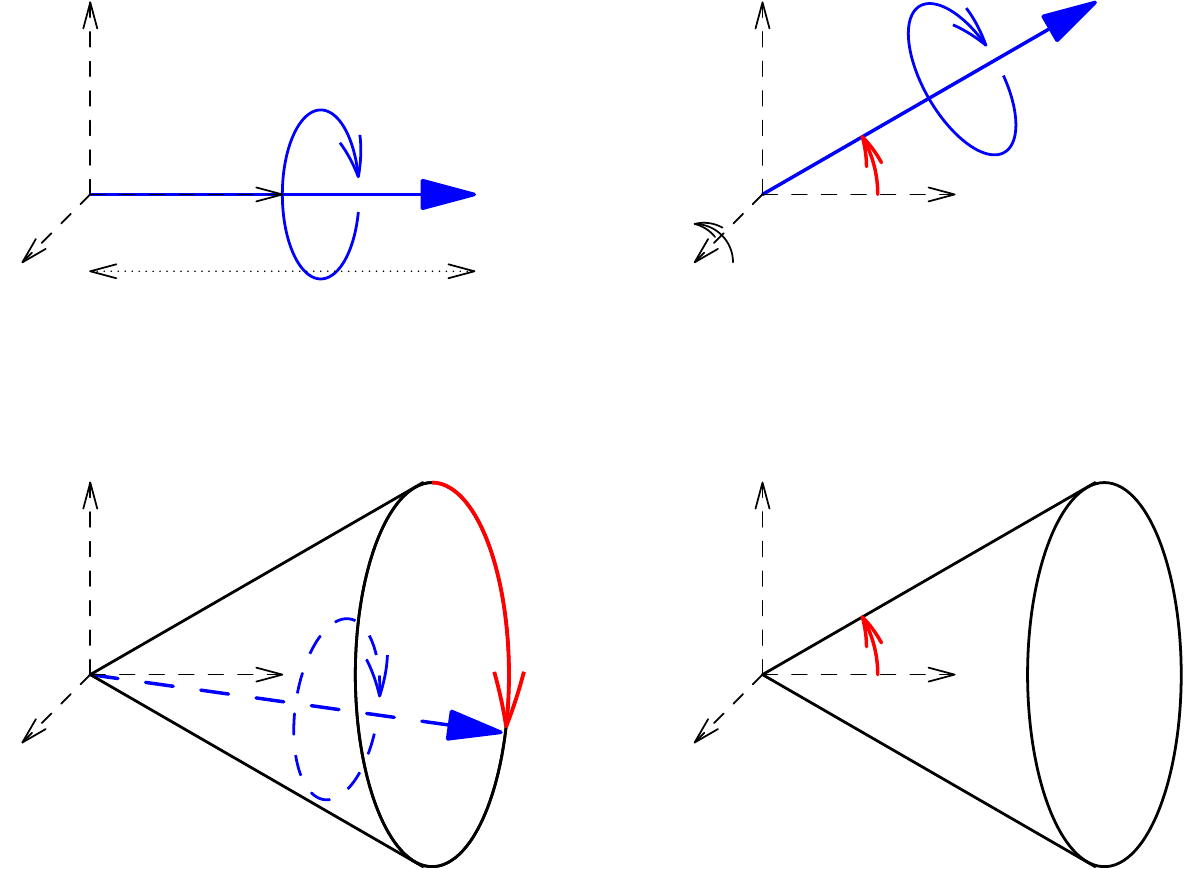}}\definecolor{ASYcolor}{gray}{0.000000}\color{ASYcolor}
\fontsize{12.000000}{14.400000}\selectfont
\usefont{\ASYencoding}{\ASYfamily}{\ASYseries}{\ASYshape}\ASYalign(-260.226526,168.899349)(-0.500000,-0.750000){$|\mathbf{p}|=k_0$}
\definecolor{ASYcolor}{gray}{0.000000}\color{ASYcolor}
\fontsize{12.000000}{14.400000}\selectfont
\ASYalignT(-312.150056,222.470526)(0.000000,-0.281266){0.800000 -0.000000 -0.000000 0.800000}{{\Large $\hat{p}_x$}}
\definecolor{ASYcolor}{gray}{0.000000}\color{ASYcolor}
\fontsize{12.000000}{14.400000}\selectfont
\ASYalignT(-287.988291,191.108761)(-0.500000,-0.781266){0.800000 -0.000000 -0.000000 0.800000}{{\Large $\hat{p}_z$}}
\definecolor{ASYcolor}{gray}{0.000000}\color{ASYcolor}
\fontsize{12.000000}{14.400000}\selectfont
\ASYalignT(-328.110907,187.439079)(-1.000000,0.290983){0.800000 -0.000000 -0.000000 0.800000}{{\Large $\hat{p}_y$}}
\definecolor{ASYcolor}{gray}{0.000000}\color{ASYcolor}
\fontsize{12.000000}{14.400000}\selectfont
\ASYalign(-293.540644,239.127585)(-0.500000,-0.250000){1)}
\definecolor{ASYcolor}{rgb}{1.000000,0.000000,0.000000}\color{ASYcolor}
\fontsize{12.000000}{14.400000}\selectfont
\ASYalign(-85.752992,204.263374)(0.000000,-0.188724){$\theta_0$}
\definecolor{ASYcolor}{gray}{0.000000}\color{ASYcolor}
\fontsize{12.000000}{14.400000}\selectfont
\ASYalign(-138.074761,239.127585)(-0.500000,-0.250000){2)}
\definecolor{ASYcolor}{gray}{0.000000}\color{ASYcolor}
\fontsize{12.000000}{14.400000}\selectfont
\ASYalign(-129.943528,171.478229)(-0.500000,-0.728418){$R_y(\theta_0)$}
\definecolor{ASYcolor}{rgb}{1.000000,0.000000,0.000000}\color{ASYcolor}
\fontsize{12.000000}{14.400000}\selectfont
\ASYalign(-200.482665,79.513538)(-1.000000,-0.378608){$\phi$}
\definecolor{ASYcolor}{gray}{0.000000}\color{ASYcolor}
\fontsize{12.000000}{14.400000}\selectfont
\ASYalign(-293.540644,100.318760)(-0.500000,-0.250000){3) $\exp(in\phi)R_z(\phi)$}
\definecolor{ASYcolor}{rgb}{1.000000,0.000000,0.000000}\color{ASYcolor}
\fontsize{12.000000}{14.400000}\selectfont
\ASYalign(-85.752992,65.454549)(0.000000,-0.188724){$\theta_0$}
\definecolor{ASYcolor}{gray}{0.000000}\color{ASYcolor}
\fontsize{12.000000}{14.400000}\selectfont
\ASYalign(-138.074761,105.871113)(-0.500000,-0.250000){4)}
 
\caption[Construction of Bessel beams of well defined helicity.]{\label{figc3:bbeam} 1) Start by taking a plane wave with well defined helicity and momentum $k_0\zhat$. 2) Rotate it by $\theta_0$ along the y axis. 3) Rotate it by $\phi$ along the $z$ axis and weigh it by $\exp(in\phi)$. Repeat this process for all $\phi$ until the whole cone is filled 4). All the plane waves have the same helicity and momentum length, so helicity and energy are well defined. All of them have the same projection onto $\hat{p}_z$: $p_z=k\cos\theta$, which is hence also well defined. The text explains in detail why $J_z$ is well defined and equal to $n$.}
	\end{center}
\end{figure}
The picture of a Bessel beam as a cone in the momentum representation is a nice geometrical image. Fig. \ref{figc3:bbeam} illustrates its construction process. In Chap. \ref{chap4} I am going to need to use them in the coordinate representation. After inserting the specified weights (\ref{eqc3:bb}) into equation (\ref{eqc3:m}), substituting the rotated plane waves by their explicit expressions (\ref{eqc3:pwhel}), changing basis from $[\xhat,\yhat,\zhat]$ to $[\rhat,\lhat,\zhat]$ and using 
\begin{equation}
\plane = \sum_m i^m J_m(\krho \rho ) \exp(i m (\arctan\frac{p_y}{p_x} - \phi)) \exp(i\kz z)
\end{equation}
before performing the integral in $d\phi$, one can finally obtain the real space expressions in cylindrical coordinates of the Bessel beams with well defined helicity ($\rho=\sqrt{x^2+y^2},\phi=\arctan\frac{y}{x},z$):
\begin{equation}
	\makebox[\textwidth][c]{$
\begin{split}
\label{eqc3:CD}
&|k,p_z,n,-\rangle\equiv\Cnkrho(\rho,\phi,z)=\sqrt{\frac{\krho}{2\pi}}i^n\exp(i(\kz z+n\phi))\times\\
&\left(\frac{i}{\sqrt{2}}\left((1+\frac{\kz}{k})J_{n+1}(\krho\rho)\exp(i\phi)\rhat+(1-\frac{\kz}{k})J_{n-1}(\krho\rho)\exp(-i\phi)\lhat \right)-\frac{\krho}{k}J_n(\krho\rho)\zhat\right),\\
&|k,p_z,n,+\rangle\equiv\Dnkrho(\rho,\phi,z)=\sqrt{\frac{\krho}{2\pi}}i^n\exp(i(\kz z+n\phi))\times\\
&\left(\frac{i}{\sqrt{2}}\left((1-\frac{\kz}{k})J_{n+1}(\krho\rho)\exp(i\phi)\rhat+(1+\frac{\kz}{k})J_{n-1}(\krho\rho)\exp(-i\phi)\lhat \right)+\frac{\krho}{k}J_n(\krho\rho)\zhat\right).\\
\end{split}
$}
\end{equation}

\subsection{Helicity and angular momentum for plane waves}\label{secc3:lambdajzpw}
The derivation (\ref{eqc3:jznolambda}) that shows that the vector in (\ref{eqc3:m}) has an angular momentum equal to $n$ is independent of the weights $c_{\pm}^{k\theta}$. In particular, it is independent of helicity. In general, helicity and angular momentum are two decoupled properties of the field. This decoupling can be appreciated by the impossibility of associating a definite angular momentum to the eigenstates of helicity depicted in the left and central parts of Fig \ref{figc3:helicity}. It is also reflected in the fact that beams of well defined $J_z$, like Bessel beams or multipoles, can be produced as arbitrary linear combinations of two modes of well defined helicity.

There is one important case where helicity and angular momentum are related: A plane wave $\pwket$. The definition of $\pwket$ as helicity eigenstates implies 
\begin{equation}
	\label{eqc3:lambda}
	\Lambda\pwket=\lambda\pwket.
\end{equation}
Let me now expand the helicity operator in (\ref{eqc3:lambda}):
\begin{equation}
	\label{eqc3:jz}
	\begin{split}
	&\Lambda\pwket=\Helfrac\pwket=\frac{\sum_{i=1}^3 J_iP_i}{|\PP|}\pwket=\\
	&\frac{J_xp_x+J_yp_y+J_zp_z}{|\pp|}\pwket=J_{\pphat}\pwket,\\
	\end{split}
\end{equation}
The second equality follows by applying the $P_i$ operators to the plane wave and obtaining the corresponding $p_i$ eigenvalues. Same with $|\PP|^{-1}$. From (\ref{eqc3:lambda}) and (\ref{eqc3:jz}), it follows that the plane wave helicity eigenstates $\pwket$ are also angular momentum ($J_{\pphat}$) eigenstates. This relationship connects helicity with the rotational properties of $\pwket$, but only along the $\pphat$ axis. 

This can also be seen in (\ref{eqc3:m}) for the case of $|(0,0,k), \lambda\rangle$. After setting $d\theta c^{k\theta}_{\pm}(\omega,\theta)=d\theta b_{\pm}^{k 0}\delta(\theta-0)$ and using the facts that $R_y(0)=I$ and $R_z(\phi)|(0,0,k),\pm\rangle=\exp(\mp\phi)|(0,0,k),\pm\rangle$, we see that the integral on $\phi$ only gives a non-zero contribution for $n=+1$ or $n=-1$, and that $b_{+}^{k 0}$ is the only remaining term when $n=1$ while $b_{-}^{k 0}$ is the only remaining term when $n=-1$. The angular momentum along the axis of the plane wave does determine its helicity, and vice versa.

For the case of $|\shortminus\pp\ \lambda\rangle$, a derivation parallel to (\ref{eqc3:jz}) leads to
\begin{equation}
\label{eqc3:-jz}
J_{\pphat}|\shortminus\pp\ \lambda\rangle=-\lambda|\shortminus\pp\ \lambda\rangle.
\end{equation}

Using (\ref{eqc3:jz}) and (\ref{eqc3:-jz}), it is easy to verify that the linear combinations 
\begin{equation}
	|\pp \ \pm\lambda\rangle + |\shortminus\pp\ \mp\lambda\rangle
\end{equation}
are also eigenstates of $J_{\pphat}$ with eigenvalue $n=\pm\lambda$.

\section{Scattering as an operator in $\mathbb{M}$}\label{secc3:scattop}
Consider the scattering situation in Fig. \ref{figc3:scatt2}, and assume harmonic time dependence $\exp(-i\omega t)$ for the fields.
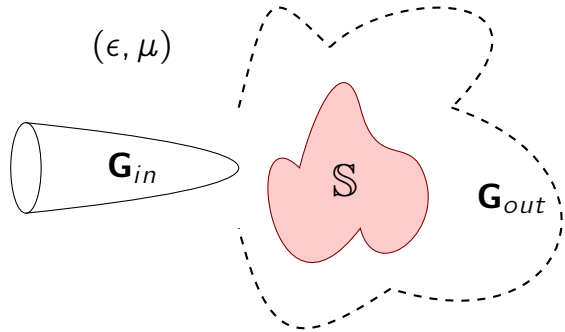
\begin{figure}[h]
	\begin{center}\begin{tikzpicture}[scale=0.8,every node/.style={scale=1.2},>=latex]
		\makeatletter{}\draw (0,0) ellipse (0.25 and 0.75);
\draw[rotate=-90](-0.75,0) parabola bend (0,3.5) (0.75,0);
\draw(1.75,0) node{$\GG_{in}$};
\filldraw[red!20!white, draw=red!50!black] (4.5,0) .. controls (4.75,1) and (5.25,2) .. (5.5,1) 
			  .. controls (5.75,0) and (6.25,0.5) .. (6.5,0)
			  .. controls (7,-1) and (5.75,-2) .. (5.5,-1) 
			  .. controls (4 ,-3 ) and (3.5,1) .. (4.5,0)--cycle;
\draw(5.25,-0.2) node{{\Large$\mathbb{S}$}};

\draw[thick,dashed] (3.5,1) .. controls (4,3) .. (5,2);
\draw[thick,dashed] (5,2) .. controls (6,3) and (9,3) .. (7,1);
\draw[thick,dashed] (7,1) .. controls (10,0) and (9,-3) .. (6,-2);
\draw[thick,dashed] (6,-2) .. controls (4.5,-3) and (4,-3) .. (3.5,-1);
\draw(8,-0.5) node{$\GG_{out}$};
\draw(1.75,2) node{$(\epsilon,\mu)$};
 
	\end{tikzpicture}\end{center}
\caption[Scattering problem.]{\label{figc3:scatt2}Scattering problem. An object $\mathbb{S}$ is embedded in an infinite isotropic and homogeneous medium with relative electric and magnetic constants $(\epsilon,\mu)$. An incident field $\GG_{in}$ impinges on the object. The interaction of $\GG_{in}$ with $\mathbb{S}$ produces a scattered field $\GG_{out}$.}
\end{figure}

$\GG_{in}$ is the field that would exist if the object $\mathbb{S}$ was not there. It is a solution of Eqs. (\ref{eqc3:iso}) and hence a member of $\mathbb{M}$ (see Sec. \ref{secc3:M}).

The interaction of $\GG_{in}$ with the charges in $\mathbb{S}$ produces induced currents $\jj$ in $\mathbb{S}$. In turn, these currents give rise to the scattered field as\footnote{In the form given in \cite[Eq. 13.1.10]{Morse1953}, equation \ref{eqc3:Gjr} has an extra term expressed as a surface integral. For the scattering problem of Fig. \ref{figc3:scatt2}, such surface can be chosen to be a sphere of infinite radius. In this case, the transversality (\cite[Chap. 10.6]{Jackson1998}) and fall off of the scattered field causes the extra term to vanish.} \cite[Eq. 50.8]{Schwinger1998}:
\begin{equation}
	\label{eqc3:Gjr}
\GG_{out}(\rr)=ik\int_{V} d\rr' G(\rr,\rr')\jj(\rr'),
\end{equation}
where $G(\rr,\rr')=\left(\mathbf{I}+\frac{\nabla\nabla}{k^2}\right)\frac{\exp{ik|\rr-\rr'|}}{|\rr-\rr'|}$ is the Green tensor of the isotropic and homogeneous medium surrounding $\mathbb{S}$, and $V$ is the volume of the object $\mathbb{S}$. 

$G(\rr,\rr')$ can be expanded by means of spherical wave functions, i.e. multipolar fields. The expansion contains two transverse and one longitudinal series of linearly independent operators \cite[Eq. 13.3.79]{Morse1953}. Since, as mentioned before, only the transverse part of the fields is of interest, I will use the expansion without the longitudinal terms (see for instance\footnote{In \cite{Wittmann1988}, the author uses $M_{j,-n}(N_{j,-n})$ instead of the hermitian conjugates in (\ref{eqc3:GMN}). The two are equivalent since the complex part of the modes is the azimuthal $\exp(in\theta)$ dependence. Other authors (e.g.  \cite[Eq. 13.3.79]{Morse1953}) use $\cos(n\theta)$ and $\sin(n\theta)$ instead of $\exp(in\theta)$; note that these modes are not eigenstates of $J_z$.} \cite[Eq. (52)-(54)]{Wittmann1988} or \cite[Eq. 4.5]{Rose1955}):
\begin{equation}
	\label{eqc3:GMN}
	G(\rr,\rr')=\sum_{j>0}\sum_{n=-j}^{j}\frac{(-1)^{n+1}}{4\pi}\left(M^3_{jn}(\rr)\left(M^1_{jn}(\rr')\right)^\dagger+N^3_{jn}(\rr)\left(N^1_{jn}(\rr')\right)^\dagger\right),
\end{equation}
for $|\rr|>|\rr'|$. When $|\rr|<|\rr'|$, the superscripts $(1,3)$ must be interchanged in the $M$ and $N$ functions. When multiplied by $\exp (-i\omega t)$, the $M_{jn}$ and $N_{jn}$ are real space representations of the multipolar states $|\omega, j(j+1),n,s\rangle$ and $|\omega,j(j+1),n,p\rangle$, respectively (see Sec. \ref{secc3:heli}). The superscripts $(1,3)$ indicate the kind of spherical radial function that they contain: Bessel or Hankel of the first kind, respectively. The main differences between $(M^{1},N^1)$ and $(M^{3},N^3)$ are that the former are finite at the origin and their behavior at infinity is that of an ``outgoing'' plus an ``incoming'' wave, while the latter are singular at the origin and at infinity contain only ``outgoing'' waves.

Since we are concerned about the field outside the scatterer, we can assume $|\rr|>|\rr'|$ for all $\rr'$ in $V$. After inserting (\ref{eqc3:GMN}) into (\ref{eqc3:Gjr}) and performing the volume $V$ integral in $\rr'$ \cite{FerCor2011}, $\GG_{out}(\rr)$ is seen to be a linear combination of the $M_{jn}^3(\rr)$ and $N_{jn}^3(\rr)$ modes. Their ``outgoing'' wave behavior at infinity makes physical sense in a scattering problem.

The $M_{jn}^3$ do not have a straightforward expansion into plane waves \cite[Eq. (46)]{Wittmann1988}. Two expansions are needed, one for $z>0$ and another one for $z<0$. The $z=0$ plane, which contains the singularity, is excluded. Additionally, both expansions contain evanescent plane waves. As mentioned before, evanescent electromagnetic modes typically exist at the interfaces of different media and exhibit a rapid decay of intensity in the direction perpendicular to the interface. The evanescent plane waves in the expansions of $M_{jn}^3$ have an imaginary eigenvalue for $P_z$. For the $z>0$ case, the sign of $p_z$ ensures that the $\exp(i p_z z)$ spatial dependence results in exponentially decaying intensity as $z$ increases. The opposite sign of $p_z$ ensures the decay of the evanescent plane waves present in the expansion for $z<0$ as $z$ decreases.

Let me now distinguish two cases. The first one, when we are interested in the scattered field at large distances of the scatterer, and the second one when we are interested in the scattered field close to the scatterer. 

In the first case, the field at each point $\rr$ is, to good approximation \cite[Chap. 3.3.4]{Mandel1995}, only dependent on the plane wave component with momentum parallel to $\rr$. Therefore, $\GG_{out}$ can be approximated as a single superposition of plane waves \cite[Chap. 7.5.2]{Tung1985}, which I will call $\GG_{out}'$. It is hence clear that $\GG_{out}'$ is a member of $\mathbb{M}$, same as $\GG_{in}$. I will now assume that the induced currents $\jj$ are a linear function of $\GG_{in}$. This means that $\GG_{out}$ (and $\GG_{out}'$) are linear functions of $\GG_{in}$. Therefore, there exists a linear operator $T$ which acts in $\mathbb{M}$ and maps the $\GG_{in}$ input vectors onto the $\GG_{out}'$ output vectors. 

At this point, I have reached firm ground in order for me to be able to apply all the machinery about Hilbert spaces contained in Sec. \ref{secc3:hilbert} to the scattering problem of Fig. \ref{figc3:scatt2}. The total field at large distances to the scatterer can be written $\GG=\GG_{in}+\GG_{out}'$. To obtain an operator relating $\GG_{in}$ with $\GG$, an identity term must be added to $T$ 
\begin{equation}
\label{eqc3:IT}
S=I+T,
\end{equation}
$S$ is typically called the {\em scattering operator} and it is the one that I will be using most of the time. 

In most of the applications of the following chapters, I will be dealing with the scattered fields at large distances of the scatterer. When we are interested in the scattered fields close to the scatterer, the situation is more complicated. In order to define a scattering operator relating $\GG_{in}$ with $\GG_{out}$, we need to pick one of the two semi-infinite spaces. If, for example, we are interested in fields at $z>0$, we would build a scattering operator $S_{z>0}$ whose resulting output field would only be valid for $z>0$. And similarly for $z<0$. Note that the additional condition of being outside the scatterer should also be met.  Provided that we take these considerations into account, it is still possible to use Hilbert spaces and the two operators $(S_{z>0},S_{z<0})$ to study the properties of scattered near-fields, including evanescent fields, by means of symmetries and conservation laws. For example: Due to translational symmetry, the evanescent wave produced by the total internal reflection of a plane wave in a planar interface must have the same transverse momentum components as the input.

To finalize this section, it is worth mentioning that absorption by the scatterer results in a non unitary operator $S$ which renders the norm of the output vectors smaller than that of the input ones. Absorption reduces the total integrated energy of the field but does not change the eigenvalues of the energy operator (the frequencies). For the latter to happen the system must be non invariant under time translations, like for example an object moving away from the source resulting in the Doppler effect. See Chap. \ref{secc4:shatlhat} for a related discussion. Independently of whether the system has losses or not, if it is invariant under the action of a given operator (transformation), the scattering of an eigenvector of such operator will only produce modes which are also eigenvectors of the operator and have the same eigenvalue as the input mode. In this respect, the applicability of symmetry arguments does not depend on whether a system has losses or not. 

\subsection{A semi-analytical technique for scattering}
The semi-analytical technique developed in \cite{FerCor2011} allows to compute the scattering operator in practical situations. 

Consider a planar multilayer structure of arbitrary number of layers, thicknesses and electric constants $\epsilon_l$, and magnetic constants $\mu_l=1$ for all layers. Consider one (or several) objects embedded in the multilayer with constants $\epsilon(\rr)$ and $\mu=1$. Note that this is a more general setting than the one depicted in Fig. \ref{figc3:scatt2}.

When particularized to the setting of Fig. \ref{figc3:scatt2}, the technique in \cite{FerCor2011} allows to obtain the operator $T$ (see (\ref{eqc3:IT})) in the plane wave basis, i.e, it allows to obtain the complex numbers
\begin{equation}
	\langle s \ \ppbar|T|\pp\ s \rangle,\ \langle s \ \ppbar|T|\pp\ p \rangle,\ \langle p \ \ppbar|T|\pp\ s \rangle,\ \langle p \ \ppbar|T|\pp\ p \rangle
\end{equation}
for all $(\pp,\ppbar)$.

The technique is based on the same ideas discussed in the previous section. The currents inside the scattering objects induced by the input field are proportional to $\left(\epsilon(\rr)-\epsilon_l\right)\EE_V(\rr)$, where $\EE_V(\rr)$ is the total field inside the volume of the scattering objects. This internal field due to a $|\pp \ s/p\rangle$ input can be obtained by means of known numerical techniques and/or commercial software packages. Once $\EE_V(\rr)$ is known, a decomposition of the Green tensor analogous to (\ref{eqc3:GMN}), but using plane waves instead of multipoles, can be used to perform the integral in $\rr'$ and obtain the desired plane wave decomposition.

In its more general setting, the Green tensor of a planar multilayer system is used instead of the one for a homogeneous and isotropic medium. The functions that expand such Green tensor are then the eigenfunctions of the multilayered structure \cite{Paulus2000}.

The technique can be extended to spherical and cylindrical multilayer structures by using the expansion of the corresponding Green's tensors in multipoles or Bessel beams. It can also be extended to incorporate both magnetic layers and magnetic scattering objects.

I have used the code that implements the technique to produce the simulation results contained in \cite{FerCor2011}, \cite{Tischler2013} and \cite{Tischler2014}. In particular, \cite{FerCor2011} contains the exact\footnote{Only limited by numerical accuracies, discretization errors, etc ..., not by simplifications in the model.} radiation diagram for a subwavelength hole in a thin metallic film under Gaussian illumination, including evanescent components.

\section{Helicity preservation and duality symmetry}\label{secc3:helpresdualsym}
It is now time for the detailed discussion on helicity and duality.

Fig. \ref{figc3:dnd} shows a graphical representation of helicity preserving and helicity non-preserving scattering in the plane wave basis.

\begin{figure}[h!]
	\begin{center}
	\makeatletter{}\def\ASYprefix{}
\newbox\ASYbox
\newdimen\ASYdimen
\long\def\ASYbase#1#2{\leavevmode\setbox\ASYbox=\hbox{#1}\ASYdimen=\ht\ASYbox\setbox\ASYbox=\hbox{#2}\lower\ASYdimen\box\ASYbox}
\long\def\ASYaligned(#1,#2)(#3,#4)#5#6#7{\leavevmode\setbox\ASYbox=\hbox{#7}\setbox\ASYbox\hbox{\ASYdimen=\ht\ASYbox\advance\ASYdimen by\dp\ASYbox\kern#3\wd\ASYbox\raise#4\ASYdimen\box\ASYbox}\put(#1,#2){#5\wd\ASYbox 0pt\dp\ASYbox 0pt\ht\ASYbox 0pt\box\ASYbox#6}}\long\def\ASYalignT(#1,#2)(#3,#4)#5#6{\ASYaligned(#1,#2)(#3,#4){
}{
}{#6}}
\long\def\ASYalign(#1,#2)(#3,#4)#5{\ASYaligned(#1,#2)(#3,#4){}{}{#5}}
\def\ASYraw#1{
currentpoint currentpoint translate matrix currentmatrix
100 12 div -100 12 div scale
#1
setmatrix neg exch neg exch translate}
 
	\makeatletter{}\setlength{\unitlength}{1pt}
\makeatletter\let\ASYencoding\f@encoding\let\ASYfamily\f@family\let\ASYseries\f@series\let\ASYshape\f@shape\makeatother{\catcode`"=12\includegraphics{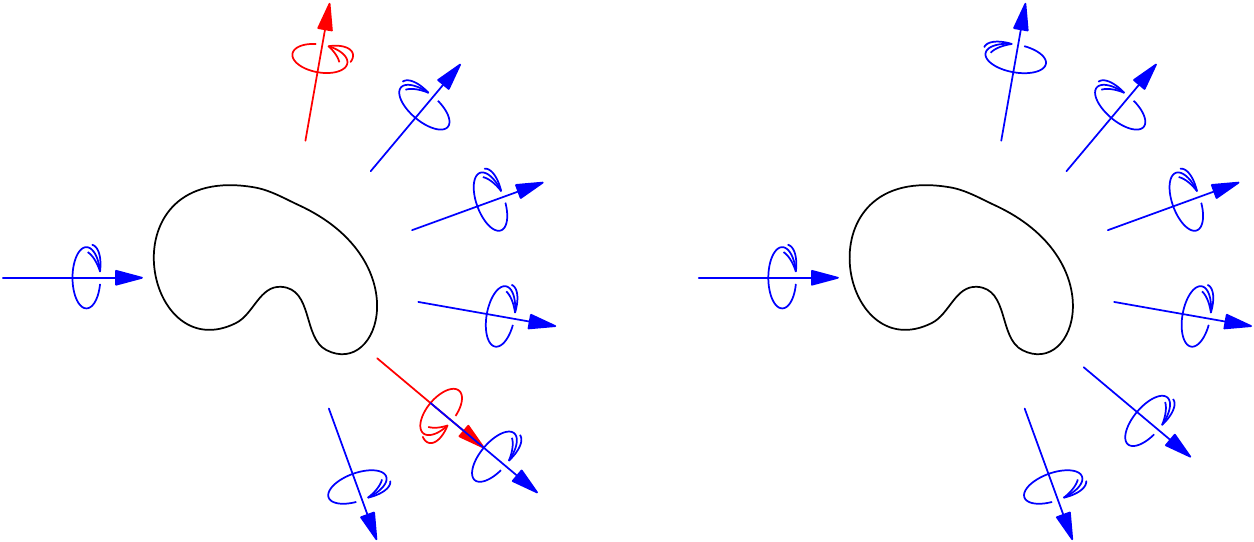}}\definecolor{ASYcolor}{gray}{0.000000}\color{ASYcolor}
\fontsize{12.000000}{14.400000}\selectfont
\usefont{\ASYencoding}{\ASYfamily}{\ASYseries}{\ASYshape}\ASYalign(-340.983523,136.203587)(-0.500000,-0.250000){a)}
\definecolor{ASYcolor}{gray}{0.000000}\color{ASYcolor}
\fontsize{12.000000}{14.400000}\selectfont
\ASYalign(-284.661531,79.881595)(-0.500000,-0.500000){Non-Dual}
\definecolor{ASYcolor}{gray}{0.000000}\color{ASYcolor}
\fontsize{12.000000}{14.400000}\selectfont
\ASYalign(-139.833550,136.203587)(-0.500000,-0.250000){b)}
\definecolor{ASYcolor}{gray}{0.000000}\color{ASYcolor}
\fontsize{12.000000}{14.400000}\selectfont
\ASYalign(-83.511558,80.686195)(-0.500000,-0.500000){Dual}
 
	\end{center}
\caption[Helicity preserving versus helicity non-preserving scatterers.]{\label{figc3:dnd}(a) The helicity of an electromagnetic field is not preserved after interaction with a non-dual symmetric object. An incoming field with well defined helicity, in this case a single plane wave of definite polarization handedness (blue), produces a scattered field that contains components of the opposite helicity (red). The helicity of the scattered field in panel (a) is not well defined because it contains plane waves of different helicities. (b) Helicity preservation after interaction with a dual symmetric object. The helicity of the scattered field is well defined and equal to the helicity of the input field.}
\end{figure}

First of all, I need to discuss a key nuance in the use of duality symmetry in light matter interactions. It is about the extra transformation of the sources that Zwanziger used to prove that the whole electrodynamic theory is invariant under the simultaneous action of the duality transformation of the fields and a similar extra transformation of the sources \cite{Zwanziger1968}. Let me write again the equations in Sec. \ref{secc2:dualhel} with $\mu_0=\epsilon_0=1$. Using real fields, Zwanziger showed that the microscopic equations with both electric and magnetic sources 
\begin{equation}
\label{eqc3:micro}
	\begin{split}
		\nabla\cdot\EE=\rho_e&,\ \nabla\cdot\HH=\rho_g,\\
	\partial_t\EE=\nabla\times\HH-\jj_e&,\partial_t\HH=-\nabla\times\EE-\jj_g.\\
	\end{split}
\end{equation}
and the Lorentz force that a field affects on a charged particle with both electric and magnetic couplings
\begin{equation}
\label{eqc3:lorentz}
	\FF=\rho_e\EE+\jj_e\times\HH+\rho_g\HH-\jj_g\times\EE,
\end{equation}
are invariant under the simultaneous action of the duality transformation of the fields
\begin{equation}
\label{eqc3:gendual}
\begin{split}
\EE&\rightarrow \EE_\theta=\EE\cos\theta  - \HH\sin\theta , \\
\HH&\rightarrow \HH_\theta=\EE\sin\theta + \HH\cos\theta ,
\end{split}
\end{equation}
and the extra transformation of the sources
\begin{equation}
\label{eqc3:sourcedual}
\begin{split}
	\begin{bmatrix}\rho_e\\\jj_e\end{bmatrix}\rightarrow \begin{bmatrix}\rho_e\\\jj_e\end{bmatrix}_\theta&=\begin{bmatrix}\rho_e\\\jj_e\end{bmatrix}\cos\theta -\begin{bmatrix}\rho_g\\\jj_g\end{bmatrix}\sin\theta , \\
	   \begin{bmatrix}\rho_g\\\jj_g\end{bmatrix}\rightarrow \begin{bmatrix}\rho_g\\\jj_g\end{bmatrix}_\theta&=\begin{bmatrix}\rho_g\\\jj_g\end{bmatrix}\cos\theta +\begin{bmatrix}\rho_e\\\jj_e\end{bmatrix}\sin\theta.\\
\end{split}
\end{equation}

Is this the kind of duality invariance that I am going to exploit in light matter interactions? No, it is not. Zwanziger answered a question about the entire theory. The question that he answered could be stated like this:

If in a universe where each fundamental particle $(s)$ has some electric $e_{s}$ and magnetic $g_{s}$ charges, you perform all the possible electrodynamic experiments and record the results: Can you then predict the results of the experiments performed in another universe where the charges of the particles are $\tilde{e}_{s}=e_{s}\cos\theta-g_{s}\sin\theta,\ \tilde{g}_{s}=g_{s}\cos\theta+e_{s}\sin\theta$? The answer, Zwanziger taught us, is yes, you can, whatever the value of $\theta$. 

My questions are going to be of a different kind. What I am interested in is the properties of scattering systems under duality transformations {\bf without the extra source transformation in (\ref{eqc3:sourcedual})}. I do not want to be able to predict what would happen if I performed the same experiment in the other universe. What I want to know is things like, for example, {\bf whether in this universe the eigenstates of helicity are preserved upon interaction with a scatterer}. Considering the extra source transformation in this context is analogous to, given an incident field, a scatterer and the corresponding scattered field, asking whether if I rotate both the incident and scattered fields by 10 degrees are they going to be solutions of a new scattering problem with the scatterer rotated by 10 degrees?: They will be, because of the invariance of physical laws under rotations, but the question does not give information about whether the scatterer preserves the eigenstates of angular momentum. In this analogy, Zwanziger's result is equivalent to the invariance of physical laws under rotation, and I am interested on whether the rotated fields will still be a solution of the original unrotated scatterer or, equivalently, if the original unrotated fields are a solution of the problem for the rotated scatterer.

Without the extra transformation of the sources (\ref{eqc3:sourcedual}), duality is not a symmetry of the microscopic equations (\ref{eqc3:micro}). The symmetry is broken in the microscopic theory. In Sec. \ref{secc3:micro}, I derive conditions on the electric and magnetic components of hypothetical elementary charges that, had they been met in Nature, would result in a dual symmetric microscopic theory. We, of course, have no control on the properties of elementary charges. Nevertheless, even though the symmetry is broken in our microscopic theory, I will also show that the duality symmetry can be restored both for the macroscopic equations and the dipolar approximation. The symmetry is restored when the constitutive relations of the macroscopic medium or the polarizability tensor of the dipolar scatterer meet certain conditions, respectively. In these two approximations to the description of light matter interactions, we can have dual symmetric helicity preserving scatterers. And we can also put the resulting theory to practical uses (see the application chapters).

Before continuing, I want to recall the key feature of the $\sqrt{2}\GG_{\pm}=\pm\EE+iZ\HH$ representation from Sec. \ref{secc3:M}. It separates the evolution equations, which were coupled for $\EE$ and $\HH$ (\ref{eqc3:iso}), into two sets of decoupled equations (\ref{eqc3:maxrs}). This is related to the discussion in Sec. \ref{subsecc2:usinghelicity} about the two helicities of the photon living separate lives in free-space. From Sec. \ref{secc3:M}, it is clear that the two helicities are also separated in an isotropic and homogeneous medium, and we could model it with two separate Hilbert spaces $\mathbb{M}_\pm$. In the scattering problem of Fig. \ref{figc3:scatt2}, a non-dual symmetric $\mathbb{S}$ couples the two spaces $\mathbb{M}_\pm$. In the next three subsections I derive the conditions that elementary charges, macroscopic media and dipolar scatterers should meet in order to maintain $\mathbb{M}_\pm$ separated from each other. 

Let me then start.
\subsection{Splitting electrodynamics at the microscopic level}
\label{secc3:micro}
As far as we know from electrodynamic experiments, we can think of all elementary particles as having only electric charge \cite{Zwanziger1968}. As I will show in this section, particles with only electric charge are source of and interact with fields of the two helicities simultaneously. This means that the microscopic equations of electrodynamics break the helicity conservation law.

Let me start the analysis by assuming a more general situation where elementary particles have both electric and magnetic charges of arbitrary value. 

With the definitions (\ref{eqc3:RS3}) $\sqrt{2}\GG_{\pm}=\pm \EE+i\HH$ the microscopic equations are:
\begin{equation}
	\label{eqc3:micrors}
	\begin{split}
		\nabla\cdot \GG_{\pm}&=\frac{1}{\sqrt{2}}\left(\pm \rho_e+i\rho_g\right)=\rho_{\pm}\\
		 i\partial_t\GG_{\pm}&=\pm\nabla\times\GG_{\pm}-\frac{i}{\sqrt{2}}(\pm\jj_e+i\jj_g)=\pm\nabla\times\GG_{\pm}-i\jj_{\pm},
	\end{split}
\end{equation}
where the last equalities contain the definitions of $\rho_{\pm}$ and $\jj_{\pm}$. 

The Lorentz force can be written:
\begin{equation}
	\label{eqc3:lorentzrs}
	\FF=\left(\rho_+\GG^*_++i\jj_+\times\GG^*_+\right)+\left(\rho_-\GG^*_--i\jj_-\times\GG^*_-\right)\equiv\FF_+ + \FF_-.
\end{equation}
To split electrodynamics into two separate uncoupled spaces, all the elementary particles should only have charges of one of the two $\pm$ kinds. According to (\ref{eqc3:micrors}) and (\ref{eqc3:lorentzrs}), under the assumption of positive energies, each of the two different kinds of $\pm$ charged particles would only interact with and produce fields of one definite helicity, opposite for each kind of particle. An elementary particle would have a single kind of $\pm$ charge if its electric and magnetic charges are related like $g_s=ie_s$ or like $g_s=-ie_s$. The charge density and current due to each particle would be either:
\begin{equation}
	\rho^s_+=0,\rho^s_-=-\sqrt{2}\rho^s_e,\jj^s_+=\mathbf{0},\jj^s_-=-\sqrt{2}\jj^s_e,
\end{equation}
or
\begin{equation}
	\rho^s_+=\sqrt{2}\rho^s_e,\rho_-=0,\jj^s_+=\sqrt{2}\jj^s_e,\jj^s_-=\mathbf{0}.
\end{equation}
Assuming that the $e_s$ are real, the particles would have purely imaginary magnetic charge $g_s=\pm i e_s$. Their ``helicity charges'' (or ``chiral charges'') $q_{\pm}$ would be real. 

In this hypothetical situation, all scatterers would preserve helicity. Upon interaction with a field of well defined helicity, only the charges of the corresponding kind present in the scatterer would react (\ref{eqc3:lorentzrs}). In turn they would produce a scattered field of the same helicity as the input field. The charges of the other kind would not react. A scatterer composed by a single kind of chiral charge would be invisible to the radiation of the opposite helicity.

Since, as it seems to be the case empirically, all elementary particles can be modeled as possessing only electric charge, equations (\ref{eqc3:micrors}) and (\ref{eqc3:lorentzrs}) show that they mix the two helicities. The particles have both kinds of helicity charge ($\pm$) simultaneously and, when reacting to the Lorentz force due to a pure helicity field, they become sources of both kinds of field at the same time. The helicity conservation law is broken at the microscopic level.

We have no control over the charges of the elementary particles. Nevertheless, we can engineer objects to effectively show duality symmetry. We can even make them invisible to fields of a particular helicity. The next two sections contain the conditions for helicity preservation (duality symmetry) in the macroscopic equations and in the dipolar approximation.

\subsection{Helicity preservation in the macroscopic equations}
The microscopic equations with sources are an exact description of classical electromagnetism. In many practical problems, though, it is impossible to consider every electron and proton of the system. For example, in the propagation of light in water. Through a series of approximations \cite[Chap. 6]{Jackson1998}, the average effect of all those microscopic charges of the medium is collected into a different set of equations where new fields $\DD$ and $\BB$ are introduced in addition to $\EE$ and $\HH$,
\begin{equation}
\label{eqc3:macro}
	\begin{split}
		\nabla\cdot\DD=0&,\ \nabla\cdot\BB=0,\\
	\partial_t\DD=\nabla\times\HH&,\partial_t\BB=-\nabla\times\EE.\\
	\end{split}
\end{equation}
and the specifics of the medium are collected in what is known as the constitutive relations which relate the two pairs of fields
\begin{equation}
\label{eqc3:consrel}
\begin{bmatrix}Z\DD\\\BB\end{bmatrix}=\begin{bmatrix}\matr{\epsilon}&\matr{\chi}\\\matr{\gamma}&\matr{\mu}\end{bmatrix}\begin{bmatrix}\EE\\Z\HH\end{bmatrix}=M\begin{bmatrix}\EE\\Z\HH\end{bmatrix}.
\end{equation}
The components of $M$ are, in the coordinate representation, integro-differential operators. Depending on the medium, $M$ can be very simple like $\DD=\epsilon\EE,\BB=\mu\HH$, for scalar electric constant $\epsilon$ and magnetic constant $\mu$, like the host medium of Fig. \ref{figc3:scatt2}. $M$ also can be very complicated, where the four 3$\times$3 blocks of $N$ are, in the coordinate representation, convolutions in time and space \cite{Agarwal1974}. The symbols $\matr{\epsilon},\matr{\chi},\matr{\gamma}$ and $\matr{\mu}$ represent whatever is needed in each case. They can be scalars or tensors, they can be constant or vary in space and frequency, they can also encode losses in the medium. The only restriction that is inherent in (\ref{eqc3:consrel}) is that the constitutive relations are linear. There are media where the constitutive relations are not linear, like in a non-linear crystal, and those are outside the scope of my thesis\footnote{Strictly speaking all media are non-linear but most of them behave linearly at low field intensities.}. Note the generality of the setup: Spatial dependency covers inhomogeneous media and boundaries between different piecewise homogeneous media, frequency dependency covers materials with different properties at different frequencies, losses can be included for example with complex $\matr{\epsilon}$ and $\matr{\mu}$, etc ... . 

Here is a key point. The description of macroscopic electromagnetism needs both the equations in (\ref{eqc3:macro}) and the constitutive relations (\ref{eqc3:consrel}). Only the two together have all the necessary information. When checking whether this or that transformation is a symmetry of macroscopic electromagnetism, one needs to verify that the transformation leaves invariant {\bf both the equations and the constitutive relations}. Alternatively, we can include (\ref{eqc3:consrel}) into (\ref{eqc3:macro}):
\begin{equation}
\label{eqc3:macrocons}
	\begin{split}
		\begin{bmatrix} \nabla\cdot & 0\\ 0&\nabla\cdot\end{bmatrix}\begin{bmatrix}Z\DD\\\BB\end{bmatrix}&=\begin{pmatrix} 0 \\ 0\end{pmatrix}\implies\begin{bmatrix} \nabla\cdot & 0\\ 0&\nabla\cdot\end{bmatrix}M\begin{bmatrix}\EE\\Z\HH\end{bmatrix}=\begin{pmatrix} 0 \\ 0\end{pmatrix},\\
		\HsixEH \begin{bmatrix}\DD\\\BB\end{bmatrix}&=\CurlsixEH\begin{bmatrix}\EE\\Z\HH\end{bmatrix}\implies\\
		\HsixEH M\begin{bmatrix}\EE\\Z\HH\end{bmatrix}&=\CurlsixEH\begin{bmatrix}\EE\\Z\HH\end{bmatrix},
	\end{split}
\end{equation}
and check the invariance of these resulting expressions. Note that the right hand sides of the first line in (\ref{eqc3:macrocons}) are two dimensional vectors with zeros in them, while the other zeros that appear in the expressions are 3$\times$3 null matrices. The different brackets are an attempt to mark this difference. 

I will now derive the conditions for helicity preservation in the Riemann-Silberstein representation (\ref{eqc3:RS3}). With 
\begin{equation}
	R=\frac{1}{\sqrt{2}}\begin{bmatrix} I & iI\\ -I&iI\end{bmatrix},
\end{equation}
I can change equations (\ref{eqc3:macrocons}) into:
\begin{equation}
	\begin{split}
		&R\begin{bmatrix} \nabla\cdot & 0\\ 0&\nabla\cdot\end{bmatrix}R^{-1}R MR^{-1}R\begin{bmatrix}\EE\\Z\HH\end{bmatrix}=\begin{pmatrix} 0\\ 0\end{pmatrix}\\
		&R\Hsix R^{-1}R MR^{-1}R\begin{bmatrix}\EE\\Z\HH\end{bmatrix}=
	 R \CurlsixEH R^{-1}R\begin{bmatrix}\EE\\Z\HH\end{bmatrix},
	\end{split}
\end{equation}
which gives
\begin{equation}
	\begin{split}
\begin{bmatrix} \nabla\cdot & 0\\ 0&\nabla\cdot\end{bmatrix}N\Gsix &=\begin{pmatrix}0\\0\end{pmatrix}\\
		\Hsix N\Gsix &=\Curlsix\Gsix,
	\end{split}
\end{equation}
with
\begin{equation}
	\label{eqc3:consrelrs}
	N=R M R^{-1}=\frac{1}{2}\begin{bmatrix} \matr{\epsilon}+\matr{\mu} - i (\matr{\chi}-\matr{\gamma} )& -\matr{\epsilon} + \matr{\mu} -i (\matr{\chi}+\matr{\gamma} )\\-\matr{\epsilon} + \matr{\mu} +i (\matr{\chi}+\matr{\gamma} )&\matr{\epsilon}+\matr{\mu} - i (\matr{\chi}-\matr{\gamma})\end{bmatrix}.
\end{equation}

In order to have uncoupled evolution equations for the two helicities, $N$ must be block diagonal, which imposes that:
\begin{equation}
	\label{eqc3:dualmacro}
	\boxed{\matr{\epsilon}=\matr{\mu},\ \matr{\chi}=-\matr{\gamma}.}
\end{equation}

It can be checked that a block diagonal structure is the necessary and sufficient condition for $N$ to be invariant under duality transformation $D_\theta^{-1}ND_\theta=N$. Then, the equations transformed by duality (\ref{eqc3:D}) have the same form for the original variables $\Gsix$ and the transformed variables $\Gsix_\theta=D_\theta \Gsix$ variables.

Conditions equivalent to (\ref{eqc3:dualmacro}) can also be found in \cite{Lindell2009}. The authors use a definition of the duality transformation for time harmonic fields which is restricted to a single value of $\theta=\pi/2$ and involves the imaginary constant $i$. Such definition is equivalent to the definition of the helicity operator. Since duality is not used as a continuous transformation generated by the helicity operator, the connection to the preservation of helicity eigenstates cannot be made with such an approach.

When (\ref{eqc3:dualmacro}) is met, the evolution equations read:
\begin{equation}
	\Hsix \begin{bmatrix} \matr{\epsilon} - i \matr{\chi}& 0\\0&\matr{\epsilon}+ i\matr{\chi}\end{bmatrix} \Gsix=c\Curlsix\Gsix.
\end{equation}

Equations (\ref{eqc3:dualmacro}) are the condition for duality symmetry (helicity preservation) for the macroscopic equations. Note that the value of $Z$ is chosen in (\ref{eqc3:consrel}). The decoupling of the $\mathbb{M}_\pm$ spaces depends on $Z$. Note also that $\mathbf{K}_\pm=\pm Z\DD+i\BB$ are the transverse vectors (and not $\GG_{\pm}$). An analogous analysis using $\mathbf{K}_\pm$ gives the same helicity preservation conditions (\ref{eqc3:dualmacro}).

Before moving to the dipolar approximation, I am going to stress an important feature of duality symmetry for the macroscopic equations: Its restoration is independent of geometrical shape.

\subsubsection{Shape independence}\label{subsecc3:shape}
In the last section I have shown that, for the macroscopic equations, helicity preservation upon scattering does not depend on the shape of the scatterer, but only on the materials of which is made. This is quite powerful. After choosing a basis, one of the four numbers will be preserved (or not) independently of the shape of the scatterer. This simplification is priceless in practical applications. This shape independence comes from the fact that the constitutive matrix $N$ can depend on space and, for instance, model a medium $\Omega$ composed of different homogeneous and isotropic domains with arbitrary boundaries. As long as (\ref{eqc3:dualmacro}) is met, the medium is dual and helicity is preserved.

In a medium like $\Omega$, $N$ has spatial discontinuities. Maybe we should worry about this discontinuities when applying the transversality condition in (\ref{eqc3:macro}), where there are spatial derivatives? It should always be possible to write $N$ so that the boundaries between two media are not sharp. Instead, $N$ can be assumed to vary very fast near them so as to go from the $(\epsilon_1,\mu_1)$ of one side to the $(\epsilon_2,\mu_2)$ of the other side in a continuous and differentiable way. There is another approach that can handle abrupt discontinuities. Appendix \ref{appc3:shape} contains a proof that uses the boundary conditions to arrive at the result that, in a piece-wise homogeneous, reciprocal, non-chiral ($\chi=\gamma=0$) and isotropic media composed of domains with sharp arbitrary boundaries, duality is restored (helicity is preserved) if and only if:
\begin{equation}
	\label{eqc3:epsmu}
	\boxed{\frac{\epsilon_i}{\mu_i}=\text{constant for all domain }i.}
\end{equation}
The proof can be found in \cite{FerCor2012p}. I include it in appendix \ref{appc3:shape} because it involves the common boundary conditions and treats the shape of the boundary explicitly. It may provide more insight than the dryer (although more general) result from (\ref{eqc3:dualmacro}). It is also the first form in which I could prove this useful result. The idea of choosing the reference $Z=\sqrt{\mu/\epsilon}$ discussed above is at the core of the proof, as one may guess by inspecting (\ref{eqc3:epsmu}).

I will now give an example of the usefulness of the result. 

\begin{figure}[h!]
 \begin{center}
	\includegraphics[scale=0.85]{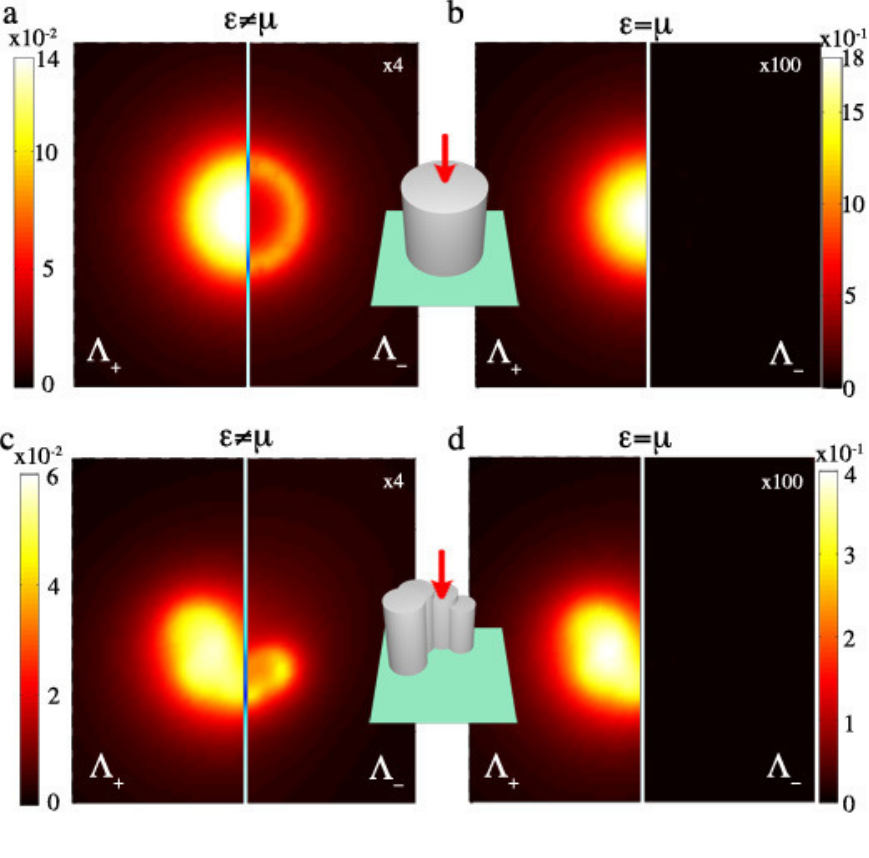}
\end{center}
\caption[Independence of helicity preservation from geometry.]{\label{figc3:newfig1} Impact of the different symmetries on the field scattered by two dielectric structures. The upper row shows the scattered intensity (in coordinate space) for a symmetric cylinder and the lower row for a panflute like shape without any rotational or spatial inversion symmetry. The displayed area is 700x700 nm. The calculation plane is perpendicular to the $z$ axis and 20 nm away from the surface of the scatterers opposite to the one where the incident field comes from. The length and diameter of the cylinder are 200 nm. The panflute is made of cylinders of different lengths and diameters, the longest one is 200 nm long and the total panflute's width is around 200 nm. In (a) and (c) the structures have $\epsilon=2.25,\ \mu=1$, while in (b) and (d) duality symmetry is enforced by setting $\epsilon=\mu=\epsilon_{\textrm{glass}}=2.25$. The incident field is a plane wave of well defined helicity equal to 1, momentum vector pointing to the positive $z$ axis and a wavelength of 633 nm.  The left half side of each subfigure corresponds to the scattered field with helicity equal to the incident plane wave; the right half is for the opposite helicity.   For color scaling purposes, the right half side is multiplied by the factor in the upper right corner. The (lack of) cylindrical symmetry of the structures results in (non-)cylindrically symmetric field patterns, which is consistent with the geometry of each case. On the other hand, both scatterers behave identically with respect to conservation of helicity, which is seen to depend exclusively on the electromagnetic properties of the material. Simulation performed by Nora Tischler.}
\end{figure}

My colleague Nora Tischler performed numerical simulations in order to illustrate the independence of helicity conservation from geometry in complex systems. The helicity change (Fig. \ref{figc3:newfig1}) for two different dielectric structures in free space was analyzed: A cylinder, which is symmetric under rotations along its axis, and a curved panflute like structure without any rotational, translational or spatial inversion symmetry. Two versions of each structure were simulated, corresponding to two different materials: the first one models the properties of silica by setting ${\epsilon=\epsilon_{\textrm{glass}}=2.25}$ and ${\mu=\mu_{\textrm{glass}}=1}$. In the second material duality is enforced (\ref{eqc3:epsmu}), by setting ${\epsilon=\mu=\epsilon_{\textrm{glass}}=2.25}$. The incident field is a circularly polarized plane wave (i.e. it has well defined helicity) with momentum vector parallel to the red arrows in the figure. Its electric field is $-(\xhat+i\yhat)/\sqrt{2}\exp(kz-\omega t)$. In the case of the cylinder, the momentum direction is aligned with the axis of the cylinder. The intensities of the two helicities ($\pm$) were computed as $|\mathbf{E}\pm i\mathbf{H}|^2$. Figure \ref{figc3:newfig1} shows that helicity is conserved independently of the spatial symmetries, whenever Eq. (\ref{eqc3:epsmu}) is fulfilled, i.e. under conditions of duality symmetry.

Equation (\ref{eqc3:epsmu}) predicts the results of Fig. \ref{figc3:newfig1} in a straightforward way. As far as I know, this cannot be done with any other existing approach.

Finally, a nice check of (\ref{eqc3:epsmu}) is to verify it in two of the few analytically solvable electromagnetic scattering problems: A planar multilayer system and a sphere. When imposing condition (\ref{eqc3:epsmu}), the Fresnel coefficients are identical for the two (TE and TM) polarizations for any plane wave impinging on the multilayer. This implies helicity preservation (Sec. \ref{secc3:helprestetm}). The same is true for the Mie coefficients representing the scattering of TE and TM multipoles off a sphere: They are identical when (\ref{eqc3:epsmu}) is met, implying preservation of the multipoles of well defined helicity. These derivations are included in Appendix \ref{appc3:fm} and in the supplementary information of \cite{FerCor2012p}.

\subsection{Helicity preservation in the dipolar approximation}
\label{secc3:dipolar}
{\em The scattering of electromagnetic waves by systems whose individual dimensions are small compared with a wavelength is a common and important occurrence. In such interactions it is convenient to think of the incident (radiation) fields as inducing electric and magnetic multipoles that oscillate in definite phase relationship with the incident wave and radiate energy in directions other than the direction of incidence. The exact form of the angular distribution is governed by the coherent superposition of multipoles induced by the fields and in general depends on the state of polarization of the incident wave. If the wavelength of the radiation is long compared to the size of the scatterer, only the lowest multipoles, usually electric and magnetic dipoles, are important}.

So starts section 10.1.A of Jackson's book \cite{Jackson1998}. It is the perfect introduction to this section. Chapters 9 and 10 of \cite{Jackson1998} contain the derivations justifying the last sentence. Typically, the dipolar approximation can be applied to scatterers that are at least ten times smaller than the wavelength. 

So, in the context of scattering (Fig. \ref{figc3:scatt2}) I am now dealing with electromagnetically small things. For example, molecules typically meet the small size requirement at optical wavelengths. As Jackson wrote, the dipolar approximation is very common in the study of light matter interactions. I will now derive the helicity preservation conditions in this approximation.

With the always present linearity assumption, the small scatterer $\mathbb{S}$ located at point $\rrp$ can be modeled as an operator $P$ that relates the incident electric $\EE$ and magnetic $\HH$ fields at point $\rrp$ with the induced electric $\ed$ and magnetic $\md$ dipoles:
\begin{equation}
\label{eqc3:M}
\begin{bmatrix}\ed(t)\\\md(t)\end{bmatrix}=P\begin{bmatrix}\EE(\rrp,t)\\\HH(\rrp,t)\end{bmatrix}=
\begin{bmatrix}\MdE & \MdH\\\MmE & \MmH\end{bmatrix}
\begin{bmatrix}\EE(\rrp,t)\\\HH(\rrp,t)\end{bmatrix}.
\end{equation}
$P$ is decomposed into its four $3\times3$ blocks, which are labeled using an obvious notation. A very common setting in the applications that use the dipolar approximation is that of polarizability tensors (the technical name for $P$) that vary with frequency. In such case, it is advantageous to work with monochromatic fields $\widehat{\FF}(\rr,\omega)\exp(-i\omega t)$ since $P$ is then constant for each frequency. For a general non-monochromatic input field the total response can be obtained by linearly decomposing it into monochromatic fields. Dropping the ``hats'' and the $\exp(-i\omega t)$ dependency from both sides: 

\begin{equation}
\label{eqc3:M2}
\begin{bmatrix}\ed(\omega)\\\md(\omega)\end{bmatrix}=P\begin{bmatrix}\EE(\rrp,\omega)\\\HH(\rrp,\omega)\end{bmatrix}=
\begin{bmatrix}\MdE(\omega) & \MdH(\omega)\\\MmE(\omega) & \MmH(\omega)\end{bmatrix}
\begin{bmatrix}\EE(\rrp,\omega)\\\HH(\rrp,\omega)\end{bmatrix}
\end{equation}
From now on, I will drop the $\omega$ and $\rrp$ as well.

I am after helicity preservation. Therefore, the field radiated by the induced dipoles must preserve the helicity of the incident field. I will first obtain the relationship that must hold between $\ed$ and $\md$ in order for their combined emission to have a well defined helicity. Then, I will find the conditions that $P$ must meet so that incident fields with well defined helicity induce dipoles which produce a scattered field with the same well defined helicity.

For the first task, I consider the field emitted by an electric dipole $\ed$ and a magnetic dipole $\md$ located at the same point in an infinite homogeneous and isotropic medium with electric and magnetic constants $(\epsilon,\mu)$. I denote by $(\Eed,\Hed)$ the fields produced by the electric dipole $\ed$ and $(\Emd,\Hmd)$ those produced by the magnetic dipole $\md$. The total fields are the sum of the fields radiated by the two dipoles
\begin{equation}
	\EE=\Eed+\Emd,\HH=\Hed+\Hmd,
\end{equation}
from which we can obtain the two helicity components:
\begin{equation}
\begin{split}
\mathbf{G}_+&=\left(\Eed+\Emd\right)+iZ\left(\Hed+\Hmd\right),\\
\mathbf{G}_-&=-\left(\Eed+\Emd\right)+iZ\left(\Hed+\Hmd\right).
\end{split}
\end{equation}
A total field with well defined helicity equal to +1 will have no component of helicity equal to -1, thus 
\begin{equation}
\label{eqc3:zero2}
	\mathbf{G}_-=-\left(\Eed+\Emd\right)+iZ\left(\Hed+\Hmd\right)=0.
\end{equation}
To solve (\ref{eqc3:zero2}), I use the relations in \cite[Chap. 9.3]{Jackson1998}. According to Jackson, a magnetic dipole $\md$ produces electric and magnetic fields $(\Emd,\Hmd)$ which are related to the electric and magnetic fields  $(\Eedbar,\Hedbar)$ produced by an auxiliary electric dipole $\edbar$ in the following way:
\begin{equation}
	\label{eqc3:ddmm}
\edbar=\frac{\md}{c},\ \Emd=-Z\Hedbar,\ \Hmd=\frac{1}{Z}\Eedbar.
\end{equation}
Note that, for now, $\edbar$ and $\ed$ are not related. Using (\ref{eqc3:ddmm}) we turn (\ref{eqc3:zero2}) into
\begin{equation}
	\label{eqc3:plus}
	\Eed-iZ\Hed = i\Eedbar+Z\Hedbar.
\end{equation}
Equation (\ref{eqc3:plus}) must be met in all points of space. The radiated fields depend linearly on the dipole vectors. The solution is hence
\begin{equation}
	\edbar=-i\ed,\text{ which means } \frac{\md}{c}=-i\ed.
\end{equation}
The corresponding steps for a well defined helicity equal to -1 result in $\edbar=i\ed$, or $\md/c=i\ed$. 

In summary:
\begin{equation}
\label{eqc3:dm}
	\ed=\pm i\frac{\md}{c}
\end{equation}
are the only two cases when an electric and magnetic dipoles at the same point produce a field with well defined helicity, respectively equal to $\pm 1$. Both types of dipole must be present for it. 

I can now advance to the last part of the program and find the conditions on the polarizability tensor $P$ under which the helicity of the incident field is preserved in the scattered field due to the induced dipoles in (\ref{eqc3:M2}). First, I will change the representations of the incident fields and the induced dipoles in equation (\ref{eqc3:M2}) in order to separate the two helicity components. For the fields, $\GG_{\pm}$ is obviously the right choice. For the dipoles, in light of (\ref{eqc3:dm}), the transformation $\mathbf{q}_{\pm}=1/\sqrt{2}\left(\ed \pm i \md/c\right)$, separates the dipolar components that produce fields with well defined helicity. The transformation matrices are
\begin{equation}
		T_1=\frac{1}{\sqrt{2}}\begin{bmatrix}I & iZ\\-I & iZ\end{bmatrix},\\
		T_2=\frac{1}{\sqrt{2}}\begin{bmatrix}I & \frac{i}{c}I\\I & -\frac{i}{c}I\end{bmatrix}.
\end{equation}
With the use of these matrices, I transform equation (\ref{eqc3:M2})
\begin{equation}
	\label{eqc3:T}
		T_2\begin{bmatrix}\ed\\\md\end{bmatrix}=T_2PT_1^{-1}T_1\begin{bmatrix}\mathbf{E}\\\mathbf{H}\end{bmatrix}
\end{equation}
into
\begin{equation}
	\label{eqc3:pg}
\begin{bmatrix}\mathbf{q}_+\\\mathbf{q}_-\end{bmatrix}=T_2PT_1^{-1}\Gsix.
\end{equation}

In light of (\ref{eqc3:pg}), the condition for helicity to be preserved is that $T_2PT_1^{-1}$ must be 3$\times$3 block diagonal, which then imposes:
\begin{equation}
\label{eqc3:dual}
\boxed{\MdE=\epsilon\MmH,\ \MmE=-\frac{\MdH}{\mu}}.
\end{equation}
When (\ref{eqc3:dual}) is met, we obtain 
\begin{equation}
	\label{eqc3:txdipoledual}
	\begin{split}
	&\begin{bmatrix}\mathbf{q}_+\\\mathbf{q}_-\end{bmatrix}=
	\begin{bmatrix}\MdE-i\sqrt{\frac{\epsilon}{\mu}}\MdH & 0\\0 & \MdE+i\sqrt{\frac{\epsilon}{\mu}}\MdH\end{bmatrix}\Gsix.
	\end{split}
\end{equation}
It is clear from the derivations that a field with well defined helicity incident upon a small scatterer whose polarizability tensor meets (\ref{eqc3:txdipoledual}) will only induce the dipole of type (\ref{eqc3:dm}) that corresponds to its helicity. The resulting scattered field radiated by such dipole will preserve the helicity of the incident field. The conclusion is that, for scatterers described by their polarizability tensors, the relations in (\ref{eqc3:dual}) are the necessary and sufficient conditions for helicity preservation, or equivalently, duality symmetry. In \cite{Karilainen2012}, the authors arrive at conditions (\ref{eqc3:dual}) as one of the necessary conditions for zero backscattering of an electrically small object. In Chap \ref{chap5}, I will clarify the relationship between zero backscattering and duality. I will also give you some examples of small realistic scatterers that meet (\ref{eqc3:dual}) in Chap. \ref{chap7}. 

Right now, I would like to pay some attention to the transformation properties of $P$ under spatial inversion.

From the fundamental definitions of the electric and magnetic dipole moments associated to a charge density $\rho(\rr)$ and corresponding current density $\jj(\rr)$ \cite[Chap. 9.2,9.3]{Jackson1998}:
\begin{equation}
	\ed=\int d\rr \ \rr\rho(\rr),\ \md=\int d\rr \ \rr\times \jj(\rr)
\end{equation}
and the transformation properties of $\rho(\rr)$ and $\jj(\rr)$ under parity, one can deduce the transformation properties of $\ed$ and $\md$ under parity. With the convention that the charge density $\rho$ is a scalar under parity and $\rr\rightarrow-\rr$, then $\ed \rightarrow -\ed$. On the other hand, since $\jj(\rr)$ does change sign under parity $\md \rightarrow \md$. Combined with the transformations of the fields $\EE\rightarrow -\EE$ and $\HH\rightarrow \HH$, we can obtain the properties of $P$ under spatial inversion:
\begin{equation}
\label{eqc3:mparity}
	\begin{bmatrix}\MdE & \MdH\\\MmE & \MmH\end{bmatrix}\stackrel{\Pi}{\rightarrow}	\begin{bmatrix}\MdE & -\MdH\\-\MmE & \MmH\end{bmatrix}.
\end{equation}

Remember that $P$ is what we have assumed fully characterizes the electromagnetic response of the scatterer $\mathbb{S}$. Relations (\ref{eqc3:mparity}) allow us to draw a few conclusions. For example: Objects that are invariant under parity, like a sphere or a cylinder, must feature a polarizability tensor $P$ with $\MdH=\MmE=0$. If parity leaves the object invariant, parity must leave $P$ invariant, which, according to (\ref{eqc3:mparity}) demands $\MdH=\MmE=0$. Chiral objects, i.e. objects that are not super imposable onto their mirror reflections \footnote{The proper definition of chirality means breaking all possible spatial inversion operations; including parity, mirror reflections and mirror reflections followed by a discrete rotation. There is more discussion about chiral objects in Chap. \ref{chap6}.}, must have non-zero $\MdH$ or $\MmE$. 

Finally, just a brief comment about the two approximations, macroscopic and dipolar. Both approximations are useful in practical cases, as you will see in the application chapters. Consider a dielectric sphere embedded in vacuum. According to the macroscopic result (\ref{eqc3:dualmacro}), only when $\epsilon_{sphere}=\mu_{sphere}$ is the system dual, regardless of the size of the sphere. But, as shown in \cite{Zambrana2013b}, the dipolar approximation trades the stringent $\epsilon_{sphere}=\mu_{sphere}$ condition for a relationship between the radius and electric constant $\epsilon_{sphere}$ (assuming $\mu_{sphere}=1$). It is important to remark that the dipolar approximation is made in order to reach Eq. (\ref{eqc3:dual}). Its use in lieu of (\ref{eqc3:dualmacro}) is limited by the electromagnetic size of the sphere.

\section{Scattering off a cone}\label{secc3:scattcone}
In this section, I will analyze the scattering off a cone by means of symmetries and conservation laws.
Let me go back to the example of the cone of Fig. \ref{figc2:conetx}, and the two transformations that leave it unchanged. Rotation along the $z$ axis $R_z(\alpha)$ and mirror reflection across the $xz$ plane, $M_\yhat$.

\begin{figure}[h!]
	\begin{center}
\makeatletter{}\def\ASYprefix{}
\newbox\ASYbox
\newdimen\ASYdimen
\long\def\ASYbase#1#2{\leavevmode\setbox\ASYbox=\hbox{#1}\ASYdimen=\ht\ASYbox\setbox\ASYbox=\hbox{#2}\lower\ASYdimen\box\ASYbox}
\long\def\ASYaligned(#1,#2)(#3,#4)#5#6#7{\leavevmode\setbox\ASYbox=\hbox{#7}\setbox\ASYbox\hbox{\ASYdimen=\ht\ASYbox\advance\ASYdimen by\dp\ASYbox\kern#3\wd\ASYbox\raise#4\ASYdimen\box\ASYbox}\put(#1,#2){#5\wd\ASYbox 0pt\dp\ASYbox 0pt\ht\ASYbox 0pt\box\ASYbox#6}}\long\def\ASYalignT(#1,#2)(#3,#4)#5#6{\ASYaligned(#1,#2)(#3,#4){
}{
}{#6}}
\long\def\ASYalign(#1,#2)(#3,#4)#5{\ASYaligned(#1,#2)(#3,#4){}{}{#5}}
\def\ASYraw#1{
currentpoint currentpoint translate matrix currentmatrix
100 12 div -100 12 div scale
#1
setmatrix neg exch neg exch translate}
 
\makeatletter{}\setlength{\unitlength}{1pt}
\makeatletter\let\ASYencoding\f@encoding\let\ASYfamily\f@family\let\ASYseries\f@series\let\ASYshape\f@shape\makeatother{\catcode`"=12\includegraphics{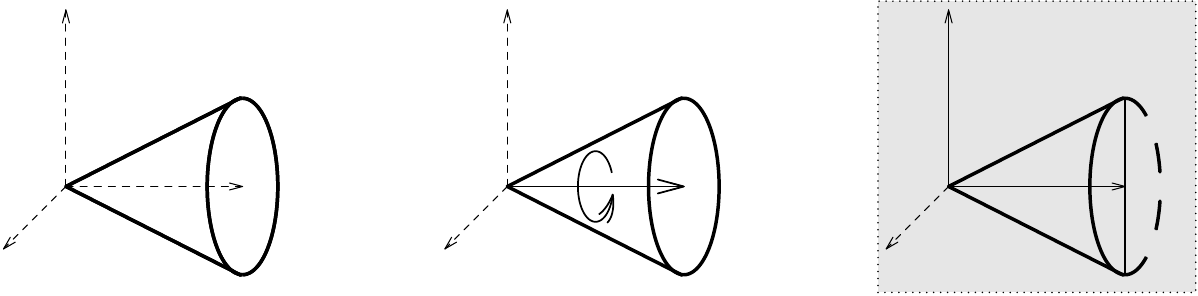}}\definecolor{ASYcolor}{gray}{0.000000}\color{ASYcolor}
\fontsize{12.000000}{14.400000}\selectfont
\usefont{\ASYencoding}{\ASYfamily}{\ASYseries}{\ASYshape}\ASYalignT(-323.250407,56.385221)(0.000000,-0.500000){0.800000 -0.000000 -0.000000 0.800000}{$\hat{x}$}
\definecolor{ASYcolor}{gray}{0.000000}\color{ASYcolor}
\fontsize{12.000000}{14.400000}\selectfont
\ASYalignT(-301.334824,27.269638)(-0.500000,-1.000000){0.800000 -0.000000 -0.000000 0.800000}{$\hat{z}$}
\definecolor{ASYcolor}{gray}{0.000000}\color{ASYcolor}
\fontsize{12.000000}{14.400000}\selectfont
\ASYalignT(-338.417113,24.394101)(-1.000000,0.218749){0.800000 -0.000000 -0.000000 0.800000}{$\hat{y}$}
\definecolor{ASYcolor}{gray}{0.000000}\color{ASYcolor}
\fontsize{12.000000}{14.400000}\selectfont
\ASYalign(-311.541057,74.246130)(-0.500000,-0.250000){0)}
\definecolor{ASYcolor}{gray}{0.000000}\color{ASYcolor}
\fontsize{12.000000}{14.400000}\selectfont
\ASYalign(-163.188005,27.269638)(-0.500000,-1.000000){$\theta$}
\definecolor{ASYcolor}{gray}{0.000000}\color{ASYcolor}
\fontsize{12.000000}{14.400000}\selectfont
\ASYalign(-183.963139,74.246130)(-0.500000,-0.250000){a)}
\definecolor{ASYcolor}{gray}{0.000000}\color{ASYcolor}
\fontsize{12.000000}{14.400000}\selectfont
\ASYalign(-56.385221,74.246130)(-0.500000,-0.250000){b)}
 
\caption[The symmetries of a cone.]{\label{figc3:conetx} The cone in (0) is left unchanged by a rotation along the $z$ axis (b), or a mirror reflection across any plane containing the $z$ axis, for example across the $xz$ plane.}
	\end{center}
\end{figure}

The task for this section is to learn as much as possible about the scattering properties of the cone by using symmetries and conservation laws. The first thing is to expand its $S$ operator in some basis. Assuming that the cone does not preserve helicity: How do we choose the basis?

The problem is that $R_z(\alpha)$ and $M_\yhat$ do not commute (Tab. \ref{tabc3:gentrans}). We cannot have a complete basis of simultaneous eigenvectors of $J_z$ and $M_\yhat$. Actually, $J_z$ anticommutes with $M_\yhat$:
\begin{equation}
	M_\yhat^{-1}J_zM_\yhat=-J_z.
\end{equation}
which means that, for an eigenstate of $J_z$, the action of $M_\yhat$ is
\begin{equation}
	M_\yhat|n\rangle=|\shortminus n\rangle,
\end{equation}
up to a phase which can be set\footnote{One gets to choose the phase resulting from the action of one of the mirror operators which contain the $\zhat$ axis \cite[Sec. 11.1.2]{Tung1985}. Once the choice is made, the rest of the operators will introduce an $n$ dependent phase.}to 1 as in the equation above. Together with the clue of mixing both helicities when the scatterer has inversion symmetries (Sec. \ref{subsecc3:tetmsym}), and the discussion on \cite[Sec. 11.1.2]{Tung1985} one may guess that
\begin{equation}
\label{eqc3:tau}
	|\tilde{n}\ \tau\rangle=\frac{1}{\sqrt{2}}\left(|n \ +\rangle+\tau|\shortminus n \ -\rangle\right),
\end{equation}
for $\tau=\pm 1$, is an interesting choice. In particular $|\tilde{n}\ \tau\rangle$ is an eigenstate of $M_\yhat$ with eigenvalue $\tau$: $M_\yhat|\tilde{n}\ \tau\rangle=\tau|\tilde{n}\ \tau\rangle.$ 

I still need two numbers to define the working basis. Energy commutes with both $R_z$ and $M_\yhat$ and will be preserved upon scattering because the cone is time translational invariant\footnote{Consequently, the eigenvalues of energy, i.e,  the frequencies $\omega$ do not change. A counter example would be a cone moving away from the source. The frequency of the scattered field would change by the Doppler effect.}. A last choice compatible with $H$, $R_z$ and $M_\yhat$ is $P_z$. $P_z$ will not be preserved by the cone since displacing it along the $z$ axis changes the system. If we complete the description of the modes in (\ref{eqc3:tau}) we find linear combinations of Bessel beams of well defined helicity:
\begin{equation}
	\label{eqc3:tau2}
	|\omega,p_z,\tilde{n}\ \tau\rangle=\frac{1}{\sqrt{2}}\left(|\omega,p_z,n \ +\rangle+\tau|\omega,p_z,\shortminus n \ -\rangle\right).
\end{equation}

I will now concentrate the discussion in the $\tilde{n}$ and $\tau$ numbers and drop $\omega$ and $p_z$ from the notation. We know that $\omega$ will be preserved and settle already for not considering the changes in $p_z$. 

The modes in (\ref{eqc3:tau2}) are eigenstates of $M_\yhat$ with eigenvalue $\tau=\pm1$ and the cone is invariant under $M_\yhat$, therefore, $\tau$ is preserved upon scattering
\begin{equation}
	\begin{split}
		&\langle \bar{\tau}\ \tilde{m}|S|\tilde{n}\ \tau\rangle=\langle \bar{\tau}\ \tilde{m}|M_\yhat^\dagger SM_\yhat|\tilde{n} \ \tau\rangle=\bar{\tau}\tau\langle \bar{\tau}\ \tilde{m}|S|\tilde{n}\ \tau\rangle\implies\\
	&\langle \bar{\tau}\ \tilde{m}|S|\tilde{n}\ \tau\rangle=0,\textrm{ unless }\tau=\bar{\tau}.
	\end{split}
\end{equation}

On the other hand, their angular momentum is not well defined, and this couples two different modes: $\tilde{n}$ and $\shortminus\tilde{n}$. 
\begin{equation}
	\label{eqc3:tosim}
	\langle \tau \ \shortminus\tilde{n}|S|\tilde{n} \ \tau\rangle\stackrel{[S,R_z]=0}{=}\frac{\tau}{2}\left(\langle -\ n|S|n\ +\rangle + \langle +\ \shortminus n|S|\shortminus n\ -\rangle\right)\stackrel{[S,M_\yhat]=0}{=}\tau\langle -\ n|S|n\ +\rangle\neq0,
\end{equation}
The two symmetries of the cone are used in (\ref{eqc3:tosim}) to relate the coupling between the $\tilde{n}$ and $\shortminus\tilde{n}$ modes to the helicity transfer coefficient for an eigenstate of $J_z$. If the cone were to have duality symmetry, this coupling would vanish.

Note also that:
\begin{equation}
	\begin{split}
		\langle \tau \ \tilde{n}|S|\tilde{n} \ \tau\rangle&=\frac{1}{2}\left(\langle +\ n|S|n\ +\rangle+\langle -\ \shortminus n|S|\shortminus n\ -\rangle\right)\\&\stackrel{[S,M_\yhat]=0}{=}\langle +\ n|S|n\ +\rangle=\langle -\ \shortminus n|S|\shortminus n\ -\rangle.
	\end{split}
\end{equation}

Consider now the transformation $T=R_z(\frac{\pi}{2n})$:
\begin{equation}
	\begin{split}
		&R_z\left(\frac{\pi}{2n}\right)|\tilde{n}\ \tau\rangle=R_z\left(\frac{\pi}{2n}\right)\frac{1}{\sqrt{2}}\left(|n \ +\rangle+\tau|\shortminus n \ -\rangle\right)=\\
	&\frac{1}{\sqrt{2}}\left(\exp\left(-i\frac{\pi}{2n}n\right)|n \ +\rangle+\tau\exp\left(i\frac{\pi}{2n}n\right)|\shortminus n \ -\rangle\right)=-i|\tilde{n}\ \shortminus\tau\rangle.
	\end{split}
\end{equation}

Therefore, $T$ interchanges the two $\tau$ modes and also leaves the cone invariant. Then, one can show that the two $\tau$ modes have the same scattering coefficients:
\begin{equation}
	\label{eqc3:equal}
	\langle \tau \ \tilde{n}|S|\tilde{n} \ \tau\rangle=\langle \tau \ \tilde{n}|T^{-1}ST|\tilde{n} \ \tau\rangle=\langle \shortminus\tau \ \tilde{n}|S|\tilde{n} \ \shortminus\tau\rangle.
\end{equation}

On the other hand, $T=R_z(\frac{\pi}{2n})$ is not going to work for $n=0$. Actually, there is no transformation that leaves the cone invariant and changes $|\tilde{0}\ \tau\rangle$ into $|\tilde{0}\ -\tau\rangle$. Then, (\ref{eqc3:equal}) does not hold and the two modes can have different scattering coefficients. If the cone had duality symmetry, these two modes would have the same scattering coefficients because:
\begin{equation}
	\langle \tau \ \tilde{0}|S|\tilde{0} \ \tau\rangle\impliesdueto{[S,D_\theta]=0}\langle \tau \ \tilde{0}|D_{\pi/2}^{-1}SD_{\pi/2}|\tilde{0} \ \tau\rangle=\langle \shortminus\tau \ \tilde{0}|S|\tilde{0} \ \shortminus\tau\rangle
\end{equation}

There is another sense in which $|\tilde{0}\ \tau\rangle$ is special: There is no other mode $|\shortminus\tilde{0}\ \tau\rangle$ to couple to. We say that the $|\tilde{0}\ \tau\rangle$ represent two invariant subspaces of dimension one. Each of them stays ``inside'' itself upon interaction with $S$. For $\tilde{n}\neq 0$, the dimensionality of the subspaces invariant under $S$ is two $|\pm\tilde{ n}\ \tau\rangle$, and would only reduce to one for a helicity preserving system.

All these findings are only based on symmetry arguments. They will hold for any system with the same symmetry properties. For example, a cylindrical hole in a layer of metal on top of a glass substrate like the one in the Preface. I will analyze such system in detail in Chap. \ref{chap4} and show you that the transmitted power of the two $|\tilde{0}\ \tau\rangle$ modes differs by more than one order of magnitude. Another example of a system with the same symmetry is the setup leading to the Stark effect: Placing an otherwise rotationally symmetric atom in a static and homogeneous electric field. The direction of the field is the only axis of rotational symmetry left. The whole system is still invariant under reflections across planes containing this axis. We should therefore expect the atom to also have a $\tau$ dependent response for $|\tilde{0}\ \tau\rangle$.

In my opinion, we have learned quite a bit about the cone without much work. Most of the applications will require just a bit more effort, but not much more. Some of them will be even simpler than the cone. 

\section{Discussion of the approach}
Studying light matter interactions by means of symmetries and conservation laws using the framework of Hilbert spaces has its virtues and its limitations. I would like to briefly comment on the ones that I have come across during my research.

The approach has important virtues. The conclusions reached by symmetry arguments are typically of very general character because the exact details of the system under consideration are not invoked. As discussed in the previous section, the scattering properties of the cone that I derived using symmetry arguments apply also to a nanohole in a metallic film or an atom in an electric field. Also, the use of abstract vectors and operators in $\mathbb{M}$ ensures that the arguments rely only on the algebraic properties of $\mathbb{M}$, common to every representation of $\mathbb{M}$. Then, the results apply to all representations of $\mathbb{M}$. In physics, this formalism is most heavily exploited in quantum mechanics. Many advances in the general theory of Hilbert spaces and linear systems have been motivated by quantum mechanics. By using the Hilbert space approach to classical electromagnetic scattering, these advances can be directly taken advantage of. Actually, the setting and results also mostly\footnote{Their direct application in single photon experiments with absorbing scatterers would require some modifications because of the finite probability that the photon is absorbed and the scattering is zero.} apply to single photon states since they have the same algebraic structure as classical fields. They do not apply to multiphoton states, which require an extension to products of spaces, i.e. $\mathbb{M}_2=\mathbb{M}\otimes\mathbb{M}$ for a two photon state, hence the limitation to linear scatterers. 

Arguably, the most serious limitation of the approach is that it is, quite often, only qualitative and not quantitative. When a system has a symmetry, the corresponding conservation law allows to make quantitative statements. When the system lacks a symmetry we cannot, in general, make quantitative statements about the changes of the non-preserved eigenvalues and eigenvectors. Other means of analysis involving the detailed description of the system, not only its symmetries, are then typically needed to obtain quantitative results.

With respect to the duality restoration conditions in (\ref{eqc3:epsmu}) and (\ref{eqc3:dualmacro}), their validity is that of the approximations implicit in the macroscopic equations. Jackson argues that the macroscopic equations are not valid for objects less than 10 nanometers in size \cite[Chap. 6.6]{Jackson1998}. The applicability of the duality condition for the dipolar limit (\ref{eqc3:dual}) is correspondingly bounded by the ratio of the size of the object to the wavelength of the radiation. 

Finally, I would like to mention again two properties of helicity, one for the fields and one for the scatterers. They captivate me because of their extreme simplicity in a normally very complex (scattering) world. 

For a field of well defined helicity and considering positive frequencies only, one of the two $\pm \EE+iZ\HH$ combinations is always zero everywhere, no matter how complicated may $\EE$ and $\HH$ be. The orthogonality of the two helicity components of the field manifests itself in a blunt way: No need to perform the integrals typically involved in the inner product. I can not think of another operator whose eigenstates have this property.

For a scatterer, the fact that it preserves or does not preserve helicity is not affected by translations or rotations, which renders this property immune to possible positioning and orientation errors in practical scenarios.

\begin{appendices}
	\chapter{Duality symmetry at media boundaries}\label{appc3:shape}
Let me consider an inhomogeneous medium $\Omega$ composed of several material domains with arbitrary geometry. I assume that each domain $i$ is homogeneous and isotropic, and fully characterized by its electric $\epsilon_i$ and magnetic $\mu_i$ constants (with $\epsilon_0=\mu_0=1$). In each domain, the constitutive relations are hence $\mathbf{B}=\mu_i\mathbf{H},\ \mathbf{D}=\epsilon_i\mathbf{E}$, and the curl equations for monochromatic fields read

\begin{equation}
\nabla \times \mathbf{H} = -i\omega\epsilon_i \mathbf{E},\
\nabla \times \mathbf{E} = i\omega\mu_i\mathbf{H}.
\end{equation}
Using $\Lambda=k^{-1}\nabla \times$ (\ref{eqc3:helnabla}) and $\omega=k_0=k/\sqrt{\epsilon_i\mu_i}$ we obtain
\begin{equation}
\Lambda \mathbf{H} = -i\sqrt{\frac{\epsilon_i}{\mu_i}} \mathbf{E},\ \Lambda \mathbf{E} = i\sqrt{\frac{\mu_i}{\epsilon_i}} \mathbf{H}.
\end{equation}
Note that to arrive at this result, the fact that the wavenumber in each medium is $k=k_0\sqrt{\epsilon_i\mu_i}$ has to be used in the expression of the helicity operator. Now, we can normalize the electric field $\mathbf{E}\rightarrow \sqrt{\frac{\epsilon_i}{\mu_i}}\mathbf{E}$, to show that inside each of the domains, we can recover the exact form of Maxwell's equations in free space. Clearly, if we want to have a consistent description for the whole medium $\Omega$, the normalization can only be done when all the different materials have the same ratio $\frac{\epsilon_i}{\mu_i}=\alpha\ \forall \ i$. In this case, the electromagnetic field equations on the whole medium $\Omega$ are invariant under the duality transformations of (\ref{eqc3:gendual}).

I need to study the matching of the fields at the interfaces between the different domains, where the material constants are discontinuous. In the absence of free currents and charges, the electromagnetic boundary conditions impose the following restrictions on the fields $\nhat \times (\mathbf{E_1}-\mathbf{E_2})=\mathbf{0}$, $\nhat \times (\mathbf{H_1}-\mathbf{H_2})=\mathbf{0}$, $\nhat \cdot (\mathbf{D_1}-\mathbf{D_2})=0$ and $\nhat \cdot (\mathbf{B_1}-\mathbf{B_2})=0$. Where $\nhat$ is the unit vector perpendicular to the interface. The boundary conditions can be seen as a real space point to point transformation of the fields. For example, at a particular point $\rr$ on the interface between domains 1 and 2, the boundary conditions may be interpreted as the following linear transformation:
\begin{equation}
\label{eqc3:T}
\begin{bmatrix} \mathbf{E_2}(\rr)\\\mathbf{H_2}(\rr)\end{bmatrix}=
\mathrm{diag}(1,1,\frac{\epsilon_1}{\epsilon_2},1,1,\frac{\mu_1}{\mu_2})
\begin{bmatrix} \mathbf{E_1}(\rr)\\\mathbf{H_1}(\rr)\end{bmatrix},
\end{equation}
where I have oriented our Cartesian reference axis so that $\zhat=\nhat$.

With the duality transformation in matrix form:
\begin{equation}
	\begin{bmatrix} \mathbf{E}_\theta\\\mathbf{H}_\theta\end{bmatrix}=\begin{bmatrix} I\cos\theta&-I\sin\theta\\I\sin\theta&I\cos\theta\end{bmatrix}\begin{bmatrix} \mathbf{E}\\\mathbf{H}\end{bmatrix}=D_{\theta}\begin{bmatrix} \mathbf{E}\\\mathbf{H}\end{bmatrix}.\nonumber
\end{equation}
It is a trivial exercise to check that the transformation matrix of (\ref{eqc3:T}) commutes with $D_\theta$ if and only if $\epsilon_1/\mu_1=\epsilon_2/\mu_2$. In such case, the fields in each of the two media can be transformed by $D_\theta$ while still meeting the boundary conditions at point $\rr$. I can now vary $\rr$ to cover all the points of the interface and repeat the same argument: The fact that $D_\theta$ does not depend on the spatial coordinates allows to reorient the reference axis as needed to follow the shape of the interface between two media ($\nhat=\zhat$). The derivation is hence independent of the shape of the interface, and we can say that the boundary conditions are invariant under duality transformations when $\epsilon_1/\mu_1=\epsilon_2/\mu_2$.
The above derivations show that both the equations and the boundary conditions in $\Omega$ are invariant under (\ref{eqc3:gendual}) when 
\begin{equation}
\label{eqc3:epsmuA}
\epsilon_i/\mu_i=\text{constant for all domain }i.
\end{equation}
The conclusion is that, independently of the shapes of each domain, a piecewise homogeneous and isotropic system has an electromagnetic response that is invariant under duality transformations if and only if all the materials have the same ratio of electric and magnetic constants. In this case, since helicity is the generator of duality transformations, the system preserves the helicity of the electromagnetic field interacting with it. 

\chapter{Multilayered systems and Mie scattering}\label{appc3:fm}
In this section I check that condition (\ref{eqc3:epsmu}) in the main text
\begin{equation}
\label{eqc3:epsmuAA}
\epsilon_i/\mu_i=\alpha \ \forall \ \textrm{domain }i,
\end{equation}
is equivalent to helicity preservation in two analytically solvable scattering problems: A planar multilayer system and a sphere. 

I will assume that the electromagnetic response of all media can be modeled with constitutive relations of the type: 
\begin{equation}
\label{eqc3:constrel}
\mathbf{B}=\mu\mathbf{H}, \ \mathbf{D}=\epsilon\mathbf{E},
\end{equation}
where $\mu$ and $\epsilon$ are scalars.

\section{Planar multilayered systems}
Planar multilayered systems are inhomogeneous systems that extend to infinity in two spatial directions, while having finite or semi-infinite domains in the third spatial direction (say $z$). These systems are best analyzed using plane waves and the Fresnel equations  for their reflection and transmission. The equations can be found, for example, in \cite[Chap. 2.8.1]{Novotny2006}, and are valid for isotropic layers. 

Let me first consider the reflection off one of the interfaces of the multilayered system in terms of the $\mathbf{s}$ and $\mathbf{p}$ polarizations. When a $\mathbf{s}$ ($\mathbf{p}$) polarized plane wave reflects on a planar interface, its energy $k$, transverse momentum $[p_x,p_y]$ and polarization character ($\mathbf{s}$ or $\mathbf{p}$) remain unchanged (see Sec. \ref{subsecc3:tetmsym}). The reflection coefficients depend on $k$, $p_{\rho}=\sqrt{p_x^2+p_y^2}$, the polarization and the electric and magnetic constants ($\epsilon,\mu$) of the two media: 
\begin{equation}
\label{eqc3:fresnel}
r^s=\frac{\mu_c p^z_c - \mu_m p^z_m}{\mu_c p^z_c + \mu_m p^z_m},\ r^p=\frac{\epsilon_c p^z_c - \epsilon_m p^z_m}{\epsilon_c p^z_c + \epsilon_m p^z_m},
\end{equation}
where $r^s,r^p$ are the reflection coefficients for the $\mathbf{s}$ and $\mathbf{p}$ polarizations, subscript $m$ refers to the initial medium, subscript $c$ refers to the second medium where the plane wave reflects from and $p^z=\sqrt{k^2-p_x^2-p_y^2}=\sqrt{k^2+p_{\rho}^2}$. 

From the results in Sec. \ref{secc3:helprestetm}, the condition for a plane wave of well defined helicity to preserve it after reflection is that $r^s=r^p$ independently of $k^2$, $p_x$ and $p_y$. It is easy to see that, in order to meet such condition in (\ref{eqc3:fresnel}), the electric and magnetic constants of both media must fulfill:

\begin{equation}
\label{eqc3:mueqeps}
\frac{\epsilon_{c}}{\mu_{c}}=\frac{\epsilon_{m}}{\mu_{m}}.
\end{equation}

By dividing all terms in $r^s$ by $\mu_m$ and all terms in $r^p$ by $\epsilon_m$, this conclusion is reached immediately.

Using the transmission coefficient formulas in \cite[Chap. 2.8.1]{Novotny2006}, it can be verified that this relation ensures the helicity conservation of the wave transmitted onto the second medium as well: by successively applying this method to the different layers we find that (\ref{eqc3:mueqeps}) is the condition that all layers must fulfill so that helicity is preserved in the whole system. By demanding helicity preservation in the multilayer we have reached Eq. (\ref{eqc3:epsmu}).

\section{Mie scattering}
 The Mie scattering theory treats the problem of a plane wave impinging on an isotropic homogeneous sphere embedded in a different isotropic and homogeneous lossless medium. The problem is solved by decomposing the incident plane wave in terms of multipolar fields, that is, waves of defined energy $k$, squared angular momentum $J^2$, angular momentum along an axis $J_z$, and parity $\Pi$. Each of these modes preserves all of its characteristics upon scattering off the sphere. It reflects off with a scattering coefficient which depends on $J^2$, $\Pi$ and the electric and magnetic constants of the sphere and surrounding medium \cite[Chap. 9.25]{Stratton1941}: 
\begin{equation}
\label{eqc3:mie}
	\begin{split}
		a_s&=\frac{\mu_mq^2j_s(qx)[xj_s(x)]'-\mu_cj_s(x)[qxj_s(qx)]'}{\mu_mq^2j_s(qx)[xh^{(1)}_s(x)]'-\mu_ch^{(1)}_s(x)[qxj_s(qx)]'},\\
  b_s&=\frac{\mu_cj_s(qx)[xj_s(x)]'-\mu_mj_s(x)[qxj_s(qx)]'}{\mu_cj_s(qx)[xh^{(1)}_s(x)]'-\mu_mh^{(1)}_s(x)[qxj_s(qx)]'},\
	\end{split}
\end{equation}

where $a_n$ and $b_n$ are the scattering coefficients for modes with $\Pi=\pm1$ and $s(s+1)$ as the eigenvalue of $J^2$, subindex $c$ refers to the sphere and subindex $m$ to the surrounding medium, $q=\frac{k_c}{k_m}=\sqrt{\frac{\epsilon_c\mu_c}{\epsilon_m\mu_m}}$, $x=k_mr=\frac{2\pi r}{\lambda_0}\sqrt{\epsilon_m\mu_m}$, $r$ is the sphere radius and $j_s(\rho)$ and $h^{(1)}_s(\rho)$ are the spherical Bessel and spherical Hankel functions of the first kind, respectively. 

Again, the condition for helicity conservation after scattering is that the two scattering coefficients for the two different parity eigenmodes must be identical. 

By dividing the numerators and denominators of $a_s$ by $\mu_c$, and those of $b_s$ by $\mu_m$, it is seen that the two expressions are equal when:

\begin{equation}
\frac{\mu_m}{\mu_c}q^2=\frac{\mu_c}{\mu_m}.
\end{equation}

Then, the condition for helicity conservation reads again,
\begin{equation}
\label{eqc3:translationsimple}
\frac{\epsilon_{c}}{\mu_{c}}=\frac{\epsilon_{m}}{\mu_{m}}.
\end{equation}
\end{appendices}
 
\makeatletter{}\chapter[SAM and OAM: A symmetry perspective]{Spin and orbital angular momentum: A symmetry perspective}
\label{chap4}

\epigraph{{\em ... and he again looked somewhat puzzled, as if I had asked him to smell a higher symmetry. But he complied courteously, and took it to his nose.}}{Oliver Sacks, ``The Man Who Mistook His Wife for a Hat and Other Clinical Tales''}

In this chapter, I use symmetries and conservation laws to pinpoint the underlying reasons for some notable effects that can be observed in focusing and scattering. Note that the action of a lens can also be understood as a scattering situation in the sense described in Sec. \ref{secc3:scattop}. The observation of optical vortices in focusing and scattering is commonly attributed to the transfer of electromagnetic spin angular momentum to electromagnetic orbital angular momentum. In this chapter, I prove that the underlying reason is very different in each case: Breaking of transverse translational symmetry in focusing, and breaking of duality symmetry in scattering. The inconsistency of the state of the art explanation can be traced back to the use of operators (spin $S_z$ and orbital $L_z$) which, in the general case, break the transversality of the fields they act on. These two operators are hence not operators in the space of transverse fields $\mathbb{M}$. On the other hand, their sum, i.e. the total angular momentum $J_z=S_z+L_z$, is. The chapter also contains a study of two transverse operators, acting within $\mathbb{M}$ which also sum to the total angular momentum $J_z$. In particular, I derive the transformations that they generate. These transformations are not rotations, as expected since these other pair of operators do not obey the angular momentum commutation rules. The transformations in question are related to frequency and helicity dependent translations.

Within the chapter, I analyze a scattering experiment by means of symmetries and conservation laws. This example shows that the framework developed in Chap. \ref{chap3} can be applied in practice in a straightforward manner.

\begin{figure}[h]
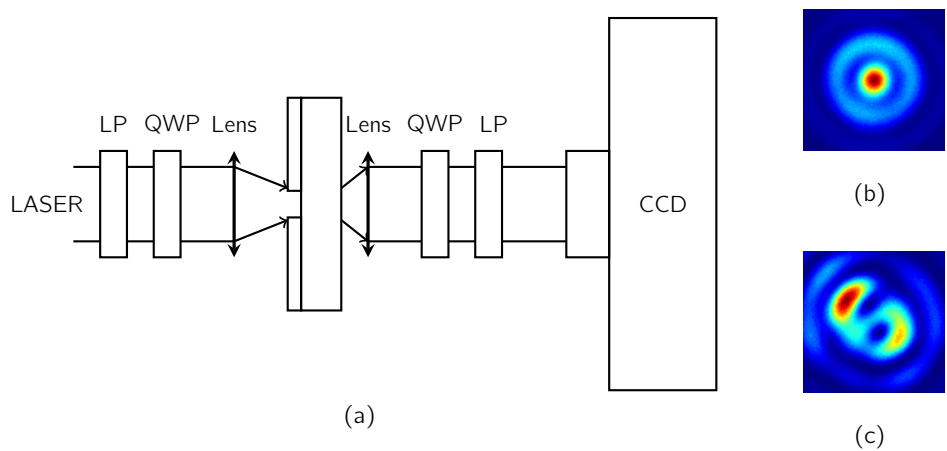

	\begin{minipage}{0.80\linewidth}
\centering
\subfloat[]{\begin{tikzpicture}[thick,scale=1.41,every node/.style={scale=0.8}]
		\makeatletter{}\draw (-0.25,0.35) -- (0,0.35);\draw (-0.5,0) node {LASER};
\draw (-0.25,-0.35) -- (0,-0.35);
\draw (0,-0.5) rectangle (0.25,0.5);\draw (0.125,0.75) node {LP};
\draw (0.25,0.35) -- (0.5,0.35);
\draw (0.25,-0.35) -- (0.5,-0.35);
\draw (0.5,-0.5) rectangle (0.75,0.5);\draw (0.675,0.75) node {QWP};
\draw (0.75,0.35) -- (1.25,0.35);
\draw (0.75,-0.35) -- (1.25,-0.35);
\draw[very thick,<->,>=stealth] (1.25,-0.5) -- (1.25,0.5);\draw (1.25,0.75) node {Lens};
\draw[->] (1.25,-0.35) -- (1.75,-0.15);
\draw[->] (1.25,0.35) -- (1.75,0.15);
\draw[shift={(0.5,0)},scale=0.25]  (5,0.5) rectangle (5.5,4);
\draw[shift={(0.5,0)},scale=0.25] (5,-0.5) rectangle (5.5,-4);
\draw[shift={(0.5,0)},scale=0.25] (5.5,-4) rectangle (7,4);
\draw[->] (2.25,0.15) -- (2.5,0.35);
\draw[->] (2.25,-0.15) -- (2.5,-0.35);
\draw[very thick,<->,>=stealth] (2.5,-0.5) -- (2.5,0.5);\draw (2.5,0.75) node {Lens};
\draw (2.5,0.35) -- (3,0.35);
\draw (2.5,-0.35) -- (3,-0.35);
\draw (3,-0.5) rectangle (3.25,0.5);\draw (3.125,0.75) node {QWP};
\draw (3.25,-0.35) -- (3.5,-0.35);
\draw (3.25,0.35) -- (3.5,0.35);
\draw (3.5,-0.5) rectangle (3.75,0.5);\draw (3.675,0.75) node {LP};
\draw (3.75,0.35) -- (4.35,0.35);
\draw (3.75,-0.35) -- (4.35,-0.35);
\draw (4.35,-0.5) rectangle (4.75,0.5);
\draw (4.75,-1.75) rectangle (5.75,1.75);\draw (5.25,0) node {CCD};
 
	\end{tikzpicture}}
\end{minipage}
	\begin{minipage}{0.15\linewidth}
		\subfloat[]{\includegraphics[scale=0.15]{direct.pdf}}\\
		\subfloat[]{\includegraphics[scale=0.15]{cross.pdf}}
	\end{minipage}
	\caption[Helicity changes in nanohole scattering: Setup and results.]{\label{figc4:setup} (a) Schematic representation of the experimental setup. The output of a He-Ne laser (with wavelength equal to 632.8 nm) is passed through a set of Linear Polarizer (LP) and Quarter Wave Plate (QWP) and focused with a microscope objective of Numerical Aperture (NA) of 0.5 onto the sample. The sample is a cylindrical hole of 400nm diameter on a 200 nm thick gold layer on top of a 1 mm glass substrate. The transmitted light is collected and collimated with another microscope objective of the same NA, analyzed with another set of QWP and LP, and imaged with a Charged Coupled Device (CCD) camera. (b) CCD image when the axis of the second LP is set to select the input polarization. (c) CCD image when the axis of the second LP is set to select the polarization orthogonal to the input one.}
\end{figure}

Figure \ref{figc4:setup} shows the experimental setup and results of a nanohole scattering experiment\footnote{This was my first (and so far only) contact with experimental physics. Assisted by our common advisor A. Prof. Molina-Terriza, my colleague Xavier Zambrana-Puyalto and I performed the experiment around June 2011 (see the Preface).}. The image in Fig. \ref{figc4:setup}-(b) was obtained when the axis of the second linear polarizer was set to select the input polarization. It shows an Airy like pattern. This is the expected result from the diffraction of a focused Gaussian laser from a sub wavelength hole. The image in Fig. \ref{figc4:setup}-(c) was obtained when the axis of the second polarizer was set to select the polarization orthogonal to the input one. It shows a very different mode from that of \ref{figc4:setup}-(b). Similar results have been reported in \cite[Fig. 4]{Gorodetski2009}, \cite[Figs. 2(c)-(d)]{Vuong2010}, and also more recently in \cite{Chimento2012,Brandao2013}. The two intensity minima of \ref{figc4:setup}-(c) correspond to two phase singularities (optical vortices) \cite{Molina2000}, and their appearance is attributed to spin to orbital angular momentum transfer \cite{Gorodetski2009,Vuong2010,Bliokh2011}. I will now summarize the argument that is commonly used to explain these experimental results.

The total angular momentum operator can be written as the sum of two operators called spin angular momentum $\SSS$ and orbital angular momentum $\LL$:
\begin{equation}
	\JJ=\SSS+\LL.
\end{equation}
Their expressions in the coordinate representation are:
\begin{equation}
	S_k=-i\epsilon_{knm},\ \LL=-i\rr\times\nabla
\end{equation}
where $\epsilon_{knm}$ is the totally antisymmetric tensor with $\epsilon_{123}=1$. Their third components are:
\begin{equation}
	S_z=\begin{pmatrix}0&-i&0\\i&0&0\\0&0&0\end{pmatrix},\ L_z=-i\partial_\theta,
\end{equation}
where $\theta = \arctan(y/x)$. 

The spin to orbital angular momentum transfer argument \cite{Bliokh2011,Marrucci2006} goes as follows. The whole setup is cylindrically symmetric, therefore $J_z=S_z+L_z$ has to be preserved. $S_z$ is associated with real space circular polarization states with eigenvalues equal to $\pm 1$. Since the cross polarization measurement resulting in \ref{figc4:setup}-(c) measures a component of changed $S_z$, say from +1 to -1, $L_z$ must pick up the difference of 2 units of angular momentum. This difference of 2 between the orbital angular momentum of the output and that of the input is what causes the appearance of the two intensity minima of Fig. \ref{figc4:setup}-(c). The minima are the locations of two phase singularities (optical vortices), which reflect the increase of the eigenvalue of $L_z$ from 0 to 2.

The conversion between spin and orbital angular momentum is used to explain phase singularities in numerical simulations of tightly focused fields as well \cite{Zhao2007,Nieminen2008,Yao2011}. A detailed discussion about spin to orbital angular momentum conversion can be found in \cite{Bliokh2011}.

From the point of view of symmetries and conservation laws, the spin to orbital angular momentum conversion argument is not valid. This is immediately clear from the fact that, when considered separately, the components of $\SSS$ and $\LL$ are not operators in $\mathbb{M}$. They break the transversality requirement and throw vectors out of the Hilbert space of transverse Maxwell fields. In general we have that, for $|\Psi\rangle$ in $\mathbb{M}$:
\begin{equation}
	(S_z+L_z)|\Psi\rangle \in \mathbb{M},\ S_z|\Psi\rangle \notin \mathbb{M}, \ L_z|\Psi\rangle \notin \mathbb{M}.
\end{equation}
Fig. \ref{figc4:breaking} illustrates the transversality breaking action.

Another way of thinking of this is to consider that, after $S_z$ or $L_z$ act on a solution of the free-space Maxwell's equations, it generally ceases being a solution. Yet another way is to observe that eigenstates of $S_z$ cannot be used to build a basis for $\mathbb{M}$. This follows from the fact that eigenstates of $S_z$ must have zero component in the real space $\zhat$ polarization and can therefore not be used to expand a general field. 
\begin{figure}[h]
	\makeatletter{}\def\ASYprefix{}
\newbox\ASYbox
\newdimen\ASYdimen
\long\def\ASYbase#1#2{\leavevmode\setbox\ASYbox=\hbox{#1}\ASYdimen=\ht\ASYbox\setbox\ASYbox=\hbox{#2}\lower\ASYdimen\box\ASYbox}
\long\def\ASYaligned(#1,#2)(#3,#4)#5#6#7{\leavevmode\setbox\ASYbox=\hbox{#7}\setbox\ASYbox\hbox{\ASYdimen=\ht\ASYbox\advance\ASYdimen by\dp\ASYbox\kern#3\wd\ASYbox\raise#4\ASYdimen\box\ASYbox}\put(#1,#2){#5\wd\ASYbox 0pt\dp\ASYbox 0pt\ht\ASYbox 0pt\box\ASYbox#6}}\long\def\ASYalignT(#1,#2)(#3,#4)#5#6{\ASYaligned(#1,#2)(#3,#4){
}{
}{#6}}
\long\def\ASYalign(#1,#2)(#3,#4)#5{\ASYaligned(#1,#2)(#3,#4){}{}{#5}}
\def\ASYraw#1{
currentpoint currentpoint translate matrix currentmatrix
100 12 div -100 12 div scale
#1
setmatrix neg exch neg exch translate}
 
	\makeatletter{}\setlength{\unitlength}{1pt}
\makeatletter\let\ASYencoding\f@encoding\let\ASYfamily\f@family\let\ASYseries\f@series\let\ASYshape\f@shape\makeatother{\catcode`"=12\includegraphics{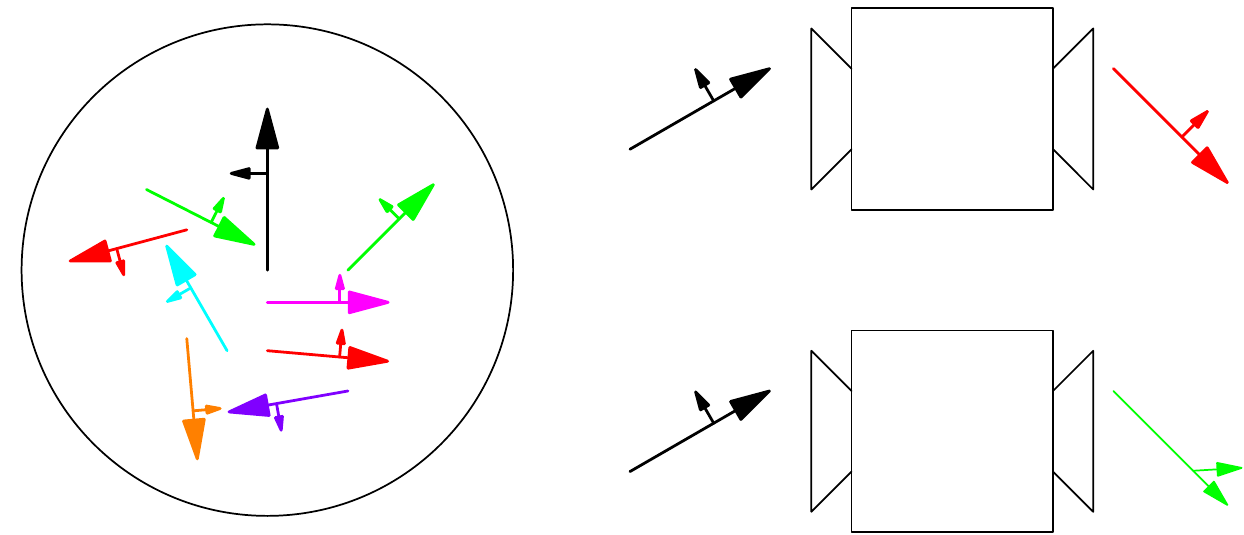}}\definecolor{ASYcolor}{gray}{0.000000}\color{ASYcolor}
\fontsize{12.000000}{14.400000}\selectfont
\usefont{\ASYencoding}{\ASYfamily}{\ASYseries}{\ASYshape}\ASYalign(-351.781429,145.838837)(-0.500000,-0.250000){a)}
\definecolor{ASYcolor}{gray}{0.000000}\color{ASYcolor}
\fontsize{12.000000}{14.400000}\selectfont
\ASYalign(-83.899694,122.544773)(-0.500000,-0.500000){{\Large $X$}}
\definecolor{ASYcolor}{gray}{0.000000}\color{ASYcolor}
\fontsize{12.000000}{14.400000}\selectfont
\ASYalign(-153.781886,145.838837)(-0.500000,-0.250000){b)}
\definecolor{ASYcolor}{gray}{0.000000}\color{ASYcolor}
\fontsize{12.000000}{14.400000}\selectfont
\ASYalign(-83.899694,29.368517)(-0.500000,-0.320001){$S_z$ or $L_z$}
\definecolor{ASYcolor}{gray}{0.000000}\color{ASYcolor}
\fontsize{12.000000}{14.400000}\selectfont
\ASYalign(-153.781886,52.662581)(-0.500000,-0.250000){c)}
 
	\caption[Hilbert space of transverse Maxwell fields.]{\label{figc4:breaking} a) Pictorial representation of the Hilbert space of transverse fields $\mathbb{M}$. The transverse character is represented by the orthogonality of the momentum (long arrows) and polarization (short arrows) of the members of $\mathbb{M}$. (b) An operator $X$ in $\mathbb{M}$ takes transverse fields and converts them into other transverse fields. Operators like the components of $\JJ$ and $\PP$ obey this rule. (c) Components of $\SSS$ or $\LL$ do not obey the transversality rule. They are not operators in $\mathbb{M}$. The resulting fields are not solutions of Maxwell's equations, in the general case.}
\end{figure}

Accordingly, $S_i$ and $L_i$ do not generate any meaningful transformation in electromagnetism. Indeed, the question of what symmetries are broken by a scatterer that cause the change in the eigenvalues of $S_z$ and $L_z$ is not well posed.

In Sec. \ref{secc4:lambdajz}, I will prove that what Fig. \ref{figc4:setup}-(c) shows is helicity changing in the scattering off the sample due to the breaking of duality symmetry. I will also prove that, in focusing, the effects attributed to spin to orbital angular momentum transfer are actually due to the fact that the lens changes the momentum of the field due to its lack of transverse translation invariance.

The rest of the chapter is organized as follows. In Sec. \ref{secc4:shatlhat} I will analyze two alternative operators, which are operators in $\mathbb{M}$, that have been shown to also sum to $\JJ$. Then, I will investigate the correlation between angular momentum and polarization. Finally, in Sec. \ref{secc4:inconsis} I will reflect on the inconsistency of the spin to orbital angular momentum explanation, and its relation with the separate consideration of $\SSS$ and $\LL$.

\section[Symmetry analysis based on $J_z$ and $\Lambda$]{Symmetry analysis based on helicity and angular momentum}\label{secc4:lambdajz}
The section of the setup in Fig. \ref{figc4:setup}-a) between the two sets of wave plates has the same symmetries as the cone in Sec. \ref{secc3:scattcone}: Rotation along the $z$ axis $R_z(\alpha)$ and mirror reflection across any plane containing the $z$ axis, for example the $xz$ plane. Instead of choosing the mirror symmetric modes of Sec. \ref{secc3:scattcone} I will use Bessel beams of well defined helicity (\ref{eqc3:CD}) to analyze the experiment:
\begin{equation}
	\makebox[\textwidth][c]{$
\begin{split}
\label{eqc4:CD}
&|k,p_z,n,-\rangle\equiv\Cnkrho(\rho,\theta,z)=\sqrt{\frac{\krho}{2\pi}}i^n\exp(i(\kz z+n\theta))\times\\
&\left(\frac{i}{\sqrt{2}}\left((1+\frac{\kz}{k})J_{n+1}(\krho\rho)\exp(i\theta)\rhat+(1-\frac{\kz}{k})J_{n-1}(\krho\rho)\exp(-i\theta)\lhat \right)-\frac{\krho}{k}J_n(\krho\rho)\zhat\right),\\
&|k,p_z,n,+\rangle\equiv\Dnkrho(\rho,\theta,z)=\sqrt{\frac{\krho}{2\pi}}i^n\exp(i(\kz z+n\theta))\times\\
&\left(\frac{i}{\sqrt{2}}\left((1-\frac{\kz}{k})J_{n+1}(\krho\rho)\exp(i\theta)\rhat+(1+\frac{\kz}{k})J_{n-1}(\krho\rho)\exp(-i\theta)\lhat \right)+\frac{\krho}{k}J_n(\krho\rho)\zhat\right).\\
\end{split}
$}
\end{equation}
The idea is to analyze the transmission through the system block by block. The input and output fields can always be expanded in the (\ref{eqc4:CD}) basis, and the symmetries of each block determine whether the eigenvalues of the four operators ($H,P_z,J_z$ and $\Lambda$) can or cannot change. In all the blocks, the self-similar component of the total field\footnote{The scattering operator $S$ in (\ref{eqc3:IT}) has a term proportional to the identity which reflects the fact that a portion of the incident field is unchanged by the scatterer.} is much smaller than the portion of the field affected by the action of the scatterer. I will ignore the self-similar portion.

I start with the Gaussian laser going through the linear polarizer and quarter wave plate (see Fig. \ref{figc4:prep}).
\begin{figure}[h]
	\centering
\begin{tikzpicture}[thick,scale=2,every node/.style={scale=1}]
\draw (-0.25,0.35) -- (0,0.35);\draw (-0.5,0) node {LASER};
\draw (-0.25,-0.35) -- (0,-0.35);
\draw (0,-0.5) rectangle (0.25,0.5);\draw (0.125,0.75) node {LP};
\draw (0.25,0.35) -- (0.5,0.35);
\draw (0.25,-0.35) -- (0.5,-0.35);
\draw (0.5,-0.5) rectangle (0.75,0.5);\draw (0.675,0.75) node {QWP};
\draw (0.75,0.35) -- (1.25,0.35);
\draw (0.75,-0.35) -- (1.25,-0.35);
\end{tikzpicture}
\caption[Setup analysis: Preparation.]{\label{figc4:prep} 632.8 nm Gaussian He-Ne laser input to a linear polarizer (LP) and a quarter wave plate (QWP). To a good approximation, the output is a linear combination of Bessel modes of well defined energy, angular momentum and helicity, with a narrow distribution of longitudinal momentum components around $p_z=k$.}
\end{figure}
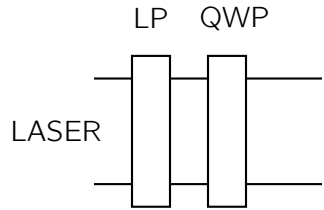
Since the input beam is collimated, the momentum is mostly along the optical axis ($z$): $ \kz\approx k$. The amplitudes of the components with small transverse momentum $\frac{\krho}{k}=\sqrt{1-\left(\frac{\kz}{k}\right)^2}<<1$ are much larger than those with large transverse momentum. 

In this case, it is useful to consider the expressions (\ref{eqc4:CD}) when $\smallkrho$:
\begin{equation}
	\label{eqc4:smallCD}
\begin{split}
\Cnkrho(\rho,\theta,z)&\approx\sqrt{\frac{\krho}{\pi}}i^{n+1}\exp(i\kz z)J_{n+1}(\krho\rho)\exp(i\theta (n+1))\rhat,\\
\Dnkrho(\rho,\theta,z)&\approx\sqrt{\frac{\krho}{\pi}}i^{n+1}\exp(i\kz z)J_{n-1}(\krho\rho)\exp(i\theta (n-1))\lhat.
\end{split}
\end{equation}

Assuming that the wave plates are set to select the real space $\rhat$ polarization, and since the Gaussian beam does not have a phase singularity $\exp(is\theta)$ in its dominant real space polarization component, the output must consist mostly of $C$ type modes with $n=-1$:

\begin{equation}
\label{eqc4:last}
\sum_{\smallkrho} c^{input}_{-1,\krho}|k,p_z,-1,-\rangle.
\end{equation}
Moving on to the microscope objective (see Fig. \ref{figc4:focusing}).
\begin{figure}[h]
	\centering
\begin{tikzpicture}[thick,scale=2,every node/.style={scale=1}]
\draw (0.75,0.35) -- (1.25,0.35);
\draw (1,0) node {\small $c_{-1}^{\smallkrho}$};
\draw (0.75,-0.35) -- (1.25,-0.35);
\draw[very thick,<->,>=stealth] (1.25,-0.5) -- (1.25,0.5);\draw (1.25,0.75) node {Lens};
\draw[->] (1.25,-0.35) -- (1.75,-0.15);
\draw[->] (1.25,0.35) -- (1.75,0.15);
\end{tikzpicture}
\caption[Setup analysis: Focusing]{\label{figc4:focusing} Microscope objective. Since it preserves energy, angular momentum and helicity (see text), its action changes only the distribution of longitudinal (transverse) momenta, increasing the weight of the smaller (higher) components. 
}
\end{figure}
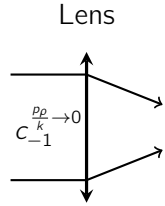

A perfect microscope objective preserves the energy and $J_z$ since its response is time invariant and the lens is cylindrically symmetric. The fact that a perfect lens preserves helicity to a very good approximation is not obvious, but has been discussed before \cite{Bliokh2011}. In \cite[App. C]{FerCor2012b}, you can find a formal proof using the language of symmetries and conservation laws. I will briefly go over the main ideas in it. 

The most commonly used approximation for modeling a high numerical aperture lens is called the aplanatic model \cite{Richards1959}. Actually, microscope objective manufacturers strive to make them behave as aplanatic lenses. Assuming monochromatic fields in the aplanatic model and a collimated input, the plane wave decomposition of the field after the lens in real space cylindrical coordinates can be written in terms of TE/TM polarization components $(\shat,\phat)$ (see \ref{secc3:shatphat}) as:
\begin{equation}
	\makebox[\textwidth][c]{$
\label{eqc4:aplanatic_us}
\EE_{out}(\rho,\phi,z)= \int_0^{\pi} \sin \thetak d\thetak \int_-\pi^{\pi} d\phik\left( g_s(\thetak, \phik )\shat_{\thetak\phik} + g_p(\thetak, \phik )\phat_{\thetak\phik}\right)\exp(i\pp\cdot\rr),
$}
\end{equation}
where $\thetak=\arccos(\kz/k)$ and $\phik=\arctan(p_y/p_x)$. The dependence from the input field comes in through the scalars
\begin{equation}
\label{eqc4:gsgp}
\begin{split}
g_s(\thetak, \phik )&= t(\thetak) \left(\pwsthetaphi{\thetak=0}{\phik}\cdot \EE_{in}( f\sin(\thetak),\phik,z=z_0) \right),\\
g_p(\thetak, \phik )&= t(\thetak) \left(\pwpthetaphi{\thetak=0}{\phik}\cdot \EE_{in}( f\sin(\thetak),\phik,z=z_0) \right),
\end{split}
\end{equation}
where $f$ is the focal number of the lens and $z_0$ a reference plane (see \cite{Richards1959} for details on the model).

Equations (\ref{eqc4:aplanatic_us})-(\ref{eqc4:gsgp}) constitute a mapping between real space polarization components of the collimated input field and momentum space polarization components of the focused output field. For my purpose here, it suffices to highlight that the key assumption is that the $t(\thetak)$ parameter in (\ref{eqc4:gsgp}) is the same for both $s$ and $p$ polarizations. This allows to prove that an aplanatic lens preserves helicity \cite[App. C]{FerCor2012b}. In essence, the proof uses the results of Sec. \ref{secc3:helprestetm} to show that an aplanatic lens preserves helicity because it treats the TE and TM components equally.

The lens thus preserves three of the four numbers that define the vectors of the chosen Bessel beam basis; namely, the eigenvalues of $H$, $J_z$ and $\Lambda$. On the other hand, it clearly changes the transverse (longitudinal) momentum distribution of the input mode: It {\em bends} the beam inwards. The distribution of amplitudes, concentrated around $\krho<<1$ at the input, shifts towards components of higher $\krho$. The eigenvalue of $P_\rho$ can change because the lens lacks transverse translational symmetry on the $xy$ plane.

In terms of the expansion coefficients, the lens can be seen as a block which transforms the input expansion coefficients into output ones like:
\begin{equation}\nonumber
c_{-1}^{\smallkrho} \xrightarrow{\rm Lens} c_{-1}^{\krho}.
\end{equation}

The optical vortices in tightly focused fields \cite{Zhao2007,Nieminen2008,Yao2011} are indeed a direct consequence of this change in momentum. Figure (\ref{figc4:bigsmall}) shows the intensities of the real space polarization components of a Bessel beam of well defined helicity as a function of the transverse $\rho$ coordinate for two $\krho$ values, one small and one large. The places where the change in $\krho$ influences the expression of the relevant Bessel beams are marked in red in (\ref{eqc4:cbigsmall}). 
\begin{equation}
	\label{eqc4:cbigsmall}
	\begin{split}
		&\mathbold{C}_{-1}^{\krho}(\rho,\theta,z)=\sqrt{\frac{\krho}{4\pi}}\exp(i\kz z)\times\\
  &{\color{red}{(1+\frac{\kz}{k})}}J_{0}({\color{red}{\krho}}\rho)\rhat+{\color{red}{(1-\frac{\kz}{k})}}J_{-2}({\color{red}{\krho}}\rho)\exp(-i2\theta)\lhat-i\sqrt{2}{\color{red}{\frac{\krho}{k}}}J_{-1}({\color{red}{\krho}}\rho)\exp(-i\theta)\zhat.
	\end{split}
\end{equation}

The value of $\krho$ in (\ref{eqc4:cbigsmall}) determines the relative weights of each real space polarization component and hence the relative weights of the different phase singularities hosted in them. For small $\krho$ values, the $\lhat$ and $\zhat$ components, which contain optical vortices, are much weaker than the dominant $\rhat$ polarization which does not have a vortex (Fig. \ref{figc4:bigsmall}-(a)). For large $\krho$ values the non-dominant components gain relative weight (Fig. \ref{figc4:bigsmall}-(b)). As the lens shifts the weight distribution towards modes with larger $\krho$ values, the strongly attenuated optical vortices of the input beam gain relative importance in the focalized beam. This, and not spin to orbital angular momentum transfer is the physical mechanism for the appearance of vortices in tightly focused beams.

\begin{figure}[htbp]
\begin{center}
\vspace{0.5cm}
			\includegraphics[width=0.45\textwidth]{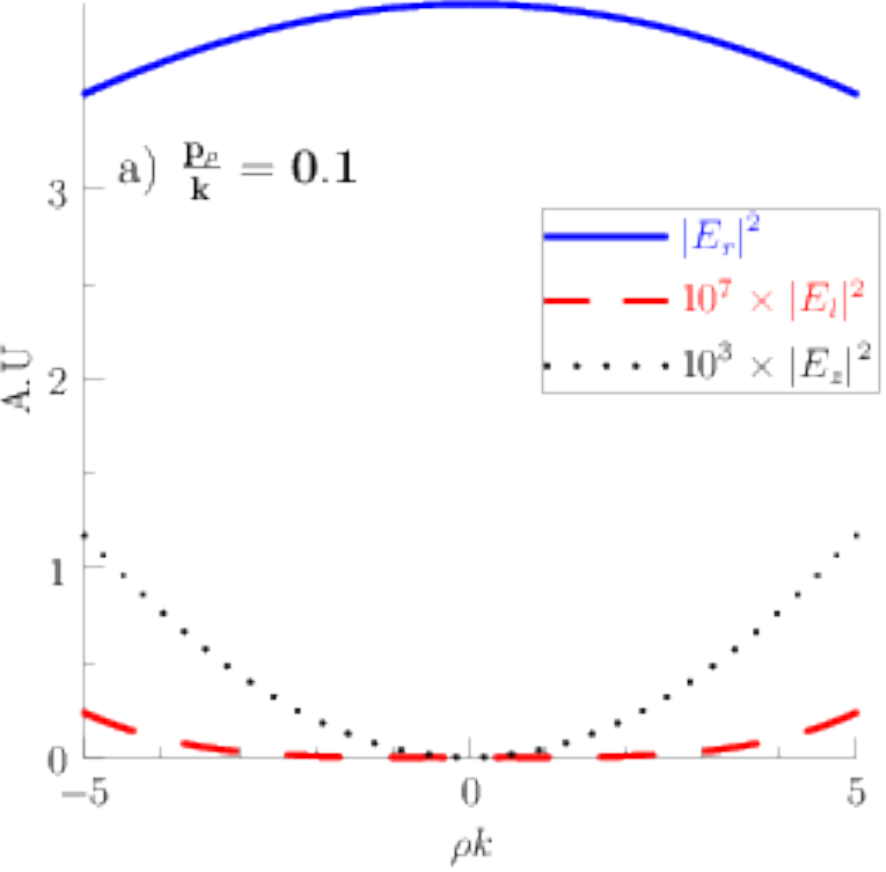}
			\includegraphics[width=0.45\textwidth]{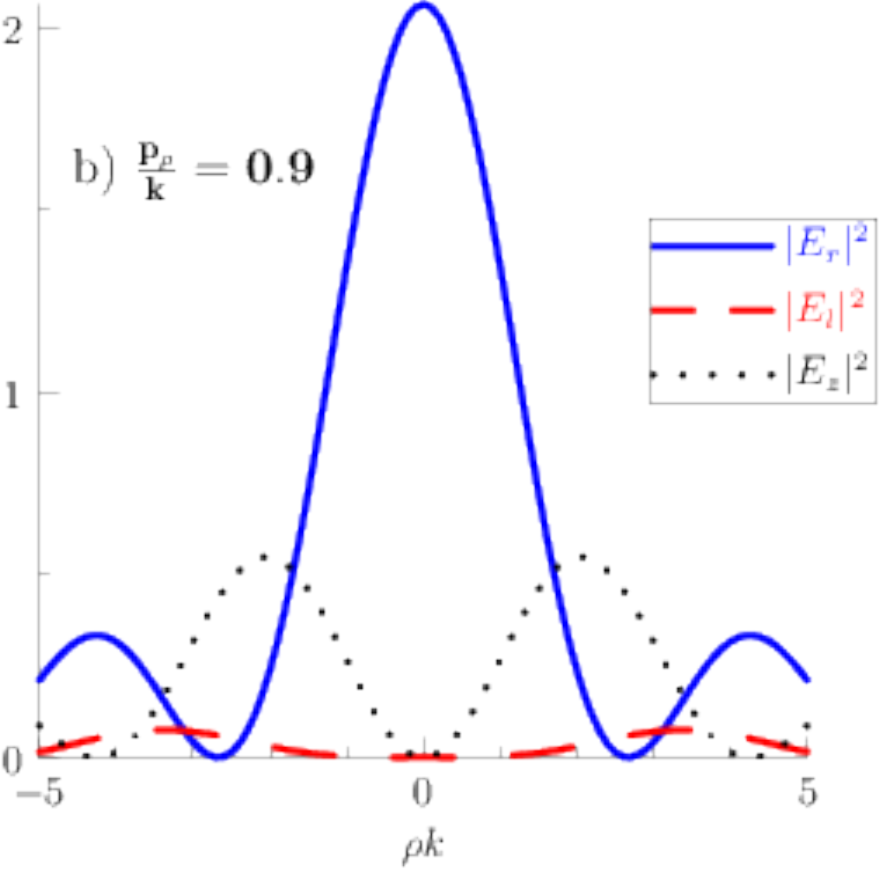}
  \end{center}
  \caption[Polarization intensities for small and large transverse momenta.]{\label{figc4:bigsmall} Normalized field intensity for the right, left and longitudinal $(\rhat,\lhat,\zhat)$ real space polarization components for two $\mathbf{C}_{-1}^{\smallkrho}$ modes, one with $\frac{\krho}{k}=0.1$ (a) and the other with $\frac{\krho}{k}=0.9$ (b). Note the scaling factors of the non-dominant polarization components in the $\frac{\krho}{k}=0.1$ case. The $\lhat$ and $\rhat$ components host optical vortices.}
\end{figure}

\begin{figure}[h]
	\centering
	\begin{tikzpicture}[thick,scale=1.5,every node/.style={scale=1}]
	\draw[->] (1.25,-0.35) -- (1.75,-0.15);
	\draw (1.3,0) node {$c_{-1}^{\krho}$};
	\draw[->] (1.25,0.35) -- (1.75,0.15);
	\draw[shift={(0.5,0)},scale=0.25]  (5,0.5) rectangle (5.5,4);
	\draw[shift={(0.5,0)},scale=0.25] (5,-0.5) rectangle (5.5,-4);
	\draw[shift={(0.5,0)},scale=0.25] (5.5,-4) rectangle (7,4);
	\draw[->] (2.25,0.15) -- (2.5,0.35);
	\draw[->] (2.25,-0.15) -- (2.5,-0.35);
	\end{tikzpicture}
\caption[Setup analysis: Interaction with the nanohole]{\label{figc4:scattering} Only energy and angular momentum are preserved in the interaction with the non-dual symmetric sample which also lacks longitudinal translational symmetry.
}
	\end{figure}
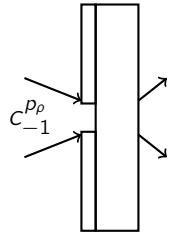

	It is now time to analyze the scattering off the sample (see Fig. \ref{figc4:scattering}). The (lack of) symmetries of the sample imply that $H$ and $J_z$ are preserved and $P_z$ is not. The fact that the collection lens collimates the beam before the measurement apparatus makes the analysis of the change in $P_z$ less important. The fourth number, the eigenvalue of helicity, is not conserved in the interaction with the sample. Assuming all materials to be nonmagnetic ($\mu_i=1 \ \forall\ i$), the relative electric constants of the glass substrate (2.25) and the gold film (-11.79+1.25i at 632.8 nm) violate the duality condition (\ref{eqc3:epsmu}). Let me assume that some portion of the incoming beam changes helicity in the interaction, and $D$ modes are created:
\begin{equation}
				c_{-1}^{\krho} \xrightarrow{\rm scattering} (c_{-1}^{q_\rho},d_{-1}^{q_\rho}).
	\end{equation}

Taking now into account the helicity preserving action of the second (collimating) microscope objective, the whole sequence of transformations before the second set of wave plates can be summarized as
\begin{equation}
	c_{-1}^{\smallkrho} \xrightarrow{\rm Lens} c_{-1}^{\krho} \xrightarrow{\rm Scattering} (c_{-1}^{\qrho},d^{\qrho}_{-1}) \xrightarrow{\rm Lens} \left(c_{-1}^{\smallqrho},d_{-1}^{\smallqrho}\right),
\end{equation}
and is graphically represented in Fig. \ref{figc4:summary}.

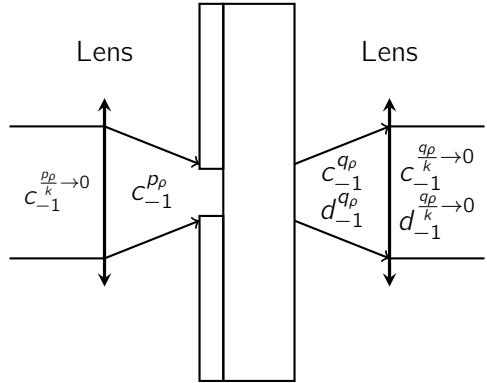
\begin{figure}[h]
	\centering
\begin{tikzpicture}[thick,scale=2.5,every node/.style={scale=1}]
\draw (0.75,0.35) -- (1.25,0.35);
\draw (1,0) node {\small $c_{-1}^{\smallkrho}$};
\draw (0.75,-0.35) -- (1.25,-0.35);
\draw[very thick,<->,>=stealth] (1.25,-0.5) -- (1.25,0.5);\draw (1.25,0.75) node {Lens};
\draw[->] (1.25,-0.35) -- (1.75,-0.15);
\draw (1.5,0) node {$c_{-1}^{\krho}$};
\draw[->] (1.25,0.35) -- (1.75,0.15);
\draw[shift={(0.5,0)},scale=0.25]  (5,0.5) rectangle (5.5,4);
\draw[shift={(0.5,0)},scale=0.25] (5,-0.5) rectangle (5.5,-4);
\draw[shift={(0.5,0)},scale=0.25] (5.5,-4) rectangle (7,4);
\draw[->] (2.25,0.15) -- (2.75,0.35);
\draw (2.5,0.1) node {$c_{-1}^{\qrho}$};
\draw (2.5,-0.1) node{ $d_{-1}^{\qrho}$};
\draw[->] (2.25,-0.15) -- (2.75,-0.35);
\draw[very thick,<->,>=stealth] (2.75,-0.5) -- (2.75,0.5);\draw (2.75,0.75) node {Lens};
\draw (2.75,0.35) -- (3.25,0.35);
\draw (3,0.13) node {$c_{-1}^{\smallqrho}$};
\draw (3,-0.13) node{ $d_{-1}^{\smallqrho}$};
\draw (2.75,-0.35) -- (3.25,-0.35);
\end{tikzpicture}
\caption[Setup analysis: End to end.]{\label{figc4:summary} Graphical representation of the block by block transformations of the expansion coefficients $c$ and $d$ corresponding to $C$ and $D$ modes with helicity eigenvalue $\mp 1$, respectively. The salient feature of the end to end effect is the creation of modes of helicity different to the input one.
}
\end{figure}
The salient feature of the overall system is the appearance of modes of different helicity due to the interaction with a non-dual symmetric scatterer. It is straightforward to show that these newly created modes are the ones that produce the two minima in the image of Fig. \ref{figc4:setup}-c). To do so, I first write the coordinate representation expressions of the corresponding Bessel modes in the limit of small transverse momenta (\ref{eqc4:smallCD}):
\begin{equation}
	\label{eqc4:CDm1}
\begin{split}
\mathbf{C}_{-1}^{\smallkrho}(\rho,\theta,z)&\approx\sqrt{\frac{\krho}{\pi}}i^{2}\exp(i\kz z)J_{0}(\krho\rho)\rhat.\\
\mathbf{D}_{-1}^{\smallkrho}(\rho,\theta,z)&\approx\sqrt{\frac{\krho}{\pi}}i^{2}\exp(i\kz z)J_{-2}(\krho\rho)\exp(-i2\theta)\lhat.
\end{split}
\end{equation}

Now, an argument parallel to the one that I have used to determine the composition of the input fields in Eq. (\ref{eqc4:last}) can be used to conclude that, when the second linear polarizer is set parallel to the first one, the $\mathbf{C}_{-1}^{\smallkrho}$ modes are dominant at the CCD; correspondingly, when it is set perpendicular to the first one, the $\mathbf{D}_{-1}^{\smallkrho}$ are dominant. The $\mathbf{C}_{-1}^{\smallkrho}$ have non-zero intensity in the center ($\rho=0$), which matches Fig. \ref{figc4:setup}-(b). The $\mathbf{D}_{-1}^{\smallkrho}$ modes have a phase singularity of order two (\ref{eqc4:CDm1}) which corresponds to the two intensity minima in Fig. \ref{figc4:setup}-(b): High order phase singularities are inherently unstable and always split into as many singularities of order one \cite{Ricci2012}.

In conclusion, the underlying reason for the vortices observed in cylindrically symmetric scattering experiments like \cite[Fig. 4]{Gorodetski2009},\cite[Figs. 2(c)-(d)]{Vuong2010} and \cite{Chimento2012,Brandao2013} is not spin to orbital angular momentum transfer, but breaking of electromagnetic duality symmetry.

\subsection{Quantitative considerations on helicity change}\label{secc4:quantitative}
In order to investigate the role of the aperture size, the experimental team in the group carried out a systematic study of the helicity conversion in a setup like the one in Fig. \ref{figc4:setup}. They obtained power conversion ratios ($\Gamma=P_{changed}/(P_{changed}+P_{unchanged})$) between roughly 0.05 and 0.4, depending on the size of the nanohole. For example, it was 0.15 for a 300 nm diameter hole. Please refer to \cite{Tischler2014} and to Fig. 3 in the experimental part of \cite{FerCor2012} for more details. These conversion ratios are mainly due to the presence of the nanohole. This can be concluded because measurements of helicity conversion due only to the two microscope objectives gave $\Gamma=10^{-4}$, which is consistent with the previously discussed helicity preservation in aplanatic lenses. Also, even though conversion through the multilayer (without the hole) was hard to measure\footnote{An intensity pattern consistent with helicity change could be seen only after summing together a large number of CCD images.} due to the small transmitted power and the extinction ratios of the linear polarizers ($\approx 5\times10^{-5}$), a simulation of the helicity conversion through the complete multilayer (without a hole) using perfect polarizers and lenses gave $\Gamma\approx 4\times10^{-4}$. 

A plausible explanation for the strong helicity conversion enhancement due to the nanohole can be obtained by combining the relationship between helicity eigenstates and eigenstates of spatial inversion operators discussed in Sec. \ref{subsecc3:tetmsym}, with the role of resonances in the transmission of light through nanoholes in metallic films \cite{Ebbesen1998,Altewischer2002,Genet2007}. The transmissivity of a nanohole in a metallic film does not follow Bethe's formula \cite{Bethe1944}. The difference is attributed to the existence of plasmonic and other resonances in the structure, which get excited by the incoming light and then re-radiate. The presence of the nanohole allows to couple propagating light modes incident on it to non-propagating modes (e.g. resonances) of the whole structure. Scattering by the nanohole produces the large transverse momentum components ($p_\rho^2>k^2$) that match those of the evanescent non-propagating modes with imaginary $p_z$ eigenvalue (see Sec. \ref{secc3:jz}). After accepting this mechanism, the helicity conversion enhancement due to the nanohole is readily explained. As can be deduced from Sec. \ref{subsecc3:tetmsym}, since the nanohole sample has mirror symmetries, the resonant modes will be linear combinations of two modes of different helicity. These TE/TM modes will not be degenerate because the structure is not dual. The incoming helicity eigenstates contain both TE and TM components, so both TE and TM resonances can be excited. After re-radiation, the outgoing mode will be an equal amplitude sum of the two helicities. Therefore, helicity conversion is greatly enhanced with respect to the case where the nanohole is not present, the large transverse momenta are not produced and the evanescent mirror symmetric modes are not excited.
\subsection{Simulation results for different input modes}\label{secc4:coneresults}
\begin{figure}[h!t]
	\begin{center}
		\makeatletter{}\def\ASYprefix{}
\newbox\ASYbox
\newdimen\ASYdimen
\long\def\ASYbase#1#2{\leavevmode\setbox\ASYbox=\hbox{#1}\ASYdimen=\ht\ASYbox\setbox\ASYbox=\hbox{#2}\lower\ASYdimen\box\ASYbox}
\long\def\ASYaligned(#1,#2)(#3,#4)#5#6#7{\leavevmode\setbox\ASYbox=\hbox{#7}\setbox\ASYbox\hbox{\ASYdimen=\ht\ASYbox\advance\ASYdimen by\dp\ASYbox\kern#3\wd\ASYbox\raise#4\ASYdimen\box\ASYbox}\put(#1,#2){#5\wd\ASYbox 0pt\dp\ASYbox 0pt\ht\ASYbox 0pt\box\ASYbox#6}}\long\def\ASYalignT(#1,#2)(#3,#4)#5#6{\ASYaligned(#1,#2)(#3,#4){
}{
}{#6}}
\long\def\ASYalign(#1,#2)(#3,#4)#5{\ASYaligned(#1,#2)(#3,#4){}{}{#5}}
\def\ASYraw#1{
currentpoint currentpoint translate matrix currentmatrix
100 12 div -100 12 div scale
#1
setmatrix neg exch neg exch translate}
 
		\makeatletter{}\setlength{\unitlength}{1pt}
\makeatletter\let\ASYencoding\f@encoding\let\ASYfamily\f@family\let\ASYseries\f@series\let\ASYshape\f@shape\makeatother{\catcode`"=12\includegraphics{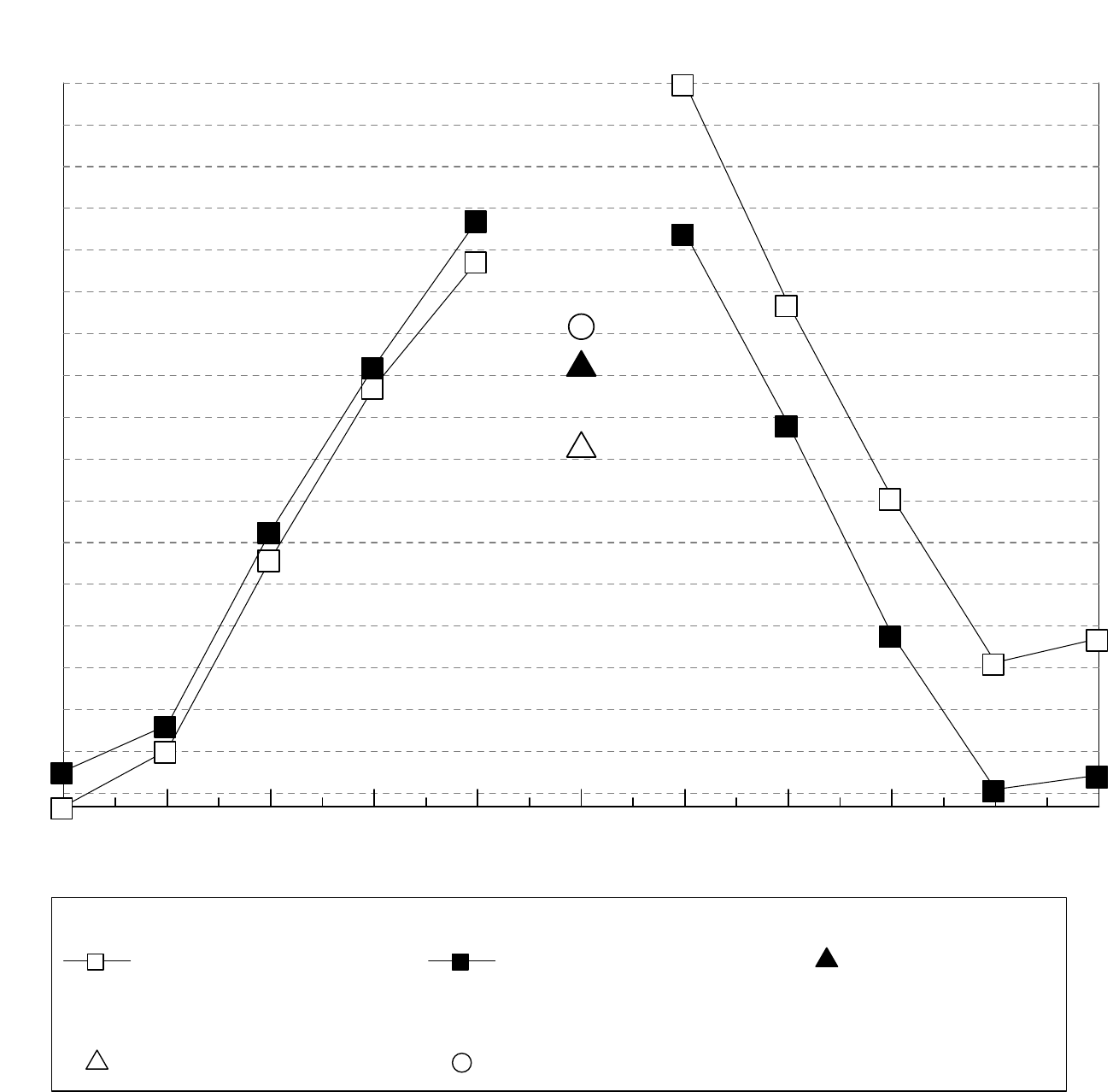}}
\begin{picture}( 376.021920, 368.463189)
\definecolor{ASYcolor}{gray}{0.000000}\color{ASYcolor}
\fontsize{12.000000}{14.400000}\selectfont
\usefont{\ASYencoding}{\ASYfamily}{\ASYseries}{\ASYshape}\ASYalignT(-359.312495,115.081301)(-1.000000,-0.404855){1.000000 -0.000000 -0.000000 1.000000}{\vphantom{$10^4$}$-8$}
\definecolor{ASYcolor}{gray}{0.000000}\color{ASYcolor}
\fontsize{12.000000}{14.400000}\selectfont
\ASYalignT(-359.312495,143.346639)(-1.000000,-0.404855){1.000000 -0.000000 -0.000000 1.000000}{\vphantom{$10^4$}$-7$}
\definecolor{ASYcolor}{gray}{0.000000}\color{ASYcolor}
\fontsize{12.000000}{14.400000}\selectfont
\ASYalignT(-359.312495,171.611977)(-1.000000,-0.404855){1.000000 -0.000000 -0.000000 1.000000}{\vphantom{$10^4$}$-6$}
\definecolor{ASYcolor}{gray}{0.000000}\color{ASYcolor}
\fontsize{12.000000}{14.400000}\selectfont
\ASYalignT(-359.312495,199.877316)(-1.000000,-0.404855){1.000000 -0.000000 -0.000000 1.000000}{\vphantom{$10^4$}$-5$}
\definecolor{ASYcolor}{gray}{0.000000}\color{ASYcolor}
\fontsize{12.000000}{14.400000}\selectfont
\ASYalignT(-359.312495,228.142654)(-1.000000,-0.404855){1.000000 -0.000000 -0.000000 1.000000}{\vphantom{$10^4$}$-4$}
\definecolor{ASYcolor}{gray}{0.000000}\color{ASYcolor}
\fontsize{12.000000}{14.400000}\selectfont
\ASYalignT(-359.312495,256.407992)(-1.000000,-0.404855){1.000000 -0.000000 -0.000000 1.000000}{\vphantom{$10^4$}$-3$}
\definecolor{ASYcolor}{gray}{0.000000}\color{ASYcolor}
\fontsize{12.000000}{14.400000}\selectfont
\ASYalignT(-359.312495,284.673331)(-1.000000,-0.404855){1.000000 -0.000000 -0.000000 1.000000}{\vphantom{$10^4$}$-2$}
\definecolor{ASYcolor}{gray}{0.000000}\color{ASYcolor}
\fontsize{12.000000}{14.400000}\selectfont
\ASYalignT(-359.312495,312.938669)(-1.000000,-0.404855){1.000000 -0.000000 -0.000000 1.000000}{\vphantom{$10^4$}$-1$}
\definecolor{ASYcolor}{gray}{0.000000}\color{ASYcolor}
\fontsize{12.000000}{14.400000}\selectfont
\ASYalignT(-359.312495,341.204007)(-1.000000,-0.500000){1.000000 -0.000000 -0.000000 1.000000}{\vphantom{$10^4$}$0$}
\definecolor{ASYcolor}{gray}{0.000000}\color{ASYcolor}
\fontsize{12.000000}{14.400000}\selectfont
\ASYalignT(-354.812495,91.976487)(-0.500000,-0.904855){1.000000 -0.000000 -0.000000 1.000000}{\vphantom{$10^4$}$-5$}
\definecolor{ASYcolor}{gray}{0.000000}\color{ASYcolor}
\fontsize{12.000000}{14.400000}\selectfont
\ASYalignT(-319.775105,91.976487)(-0.500000,-0.904855){1.000000 -0.000000 -0.000000 1.000000}{\vphantom{$10^4$}$-4$}
\definecolor{ASYcolor}{gray}{0.000000}\color{ASYcolor}
\fontsize{12.000000}{14.400000}\selectfont
\ASYalignT(-284.737715,91.976487)(-0.500000,-0.904855){1.000000 -0.000000 -0.000000 1.000000}{\vphantom{$10^4$}$-3$}
\definecolor{ASYcolor}{gray}{0.000000}\color{ASYcolor}
\fontsize{12.000000}{14.400000}\selectfont
\ASYalignT(-249.700325,91.976487)(-0.500000,-0.904855){1.000000 -0.000000 -0.000000 1.000000}{\vphantom{$10^4$}$-2$}
\definecolor{ASYcolor}{gray}{0.000000}\color{ASYcolor}
\fontsize{12.000000}{14.400000}\selectfont
\ASYalignT(-214.662935,91.976487)(-0.500000,-0.904855){1.000000 -0.000000 -0.000000 1.000000}{\vphantom{$10^4$}$-1$}
\definecolor{ASYcolor}{gray}{0.000000}\color{ASYcolor}
\fontsize{12.000000}{14.400000}\selectfont
\ASYalignT(-179.625545,91.976487)(-0.500000,-1.000000){1.000000 -0.000000 -0.000000 1.000000}{\vphantom{$10^4$}$0$}
\definecolor{ASYcolor}{gray}{0.000000}\color{ASYcolor}
\fontsize{12.000000}{14.400000}\selectfont
\ASYalignT(-144.588155,91.976487)(-0.500000,-1.000000){1.000000 -0.000000 -0.000000 1.000000}{\vphantom{$10^4$}$1$}
\definecolor{ASYcolor}{gray}{0.000000}\color{ASYcolor}
\fontsize{12.000000}{14.400000}\selectfont
\ASYalignT(-109.550765,91.976487)(-0.500000,-1.000000){1.000000 -0.000000 -0.000000 1.000000}{\vphantom{$10^4$}$2$}
\definecolor{ASYcolor}{gray}{0.000000}\color{ASYcolor}
\fontsize{12.000000}{14.400000}\selectfont
\ASYalignT(-74.513375,91.976487)(-0.500000,-1.000000){1.000000 -0.000000 -0.000000 1.000000}{\vphantom{$10^4$}$3$}
\definecolor{ASYcolor}{gray}{0.000000}\color{ASYcolor}
\fontsize{12.000000}{14.400000}\selectfont
\ASYalignT(-39.475985,91.976487)(-0.500000,-1.000000){1.000000 -0.000000 -0.000000 1.000000}{\vphantom{$10^4$}$4$}
\definecolor{ASYcolor}{gray}{0.000000}\color{ASYcolor}
\fontsize{12.000000}{14.400000}\selectfont
\ASYalignT(-4.438595,91.976487)(-0.500000,-1.000000){1.000000 -0.000000 -0.000000 1.000000}{\vphantom{$10^4$}$5$}
\definecolor{ASYcolor}{gray}{0.000000}\color{ASYcolor}
\fontsize{12.000000}{14.400000}\selectfont
\ASYalignT(-179.625545,76.365092)(-0.500000,-1.000000){1.250000 -0.000000 -0.000000 1.250000}{$\tilde{n}$}
\definecolor{ASYcolor}{gray}{0.000000}\color{ASYcolor}
\fontsize{12.000000}{14.400000}\selectfont
\ASYalignT(-179.625545,355.336676)(-0.500000,-0.166665){1.250000 -0.000000 -0.000000 1.250000}{$\Gamma=\log_{10}\left({P/P_{|\tilde{n}=1 \ \tau\rangle\rightarrow|\tilde{n}=1 \ \tau\rangle}}\right)$}
\end{picture}\kern -376.021920pt
\definecolor{ASYcolor}{gray}{0.000000}\color{ASYcolor}
\fontsize{12.000000}{14.400000}\selectfont
\usefont{\ASYencoding}{\ASYfamily}{\ASYseries}{\ASYshape}\ASYalignT(-329.528120,44.417955)(0.000000,-0.069445){0.750000 -0.000000 -0.000000 0.750000}{$\begin{array}{l}|\tilde{n}\ \tau\rangle\rightarrow|\tilde{n}\ \tau\rangle\\|n\ +\rangle\rightarrow|n\ +\rangle\\|{-}n \ -\rangle\rightarrow|{-}n \ -\rangle\end{array}$}
\definecolor{ASYcolor}{gray}{0.000000}\color{ASYcolor}
\fontsize{12.000000}{14.400000}\selectfont
\ASYalignT(-206.057381,44.417955)(0.000000,-0.104167){0.750000 -0.000000 -0.000000 0.750000}{$\begin{array}{l}|\tilde{n}\ \tau\rangle\rightarrow|{-}\tilde{n}\ \tau\rangle \\ |n\ +\rangle  \rightarrow|n\ -\rangle\end{array}$}
\definecolor{ASYcolor}{gray}{0.000000}\color{ASYcolor}
\fontsize{12.000000}{14.400000}\selectfont
\ASYalignT(-82.586642,44.417955)(0.000000,-0.104167){0.750000 -0.000000 -0.000000 0.750000}{$\begin{array}{l}|0\ +\rangle\rightarrow|0\ -\rangle\\|0\ -\rangle\rightarrow|0\ +\rangle\end{array}$}
\definecolor{ASYcolor}{gray}{0.000000}\color{ASYcolor}
\fontsize{12.000000}{14.400000}\selectfont
\ASYalignT(-329.528120,9.766312)(0.000000,-0.250000){0.750000 -0.000000 -0.000000 0.750000}{$|0\ \tau=1\rangle\rightarrow|0\ \tau=1\rangle$}
\definecolor{ASYcolor}{gray}{0.000000}\color{ASYcolor}
\fontsize{12.000000}{14.400000}\selectfont
\ASYalignT(-206.057381,9.766312)(0.000000,-0.250000){0.750000 -0.000000 -0.000000 0.750000}{$|0\ \tau=-1\rangle\rightarrow|0\ \tau=-1\rangle$}
 
	\end{center}
	\caption[Scattering off a cone.]{\label{figc4:hole_theory} Nanohole transmitted scattered power in normalized logarithmic scale for different input modes. Both transmission of the original modes and conversion between modes coupled by the sample are shown. In the legend, $|a\rangle\rightarrow|b\rangle$ indicates transmittivity from input $|a\rangle$ to output $|b\rangle$ modes. The values are normalized to the transmittivity of the $|\tilde{n}=1\ \tau\rangle$ mode: $\Gamma=\log_{10}({P/P_{|\tilde{n}=1 \ \tau\rangle\rightarrow|\tilde{n}=1 \ \tau\rangle}})$. The legend indicates the different pairs of modes for which the transmissivities are equal due to symmetry reasons (see Sec. \ref{secc3:scattcone}). The values of $\lambda=\pm1$ are indicated by isolated $\pm$, while the values of $\tau=\pm1$ are indicated explicitly. The case $\tilde{n}=0$ is special. The asymmetry around $\tilde{n}=0$ is consistent with the fact that the system does not have a symmetry connecting the $|\tilde{n}\ \tau\rangle$ modes with the $|-\tilde{n}\ \bar{\tau}\rangle$ modes. See the text for a detailed explanation.}
\end{figure}
All the experimental work contained in \cite{FerCor2012} was done using either $|n=1,\lambda=1\rangle$ or $|n=-1,\lambda=-1\rangle$ as input modes. I have used the method in \cite{FerCor2011} to obtain simulation results for other input modes: $|n\ \lambda\rangle$ modes with different $n$ and the mirror symmetric modes $|\tilde{n}\ \tau\rangle=1/\sqrt{2}\left(|n\ +\rangle+\tau|-n\ -\rangle\right)$ introduced in Sec. \ref{secc3:scattcone}. The cone scattering results contained in Sec. \ref{secc3:scattcone} are relevant for the nanohole sample described in Fig. \ref{figc4:setup} because the two systems have and lack the same symmetries. They have $R_z(\alpha)$ symmetry and any mirror plane of symmetry containing the $z$ axis. They lack the three translational symmetries, $M_\zhat$ and duality symmetry.

Fig. \ref{figc4:hole_theory} shows the transmitted scattered power in logarithmic scale as a function of $\tilde{n}$. For each $|\tilde{n} \ \tau\rangle$ case, the input mode is a linear combination of Bessel beams with a 632.8 nm wavelength and $p_z=k\cos(\pi/6)$:
\begin{equation}
	\frac{1}{\sqrt{2}}\left(|\omega\ p_z \ n \ +\rangle +\tau |\omega \ p_z \ \shortminus n \ -\rangle\right)
\end{equation}

The output is collected with a microscopic objective of NA=0.9. The sample is a hole of 300 nm of diameter in a 200 nm thick gold layer on top of a 300 nm glass layer\footnote{Numerical limitations preclude the simulation of a realistic (much thicker) glass layer.}. The magnetic constants of all layers are set to unity and the electric constants of the gold and glass to -11.79+1.25i and 2.25, respectively. All the transmissivity information for both $|\tilde{n} \ \tau\rangle$ and $|n \ \lambda\rangle$, including their respective power conversion to the $|-\tilde{n}\ \tau\rangle$ and $|n\ \shortminus\lambda\rangle$ modes, can be read off the graphs. The legend indicates the equivalences between the results for $|\tilde{n}\ \tau\rangle$ modes and $|n\ \lambda\rangle$ modes, which are immediate from the symmetry analysis in Sec. \ref{secc3:scattcone}.

The special character of the zero angular momentum modes is seen in Fig. \ref{figc4:hole_theory}. According to the results of Sec. \ref{secc3:scattcone}, $\tilde{n}=0$ is the only case where the transmissivity of the $|\tilde{n}\ \tau\rangle$ modes can depend on $\tau$. In this case, the transmitted power is 30 times larger for the $\tau=-1$ than for the $\tau=1$ case. This is consistent with the fact that the surface plasmonic modes are TM modes \cite[Chap. 2.2]{Maier2007}, i.e. the subtraction of two otherwise identical modes of opposite helicity (Sec. \ref{secc3:tetmhelhel}).  Differences in the transmissivity of the two modes have been reported in the literature for nanoholes \cite{Muller2007} and also for gratings with the same symmetries \cite{Lerman2011}.

It is interesting to note that the $|0 \ \tau\rangle$ modes are the only ones that have both well defined mirror symmetry and angular momentum, the two symmetries of the sample.

It is also worth mentioning that, in the laboratory, it is easier to use the $|n\ \lambda\rangle$ modes than the $|\tilde{n}\ \tau\rangle$ modes. Thanks to the equivalences in Sec. \ref{secc3:scattcone}, the information about the scattering of one set of modes can be obtained using the other.

The results for $\tilde{n}<0$ show more power in converted modes than on the original modes. This suggests that other mechanisms of helicity conversion besides the one outlined in Sec. \ref{secc4:quantitative} may exists: The coupling to resonant modes of mixed helicity can at most produce and equal amount of converted and non-converted power upon re-radiation. Systems where the power of the helicity converted component is larger than that of the conserved helicity component do exist \cite{Zambrana2013b}. In the examples provided in \cite{Zambrana2013b}, the helicity conversion ratio was seen to depend on the input mode.
In any case, since the results do not show the reflected power, a definitive assessment cannot be made of whether the total scattered power of the changed helicity component is larger than that of the preserved helicity component.

\section[The split of angular momentum in two terms]{The split of angular momentum in two terms}\label{secc4:shatlhat}
In electromagnetism, the separation of the total angular momentum in two components has its basis in the following fact. In the coordinate representation, the average of the total angular momentum of the field can be computed as the sum of two averages  \cite[Chap. XXI \S23]{Messiah1958}, \cite[probl. 7.27]{Jackson1998}: 

\begin{align}
\label{eqc4:jtrans}
\langle {\JJ} \rangle&=\int d\rr\ \rr\times (\EE_t\times\BB)\\
\label{eqc4:OAMSAM}
&=\int d\rr\ (\EE_t\times\mathbf{A}_t)+\int d\rr \ \sum_{i=1}^3 E_t^i(\rr\times\nabla) A^i_t,
\end{align}
where $\EE_t$ is the transverse part of the electric field and $\mathbf{A}_t$ the transverse part of the vector potential\footnote{I have not used the vector potential in my thesis. Its relationship with the fields is: $\HH=\nabla\times\mathbf{A}$, $\EE=-\partial_t\mathbf{A}+\nabla\phi$, where $\phi$ is the scalar potential. The scalar and vector potentials form a four-vector $A_\mu=(\phi,\mathbf{A})$.}.

The use of the word {\em average} is purposeful. Typically, Eq. (\ref{eqc4:jtrans}) is just called the total angular momentum of the field. I will now explain why the qualification {\em average} is appropriate. Consider a vector in $\mathbb{M}$ expanded in an orthonormal basis of eigenvectors of $J_z$
\begin{equation}
	|\Phi\rangle=\sum_{\nu,n}\alpha(\nu,n)|\nu,n\rangle,
\end{equation}
where $\nu$ contains eigenvalues of other operators.
It is straightforward to prove that the third component of the vector $\langle {\JJ} \rangle$ in Eq. (\ref{eqc4:jtrans}) is nothing but the real space calculation of the quantity:
\begin{equation}
	\label{eqc4:average}
	\langle \Phi|J_z|\Phi \rangle=\sum_{\nu,n} n |\alpha_{\nu,n}|^2.
\end{equation}
This can be seen in \cite[\S 9, p. 151]{Birula1975} and \cite[Eq. 9.143]{Jackson1998} after the proper normalizations are taken into account.

Equation (\ref{eqc4:average}) is the weighted average of the different $J_z$ eigenvalues of the modes present in the expansion of $|\Phi\rangle$. The weights are the squared norms of the expansion coefficients. Equation (\ref{eqc4:average}) is a sensible definition of the average third component of the angular momentum of $|\Phi\rangle$. Clearly, modes with different angular momentum content can result in the same average value of Eq. (\ref{eqc4:jtrans}). In quantum mechanics, expression (\ref{eqc4:average}) is the average $J_z$ of the state $|\Phi\rangle$.

The same considerations apply to the real space integrals used to compute the energy, the momentum, and other properties of the fields. They are also average values in the sense of (\ref{eqc4:average}). To verify this, it suffices to expand a field in a basis of eigenvectors of the corresponding operator to reach an expression similar to (\ref{eqc4:average}). Plane waves can be used for the energy and momentum cases. Those integrals can be found in \cite[\S 9, Eq. 33(a/b)]{Birula1975}. For example, it is appropriate to say that the scalar typically referred to as the energy of the classical electromagnetic field is its average frequency. Again, the average contains limited information: A field with an average frequency equal to the transition energy of a system, may or may not be able to excite such transition depending on whether the field actually contains a component with frequency equal to the average.

Returning to the main discussion, the identification of the two parts of equation (\ref{eqc4:OAMSAM}) with spin and orbital angular momenta is tempting due to the appearance of the operator $\LL=-i\mathbf{r}\times\nabla$ and the relationship of the cross-product with the spin-1 matrices representing $\mathbf{S}$. But, when the standard second quantization techniques are used to obtain expressions for the two operators, they are found to not obey the commutation relations that define angular momenta. The resulting Fock space operators are not\footnote{Note that these two vectors of operators do obey the commutation relations of angular momenta.} $\SSS$ and $\LL$. After the seminal work of Van Enk and Nienhuis in \cite{VanEnk1994}, several authors have studied the properties of these ``other'' operators \cite{Jauregui2005,Barnett2010,Bliokh2010}. 

I will call these operators $\Shat$ and $\Lhat$, for the purpose of distinguishing them from $\SSS$ and $\LL$. To summarize, the situation is the following one. Both pairs of vector operators sum up to $\JJ$:
\begin{equation}
	\JJ=\SSS+\LL=\Shat+\Lhat.
\end{equation}

$\SSS$ and $\LL$ are angular momenta operators but, due to their transversality violation (see Fig. \ref{figc4:breaking}), they are proper operators in the Hilbert space of transverse fields $\mathbb{M}$. $\Shat$ and $\Lhat$ are operators in $\mathbb{M}$, but they are not angular momenta. They will not be directly useful in questions regarding the rotational properties of fields and scatterers. The question is then: What can they be used for?

I will now study $\Shat$ and $\Lhat$, starting by deriving what the transformations that they generate are.

The expressions for the $\Shat$ operators in the Fock space representation of quantized modes with well defined momentum $\pp$ and helicity $\lambda=\pm$ can be found in \cite[chap XXI. prob. 7]{Messiah1958},\cite[Chap. 10.6.3]{Mandel1995}, \cite{VanEnk1994}:
\begin{equation}
	\label{eqc4:second}
	\Shat_{F}=\int d{\pp}\left(\hat{n}_{\pp,+}-\hat{n}_{\pp,-}\right)\frac{\pp}{|\pp|},
\end{equation}
where the $\hat{n}_{\pp,\pm}$ are the number operators. 

The expression for the $\Shat$ operator in the momentum representation of classical fields can be deduced from \cite[eq. 6]{Bliokh2010} to be:
\begin{equation}
	\label{eqc4:first}
	\Shat_{m}=\int{d\pp}\frac{\pp}{|\pp|}\left(|\pp\ +\rangle\langle +\ \pp|-|\pp\ -\rangle\langle - \pp|\right).
\end{equation}
In both (\ref{eqc4:second}) and (\ref{eqc4:first}), $\pp$ are numbers: The three momenta eigenvalues of the modes on which the operators act.

The commutation relations between $\Shat$ and $\Lhat$ were found to be exactly the same in the two representations of Eqs. (\ref{eqc4:second}) and (\ref{eqc4:first}) (\cite{VanEnk1994b,Bliokh2010}), reflecting the fact that they represent the same algebraic structure. With $\varepsilon_{jkl}$ denoting the totally antisymmetric tensor with $\varepsilon_{123}=1$, the commutation relations read
\begin{equation}
	\label{eqc4:commu}
	\begin{split}
		[\hat{S}_j,\hat{S}_k]=0&,\ [\hat{L}_j,\hat{L}_k]=i\sum_l\varepsilon_{jkl}(\hat{L}_l-\hat{S}_l),\\
		  [\hat{S}_j,\hat{L}_k]&=i\sum_l\varepsilon_{jkl}\hat{S}_l.
	\end{split}
\end{equation}

These are different from the commutation relations that define angular momentum operators: $[J_j,J_k]=i\sum_l\varepsilon_{jkl}J_l$. Clearly, neither $\Shat$ nor $\Lhat$ are angular momenta. They do not generate rotations and, consequently, their eigenstates are not necessarily preserved upon interaction with rotationally symmetric systems. On the other hand, they may be preserved by systems without rotational symmetry. I will later give examples of both these cases.

With the definition
\begin{equation}
	\label{eqc4:shat}
	\Shat=\Lambda\frac{\PP}{|\PP|},
\end{equation}
which implies $\Lhat=\JJ-\Lambda\PP/|\PP|$, $\Lhat$ and $\Shat$ meet the commutation relations in (\ref{eqc4:commu}). This can be verified using that $\Lambda$ commutes with both $\JJ$ and $\PP$. For vectors in $\mathbb{M}$, the action of $\Shat$ on the modes $\pwket$ of well defined momentum and helicity is the same as the one produced by (\ref{eqc4:second}) and (\ref{eqc4:first}) in their particular representations. I can therefore take $\Shat$ in (\ref{eqc4:shat}) as the representation-independent expression of the operators in (\ref{eqc4:second}) and (\ref{eqc4:first}).

In order to further understand $\Shat$ and $\Lhat$, and to be able to use them in the context of symmetries and conservation laws in light matter interactions, I wish to obtain more insight into the exact action of the transformation $\exp(-i\boldsymbol\beta\cdot\Lambda\PP/|\PP|)$, where $\boldsymbol\beta$ is a real vector. I will try its action on simultaneous eigenstates of $\Lambda$ and $|\PP|$. The test modes are hence monochromatic modes with well defined helicity $\lambda$ and frequency $\omega=|\pp|$ (in units of $c=1$), which I denote by $\Field$, ignoring any other well defined quantity. Note that $\Field$ can be a superposition of modes with different momentum directions. For simplicity, I study first a single component of $\Shat$. I take the first component of $\Shat$, use it to generate the corresponding continuous transformation with a real scalar parameter $\beta_x$ and apply such transformation to $\Field$.  I then manipulate this expression using the Taylor expansion of the exponential, the fact that helicity and momentum commute, that $\Lambda^2=I$ for transverse electromagnetic fields (Sec. \ref{secc3:M}), and then substitute the operators $\Lambda$ and $|\PP|$ by their eigenvalues $\lambda$ and $\omega$:
{\small
\begin{eqnarray}\nonumber
\label{eqc4:deriv}
&&\exp\Gen\Field=\\\nonumber
&&\sum_{k=0}^\infty \left[\frac{\Gen^{2k}}{(2k)!}+\frac{\Gen^{2k+1}}{(2k+1)!}\right]\Field=\\\nonumber
&&\sum_{k=0}^\infty\left[\frac{\Gentwo^{2k}}{(2k)!}+\frac{\Gentwo^{2k+1}\Lambda}{(2k+1)!}\right]\Lambda^{2k}\Field=\\\nonumber
&&\sum_{k=0}^\infty\frac{\Gentwo^{2k}}{(2k)!}\Field+\lambda\sum_{k=1}^\infty \frac{\Gentwo^{2k+1}}{(2k+1)!}\Field=\\
&&\exp\Genthree\Field.
\end{eqnarray}
}
The final expression in (\ref{eqc4:deriv}) is a translation along the $x$ axis with displacement $\lambda\beta_x/\omega$. For a fixed value of $\beta_x$, the magnitude of the translation depends on the frequency of the field. The direction of the translation, i.e. whether it is towards larger or smaller $x$ values, depends on the helicity of the field. 

Since the components of $\Shat$ commute, the derivation in (\ref{eqc4:deriv}) can be easily extended to the case of $\exp(-i\boldsymbol\beta\cdot\Lambda\PP/|\PP|)$, resulting in:
\begin{equation}
	\label{eqc4:trans}
	\exp(-i\boldsymbol\beta\cdot\Lambda\PP/|\PP|)\Field=\exp(-i(\lambda/\omega)\boldsymbol\beta\cdot\PP)\Field.
\end{equation}
Equation (\ref{eqc4:trans}) is a translation along the direction of the unitary vector $\hat{\boldsymbol\beta}=\boldsymbol\beta/|\boldsymbol\beta|$ by a displacement equal to $\lambda|\boldsymbol\beta|/\omega$. The value of helicity $\lambda$ controls the direction of the translation along the $\hat{\boldsymbol\beta}$ axis, and $|\boldsymbol\beta|/\omega$ its absolute value.

The particular case of monochromatic fields in the coordinate representation is worth examining because the action of $\exp(-i\boldsymbol\beta \cdot\Lambda\PP/|\PP|)$ as a helicity dependent translation can be very clearly seen. For a monochromatic field of well defined helicity $\mathbf{F}^{\omega}_{\pm}(x,y,z,t)=\mathbf{\widehat{F}}_{\pm}(x,y,z)\exp(-i\omega t)$, it follows from (\ref{eqc4:trans}), that the action of $\exp(-i\boldsymbol\beta \cdot\Lambda\PP/|\PP|)$ is
\begin{equation}
\label{eqc4:F}
\begin{split}
	&\exp(-i\boldsymbol\beta\cdot\Lambda\PP/|\PP|)\mathbf{\widehat{F}}_{\pm}(x,y,z)\exp(-i\omega t)=\\
 &\mathbf{\widehat{F}}_{\pm}(x\mp \beta_x/\omega,y\mp \beta_y/\omega ,z\mp \beta_z/\omega ,t)\exp(-i\omega t),
\end{split}
\end{equation}
where the anticipated spatial translation is explicitly seen in the displacements of the cartesian coordinates.

\begin{figure}[h!]
\begin{center}
	\makeatletter{}\def\ASYprefix{}
\newbox\ASYbox
\newdimen\ASYdimen
\long\def\ASYbase#1#2{\leavevmode\setbox\ASYbox=\hbox{#1}\ASYdimen=\ht\ASYbox\setbox\ASYbox=\hbox{#2}\lower\ASYdimen\box\ASYbox}
\long\def\ASYaligned(#1,#2)(#3,#4)#5#6#7{\leavevmode\setbox\ASYbox=\hbox{#7}\setbox\ASYbox\hbox{\ASYdimen=\ht\ASYbox\advance\ASYdimen by\dp\ASYbox\kern#3\wd\ASYbox\raise#4\ASYdimen\box\ASYbox}\put(#1,#2){#5\wd\ASYbox 0pt\dp\ASYbox 0pt\ht\ASYbox 0pt\box\ASYbox#6}}\long\def\ASYalignT(#1,#2)(#3,#4)#5#6{\ASYaligned(#1,#2)(#3,#4){
}{
}{#6}}
\long\def\ASYalign(#1,#2)(#3,#4)#5{\ASYaligned(#1,#2)(#3,#4){}{}{#5}}
\def\ASYraw#1{
currentpoint currentpoint translate matrix currentmatrix
100 12 div -100 12 div scale
#1
setmatrix neg exch neg exch translate}
 
	\makeatletter{}\setlength{\unitlength}{1pt}
\makeatletter\let\ASYencoding\f@encoding\let\ASYfamily\f@family\let\ASYseries\f@series\let\ASYshape\f@shape\makeatother{\catcode`"=12\includegraphics{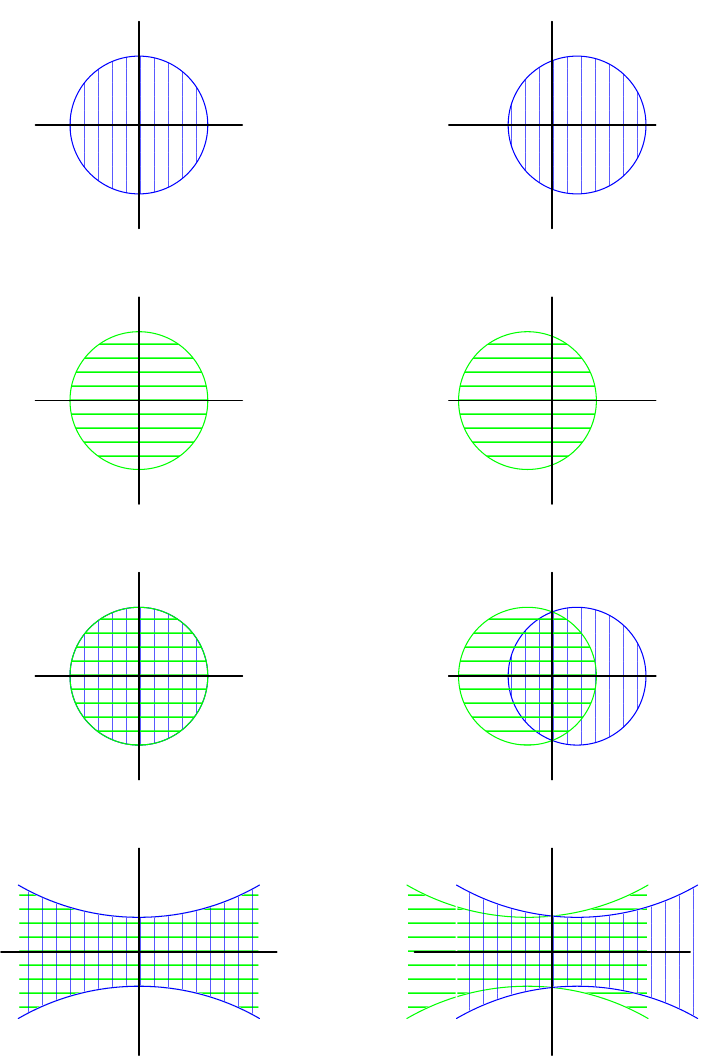}}\definecolor{ASYcolor}{gray}{0.000000}\color{ASYcolor}
\fontsize{12.000000}{14.400000}\selectfont
\usefont{\ASYencoding}{\ASYfamily}{\ASYseries}{\ASYshape}\ASYalign(-117.411902,287.608609)(0.000000,0.000000){$\lambda=1$}
\definecolor{ASYcolor}{gray}{0.000000}\color{ASYcolor}
\fontsize{12.000000}{14.400000}\selectfont
\ASYalign(-191.104540,291.591994)(0.000000,0.250000){{ a)}}
\definecolor{ASYcolor}{gray}{0.000000}\color{ASYcolor}
\fontsize{12.000000}{14.400000}\selectfont
\ASYalign(-158.183037,299.004875)(0.000000,-0.188891){$y$}
\definecolor{ASYcolor}{gray}{0.000000}\color{ASYcolor}
\fontsize{12.000000}{14.400000}\selectfont
\ASYalign(-131.907644,265.529481)(-0.500000,-1.000000){$x$}
\definecolor{ASYcolor}{gray}{0.000000}\color{ASYcolor}
\fontsize{12.000000}{14.400000}\selectfont
\ASYalign(-71.602965,291.591994)(0.000000,0.250000){{ b)}}
\definecolor{ASYcolor}{gray}{0.000000}\color{ASYcolor}
\fontsize{12.000000}{14.400000}\selectfont
\ASYalign(-191.104540,211.924278)(0.000000,0.250000){{ c)}}
\definecolor{ASYcolor}{gray}{0.000000}\color{ASYcolor}
\fontsize{12.000000}{14.400000}\selectfont
\ASYalign(-123.386981,207.940892)(0.000000,0.107143){$\lambda=-1$}
\definecolor{ASYcolor}{gray}{0.000000}\color{ASYcolor}
\fontsize{12.000000}{14.400000}\selectfont
\ASYalign(-71.602965,211.924278)(0.000000,0.250000){{ d)}}
\definecolor{ASYcolor}{gray}{0.000000}\color{ASYcolor}
\fontsize{12.000000}{14.400000}\selectfont
\ASYalign(-191.104540,132.256561)(0.000000,0.250000){{ e)}}
\definecolor{ASYcolor}{gray}{0.000000}\color{ASYcolor}
\fontsize{12.000000}{14.400000}\selectfont
\ASYalign(-71.602965,132.256561)(0.000000,0.250000){{ f)}}
\definecolor{ASYcolor}{gray}{0.000000}\color{ASYcolor}
\fontsize{12.000000}{14.400000}\selectfont
\ASYalign(-191.104540,52.588845)(0.000000,0.250000){{ g)}}
\definecolor{ASYcolor}{gray}{0.000000}\color{ASYcolor}
\fontsize{12.000000}{14.400000}\selectfont
\ASYalign(-121.949179,26.526331)(-0.500000,-1.000000){$z$}
\definecolor{ASYcolor}{gray}{0.000000}\color{ASYcolor}
\fontsize{12.000000}{14.400000}\selectfont
\ASYalign(-158.183037,60.001725)(0.000000,-0.188891){$y$}
\definecolor{ASYcolor}{gray}{0.000000}\color{ASYcolor}
\fontsize{12.000000}{14.400000}\selectfont
\ASYalign(-71.602965,52.588845)(0.000000,0.250000){{ h)}}
 
\end{center}
\caption[The action of $\Shat$.]{The diagrams on the left represent the transverse (a,c,e) and longitudinal (g) intensity patterns of Gaussian-like monochromatic fields with different helicity content. The diagrams on the right show the effect that the application of transformations generated by $\Shat$ have on these fields. (a-f) Effect of $\exp(-i\beta_x \hat{S}_x)$ on the transverse intensity pattern for (a, b) a field of well defined helicity equal to one, (c, d) a field of well defined helicity equal to minus one, (e, f) a field containing both helicity components. (g, f) show the effect of $\exp(-i\beta_z \hat{S}_z)$ on the longitudinal intensity pattern of a field containing both helicity components.}
\label{fig:gauss}
\end{figure}

Equations (\ref{eqc4:deriv}), (\ref{eqc4:trans}) and (\ref{eqc4:F}), provide physical insight about the transformations generated by $\Shat$. Figure \ref{fig:gauss} depicts the helicity dependent displacement experienced by a monochromatic Gaussian-like field upon application of $\exp(-i\beta_x \hat{S}_x)$ or $\exp(-i\beta_z \hat{S}_z)$.

I now consider the other part of the split. The transformations generated by $\Lhat$ have a straightforward interpretation in relation with the transformations generated by $\Shat$. Since $\JJ=\Lhat+\Shat$:
\begin{equation}
	\Lhat=\JJ-\Shat=\JJ-\Lambda\frac{\PP}{|\PP|}.
\end{equation}
Since rotations and translations along the same axis commute and helicity commutes with all rotations and translations (Tab. \ref{tabc3:gentrans}), the transformations generated by $\Lhat$ are trivially separated into those generated by $\Shat$ and those generated by $\JJ$. Referring again to the example of monochromatic fields, each component of $\Lhat$, $\hat{L}_i$, generates a helicity dependent translation along the $i$-axis followed by a rotation around the same axis. The order in which the two operations are applied does not matter.

With the insight gained up to this point, I can now make some qualitative considerations about light matter interactions using $\Shat$ and $\Lhat$. 

As discussed in Sec. \ref{secc4:lambdajz}, an aplanatic lens preserves helicity and the component of angular momentum along its axis, say $J_z$. It does not preserve either $\hat{S}_z$ or $\hat{L}_z$ because the lensing action changes $P_z$. The lens is thus a cylindrically symmetric system that does not preserve either $\hat{S}_z$ nor $\hat{L}_z$. On the other hand, the natural modes of a straight electromagnetic waveguide of arbitrary cross-section (see Fig. \ref{figc4:wg}) will be eigenstates of $P_z$, and, if all the materials have the same ratio of electric and magnetic constants, they will be eigenstates of helicity (Eq. \ref{eqc3:epsmu}). Therefore, $\hat{S}_z$ can be used to classify the eigenmodes of a non cylindrically symmetric system. These two examples illustrate the fact that $\Shat$ and $\Lhat$ are not related to rotations.
\begin{figure}[h!]
	\centering
	\makeatletter{}\def\ASYprefix{}
\newbox\ASYbox
\newdimen\ASYdimen
\long\def\ASYbase#1#2{\leavevmode\setbox\ASYbox=\hbox{#1}\ASYdimen=\ht\ASYbox\setbox\ASYbox=\hbox{#2}\lower\ASYdimen\box\ASYbox}
\long\def\ASYaligned(#1,#2)(#3,#4)#5#6#7{\leavevmode\setbox\ASYbox=\hbox{#7}\setbox\ASYbox\hbox{\ASYdimen=\ht\ASYbox\advance\ASYdimen by\dp\ASYbox\kern#3\wd\ASYbox\raise#4\ASYdimen\box\ASYbox}\put(#1,#2){#5\wd\ASYbox 0pt\dp\ASYbox 0pt\ht\ASYbox 0pt\box\ASYbox#6}}\long\def\ASYalignT(#1,#2)(#3,#4)#5#6{\ASYaligned(#1,#2)(#3,#4){
}{
}{#6}}
\long\def\ASYalign(#1,#2)(#3,#4)#5{\ASYaligned(#1,#2)(#3,#4){}{}{#5}}
\def\ASYraw#1{
currentpoint currentpoint translate matrix currentmatrix
100 12 div -100 12 div scale
#1
setmatrix neg exch neg exch translate}
 
	\makeatletter{}\setlength{\unitlength}{1pt}
\makeatletter\let\ASYencoding\f@encoding\let\ASYfamily\f@family\let\ASYseries\f@series\let\ASYshape\f@shape\makeatother{\catcode`"=12\includegraphics{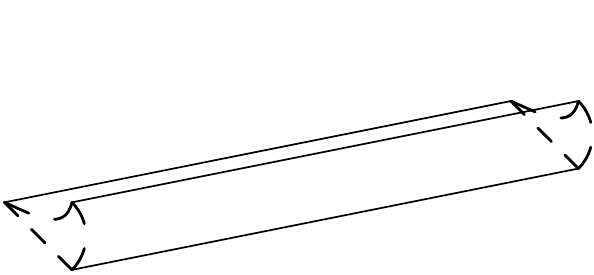}}\definecolor{ASYcolor}{gray}{0.000000}\color{ASYcolor}
\fontsize{12.000000}{14.400000}\selectfont
\usefont{\ASYencoding}{\ASYfamily}{\ASYseries}{\ASYshape}\ASYalign(-87.379642,23.312541)(-0.500000,-0.250000){$(\epsilon_1,\mu_1)$}
\definecolor{ASYcolor}{gray}{0.000000}\color{ASYcolor}
\fontsize{12.000000}{14.400000}\selectfont
\ASYalign(-87.379642,68.696808)(-0.500000,-0.112509){$(\epsilon_2,\mu_2)$ with
		${\frac{\epsilon_1}{\mu_1}=\frac{\epsilon_2}{\mu_2}}$}
 
	\caption[Example of using $\Shat$ in light matter interactions: Waveguide.]{\label{figc4:wg} Infinitely long waveguide of arbitrary cross section. The indicated relationship between the electric and magnetic constants of the waveguide $(\epsilon_1,\mu_1)$ and those of the medium $(\epsilon_2,\mu_2)$ make the system dual symmetric (Eq. \ref{eqc3:epsmu}). The dual symmetric waveguide is also translationally invariant along its axis and time invariant. Its eigenmodes can therefore be classified using eigenvalues of $\Lambda$, $P_z$ and $H$. They can also be classified using $\hat{S}_z$, showing that $\Shat$ is not connected with rotational symmetry.}
\end{figure}

I now discuss an application of one kind of simultaneous eigenstates of $\Shat$. Since the three components of $\Shat$ commute, there exist modes with simultaneously well defined values for the three of them. The eigenvalue of one more independent commuting operator is needed to completely specify an electromagnetic field. Choosing helicity results in a plane wave of well defined helicity $\pwket$. Choosing parity, which commutes with $\Shat$ since it simultaneously flips the sign of both helicity and momentum, results in a so called standing or stationary wave. 
\begin{equation}
	\frac{1}{\sqrt{2}}\left(|\pp \ +\rangle \pm |\shortminus\pp \ -\rangle\right).
\end{equation}
In the coordinate representation, the electric field of such a mode reads, for $\pp=p \mathbf{\hat{z}}$: 
\begin{equation}
\label{eqc4:standing}
\left(\xhat+i\yhat\right)\begin{bmatrix}\cos(\pp\cdot\rr)\\i\sin(\pp\cdot\rr)\end{bmatrix}\exp\left(-i\omega t\right).
\end{equation}
Fields similar to those in (\ref{eqc4:standing}) were recently predicted to achieve an enhanced interaction with chiral molecules \cite{Tang2010}. This points towards a role for simultaneous eigenstates of $\Shat$ and parity in the study of the interactions of light with chiral molecules. On the other hand, it should be noted that the name ``superchiral fields'' used in that work can be misleading because of the fact that, while chiral objects undergo a fundamental change after a parity transformation, the fields in (\ref{eqc4:standing}) are eigenstates of parity, and therefore stay invariant after a parity transformation.

To finalize, I include one other instance where the $\Shat$ operator plays a role. 

\subsection{$\Shat$ and the Pauli-Lubanski four-vector}
The $\Shat$ operators are related to the spatial part of a well known object in relativistic field theory: The Pauli-Lubanski four-vector $W_\mu$. The length of the Pauli-Lubanski four vector $W_\mu W^{\mu}$ is one of the Poincar\'e invariants used to classify elementary particles \cite[Chap. 10.4.3]{Tung1985}. It is known \cite[expr. 6.6.6]{Penrose1986} that for a massless field: $W_\mu=\Lambda P_\mu$. For the space components $W_k$ $k=1,2,3$, we have then that:
\begin{equation}
	\label{eqc4:sw}
	W_k=\Lambda P_k=\Lambda P_k\frac{|\PP|}{|\PP|}=\hat{S}_k |\PP|=\hat{S}_k H,
\end{equation}
where the third equality follows from the definition in (\ref{eqc4:shat}) and the fourth from the assumption of positive frequencies which selects the $H=|\PP|$ option and discards the $H=-|\PP|$ from the massless condition $H^2=|\PP|^2$. As far as I know, relationship (\ref{eqc4:sw}) has not been reported previously. 

I also note that the four-vector operator $(\hat{X},\mathbf{\hat{\Pi}})$ defined in \cite{Bliokh2011b}, with time component equal to the ``chirality'' ($\hat{X}$) and space component equal to the ``chiral momentum'' ($\mathbf{\hat{\Pi}}$), which, in the present notation would be $\hat{X}\equiv \Lambda H$ and $\mathbf{\hat{\Pi}}\equiv \Shat H$, is exactly the Pauli Lubanski four-vector $(\hat{X},\mathbf{\hat{\Pi}})\equiv(\Lambda H,\Shat H)=W_\mu$.

\section{Angular momentum and polarization}\label{secc4:polam}
I will now study a common interpretation of the split of the total angular momentum. In the literature, the polarization of a field is often considered to be a contributor to its angular momentum \cite{Oneil2002,Padgett2011,Cameron2012}. In general, though, the polarization of the field is completely decoupled from its angular momentum, as I will now show. 

It is possible to give an argument for such decoupling from the constructive way of generating solutions of the vectorial Helmholtz equation from solutions of the scalar vector equations (\cite[Sec. 13.1]{Morse1953}, \cite[Chap. VII]{Stratton1941}) considered in Chap. \ref{secc3:heli}. The argument is that angular momentum is determined by a scalar function, which gives rise to two transverse orthogonally polarized fields. Making arbitrary linear combinations of those two modes will maintain the same angular momentum but vary the polarization degree of freedom through its complete range of possible values. Polarization cannot affect angular momentum. The same is true for the other properties that the vectorial mode inherits from the scalar solution.

I will now give a formal proof of the idea. Consider the construction (\ref{eqc3:m}) of a general\footnote{Evanescent components can be included by a change in the integration limits of $\theta$. See the discussion around (\ref{eqc3:m}). I will not include them here to avoid cluttering the notation.} mode of well defined $J_z$:
\begin{equation}	
\begin{split}
	\label{eqc4:m}
	&|\Phi_n\rangle=\Longint\exp(in\phi)R_z(\phi)R_y(\theta)\\	
	&\left(c_+^{k\theta}|(0,0,k),+\rangle+c_-^{k\theta}|(0,0,k),-\rangle\right).
	\end{split}
\end{equation}

The mode in (\ref{eqc4:m}) is generated by a linear superposition of plane wave modes. Each plane wave is initially built as a linear superposition of two plane waves of well defined helicity ($\pm$) and initial momentum aligned with the positive $z$-axis, $\pp=(0,0,|\pp|=k)$. The complex coefficients of the linear superposition are $c_{\pm}^{k\theta}$. As discussed in Sec. \ref{secc3:jz}, the rotation $R_z(\phi)R_y(\theta)$ preserves the two helicity components and does not imprint any additional phases on them. Therefore, the ensemble of $c_{\pm}^{k\theta}$ completely determine the polarization of $|\Phi_n\rangle$.

Crucially, the steps (\ref{eqc3:jznolambda}) in the proof that $|\Phi_n\rangle$ is an eigenstates of $J_z$ with eigenvalue $n$ are independent of $c_{\pm}^{k\theta}$, that is, independently of polarization.

The argument holds for arbitrarily small non-null values of $\theta$. It applies for electromagnetic fields that fall within the paraxial approximation \cite{Lax1975}. The case of a single plane wave is different, as discussed in Sec. \ref{secc3:lambdajzpw}. The angular momentum along the axis of the plane wave does determine its helicity, and vice versa.

\section{Discussion about the separation of $\JJ$}
\label{secc4:inconsis}
Caution against the separate consideration of $\SSS$ and $\LL$ for the electromagnetic field can be found in reference books like \cite[\S 16]{Berestetskii1982} and \cite[p. 50]{Cohen1997}. In 1992, Allen {\em et al.} published a seminal paper \cite{Allen1992}, where solutions of the paraxial equation are used to argue the separate observability of $S_z$ and $L_z$ and propose an experimental setup to measure $L_z$. 

To this date, there has not been such experimental observation.

I have strong reservations about the validity of the conclusions reached in \cite{Allen1992}. The article uses the properties of solutions of the paraxial equation, which are not solutions of Maxwell's equations, to identify observable properties of electromagnetic fields which, collimated or not, {\bf are} solutions of Maxwell's equations. The properties of the two kinds of solutions and the algebraic structure of the Hilbert spaces that they belong to are different. For example, in the paraxial equation, the scalar and polarization parts of the solution can be separately specified. It is then clear than two different generators of rotations, one for the scalar part ($L_z$) and one for the polarization part ($S_z$), can exist. Real electromagnetic fields have a single such kind of observable: $J_z$. Nevertheless, one of the claims in \cite{Allen1992} is that a paraxial photon has two distinct observables connected to its rotational properties: $L_z$ and $S_z$. This amounts to suggesting that new observables for the field arise under the paraxial approximation. Physical reasons supporting such claim can not be found in \cite{Allen1992} nor anywhere else, as far as I know. I have also been unable to find a systematic discussion on the domain of validity of the claimed $L_z$ and $S_z$ separation. Such discussion should necessarily contain a measure of the separate observability as a function of the parameters involved in the approximation.

In any case, what seems to be uncontroversial is that, outside the paraxial regime, separate consideration of $S_z$ and $L_z$ is meaningless. Several efforts have been undertaken to rigorously extend the separation in the paraxial regime to the non-paraxial regime, but have encountered fundamental difficulties \cite{Barnett1994,Allen1999}. Accordingly, experiments with material particles trapped in the center of a beam \cite{He1995,Simpson1997,Friese1996} have shown that the rotation rates of the particles depend only on the total angular momentum $J_z$. Often the experimental results in \cite{Oneil2002} are interpreted as showing separate measurement of $S_z$ and $L_z$. In that work, the particles were trapped away from the center of the beam. Since $J_z$ does not commute with $P_x$ or $P_y$ (Tab. \ref{tabc3:gentrans}), even if the beam is an eigenstate of $J_z$ with respect to its axis, a decomposition in $J_z$ eigenstates with respect to an off axis point will contain not one, but several modes with different eigenvalues. Such multiplicity of modes is not considered in \cite{Oneil2002}. It is also important to note that all the cited experiments were carried out using a microscope objective with high numerical aperture. In the strongly non-paraxial regime of experiments like \cite{Oneil2002}, the separation of $S_z$ and $L_z$ has no theoretical support at all. The observation of two distinct kinds of angular momentum transfer from the electromagnetic field to other objects, one due solely to $L_z$ and the other due solely to $S_z$, would indeed be extraordinary.

As mentioned above, there is no evidence of such observation in the paraxial regime either.

Unfortunately, the idea that light possesses two distinct forms of angular momentum, spin an orbital, is deeply entrenched in the community, as an online search quickly shows. The assessment of whether the relevant electromagnetic beams fit within the paraxial approximation is rarely made. This then leads to inconsistent theories like spin to orbital angular momentum transfer. As discussed, the appearance of optical vortices in focusing and scattering is commonly explained as two instances of such transfer. As I have shown in Sec. \ref{secc4:lambdajz}, the appearance of the vortices is in fact due to the breaking of two different symmetries: Transverse translational symmetry in focusing and duality symmetry in scattering. There are other notable phenomena that are explained by spin to orbit, for example, the spin Hall effect of light \cite{Bliokh2008,Shitrit2011,Luo2012,Yin2013} or the action of the so called ``q-plates'' \cite{Marrucci2006}. Given that Maxwell's fields do not have spin or orbital angular momentum, the symmetry analysis of any instance where the transfer explanation is used should result in new insights.

The approach taken in this thesis to light matter interactions results in predictions which can be experimentally verified \cite{FerCor2013c,FerCor2012c} (Chaps. \ref{chap5} and \ref{chap6}), answers the question of {\em why} there are observable changes \cite{FerCor2012b} (this chapter), {\em why} do notable phenomena happen \cite{FerCor2012p,Zambrana2013} (Chap. \ref{chap5}), and provides sufficient insight for engineering purposes \cite{FerCor2012c,FerCor2013c,FerCor2013} (Chaps. \ref{chap5}-\ref{chap7}). The separate consideration of $\SSS$ and $\LL$ has none of these features. For example: If the sample in Fig. \ref{figc4:setup} was approximately dual like the small spheres studied in \cite{Zambrana2013b}, there will be no significant intensity in the image of the cross polarized measurement. This prediction cannot be reached using $\SSS$ and $\LL$.

\makeatletter{}\chapter[Forward and backward scattering]{Forward and backward scattering off systems with discrete rotational symmetries}
\label{chap5}
\epigraph{There is a noticeable general difference between the sciences and mathematics on the one hand, and the humanities and social sciences on the other. [...] In the former, the factors of integrity tend to dominate more over the factors of ideology. It's not that scientists are more honest people. It's just that nature is a harsh taskmaster.}{Noam Chomsky}
In this chapter, I will use symmetries and conservation laws to study the electromagnetic forward and backward scattering properties of linear systems with discrete rotational symmetries $R_z(2\pi/n)$ for $n=1,2,3,\ldots$ . I will show that the scattering coefficients are restricted for systems with symmetries of degree $n\ge3$: Along the axis of symmetry, forward scattering can only be helicity preserving while backward scattering can only be helicity flipping. These restrictions do not exist for systems with symmetries of degree $n=2$ or the trivial $n=1$. These results depend only on the discrete rotational symmetry properties of the scatterer. In particular, they do not depend on the (lack of) duality symmetry of the scatterer\footnote{As I will show, this is because only two particular scattering directions are considered. The field scattered in other directions is disregarded.}.

I will also show that, if in addition to the discrete rotational symmetry of degree $n\ge3$, the system has electromagnetic duality symmetry, it will exhibit zero backscattering. 

When $n\rightarrow \infty$, and the system reaches cylindrical symmetry, these results provide the symmetry reasons underpinning the prediction in \cite{Kerker1983} of zero backscattering for vacuum embedded spheres with $\epsilon=\mu$. In the same article, the authors also find that upon scattering of such spheres, the state of polarization of an incident plane wave is preserved independently of the scattering angle. In this case, the underlying symmetry reasons can be seen to be the simultaneous duality and mirror symmetries exhibited by the spheres.

These results could have applications in the engineering of structures for polarization control and also in reducing the backscattering from solar cells. The zero backscattering effect can also be used for testing the performance of devices designed to achieve duality symmetry.

\section{Description of the problem}
In this chapter, the scattering problem that I am interested in has the following characteristics
\begin{itemize}
\item The input field is a single plane wave with momentum vector pointing to the positive $z$ direction.
\item Only two scattering directions are of interest:
	\begin{itemize}
	\item Forward: The scattered plane wave whose momentum is parallel to the one of the input plane wave ($\zhat$).
	\item Backward: The scattered plane wave whose momentum is anti parallel to the one of the input plane wave ($-\zhat$).
	\end{itemize}
	In general, there is scattering in other directions, but they are not considered in this chapter.
\item The scatterers have discrete rotational symmetries, that is, they are invariant under a rotation by a discrete angle $2\pi/n$, where $n=2,3 \ldots$. See Fig. \ref{figc5:discrot}.
\end{itemize}
Note that this setting excludes the input used in the experimental setup of Fig. \ref{figc4:setup}: The decomposition of such focused field in the plane wave basis contains plane waves whose momentum is not aligned with the $z$ direction.
\begin{figure}[h]
\begin{center}
	\makeatletter{}\def\ASYprefix{}
\newbox\ASYbox
\newdimen\ASYdimen
\long\def\ASYbase#1#2{\leavevmode\setbox\ASYbox=\hbox{#1}\ASYdimen=\ht\ASYbox\setbox\ASYbox=\hbox{#2}\lower\ASYdimen\box\ASYbox}
\long\def\ASYaligned(#1,#2)(#3,#4)#5#6#7{\leavevmode\setbox\ASYbox=\hbox{#7}\setbox\ASYbox\hbox{\ASYdimen=\ht\ASYbox\advance\ASYdimen by\dp\ASYbox\kern#3\wd\ASYbox\raise#4\ASYdimen\box\ASYbox}\put(#1,#2){#5\wd\ASYbox 0pt\dp\ASYbox 0pt\ht\ASYbox 0pt\box\ASYbox#6}}\long\def\ASYalignT(#1,#2)(#3,#4)#5#6{\ASYaligned(#1,#2)(#3,#4){
}{
}{#6}}
\long\def\ASYalign(#1,#2)(#3,#4)#5{\ASYaligned(#1,#2)(#3,#4){}{}{#5}}
\def\ASYraw#1{
currentpoint currentpoint translate matrix currentmatrix
100 12 div -100 12 div scale
#1
setmatrix neg exch neg exch translate}
 
	\makeatletter{}\setlength{\unitlength}{1pt}
\makeatletter\let\ASYencoding\f@encoding\let\ASYfamily\f@family\let\ASYseries\f@series\let\ASYshape\f@shape\makeatother{\catcode`"=12\includegraphics{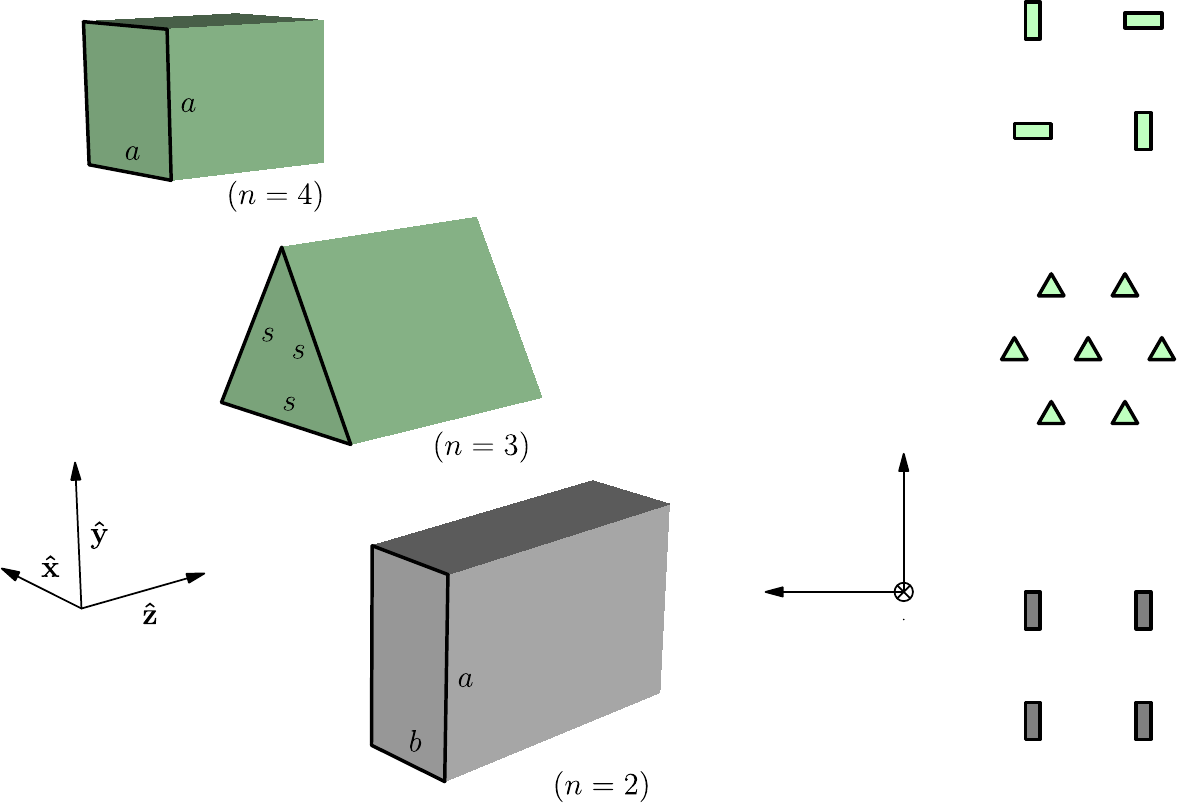}}\definecolor{ASYcolor}{gray}{0.000000}\color{ASYcolor}
\fontsize{12.000000}{14.400000}\selectfont
\usefont{\ASYencoding}{\ASYfamily}{\ASYseries}{\ASYshape}\ASYalignT(-98.788313,64.358871)(-0.500000,0.000000){0.750000 -0.000000 -0.000000 0.750000}{$\mathbf{\hat{x}}$}
\definecolor{ASYcolor}{gray}{0.000000}\color{ASYcolor}
\fontsize{12.000000}{14.400000}\selectfont
\ASYalignT(-75.203617,80.743566)(0.000000,-0.281251){0.750000 -0.000000 -0.000000 0.750000}{$\mathbf{\hat{y}}$}
\definecolor{ASYcolor}{gray}{0.000000}\color{ASYcolor}
\fontsize{12.000000}{14.400000}\selectfont
\ASYalignT(-78.803617,52.764993)(-0.500000,-0.500000){0.750000 -0.000000 -0.000000 0.750000}{$\mathbf{\hat{z}}$}
 
	\caption[Systems with discrete rotational symmetries $R_z(2\pi/n)$.]{\label{figc5:discrot} Systems with discrete rotational symmetries $R_z(2\pi/n)$ of degree $n=2,3,4$. Left panel: Prisms. Right panel: Unit cells of two dimensional arrays. Note that the coordinate axes have different orientation in the two panels.}
\end{center}
\end{figure}

\section{Derivations and results}\label{secc5:xixa}
The first part of the analysis does not involve the duality symmetry. It only involves the discrete rotational symmetries. Nevertheless, helicity is still relevant because the problem is restricted to plane waves with momentum parallel or anti parallel to a fixed axis. Let me particularize the results obtained in (\ref{eqc3:jz}) and (\ref{eqc3:-jz}) to the case of plane waves with momentum $p\zhat$ or $-p\zhat$, where $p=|\pp|$:
\begin{equation}
J_z\pmz=\pm\lambda\pmz.
\end{equation}
In the case of $\z$, the eigenvalues of $\Lambda$ and $J_z$ coincide, while for $\mz$, they have opposite sign. Accordingly, the properties of $\pmz$ under rotations along the $z$ axis are:
\begin{equation}
	\label{eqc5:ralpha}
	R_z(\alpha)\pmz=\exp(\mp i\alpha\lambda)\pmz.\\
\end{equation}
I will also need their hermitian conjugate version:
\begin{equation}
	\label{eqc5:ralphaherm}
		\pmzbra R_z(\alpha)^{-1}=\left(R_z(\alpha)\pmz\right)^{\dagger}\duetoref{eqc5:ralpha}\exp(\pm i\alpha\lambda)\pmzbra.
\end{equation}

The helicity dependent forward ($\tau_f^{\lambda\bar{\lambda}}$) and backward ($\tau_b^{\lambda\bar{\lambda}}$) scattering coefficients are the following matrix elements of the scattering operator $S$:
\begin{equation}
	\tau_f^{\lambda\bar{\lambda}}=\zbra S \z,\ \tau_b^{\lambda\bar{\lambda}}=\mzbra S \z,
\end{equation}
where $\lambda(\bar{\lambda})$ is the helicity of the input(scattered) plane wave. 

I will now assume that the scattering system has a discrete rotational symmetry  $2\pi/n$ along the $z$ axis as in Fig. \ref{figc5:discrot}. What this symmetry means for the scattering operator is that $S$ is invariant under a transformation by $R_z(2\pi/n)$:
\begin{equation}
\label{eqc5:rsr}
	R_z^{-1}(2\pi/n)SR_z(2\pi/n)=S.
\end{equation}
All is now ready. Let me first study the forward scattering coefficient $\tau_f^{\lambda\bar{\lambda}}$:
\begin{equation}
\label{eqc5:f}
	\begin{split}
		&\tau_f^{\lambda\bar{\lambda}}=\zbra S \z\stackrel{(\ref{eqc5:rsr})}{=}\zbra R_z^{-1}(2\pi/n)SR_z(2\pi/n) \z\\
		&\stackrel{(\ref{eqc5:ralpha}),(\ref{eqc5:ralphaherm})}{=}\exp(-i(\lambda-\bar{\lambda})\frac{2\pi}{n})\zbra S \z=\exp(-i(\lambda-\bar{\lambda})\frac{2\pi}{n})\tau_f^{\lambda\bar{\lambda}}.
	\end{split}
\end{equation}
For helicity preserving scattering $\lambda=\bar{\lambda}$, (\ref{eqc5:f}) results in the trivial statement $\tau_f^{\lambda\lambda}=\tau_f^{\lambda\lambda}$, which contains no information. But, for helicity changing scattering $\lambda=-\bar{\lambda}$ (with $\lambda=\pm1$), there are only two ways to meet $\tau_f^{\lambda,-\lambda}=\tau_f^{\lambda,-\lambda}\exp(\pm i 4\pi/n)$. One is that $\tau_f^{\lambda,-\lambda}=0$, and the other is that there exist an integer $k$ such that:
\begin{equation}
\label{eqc5:ff}
	\frac{4\pi}{n}=2\pi k \implies \frac{2}{n}=k,
\end{equation}
This second way is only possible if $n=1$ or $n=2$. It can not happen for $n\ge 3$, which then forces $\tau_f^{\lambda,-\lambda}=0$. This means that there is no component of changed helicity in the forward scattering direction of a system with a discrete rotational symmetry $R_z(2\pi/n)$ with $n\ge 3$. The forward scattering direction can only contain the preserved helicity component. Note that the derivation does not involve the duality properties of the scatterer, i.e. its general helicity preservation properties.

Let me now turn to the backward scattering coefficient. 
\begin{equation}
\begin{split}
	\label{eqc5:back}
	&\tau_b^{\lambda\bar{\lambda}}=\mzbra S \z\stackrel{(\ref{eqc5:rsr})}{=}\mzbra R_z^{-1}(2\pi/n)SR_z(2\pi/n) \z\\
	&\stackrel{(\ref{eqc5:ralpha}),(\ref{eqc5:ralphaherm})}{=}\exp(-i(\lambda+\bar{\lambda})\frac{2\pi}{n})\mzbra S \z=\exp(-i(\lambda+\bar{\lambda})\frac{2\pi}{n})\tau_b^{\lambda\bar{\lambda}}.
	\end{split}
\end{equation}
The situation is now reversed with respect to the helicity. For helicity flipping backscattering $\lambda=-\bar{\lambda}$, the result is trivial $\tau_b^{\lambda,-\lambda}=\tau_b^{\lambda,-\lambda}$. For helicity preserving backscattering $\lambda=\bar{\lambda}$, one can go through the same arguments as before to conclude that: There is no preserved helicity component in the backward scattering direction of a system with a discrete rotational symmetry $R_z(2\pi/n)$ with $n\ge3$. For such case, the backward scattering direction can only contain the changed helicity component.

In summary, for $n=1,2$, the two helicities are possible in both forward and backward scattering. For $n\ge3$, forward scattering can only be helicity preserving and backward scattering can only be helicity flipping. Fig. \ref{figc5:nyaca} illustrates these results. It is worth recalling that this analysis is only valid for the forward and backward scattering directions: The derivations do not apply to other scattering directions. In those directions, both preserved and flipped helicity are expected to be present, except, of course, for dual symmetric scatterers where helicity changes are forbidden. 
\begin{figure}[h!]
\begin{center}
	\makeatletter{}\def\ASYprefix{}
\newbox\ASYbox
\newdimen\ASYdimen
\long\def\ASYbase#1#2{\leavevmode\setbox\ASYbox=\hbox{#1}\ASYdimen=\ht\ASYbox\setbox\ASYbox=\hbox{#2}\lower\ASYdimen\box\ASYbox}
\long\def\ASYaligned(#1,#2)(#3,#4)#5#6#7{\leavevmode\setbox\ASYbox=\hbox{#7}\setbox\ASYbox\hbox{\ASYdimen=\ht\ASYbox\advance\ASYdimen by\dp\ASYbox\kern#3\wd\ASYbox\raise#4\ASYdimen\box\ASYbox}\put(#1,#2){#5\wd\ASYbox 0pt\dp\ASYbox 0pt\ht\ASYbox 0pt\box\ASYbox#6}}\long\def\ASYalignT(#1,#2)(#3,#4)#5#6{\ASYaligned(#1,#2)(#3,#4){
}{
}{#6}}
\long\def\ASYalign(#1,#2)(#3,#4)#5{\ASYaligned(#1,#2)(#3,#4){}{}{#5}}
\def\ASYraw#1{
currentpoint currentpoint translate matrix currentmatrix
100 12 div -100 12 div scale
#1
setmatrix neg exch neg exch translate}
 
	\makeatletter{}\setlength{\unitlength}{1pt}
\makeatletter\let\ASYencoding\f@encoding\let\ASYfamily\f@family\let\ASYseries\f@series\let\ASYshape\f@shape\makeatother{\catcode`"=12\includegraphics{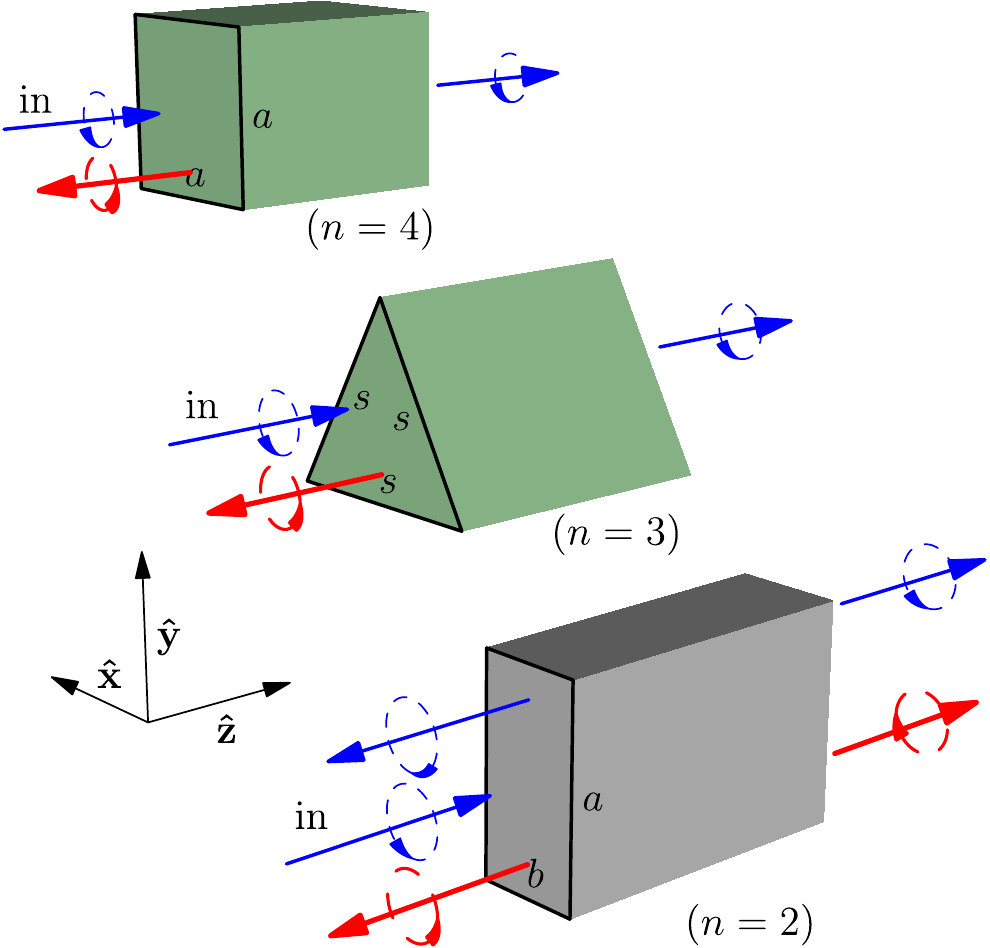}} 
\caption[Forward and backward scattering.]{Forward and backward scattering produced by input plane waves of well defined helicity impinging on structures with discrete rotational symmetries $R_z(2\pi/n)$: Gray rectangular prism ($n=2$), green equilateral triangular prism ($n=3$) and green square prism ($n=4$). Plane waves of positive helicity (left handed polarization) are blue. Plane waves of negative helicity (right handed polarization) are red. The input plane waves, labeled as ``in'', have positive helicity and momentum aligned with the positive $z$ axis. The text shows that for $n\ge 3$, the forward scattering can only contain components with the same helicity as the input and the backward scattering can only contain components with helicity opposite to the input one. In the figure, this is reflected in the restricted helicity components drawn in forward and backward scattering for the square ($n=4$) and triangular ($n=3$) prisms. No such restrictions apply to the rectangular prism ($n=2$): Any helicity is allowed in both forward and backward scattering. \label{figc5:nyaca}}
\end{center}
\end{figure}
Two very clear examples for $n=4$ can be found in \cite{Decker2010} and \cite{Kaschke2012}. In \cite{Decker2010} the authors design an array of split ring resonators which has $R_z(2\pi/4)$ symmetry. Their forward scattering results show very small ($<10^{-5}$) conversion ratios between the two circular polarizations. 
In \cite{Kaschke2012}, the authors study the forward and backward scattering properties of their design, which consist of four gold helices set in a square, specifically arranged to have $R_z(2\pi/4)$ symmetry. This structure is compared with a single helix, which lacks rotational symmetries (except the trivial $n=1$). The authors numerically analyze several cases where the helices have different number of turns and use both a lossless and a lossy model for the response of the gold. Their results show zero circular polarization conversion ratio in forward scattering for the $n=4$ structure, regardless of the number of helix turns and the loss model. The single helix shows non-zero conversion in both lossless and lossy cases.

The results from the two references are consistent with the restrictions obtained in (\ref{eqc5:f}) and (\ref{eqc5:ff}): Helicity cannot change in forward scattering when $n\ge 3$. Since the fields are plane waves, helicity preservation in forward scattering translates in preservation of the real space circular polarizations ($\rhat,\lhat$). This can be seen in Tab. \ref{tabc5:helreal}, which contains the explicit expressions of $\pmz$ in the coordinate representation. 

\begin{table}[h]
\begin{center}\begin{tabular}{rcc} \toprule
	&	$\lambda=1$ & $\lambda=-1$ \\ \midrule
$p\zhat\textrm{\hspace{0.28cm}}$&$-\lhat\exp(i(pz-\omega t))$ & $\rhat\exp(i(pz-\omega t))$\\
$\shortminus p\zhat\textrm{\hspace{0.3cm}}$   &$\rhat\exp(i(\shortminus pz-\omega t))$ & $-\lhat\exp(i(\shortminus pz-\omega t))$\\
	\bottomrule
\end{tabular}\end{center}
\caption[Plane waves of well defined helicity in the coordinate representation.]{\label{tabc5:helreal} Expressions for monochromatic plane waves of well defined helicity ($\lambda=\pm 1$) in the real space representation of electromagnetic fields. The momentum of the plane waves is either parallel or anti-parallel to the $z$ axis $\pm p\zhat$. The real space polarization vectors in the expressions are $\lhat=(\xhat+i\yhat)/\sqrt{2}$ and $\rhat=(\xhat-i\yhat)/\sqrt{2}$. In order to have the same handedness with respect to its momentum vector, the real space polarization vector must change when the momentum changes sign.}
\end{table}

The backward scattering analysis in \cite{Kaschke2012} shows preservation of the real space circular polarizations ($\rhat,\lhat$) for the $n=4$ structure as opposed to polarization conversion for the $n=1$ structure. Again, these findings do not depend on the number of helix turns and loss model. These results are consistent with the result in (\ref{eqc5:back}): Helicity cannot be preserved in backward scattering when $n\ge 3$. In backward scattering, helicity change translates into preservation of the real space circular polarization vectors: See in Tab. \ref{tabc5:helreal} how plane waves of opposite helicity and momentum have the same real space polarization vector.

One of the properties of the symmetry arguments that I have used in the derivations is that they are independent of factors like the wavelength of the illumination, the number of turns of the helices, the loss model, or whether the system is an array, a square arrangement of four helices or a single helix. 

The restrictions on the forward and backward helicity scattering coefficients for $n\ge 3$ agree with the results in \cite{Hu1987}. In that work, the authors study the consequences that different geometrical symmetries of the scatterer, including discrete rotational symmetries, have on the forward and backward scattering Mueller and Jones matrices. The results are also in agreement with the findings contained in \cite{Menzel2010,Bai2012}. In these references, the authors study the forward subscattering matrices of two dimensional planar arrays with different kinds of symmetry. In particular, they find that the matrices of systems with discrete rotational symmetries with degree $n=\{3,4,6\}$ have eigenstates that coincide with the two circular polarizations.

Before involving the duality symmetry in the discussion, it is worth considering the results in \cite{Hopkins2013}. The authors show that the extinction, absorption and scattering cross sections of nanoparticle clusters with $n\ge 3$ discrete rotational symmetry are independent of the linear polarization angle of the input plane wave. Using the formalism of this paper, the polarization independence of the extinction cross section can be recovered for systems with $n\ge 3$ featuring a mirror plane of symmetry ($M$) containing the symmetry axis. For a given input polarization, the optical theorem \cite[Chap. 10.11]{Jackson1998} states that the total extinction cross section is proportional to the imaginary part of the co-polarization forward scattering coefficient. Let me show that the mirror symmetry forces the forward scattering coefficient to be identical for both helicities:

\begin{equation}
\label{eq:ref}
\tau_f^{\lambda\lambda}=\langle \lambda \ p\zhat|S|p\zhat\ \lambda\rangle=\langle  \lambda \ p\zhat|M^{\dagger}SM|p\zhat \ \lambda\rangle=\langle -\lambda \ p\zhat|S|p\zhat \ -\lambda\rangle=\tau_f^{-\lambda-\lambda},
\end{equation}
where the second equality follows from the invariance of the scatterer under the mirror reflection, and the third one from the fact that the momentum $p\zhat$ is left unchanged since the mirror plane contains the $z$ axis, but helicity flips sign under any spatial inversion.

Together with the inherent helicity preservation of the $n\ge3$ system, which means that $\tau_f^{\lambda,-\lambda}=0$, Eq. (\ref{eq:ref}) implies that, in the helicity basis, the 2x2 Jones matrix is proportional to the identity. It is therefore the same in all polarization bases, in particular, in the linear polarization basis. The extinction cross section will hence be independent of the polarization. It is interesting to note that, while the structures considered in \cite{Hopkins2013} are mirror symmetric, the ones in \cite{Decker2010} and \cite{Kaschke2012} are not, so the above conclusion does not apply to them.

Consider now the following question: What happens if, on top of a discrete rotational symmetry with $n\ge3$, the scatterer has duality symmetry? In this case, there will not be any scattering in the backwards direction at all:  Since $n\ge3$, only the changed helicity component is possible, but, because of duality symmetry (helicity preservation) helicity cannot change upon scattering. The solution is that $\tau_b^{\lambda\bar{\lambda}}=0$ for all $(\lambda,\bar{\lambda})$. The system exhibits zero backscattering. See Fig. \ref{figc5:zbs}.

\begin{figure}[h!]
\begin{center}
	\makeatletter{}\def\ASYprefix{}
\newbox\ASYbox
\newdimen\ASYdimen
\long\def\ASYbase#1#2{\leavevmode\setbox\ASYbox=\hbox{#1}\ASYdimen=\ht\ASYbox\setbox\ASYbox=\hbox{#2}\lower\ASYdimen\box\ASYbox}
\long\def\ASYaligned(#1,#2)(#3,#4)#5#6#7{\leavevmode\setbox\ASYbox=\hbox{#7}\setbox\ASYbox\hbox{\ASYdimen=\ht\ASYbox\advance\ASYdimen by\dp\ASYbox\kern#3\wd\ASYbox\raise#4\ASYdimen\box\ASYbox}\put(#1,#2){#5\wd\ASYbox 0pt\dp\ASYbox 0pt\ht\ASYbox 0pt\box\ASYbox#6}}\long\def\ASYalignT(#1,#2)(#3,#4)#5#6{\ASYaligned(#1,#2)(#3,#4){
}{
}{#6}}
\long\def\ASYalign(#1,#2)(#3,#4)#5{\ASYaligned(#1,#2)(#3,#4){}{}{#5}}
\def\ASYraw#1{
currentpoint currentpoint translate matrix currentmatrix
100 12 div -100 12 div scale
#1
setmatrix neg exch neg exch translate}
 
	\makeatletter{}\setlength{\unitlength}{1pt}
\makeatletter\let\ASYencoding\f@encoding\let\ASYfamily\f@family\let\ASYseries\f@series\let\ASYshape\f@shape\makeatother{\catcode`"=12\includegraphics{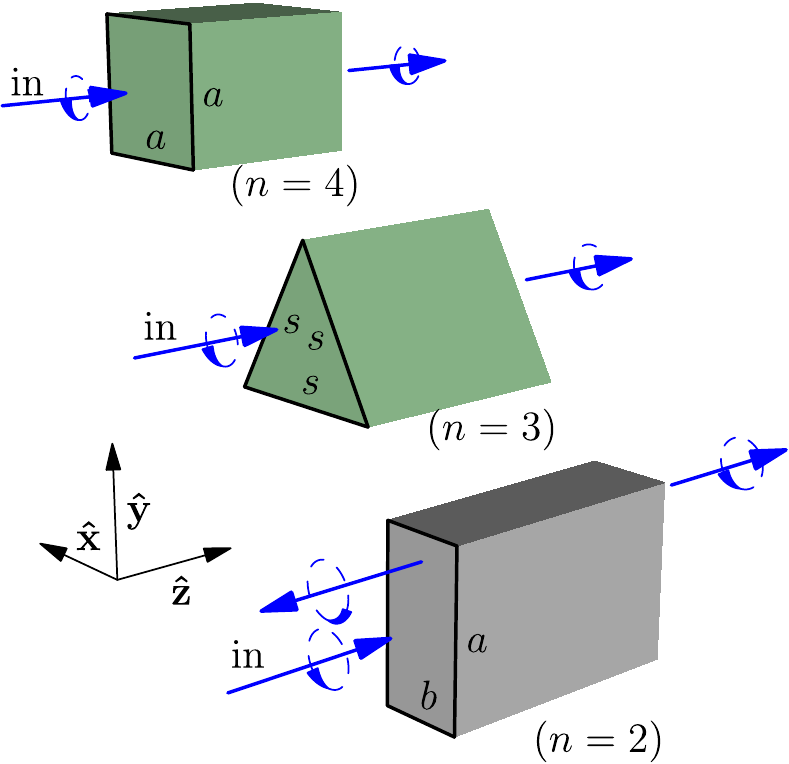}} 
\caption[Zero backscattering.]{\label{figc5:zbs} If, besides the discrete rotational symmetry, the scatterers also have electromagnetic duality symmetry, all the red right handed plane waves will disappear from Fig. \ref{figc5:nyaca} because duality enforces helicity preservation in all scattering directions. Therefore, dual objects with discrete rotational symmetries with $n\ge 3$ will exhibit zero backscattering.}
\end{center}
\end{figure}

In \cite{Lindell2009,Karilainen2012}, the authors study zero backscattering from dual systems with $R_z(2\pi/4)$ symmetry. In this chapter, the consideration of the connection between helicity and duality, and the relationship between helicity and angular momentum for plane waves, allows to conclude that the same zero backscattering effect exists for $R_z(2\pi/3),R_z(2\pi/4),R_z(2\pi/5),R_z(2\pi/6),...,$ etc, symmetric scatterers.

A direct application of the results contained in this section is the design of a planar array exhibiting zero backscattering. One requirement is to arrange the array inclusions so that the system has a discrete rotational symmetry of degree $n\ge 3$. The other one is to make the inclusions be dual symmetric scatterers. If the inclusions are small enough, the dipolar duality conditions in (\ref{eqc3:dual}) are the design goal.

Recently, solar cells of semiconductor nanowires arranged in a square lattice have been shown to achieve significant efficiencies \cite{Wallentin2013}. According to the results of this section, investigating the duality properties of the nanowires could lead to insights for reducing their reflection of normally incident light.

\section{Scattering off magnetic spheres}\label{secc5:kerker}
In 1983 Kerker {\em et. al.} \cite{Kerker1983} reported several unusual scattering effects for magnetic spheres (with magnetic constant $\mu\neq1$). One of them was the fact that a plane wave impinging on a vacuum embedded sphere with $\epsilon={\mu}$ does not produce any backscattered field. This effect, which has been referred to as an anomaly \cite{Nieto2011}, can be easily understood using the results from Sec. \ref{secc5:xixa}.

First, a vacuum embedded sphere with $\epsilon=\mu$ meets the macroscopic duality condition (\ref{eqc3:epsmu}). It is therefore dual and preserves helicity. Second, a sphere has cylindrical symmetry, which is the limiting case of a discrete rotational symmetry $R_z(2\pi/n)$ when $n$ tends to infinity. The results from Sec. \ref{secc5:xixa} apply directly to Kerker's setup: The sphere is a dual object with a discrete rotational symmetry of degree $n\ge 3$ and will therefore exhibit zero backscattering due to symmetry reasons. 

It is worth mentioning that already in \cite{Wagner1963}, the author uses Maxwell's equations to derive that $\epsilon(\rr)={\mu}(\rr)$ and cylindrical symmetry are sufficient conditions for zero backscattering.

Going back to \cite{Kerker1983}, the authors found that, upon scattering off a vacuum embedded sphere with $\epsilon=\mu$, the state of polarization of light is preserved independently of the scattering angle. I will now show that the root cause of such interesting phenomenon is the simultaneous invariance of the system with respect to duality transformations, due to the material of the sphere, and any mirror reflection through a plane containing the origin of coordinates, due to the geometry of the sphere. This can be seen using the results of Sec. \ref{secc3:tetmhel} as follows.

Let me take any pair of input and output plane waves with momenta $\pp$ and $\ppbar$, respectively. Since the sphere is invariant upon a reflection across the plane defined by the two momenta, the TE/TM character of the input plane wave will be preserved in the output plane wave. This follows from the relationship between spatial inversion operators and the TE/TM character of electromagnetic fields that I discussed in Sec. \ref{subsecc3:tetmsym}. Consequently, the 2x2 scattering submatrix between the two plane waves must be diagonal in the TE/TM basis ($\uparrow\downarrow$):
\begin{equation}
\label{eqc5:kerker2}
		S_\pp^{\ppbar}(\uparrow\downarrow)=\begin{bmatrix}\alpha_\pp^{\ppbar}&0\\0&\gamma_\pp^{\ppbar}\end{bmatrix}.
\end{equation}
Additionally, the sphere is a dual object which preserves helicity. According to (\ref{eqc3:prestetm}), this forces the diagonal terms of the (\ref{eqc5:kerker2}) to be equal. Then:
\begin{equation}
\label{eqc5:kerker23}
		S_\pp^{\ppbar}(\uparrow\downarrow)=\alpha_\pp^{\ppbar}\begin{bmatrix}1&0\\0&1\end{bmatrix}.
\end{equation}
The subscattering matrix is proportional to the identity. The polarization of the output plane wave is identical to the one of the input plane wave. This argument works for all $(\pp,\ppbar)$ and explains the preservation of the state of polarization after scattering off a sphere with $\epsilon=\mu$.
 
\makeatletter{}\chapter{Optical activity}
\label{chap6}
\epigraph{{\em In science, it is not speed that is the most important. It is the dedication, the commitment, the interest and the will to know something and to understand it. These are the things that come first.}}{Eugene Wigner}
An object which cannot be superimposed onto its mirror image is said to be chiral. Chirality is entrenched in nature. For instance, some interactions among fundamental particles are not equivalent to their mirrored versions \cite{Wu1957}. Also, the DNA, and many aminoacids, proteins and sugars are chiral. The understanding and control of chirality has become important in many scientific disciplines. In chemistry, the control of the chiral phase (left or right) of the end product of a reaction is crucial, since the two versions can have very different properties. In nanoscience and nanotechnology, chirality plays an increasingly important role \cite{Noguez2011,Zhang2005}. 

The chirality of electromagnetic fields is mapped onto its helicity. Since electromagnetic fields are routinely used to interact with matter at the nano, meso, molecular and atomic scales, it is not surprising that the interaction between chiral light and chiral matter has become an important subject of study for practical and also fundamental reasons. The subject itself is quite old \cite{Biot1815,Pasteur1848} and, from the beginning, has always been associated with the rotation of the linear polarization of light, an effect that occurs for example upon propagation through a solution of chiral molecules. A comprehensive theoretical study of optical activity based in symmetry principles can be found in \cite{Barron2004}, and the modern theoretical and computational methods for optical activity calculations are reviewed in \cite{Autschbach2012}. Recently, the powerful techniques of group representation theory have been used to study optical activity effects \cite{Saba2011,Turner2013,Saba2013}.

In this chapter, I study the rotation of the plane of linear polarization from the point of view of symmetries and conservation laws. In Sec. \ref{secc6:nec} I derive two necessary conditions for a scatterer to rotate linear polarization states: Helicity preservation and breaking of a mirror symmetry. In Sec. \ref{secc6:moa} I investigate how a solution of chiral molecules meets those two conditions in the forward scattering direction. The random orientation of the molecules in the solution is the crucial factor. This randomness effectively endows the solution with cylindrical symmetry, which leads to helicity preservation by the solution in the forward scattering direction (see Sec. \ref{secc5:kerker}). The fact that this preservation is ``automatic'' may be the reason why its role is not commonly considered in optical activity. Its importance is evident when considering non-forward scattering directions in a solution: The geometric argument leading to helicity preservation in forward scattering does not apply to non forward scattering directions. In Sec. \ref{secc6:aoa} I discuss the design of structures for achieving artificial optical activity using the results from the previous sections. Finally, Sec. \ref{secc6:fad} is devoted to another well known means of polarization rotation: The Faraday effect. I analyze the difference between the Faraday effect and molecular optical activity in terms of the space and time inversion symmetries. 

\section{Necessary conditions for polarization rotation}\label{secc6:nec}
Consider the scatterer $\mathbb{S}$ in Fig. \ref{figc6:challenge} and assume the following conditions. Upon excitation by an incident plane wave with momentum $\pp$ and linear polarization $\alpha$, the scattered component with momentum $\ppbar$ has linear polarization $\alpha+\beta$, where $\beta$ is independent of $\alpha$. The polarization angles are measured with respect to the momentum dependent direction set by the polarization vector of the TE component in each plane wave (see Eq. (\ref{eqc3:pwhel0})).
\begin{figure}[h]
	\begin{center}\begin{tikzpicture}[very thick,scale=1,every node/.style={scale=1},>=latex]
		\makeatletter{}\useasboundingbox (0,-2) rectangle (10,3);
\draw [line width=0.75mm,->,>=stealth,shift={(1.5,0)},rotate=0,scale=1] (0,0) -- (2,0);
\draw(2.75,-0.5) node{$|\mathbf{p},\alpha\rangle$};
\draw (4.5,0) .. controls (4.75,1) and (5.25,2) .. (5.5,1) .. controls (5.75,0) and (6.25,0.5) .. (6.5,0);
\draw (6.5,0) .. controls (7,-1) and (5.75,-2) .. (5.5,-1) .. controls (4 ,-3 ) and (3.5,1) .. (4.5,0);
\draw [line width=0.75mm,->,>=stealth,shift={(6.5,1)},rotate=60,scale=1] (0,0) -- (2,0);
\node [shift={(7.5,1.5)},rotate=60] at(0,0) {$|\mathbf{\bar{p}},\alpha+\beta\rangle$};
\draw(5.25,-0.25) node{{\Large $\mathbb{S}$}};

	\end{tikzpicture}\end{center}
	\caption[Polarization rotation in a general scattering direction.]{\label{figc6:challenge} The linear polarization of the input plane wave ($|\pp,\alpha\rangle=\exp(i\alpha)|\pp\ +\rangle+\exp(-i\alpha)|\pp\ -\rangle$) is rotated by the scatterer in the figure in the following way: For the scattered plane wave with momentum $\ppbar$, the plane of linear polarization has rotated by an angle $\beta$ with respect to the input. $\beta$ is independent of $\alpha$.}
\end{figure}
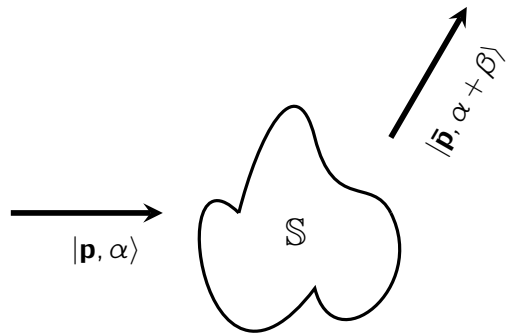
In a generalization of this transformation, the polarization is allowed to become elliptical, and the degree of ellipticity is independent of $\alpha$ as well. This is illustrated in Fig. \ref{figc6:polrot}. Imagine that the scatterer $\mathbb{S}$ behaves in such way, i.e., as a ``generalized'' polarization rotator for the $(\pp,\ppbar)$ input/output plane waves: What can be said about the symmetries of $\mathbb{S}$?
\begin{figure}[h]
	\centering
	\begin{center}\begin{tikzpicture}[thick,scale=0.5,every node/.style={scale=1},>=latex]
		\makeatletter{}\draw[->] (-4.5,0) -- (4.5,0);\draw[->] (0,-4.5) -- (0,4.5);\draw[line width=0.5mm,<->,rotate=25] (-4,0) -- (4,0);\draw[->] (1.4,0) arc (0:25:1.4);\draw[rotate=10] (1.9,0) node{$\alpha$} ;
\draw(6,0) node[scale=2]{$\Longrightarrow$};
\draw[->,xshift=12cm] (-4.5,0) -- (4.5,0);
\draw[->,xshift=12cm] (0,-4.5) -- (0,4.5);\draw[dashed,xshift=12cm,rotate=25+50] (-4,0) -- (4,0);\draw[line width=0.5mm,xshift=12cm,rotate=25+50] (0,0) ellipse [x radius=3.45cm,y radius=2.25cm];
\draw[->,xshift=12cm] (1.4,0) arc (0:25:1.4);\draw[xshift=12cm,rotate=10] (1.9,0) node{$\alpha$} ;\draw[->,xshift=12cm,rotate=25,red] (1.4,0) arc (0:50:1.4);\draw[xshift=12cm,rotate=50] (1.9,0) node{$\beta$} ;

	\end{tikzpicture}\end{center}
	\caption[Generalization of linear polarization rotation.]{\label{figc6:polrot} Generalization of linear polarization rotation. The output polarization is elliptical, with the restriction that both the ellipticity and the angle of rotation $\beta$ are independent of the input polarization angle $\alpha$.}
\end{figure}
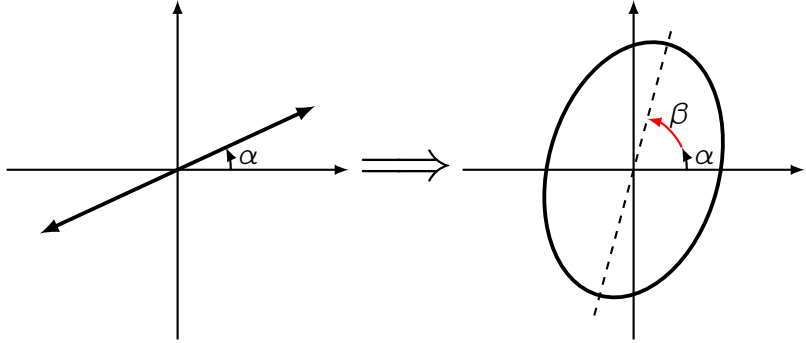

To start I consider a general 2x2 sub-scattering matrix for $(\pp,\ppbar)$ in the helicity basis
\begin{equation}
	S_\pp^{\ppbar}=\begin{bmatrix}a&b\\c&d\end{bmatrix}
\end{equation}
and impose the type of transformation illustrated by Fig. \ref{figc6:polrot}. A linear polarization state\footnote{The superscripted symbol $^{T}$ denotes transposition.} $[ \exp(i\alpha)\, ,\, \exp(-i\alpha)]^{T}/\sqrt{2}$ is transformed into a new state $[ f_+\,\ f_-]^{T}$:
\begin{equation}
\label{eqc6:tx}
\begin{bmatrix}f_+\\f_-\end{bmatrix}=\begin{bmatrix}a & b\\c&d\\\end{bmatrix}
\frac{1}{\sqrt{2}}\begin{bmatrix} \exp(i\alpha)\\ \exp(-i\alpha)\end{bmatrix}
=\frac{1}{\sqrt{2}}
\begin{bmatrix}a\exp (i\alpha)+b\exp (-i\alpha)\\c\exp (i\alpha)+d\exp (-i\alpha) \end{bmatrix}.
\end{equation}
The angle of the major ellipse axis with respect to the horizontal axis is $\theta=\frac{1}{2}\arg{\left({f_+}{f_-}^*\right)}$. According to the specification, it must be that 
\begin{equation}
2\theta=2(\alpha+\beta) \text{ for all } \alpha,
\end{equation}
which then forces 

\begin{equation}
{f_+}{f_-}^*=\eta \exp(i2(\alpha+\beta)),
\end{equation}

where $\eta$ is a real number. Using (\ref{eqc6:tx}):
\begin{equation}
\label{eqc6:eler}
	{f_+}{f_-}^*=ac^*+ad^*\exp(i2\alpha)+bc^*\exp(-i2\alpha)+bd^*=\eta \exp(i2(\alpha+\beta)),
\end{equation}
which must be valid for all $\alpha$ and hence imposes $b=c=0$ and gives $2\beta=2(\arg{a}-\arg{d})$. The most general matrix which meets the requirement is hence diagonal in the helicity basis
\begin{equation}
S_\pp^{\ppbar}=\begin{bmatrix}\aaa& 0\\0&\ddd \end{bmatrix}
=\begin{bmatrix}|\aaa|\exp(i\arg\aaa) & 0\\0&|\ddd|\exp(i\arg\ddd) \end{bmatrix}
.
\end{equation}

The conclusion is that the specified transformation needs helicity preservation. Helicity preservation will happen if $\mathbb{S}$ has duality symmetry. As shown in Chap. \ref{chap5}, it will also happen if $\pp=\ppbar$ and $\mathbb{S}$ has a discrete rotational symmetry of degree higher than 3 along the $\pp$ axis.

Consider now the mirror operation $M_{\pp\ppbar}$ across the plane defined by the two vectors $(\pp,\ppbar)$ and assume that the system possesses this mirror symmetry: $M_{\pp\ppbar}^{-1}SM_{\pp\ppbar}=S$. This particular mirror reflection leaves the momentum vectors invariant because they are contained in the reflection plane and, since any spatial inversion (parity) flips the helicity value, the plane wave states transform as 
\begin{equation}
M_{\pp\ppbar}|\pp,\pm\rangle=|\pp,\mp\rangle,M_{\pp\ppbar}|\ppbar,\pm\rangle=|\ppbar,\mp\rangle. 
\end{equation}
Using these transformation properties and the fact that the mirror operator is unitary ($M_{\pp\ppbar}^{-1}=M_{\pp\ppbar}^{\dagger}$), we can see that, if the system is invariant under this mirror transformation, the angle of rotation $\beta_{\pp}^{\ppbar}$ is equal to zero because $\aaa=\ddd$:
\begin{equation}\nonumber
	\begin{split}
		\aaa&=\langle +,\ppbar|S|\pp,+\rangle=\langle +,\ppbar|M_{\pp\ppbar}^{\dagger}SM_{\pp\ppbar}|\pp,+\rangle\\&=\langle -,\ppbar|S|\pp,-\rangle=\ddd \Rightarrow \beta_{\pp}^{\ppbar}=\arg \aaa -\arg\ddd=0.
	\end{split}
\end{equation}

Therefore, in order for $\mathbb{S}$ to perform the generalized polarization rotation, it must break (lack) the $M_{\pp\ppbar}$ mirror symmetry. For the $\pp=\ppbar$ case, there are infinitely many mirror planes defined by $(\pp,\pp)$. To avoid $a_{\pp}^{\pp}=d_{\pp}^{\pp}$, the scatterer must break all of them.

In conclusion, helicity preservation and breaking of the $M_{\pp\ppbar}$ mirror symmetries are necessary conditions for the rotation of linear polarization on the $(\pp,\ppbar)$ input/output directions.

\section{Molecular optical activity}\label{secc6:moa}
The study of the phenomenon of polarization rotation is an old scientific endeavor. In 1811, Arago discovered that the plane of linear polarization rotates upon propagation through a quartz crystal. Around 1815, Biot discovered that when light propagates through a solution of certain types of molecules, its linear polarization rotates as well \cite{Biot1815}. Commonly referred to as molecular optical activity, the study of its root causes has a long history \cite{Ingold1967,Oloane1980,Barron2004}. In 1848, Pasteur identified the absence of mirror planes of symmetry of the molecule as a necessary condition \cite{Pasteur1848}. He called it ``dissym\'etrie mol\'eculaire'' and by it Pasteur meant non-superimposability of the molecule and its mirror image, in other words: Chirality. Nowadays, this necessary condition is assumed to also be sufficient, and the exceptions to the rule are explained by other means \cite[sec. II.G]{Oloane1980}, \cite[Chap. 2.6]{Bishop1993}. Nevertheless, in his seminal work \cite{Condon1937}, Condon posed a still unresolved question: {\em ``The generality of the symmetry argument is also its weakness. It tells us that two molecules related as mirror images will have equal and opposite rotatory powers, but it does not give us the slightest clue as to what structural feature of the molecule is responsible for the activity. Any pseudoscalar associated with the structure might be responsible for the activity and the symmetry argument would be unable to distinguish between them.''.}

Condon's question suggests that the chirality of the molecule is not the whole story in optical activity. Section \ref{secc6:nec} shows that helicity preservation is a necessary condition. I will come back to his question later in this chapter.

The results of Sec. \ref{secc6:nec} show that helicity preservation is a necessary condition for optical activity, which is at odds with the common understanding of molecular optical activity, that is, that chirality of the molecule is the only necessary and sufficient condition. I will now show that in molecular optical activity, the randomness of the orientations of the molecules in the solution is the key factor that reconciles the results of Sec. \ref{secc6:nec} with the common view. The idea is that the solution acquires an effective cylindrical symmetry due to the randomness. The cylindrical symmetry implies the ``automatic'' preservation of helicity in the forward scattering direction (results in Secs. \ref{secc5:xixa} and \ref{secc5:xixa}).   

In molecular optical activity, the measurements are performed in the forward scattering direction. This is the special case $\pp=\ppbar$ (see Fig. \ref{figc6:fw}). As proved in Chap. \ref{chap5}, when only the forward scattering direction is considered, helicity preservation can happen independently of whether the scatterer has duality symmetry. It occurs for systems with discrete rotational symmetries of degree $n\ge 3$, and in particular for cylindrical symmetry ($n\rightarrow\infty$). I will now prove that, due to the randomness of the solution:
\begin{itemize}
	\item I) A solution of molecules is, effectively, rotationally symmetric. It has hence cylindrical symmetry along any axis, in particular $\pp=\ppbar$. 
	\item II) For the solution to break any mirror symmetry, the individual particle must be chiral.
\end{itemize}

\begin{figure}[h]
\begin{center}\begin{tikzpicture}[very thick,scale=1,every node/.style={scale=1},>=latex]
	\makeatletter{}\useasboundingbox (0,-2) rectangle (10,2);
\draw [line width=0.75mm,->,>=stealth,shift={(1.5,0)},rotate=0,scale=1] (0,0) -- (2,0);
\draw (4.5,0) .. controls (4.75,1) and (5.25,2) .. (5.5,1) .. controls (5.75,0) and (6.25,0.5) .. (6.5,0);
\draw (6.5,0) .. controls (7,-1) and (5.75,-2) .. (5.5,-1) .. controls (4 ,-3 ) and (3.5,1) .. (4.5,0);
\draw [line width=0.75mm,->,>=stealth,shift={(7,0)},rotate=0,scale=1] (0,0) -- (2,0);
\draw(5.25,-0.25) node{{\Large $\mathbb{S}$}};
 
\end{tikzpicture}\end{center}
\caption[Polarization rotation in forward scattering.]{\label{figc6:fw} The rotation of linear polarization observed in molecular optical activity is measured in the forward scattering direction. This direction is special w.r.t helicity preservation between input and output. As shown in Chap. \ref{chap5}, rotational symmetries can result in the preservation of helicity independently of whether the scatterer has duality symmetry.}
\end{figure}
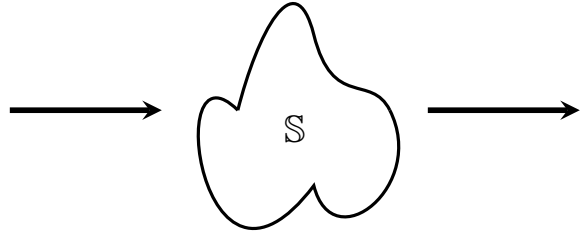
To prove these two points, I will make use of the theory of independent random scattering to study the Mueller matrix of the solution. The Mueller matrix relates the input Stokes parameters with the output Stokes parameters \cite[Chap. 3.2]{Tsang2004}. I will assume a mixture containing a large number of randomly oriented scattering particles immersed in an isotropic and homogeneous medium. I will also assume that the mixture has a linear response and that it contains only one kind of particle. 

The theory of independent random scattering \cite[1.21, 4.22]{Hulst1957}, \cite[Chap. 3.2]{Tsang2004} is typically used to approximately describe electromagnetic propagation in a random solution of small scattering particles. It is exact when the individual particles are sufficiently separated\footnote{A condition on the standard deviation (SD) of the random distance $d_{ij}$ between two particles which ensures sufficient separation can be found in \cite[expr. 3.1.13]{Tsang2004}: $SD(d_{ij})\ge \frac{\nu}{4}$, where $\nu$ is the wavelength. In \cite[Chap. 1.21]{Hulst1957}, the condition for applying independent scattering is given in terms of the radius of the particles $R$:  $d_{ij}>>3R$.} and the number of particles tends to infinity. In this case, the Mueller matrix of the total solution $L_S(\pp,\ppbar)$ can be computed as the average sum of the Mueller matrices for all possible orientations of the individual particle. If $f(\cdot)$ is the function that converts a 2x2 scattering matrix to its corresponding Mueller matrix \footnote{Formula A4.12 in \cite[App. IV]{Fujiwara2007} reads $f(N)=A(N\otimes N^*)A^{-1}$, where $N$ is the 2$\times$2 Jones matrix and $\otimes$ denotes the Kronecker matrix product. For the circular polarization basis $A=\begin{bmatrix}1&0&0&1\\0&1&1&0\\0&i&-i&0\\1&0&0&-1\end{bmatrix}$.} we have that
\begin{equation}
\label{eqc6:L}
L_S(\pp,\ppbar)=n_0 \int dR\ f(S_u^R(\pp,\ppbar))=n_0 \int dR\ f(\langle\bar{\lambda},\ppbar|R^{\dagger} S_u R|\pp,\lambda\rangle),
\end{equation}
where $n_0$ is the density of particles per unit volume, $\int dR$ indicates the sum over all possible rotations and $S_u^R(\pp,\ppbar)$ is the 2x2 scattering matrix of a $R$-rotated version of the individual particle with coefficients $\langle\bar{\lambda},\ppbar|R^{\dagger}S_uR|\pp,\lambda\rangle$. Note that $S_u$ denotes the scattering operator of the individual particle, while $S$ denotes the scattering operator of the solution as a whole. It is important to note that due to the integral over all rotations, equation (\ref{eqc6:L}) is only exact in the limit of infinite number of randomly oriented particles. From now on, I will take (\ref{eqc6:L}) as an effective response for the mixture and comment on which of the obtained results explicitly rely on the $\int dR$ average and which do not.

Let me start with statement II) concerning mirror symmetry and assume that the solution breaks one given mirror symmetry: 
\begin{equation}
\exists \ \vhat \textrm{ such that } [S,M_{\vhat}]\neq 0.
\end{equation}
The Mueller matrix of the mirror system can be written\footnote{The right hand side of (\ref{eqc6:LM}) can be interpreted as the response of a solution of a particle which is the mirror image version of the original particle. In the limit of independent scattering from an infinite number of particles, such response is identical to that of the mirror image of a solution of the original particle.}
\begin{equation}
\label{eqc6:LM}
L_{M_{\vhat}^\dagger SM_{\vhat}}(\pp,\ppbar)=n_0 \int dR\ f(\langle\bar{\lambda},\ppbar|R^{\dagger}M_{\vhat}^\dagger S_uM_{\vhat}R|\pp,\lambda\rangle).
\end{equation}

Lack of the mirror plane of symmetry $M_{\vhat}$ for the mixture implies that $L_S(\pp,\ppbar)\neq L_{M_{\vhat}^\dagger SM_{\vhat}}(\pp,\ppbar)$ for at least one pair $(\pp,\ppbar)$. 

Now, let me assume that the individual particle possesses a symmetry of the rotation-reflection kind: $Q_{\what}^m=M_{\what}R_{\what}\left(\frac{2\pi}{m}\right)$. When we assume any of these symmetries for $S_u$, the argument of $f(\cdot)$ in (\ref{eqc6:LM}) can be written\footnote{To obtain such result, substitute $M_{\vhat}^{\dagger}S_uM_{\vhat}=M_{\vhat}^{\dagger}R_{\what}^{\dagger}\left(\frac{2\pi}{m}\right) R_{\what}^{\dagger}(\pi)\Pi^{\dagger}S_u \Pi R_{\what}(\pi)R_{\what}\left(\frac{2\pi}{m}\right)M_{\vhat}$, use the facts that any mirror symmetry is the product of parity and a rotation ($M_\vhat=R_\vhat(\pi)\pi$), that the parity operator $\Pi$ commutes with any rotation and that $\Pi^2$ is the identity, and group all fixed rotations (which depend on $\vhat,\what$) into rotation $\tilde{R}$.} as $\langle\bar{\lambda},\ppbar|R^{\dagger}\tilde{R}^\dagger S_u \tilde{R} R|\pp,\lambda\rangle$, where $\tilde{R}$ is a fixed rotation which depends on $\vhat$ and $\what$. Then:
\begin{equation}
\label{eqc6:ms}
		L_{M_{\vhat}^\dagger SM_{\vhat}}(\pp,\ppbar)=n_0 \int dR\ f(\langle\bar{\lambda},\ppbar|R^{\dagger}\tilde{R}^\dagger S_u \tilde{R} R|\pp,\lambda\rangle)
=L_S(\pp,\ppbar),\ \forall \ (\pp,\ppbar).
\end{equation}

The second equality follows from the fact that, when $R$ covers all possible rotations once, $\tilde{R}R$ also covers all possible rotations once and the result of the integral is always the same, independently of $\tilde{R}$ (including the case of the identity $\tilde{R}=I$). This is an application of the re-arrangement lemma from group theory \cite[Chap. 2]{Tung1985}. Equation (\ref{eqc6:ms}) means that if the individual particle has any $Q_{\what}^m$ symmetry, the solution is mirror symmetric across any plane. Therefore it does not meet one of the necessary conditions from Sec. \ref{secc6:nec}.

In order for the solution to meet such condition, the individual particle can not have any of the $Q_{\what}^m$ symmetries. The lack of all the $Q_{\what}^m$ symmetries is the exact definition of chirality \cite{Bishop1993}. The molecule must therefore be chiral. A solution of chiral molecules will break all mirror symmetries. Due to disorder, for the solution to break one mirror symmetry, it must break them all. Since this result needs the averaging over random orientations, it will not apply to an ordered system. For instance, an ensemble of non-chiral oriented molecules can easily lack one mirror plane of symmetry without lacking them all. In \cite{Barron1972}, Barron mentions that there are non-chiral ordered systems which have been empirically shown to be optically active.

At this point, it is interesting to note that a helicity eigenstate meets the definition of chirality, i.e, lacks all the $Q_{\what}^m$ symmetries, in the sense that $
Q_{\what}^m|\eta, \lambda\rangle$  will always be orthogonal to $|\eta, \lambda\rangle$ because helicity will flip independently of what the other three numbers ($\eta$) represent. It is hence appropriate to say that the chiral character of the electromagnetic field is represented by its helicity.

With respect to cylindrical symmetry (statement I)), using again the re-arrangement lemma (which needs the $\int dR$ averaging) one can show that $L_S(\pp,\ppbar)=L_{\tilde{R}^\dagger S\tilde{R}}(\pp,\ppbar)$, for any rotation $\tilde{R}$. 
\begin{equation}
		L_{\tilde{R}^\dagger S\tilde{R}}(\pp,\ppbar)=\int dR\ f(\langle\bar{\lambda},\ppbar|R^{\dagger}\tilde{R}^\dagger S_u \tilde{R}R|\pp,\lambda\rangle)=L_S(\pp,\ppbar).
\end{equation}
Such effective rotational symmetry implies conservation of the angular momentum along any axis and ensures helicity preservation in the forward scattering direction. Having used the average over all possible rotations, this conclusion will not apply to ordered systems or systems with a small number of particles. For example, the result does not apply to an ensemble of oriented molecules. The acquisition of effective rotational symmetry due to orientation randomness and its breaking by an ordered sample is illustrated in Fig. \ref{figc6:helices}. 

\begin{figure}[h!]
\begin{center}
\subfloat{
	\makeatletter{}\def\ASYprefix{}
\newbox\ASYbox
\newdimen\ASYdimen
\long\def\ASYbase#1#2{\leavevmode\setbox\ASYbox=\hbox{#1}\ASYdimen=\ht\ASYbox\setbox\ASYbox=\hbox{#2}\lower\ASYdimen\box\ASYbox}
\long\def\ASYaligned(#1,#2)(#3,#4)#5#6#7{\leavevmode\setbox\ASYbox=\hbox{#7}\setbox\ASYbox\hbox{\ASYdimen=\ht\ASYbox\advance\ASYdimen by\dp\ASYbox\kern#3\wd\ASYbox\raise#4\ASYdimen\box\ASYbox}\put(#1,#2){#5\wd\ASYbox 0pt\dp\ASYbox 0pt\ht\ASYbox 0pt\box\ASYbox#6}}\long\def\ASYalignT(#1,#2)(#3,#4)#5#6{\ASYaligned(#1,#2)(#3,#4){
}{
}{#6}}
\long\def\ASYalign(#1,#2)(#3,#4)#5{\ASYaligned(#1,#2)(#3,#4){}{}{#5}}
\def\ASYraw#1{
currentpoint currentpoint translate matrix currentmatrix
100 12 div -100 12 div scale
#1
setmatrix neg exch neg exch translate}
 
	\makeatletter{}\setlength{\unitlength}{1pt}
\makeatletter\let\ASYencoding\f@encoding\let\ASYfamily\f@family\let\ASYseries\f@series\let\ASYshape\f@shape\makeatother{\catcode`"=12\includegraphics{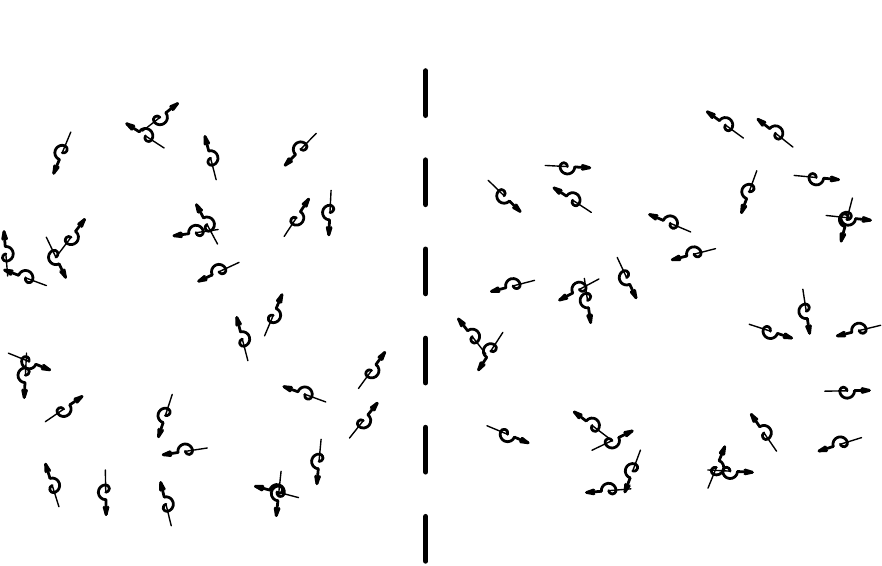}}\definecolor{ASYcolor}{gray}{0.000000}\color{ASYcolor}
\fontsize{12.000000}{14.400000}\selectfont
\usefont{\ASYencoding}{\ASYfamily}{\ASYseries}{\ASYshape}\ASYalign(-131.714126,155.365262)(-0.500000,-0.250001){{\small a) Disordered sample. 90 deg. rotation.}}
 
}
\\
\subfloat{
	\makeatletter{}\def\ASYprefix{}
\newbox\ASYbox
\newdimen\ASYdimen
\long\def\ASYbase#1#2{\leavevmode\setbox\ASYbox=\hbox{#1}\ASYdimen=\ht\ASYbox\setbox\ASYbox=\hbox{#2}\lower\ASYdimen\box\ASYbox}
\long\def\ASYaligned(#1,#2)(#3,#4)#5#6#7{\leavevmode\setbox\ASYbox=\hbox{#7}\setbox\ASYbox\hbox{\ASYdimen=\ht\ASYbox\advance\ASYdimen by\dp\ASYbox\kern#3\wd\ASYbox\raise#4\ASYdimen\box\ASYbox}\put(#1,#2){#5\wd\ASYbox 0pt\dp\ASYbox 0pt\ht\ASYbox 0pt\box\ASYbox#6}}\long\def\ASYalignT(#1,#2)(#3,#4)#5#6{\ASYaligned(#1,#2)(#3,#4){
}{
}{#6}}
\long\def\ASYalign(#1,#2)(#3,#4)#5{\ASYaligned(#1,#2)(#3,#4){}{}{#5}}
\def\ASYraw#1{
currentpoint currentpoint translate matrix currentmatrix
100 12 div -100 12 div scale
#1
setmatrix neg exch neg exch translate}
 
	\makeatletter{}\setlength{\unitlength}{1pt}
\makeatletter\let\ASYencoding\f@encoding\let\ASYfamily\f@family\let\ASYseries\f@series\let\ASYshape\f@shape\makeatother{\catcode`"=12\includegraphics{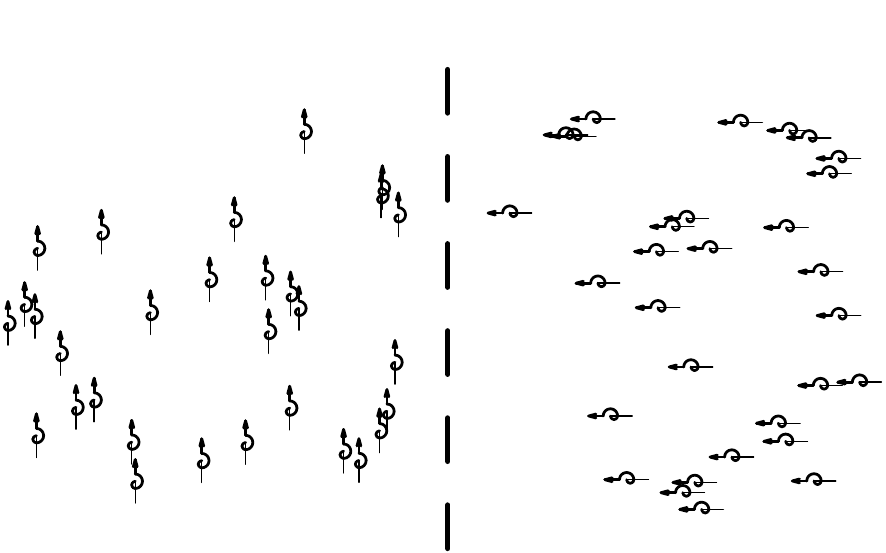}}\definecolor{ASYcolor}{gray}{0.000000}\color{ASYcolor}
\fontsize{12.000000}{14.400000}\selectfont
\usefont{\ASYencoding}{\ASYfamily}{\ASYseries}{\ASYshape}\ASYalign(-125.512546,151.814971)(-0.500000,-0.250001){{\small b) Ordered sample. 90 deg. rotation.}}
 
}
\end{center}
\caption[Effective rotational symmetry induced by disorder.]{Illustration of the effective rotational symmetry acquired by a random solution. The left parts of the figures are the initial mixtures and the right parts are the rotated versions of the initial mixtures. Even though the limit of infinite number of particles cannot be graphically illustrated, it can already be perceived in the figures that, after a rotation, the light scattering properties of an ordered sample (b) should significantly change, while those of a disordered sample (a) should not; in fact, under the assumptions made in the text, they do not change at all when the number of particles tends to infinity. Note that the effective rotational symmetry in (a) is acquired independently of any property of the individual particle. In particular, it does not matter whether the particle is chiral or not.}
\label{figc6:helices}
\end{figure}
These results show that the forward scattering direction of a random solution of any chiral molecule meets the two conditions for polarization rotation derived in Sec. \ref{secc6:nec}. Due to randomness, the chirality of the molecule is necessary and sufficient for breaking any mirror symmetry and the helicity preservation in forward scattering comes from effective cylindrical symmetry. 

The question is then: What happens in non-forward ($\pp\neq\ppbar$) scattering directions? The solution still lacks all mirror planes of symmetry, in particular $M_{\pp\ppbar}$, but, crucially, the geometrical argument that leads to helicity preservation in forward scattering (see Chap. \ref{chap5}) does not work for non-forward scattering directions. One has therefore no apparent reason to expect helicity preservation in a non-forward measurement. Actually, according to Gell-Mann's totalitarian principle, ``everything that is not forbidden is mandatory'', one should expect that helicity is not preserved. This principle does not seem to be a proved result of theoretical physics, but rather a conjecture. Nevertheless, the principle is met for helicity non-preservation in non-forward scattering, as my colleagues Alex Barbara and Dr. Xavier Vidal have shown in a series of experiments. Table \ref{tabc6:fw} and Fig. \ref{figc6:nfw} show the results of one of their experiments. The sample consisted of a solution of maltose, which is a chiral sugar. They performed measurements at two scattering directions, forward (results in Tab. \ref{tabc6:fw}) and at a scattering angle of 90 degrees (results in Fig. \ref{figc6:nfw}). The conclusion that can be extracted from the two measurements is that helicity is preserved in forward scattering but is not preserved at 90 degrees. 

The forward scattering helicity preservation can be seen in the measurements of the input and output Stokes parameters contained in Tab. \ref{tabc6:fw}. The sample was illuminated by a collimated Gaussian laser which was prepared\footnote{See Eq. (\ref{eqc4:smallCD}) and the discussion around it for the preparation of approximate helicity eigenstates in the collimated case.} as an approximate helicity eigenstate with eigenvalue equal to either +1 or -1. The measurements of the Stokes parameters reveal that helicity is preserved.
\begin{table}[h!]\small
	\makebox[\textwidth][c]{
		\begin{tabular}{lcccc} \toprule
 & $S_1$ & $S_2$ & $S_3$&  DOP(\%) \\ \midrule
Input ($\lambda=1$) & $-5\e{-5}\pm3.2\e{-3}$ & $-2.85\e{-3}\pm4.18\e{-3}$ & $0.999992\pm7\e{-6}$ &$99.850\pm0.094$ \\
Output & $-1.7\e{-3}\pm4\e{-3}$ & $3.5\e{-3}\pm6.8\e{-3}$ & $0.999986\pm2.6\e{-5}$ & $99.796\pm0.072$\\\midrule
Input ($\lambda=-1$) & $-3.01\e{-3}\pm3.5\e{-3}$ & $-7.8\e{-3}\pm3.5\e{-3}$ & $-0.999962\pm 2.9\e{-5}$ & $99.310\pm0.068$\\
Output & $-6.7\e{-3}\pm3.6\e{-3}$ & $-1.19\e{-2}\pm3.7\e{-3}$ & $-0.999903\pm4.9\e{-5}$ & $99.23\pm0.066$\\
\end{tabular}
}

\caption[Maltose forward scattering polarimetric measurements.]{\label{tabc6:fw} Forward scattering polarimetric measurements off a maltose solution. The Stokes parameters and the degree of polarization (DOP) of the input and output beams were measured with a polarimeter. The two first data rows correspond to an input of positive helicity and the two last rows to an input of negative helicity. The preservation of $S_3$ after propagation of the solution shows that helicity is preserved by the sample in the forward scattering direction. Experiment performed by Alex Barbara and Dr. Xavier Vidal.}
\end{table}

On the other hand, the measurements at 90 degrees in Fig. \ref{figc6:nfw} show that helicity was not preserved. The power measurements at 90 degrees were done using a linear polarizer which was rotated at 5 degree steps. If the input helicity had been preserved, such measurement would have given a flat curve with respect to the polarizer orientation.

\begin{figure}[h!]
\begin{center}
\makeatletter{}\def\ASYprefix{}
\newbox\ASYbox
\newdimen\ASYdimen
\long\def\ASYbase#1#2{\leavevmode\setbox\ASYbox=\hbox{#1}\ASYdimen=\ht\ASYbox\setbox\ASYbox=\hbox{#2}\lower\ASYdimen\box\ASYbox}
\long\def\ASYaligned(#1,#2)(#3,#4)#5#6#7{\leavevmode\setbox\ASYbox=\hbox{#7}\setbox\ASYbox\hbox{\ASYdimen=\ht\ASYbox\advance\ASYdimen by\dp\ASYbox\kern#3\wd\ASYbox\raise#4\ASYdimen\box\ASYbox}\put(#1,#2){#5\wd\ASYbox 0pt\dp\ASYbox 0pt\ht\ASYbox 0pt\box\ASYbox#6}}\long\def\ASYalignT(#1,#2)(#3,#4)#5#6{\ASYaligned(#1,#2)(#3,#4){
}{
}{#6}}
\long\def\ASYalign(#1,#2)(#3,#4)#5{\ASYaligned(#1,#2)(#3,#4){}{}{#5}}
\def\ASYraw#1{
currentpoint currentpoint translate matrix currentmatrix
100 12 div -100 12 div scale
#1
setmatrix neg exch neg exch translate}
 
\makeatletter{}\setlength{\unitlength}{1pt}
\makeatletter\let\ASYencoding\f@encoding\let\ASYfamily\f@family\let\ASYseries\f@series\let\ASYshape\f@shape\makeatother{\catcode`"=12\includegraphics{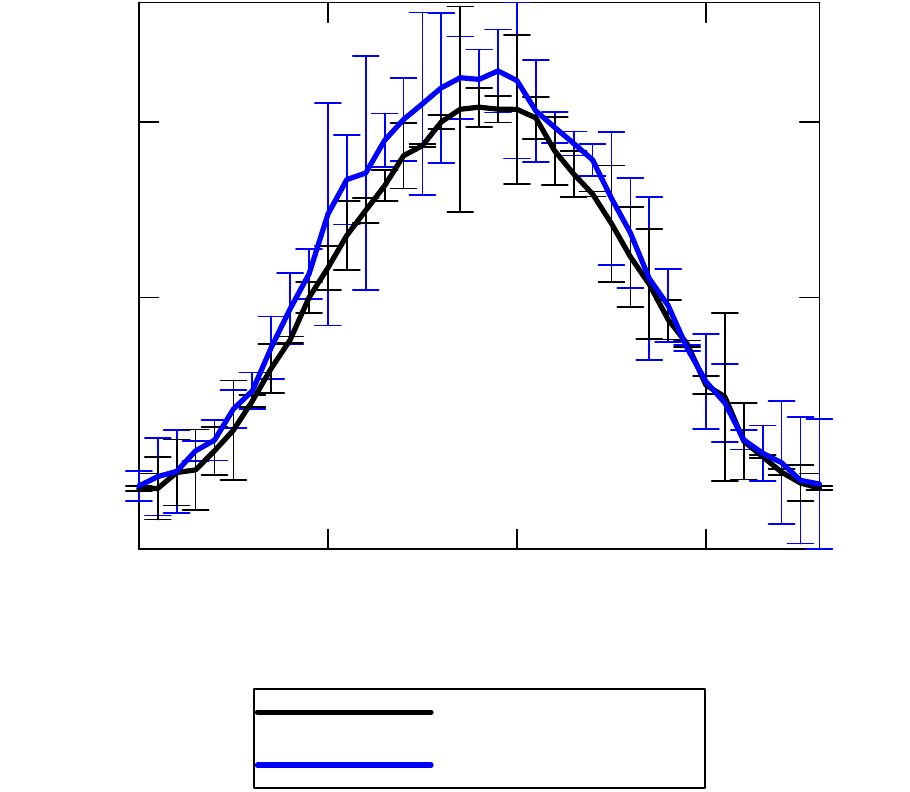}}\definecolor{ASYcolor}{gray}{0.000000}\color{ASYcolor}
\fontsize{12.000000}{14.400000}\selectfont
\usefont{\ASYencoding}{\ASYfamily}{\ASYseries}{\ASYshape}\ASYalign(-227.080853,91.273895)(-1.000000,-0.500000){\vphantom{$10^4$}$50$}
\definecolor{ASYcolor}{gray}{0.000000}\color{ASYcolor}
\fontsize{12.000000}{14.400000}\selectfont
\ASYalign(-227.080853,142.119204)(-1.000000,-0.500000){\vphantom{$10^4$}$100$}
\definecolor{ASYcolor}{gray}{0.000000}\color{ASYcolor}
\fontsize{12.000000}{14.400000}\selectfont
\ASYalign(-227.080853,192.964514)(-1.000000,-0.500000){\vphantom{$10^4$}$150$}
\definecolor{ASYcolor}{gray}{0.000000}\color{ASYcolor}
\fontsize{12.000000}{14.400000}\selectfont
\ASYalignT(-249.806798,148.371943)(-0.500000,0.249999){0.000000 -1.000000 1.000000 0.000000}{{\small Measured power (a.u)}}
\definecolor{ASYcolor}{gray}{0.000000}\color{ASYcolor}
\fontsize{12.000000}{14.400000}\selectfont
\ASYalign(-222.580853,64.872776)(-0.500000,-1.000000){\vphantom{$10^4$}$0$}
\definecolor{ASYcolor}{gray}{0.000000}\color{ASYcolor}
\fontsize{12.000000}{14.400000}\selectfont
\ASYalign(-167.934437,64.872776)(-0.500000,-1.000000){\vphantom{$10^4$}$50$}
\definecolor{ASYcolor}{gray}{0.000000}\color{ASYcolor}
\fontsize{12.000000}{14.400000}\selectfont
\ASYalign(-113.288022,64.872776)(-0.500000,-1.000000){\vphantom{$10^4$}$100$}
\definecolor{ASYcolor}{gray}{0.000000}\color{ASYcolor}
\fontsize{12.000000}{14.400000}\selectfont
\ASYalign(-58.641607,64.872776)(-0.500000,-1.000000){\vphantom{$10^4$}$150$}
\definecolor{ASYcolor}{gray}{0.000000}\color{ASYcolor}
\fontsize{12.000000}{14.400000}\selectfont
\ASYalign(-124.217305,50.261381)(-0.500000,-0.750001){{\small Orientation of the measurement polarizer (degrees)}}
\definecolor{ASYcolor}{gray}{0.000000}\color{ASYcolor}
\fontsize{12.000000}{14.400000}\selectfont
\ASYalign(-134.519251,22.252736)(0.000000,-0.281250){{\small Input $\lambda_{in}=1$}}
\definecolor{ASYcolor}{rgb}{0.000000,0.000000,1.000000}\color{ASYcolor}
\fontsize{12.000000}{14.400000}\selectfont
\ASYalign(-134.519251,6.970040)(0.000000,-0.281250){{\small Input $\lambda_{in}=-1$}}
 
\end{center}
\caption[Measurements at 90 degree scattering off a maltose solution.]{\label{figc6:nfw} Linear polarization power at 90 degrees scattering angle off a solution of maltose as a function of the measurement polarizer orientation. The error bars denote 95\% confidence intervals. Since the collimated inputs were helicity eigenstates to a good approximation, both curves should be flat in the case of a helicity preserving scattering. The shape of the curves implies some degree of helicity changing by the solution of maltose at 90 degrees scattering angle. Experiment performed by Alex Barbara and Dr. Xavier Vidal.}
\end{figure}

\subsection{Disorder and duality symmetry}\label{secc6:disorder}
In this section I provide some analytical and numerical evidence that helicity is not preserved by a random mixture of non-dual particles. It indicates that duality of the particle is required for duality of the solution. Unlike in the case of forward scattering, randomness does not induce helicity preservation in a general scattering direction.

I will use the Mueller matrix formalism to study random mixtures of different kinds of particles and provide analytical and numerical evidence that helicity is only preserved for all $(\pp,\ppbar)$ when the individual particles preserve helicity. 

In general, equation (\ref{eqc6:L}) must be evaluated numerically. I later provide numerically obtained values for mixtures of spherical, conical and helical particles. For the simple case of small (w.r.t. the wavelength) spherical particles with relative electric constant $\epsilon$ and relative magnetic constant $\mu=1$, there is an analytical expression for $S_u(\pp,\ppbar)$ \cite{Tsang2004} which allows to obtain $L_S(\pp,\ppbar)$ also analytically,
\begin{equation}
\label{eqc6:sphere}
\begin{split}
&L_S(\pp,\ppbar)=\delta_{\pp\ppbar}I_{4\text{x}4}+n_0 k^2a^3\frac{\epsilon -1}{4\pi(\epsilon+2)}\times\\
&\times\begin{bmatrix}
\cos^2(\dtheta)+1 & \cos^2(\dtheta)-1 & 0 & 0\\
\cos^2(\dtheta)-1 & \cos^2(\dtheta)+1 & 0 & 0\\
0&0& 2\cos(\dtheta)&0\\
0&0&0& 2\cos(\dtheta)\end{bmatrix},
\end{split}
\end{equation}
where $k$ is the wavenumber, $a$ is the radius of the sphere and $\dtheta$ is the angle between the input and output momentum vectors. The first term is the $4\times 4$ identity matrix, which, as indicated by the Kronecker delta $\delta_{\pp\ppbar}$ is only added when $\pp=\ppbar$. It represents the contribution of the original input plane wave (\ref{eqc3:IT}).

For a helicity preserving system, the two Stokes vectors of well defined helicity $\begin{bmatrix} 1&0&0&\pm1\end{bmatrix}$ must be eigenvectors of the Mueller matrix of the system $L$. This restricts the matrix coefficients $L_{ij}$:
\begin{equation}
\label{eqc6:helpres}
L_{11}=L_{44},\ L_{14}=L_{41},\ L_{21}=L_{31}=L_{24}=L_{34}=0.
\end{equation}

For general $\dtheta$, matrix (\ref{eqc6:sphere}) violates the helicity preserving conditions (\ref{eqc6:helpres}). Therefore, in general, a solution of small non-dual spheres does not preserve helicity.

Nora Tischler, Dr. Xavier Vidal and I designed and performed simulations to investigate whether the conclusions reached for small spheres also hold for mixtures of other kinds of particles and sizes. We numerically computed the rotational average (\ref{eqc6:L}) for small conical, and helical particles and for spheres of different sizes, with $\epsilon=2.25$ and $\mu=1$ immersed in vacuum.

To measure the degree of helicity transformation in each case we used the following metric on the resulting Mueller matrices:
\begin{equation}
\label{eqc6:gamma}
\begin{split}
\Gamma = & \frac{\left(L_{11}+L_{14}-\left(L_{41}+L_{44}\right)\right)^2}{2\left(L_{11}+L_{14}-\left(L_{41}+L_{44}\right)\right)^2+\left(L_{11}+L_{14}+\left(L_{41}+L_{44}\right)\right)^2}+\\
&\frac{\left(L_{11}-L_{14}+\left(L_{41}-L_{44}\right)\right)^2}{2\left(L_{11}-L_{14}+\left(L_{41}-L_{44}\right)\right)^2+\left(L_{11}-L_{14}-\left(L_{41}-L_{44}\right)\right)^2}.
\end{split}
\end{equation}
The first (second) line in (\ref{eqc6:gamma}) is the relative helicity change affected by the Mueller matrix on a Stokes vector of well defined positive (negative) helicity. $\Gamma=0$ for a helicity preserving Mueller matrix (\ref{eqc6:helpres}), and $\Gamma=1$ for a perfect helicity flipping Mueller matrix.

\begin{figure}[h!]
\begin{flushleft}
\subfloat{
	\makeatletter{}\def\ASYprefix{}
\newbox\ASYbox
\newdimen\ASYdimen
\long\def\ASYbase#1#2{\leavevmode\setbox\ASYbox=\hbox{#1}\ASYdimen=\ht\ASYbox\setbox\ASYbox=\hbox{#2}\lower\ASYdimen\box\ASYbox}
\long\def\ASYaligned(#1,#2)(#3,#4)#5#6#7{\leavevmode\setbox\ASYbox=\hbox{#7}\setbox\ASYbox\hbox{\ASYdimen=\ht\ASYbox\advance\ASYdimen by\dp\ASYbox\kern#3\wd\ASYbox\raise#4\ASYdimen\box\ASYbox}\put(#1,#2){#5\wd\ASYbox 0pt\dp\ASYbox 0pt\ht\ASYbox 0pt\box\ASYbox#6}}\long\def\ASYalignT(#1,#2)(#3,#4)#5#6{\ASYaligned(#1,#2)(#3,#4){
}{
}{#6}}
\long\def\ASYalign(#1,#2)(#3,#4)#5{\ASYaligned(#1,#2)(#3,#4){}{}{#5}}
\def\ASYraw#1{
currentpoint currentpoint translate matrix currentmatrix
100 12 div -100 12 div scale
#1
setmatrix neg exch neg exch translate}
 
	\makeatletter{}\setlength{\unitlength}{1pt}
\makeatletter\let\ASYencoding\f@encoding\let\ASYfamily\f@family\let\ASYseries\f@series\let\ASYshape\f@shape\makeatother{\catcode`"=12\includegraphics{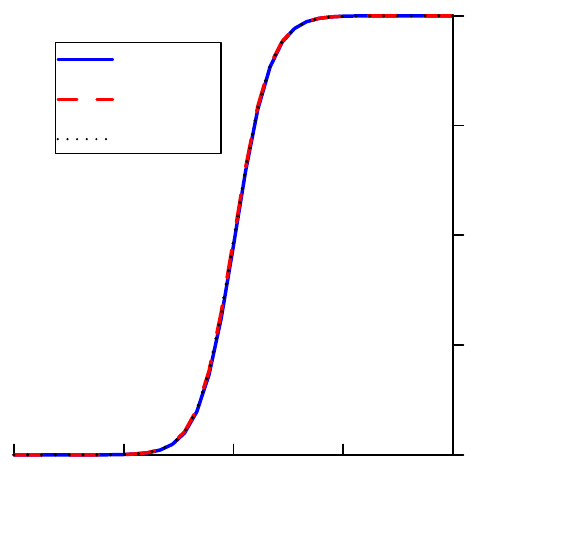}}\definecolor{ASYcolor}{gray}{0.000000}\color{ASYcolor}
\fontsize{8.000000}{9.600000}\selectfont
\usefont{\ASYencoding}{\ASYfamily}{\ASYseries}{\ASYshape}\ASYalign(-28.655180,27.302780)(0.000000,-0.500000){\vphantom{$10^4$}$0$}
\definecolor{ASYcolor}{gray}{0.000000}\color{ASYcolor}
\fontsize{8.000000}{9.600000}\selectfont
\ASYalign(-28.655180,59.033174)(0.000000,-0.500000){\vphantom{$10^4$}$0.25$}
\definecolor{ASYcolor}{gray}{0.000000}\color{ASYcolor}
\fontsize{8.000000}{9.600000}\selectfont
\ASYalign(-28.655180,90.763567)(0.000000,-0.500000){\vphantom{$10^4$}$0.5$}
\definecolor{ASYcolor}{gray}{0.000000}\color{ASYcolor}
\fontsize{8.000000}{9.600000}\selectfont
\ASYalign(-28.655180,122.493961)(0.000000,-0.500000){\vphantom{$10^4$}$0.75$}
\definecolor{ASYcolor}{gray}{0.000000}\color{ASYcolor}
\fontsize{8.000000}{9.600000}\selectfont
\ASYalign(-28.655180,154.224355)(0.000000,-0.500000){\vphantom{$10^4$}$1$}
\definecolor{ASYcolor}{gray}{0.000000}\color{ASYcolor}
\fontsize{12.000000}{14.400000}\selectfont
\ASYalign(-8.842725,90.763567)(0.000000,-0.500000){$\Gamma$}
\definecolor{ASYcolor}{gray}{0.000000}\color{ASYcolor}
\fontsize{8.000000}{9.600000}\selectfont
\ASYalign(-160.988007,24.302780)(-0.500000,-1.000000){\vphantom{$10^4$}$0$}
\definecolor{ASYcolor}{gray}{0.000000}\color{ASYcolor}
\fontsize{8.000000}{9.600000}\selectfont
\ASYalign(-129.257613,24.302780)(-0.500000,-1.000000){\vphantom{$10^4$}$0.25$}
\definecolor{ASYcolor}{gray}{0.000000}\color{ASYcolor}
\fontsize{8.000000}{9.600000}\selectfont
\ASYalign(-97.527219,24.302780)(-0.500000,-1.000000){\vphantom{$10^4$}$0.5$}
\definecolor{ASYcolor}{gray}{0.000000}\color{ASYcolor}
\fontsize{8.000000}{9.600000}\selectfont
\ASYalign(-65.796826,24.302780)(-0.500000,-1.000000){\vphantom{$10^4$}$0.75$}
\definecolor{ASYcolor}{gray}{0.000000}\color{ASYcolor}
\fontsize{8.000000}{9.600000}\selectfont
\ASYalign(-34.066432,24.302780)(-0.500000,-1.000000){\vphantom{$10^4$}$1$}
\definecolor{ASYcolor}{gray}{0.000000}\color{ASYcolor}
\fontsize{12.000000}{14.400000}\selectfont
\ASYalign(-97.527218,12.911665)(-0.500000,-0.719729){{\small $\chi_{\mathbf{p\bar{p}}}/\pi$}}
\definecolor{ASYcolor}{rgb}{0.000000,0.000000,1.000000}\color{ASYcolor}
\fontsize{12.000000}{14.400000}\selectfont
\ASYalignT(-129.842306,141.532197)(0.000000,-0.281251){0.700000 -0.000000 -0.000000 0.700000}{Spheres}
\definecolor{ASYcolor}{rgb}{1.000000,0.000000,0.000000}\color{ASYcolor}
\fontsize{12.000000}{14.400000}\selectfont
\ASYalignT(-129.842306,130.050347)(0.000000,-0.500000){0.700000 -0.000000 -0.000000 0.700000}{Cones}
\definecolor{ASYcolor}{gray}{0.000000}\color{ASYcolor}
\fontsize{12.000000}{14.400000}\selectfont
\ASYalignT(-129.842306,118.568496)(0.000000,-0.500000){0.700000 -0.000000 -0.000000 0.700000}{Helices}
 
}
\subfloat{
	\makeatletter{}\def\ASYprefix{}
\newbox\ASYbox
\newdimen\ASYdimen
\long\def\ASYbase#1#2{\leavevmode\setbox\ASYbox=\hbox{#1}\ASYdimen=\ht\ASYbox\setbox\ASYbox=\hbox{#2}\lower\ASYdimen\box\ASYbox}
\long\def\ASYaligned(#1,#2)(#3,#4)#5#6#7{\leavevmode\setbox\ASYbox=\hbox{#7}\setbox\ASYbox\hbox{\ASYdimen=\ht\ASYbox\advance\ASYdimen by\dp\ASYbox\kern#3\wd\ASYbox\raise#4\ASYdimen\box\ASYbox}\put(#1,#2){#5\wd\ASYbox 0pt\dp\ASYbox 0pt\ht\ASYbox 0pt\box\ASYbox#6}}\long\def\ASYalignT(#1,#2)(#3,#4)#5#6{\ASYaligned(#1,#2)(#3,#4){
}{
}{#6}}
\long\def\ASYalign(#1,#2)(#3,#4)#5{\ASYaligned(#1,#2)(#3,#4){}{}{#5}}
\def\ASYraw#1{
currentpoint currentpoint translate matrix currentmatrix
100 12 div -100 12 div scale
#1
setmatrix neg exch neg exch translate}
 
	\makeatletter{}\setlength{\unitlength}{1pt}
\makeatletter\let\ASYencoding\f@encoding\let\ASYfamily\f@family\let\ASYseries\f@series\let\ASYshape\f@shape\makeatother{\catcode`"=12\includegraphics{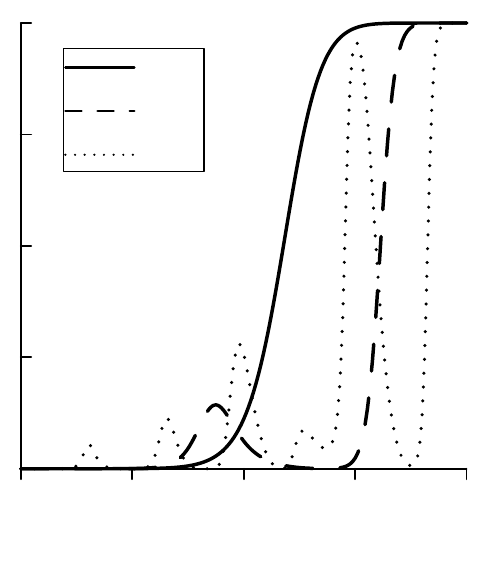}}\definecolor{ASYcolor}{gray}{0.000000}\color{ASYcolor}
\fontsize{12.000000}{14.400000}\selectfont
\usefont{\ASYencoding}{\ASYfamily}{\ASYseries}{\ASYshape}\ASYalign(-136.581399,29.714030)(-1.000000,-0.500000){\vphantom{$10^4$}$ $}
\definecolor{ASYcolor}{gray}{0.000000}\color{ASYcolor}
\fontsize{12.000000}{14.400000}\selectfont
\ASYalign(-136.581399,61.927841)(-1.000000,-0.500000){\vphantom{$10^4$}$ $}
\definecolor{ASYcolor}{gray}{0.000000}\color{ASYcolor}
\fontsize{12.000000}{14.400000}\selectfont
\ASYalign(-136.581399,94.141652)(-1.000000,-0.500000){\vphantom{$10^4$}$ $}
\definecolor{ASYcolor}{gray}{0.000000}\color{ASYcolor}
\fontsize{12.000000}{14.400000}\selectfont
\ASYalign(-136.581399,126.355463)(-1.000000,-0.500000){\vphantom{$10^4$}$ $}
\definecolor{ASYcolor}{gray}{0.000000}\color{ASYcolor}
\fontsize{12.000000}{14.400000}\selectfont
\ASYalign(-136.581399,158.569274)(-1.000000,-0.500000){\vphantom{$10^4$}$ $}
\definecolor{ASYcolor}{gray}{0.000000}\color{ASYcolor}
\fontsize{8.000000}{9.600000}\selectfont
\ASYalign(-132.081399,24.302780)(-0.500000,-1.000000){\vphantom{$10^4$}$0$}
\definecolor{ASYcolor}{gray}{0.000000}\color{ASYcolor}
\fontsize{8.000000}{9.600000}\selectfont
\ASYalign(-99.867588,24.302780)(-0.500000,-1.000000){\vphantom{$10^4$}$0.25$}
\definecolor{ASYcolor}{gray}{0.000000}\color{ASYcolor}
\fontsize{8.000000}{9.600000}\selectfont
\ASYalign(-67.653777,24.302780)(-0.500000,-1.000000){\vphantom{$10^4$}$0.5$}
\definecolor{ASYcolor}{gray}{0.000000}\color{ASYcolor}
\fontsize{8.000000}{9.600000}\selectfont
\ASYalign(-35.439966,24.302780)(-0.500000,-1.000000){\vphantom{$10^4$}$0.75$}
\definecolor{ASYcolor}{gray}{0.000000}\color{ASYcolor}
\fontsize{8.000000}{9.600000}\selectfont
\ASYalign(-3.226155,24.302780)(-0.500000,-1.000000){\vphantom{$10^4$}$1$}
\definecolor{ASYcolor}{gray}{0.000000}\color{ASYcolor}
\fontsize{12.000000}{14.400000}\selectfont
\ASYalign(-67.653776,12.911665)(-0.500000,-0.719729){{\small $\chi_{\mathbf{p\bar{p}}}/\pi$}}
\definecolor{ASYcolor}{gray}{0.000000}\color{ASYcolor}
\fontsize{12.000000}{14.400000}\selectfont
\ASYalignT(-96.758945,145.683750)(0.000000,-0.250000){0.700000 -0.000000 -0.000000 0.700000}{$\lambda_0/2$}
\definecolor{ASYcolor}{gray}{0.000000}\color{ASYcolor}
\fontsize{12.000000}{14.400000}\selectfont
\ASYalignT(-96.758945,133.081860)(0.000000,-0.322369){0.700000 -0.000000 -0.000000 0.700000}{$\lambda_0$}
\definecolor{ASYcolor}{gray}{0.000000}\color{ASYcolor}
\fontsize{12.000000}{14.400000}\selectfont
\ASYalignT(-96.758945,120.479970)(0.000000,-0.322369){0.700000 -0.000000 -0.000000 0.700000}{$2\lambda_0$}
 
}

\end{flushleft}
\caption[Helicity transformation in solutions of different particles.]{\label{figc6:final} Helicity transformation metric $\Gamma(\dtheta)$ results obtained from the numerical computation of the Mueller matrices of solutions (in vacuum) of different kinds of particles with $\epsilon=2.25$ and $\mu=1$. $\Gamma=0$ corresponds to helicity preservation and $\Gamma=1$ to perfect helicity flipping. A helicity preserving solution should exhibit $\Gamma=0$ for all $\dtheta$. (a) Spheres, cones and helices of dimension $\approx \lambda_0/6$. The wavelength of light was $\lambda_0=$632 nm. (b) Spheres with diameters $\approx [\lambda_0/2,\lambda_0,2\lambda_0]$. All cases break helicity preservation. In the small particle case (a), the results are very similar independently of the kind of particle (small differences not visible in the figure). For larger sizes (b), there is a more complex behavior of $\Gamma(\dtheta)$ with oscillations where $\Gamma$ is close to zero for some angles. This behavior may be related to the excitation of higher (than dipole) multipolar moments. }
\end{figure}

Figure \ref{figc6:final}-(a) plots $\Gamma$ as a function of the relative angle between the input and output momenta ($\dtheta$) for spheres, cones and helices of dimensions $\approx \lambda_0/6$, where $\lambda_0$ is the wavelength. The individual scattering matrices $S^R_u(\pp,\ppbar)$ for the computation of (\ref{eqc6:L}) were obtained by illuminating the single scatterer with plane waves of defined helicity from all input directions described by polar and azimuthal angles $(\theta,\phi)$. For each input direction, the scattered far field in all output directions $(\bar{\theta},\bar{\phi})$ was then computed. For the sphere, the computation was done analytically by applying the Mie scattering theory. For the cylinder and the helix, a commercial finite elements package (COMSOL) was used in which far field calculations are made with the Stratton-Chu formula. Given an input output momenta pair $(\theta,\phi),(\bar{\theta},\bar{\phi})$, the helicity scattering coefficients are asymptotically proportional to the projection of the far field for direction $(\bar{\theta},\bar{\phi})$ onto the corresponding polarization vectors (see (\ref{eqc3:pwhel})). For the sphere, the integral $\int dR$ in (\ref{eqc6:L}) is trivial. For the cone and the helix, it must be numerically computed. All angles were discretized using 5 degrees intervals. Fig. \ref{figc6:final}-(b) plots $\Gamma(\dtheta)$ for spheres of diameters $\approx [\lambda_0/2,\lambda_0,2\lambda_0]$. All cases show that the solutions do not preserve helicity. This may indicate that, in general, for the solution to preserve helicity, the individual particles must preserve helicity. Disorder seems to help only in the forward scattering direction. Note how the helicity preservation properties are independent of the geometrical properties of the particles. Spheres have all mirror planes of symmetry, helices lack them all and cones have some but not all. In all cases $\Gamma(0)=0$ and $\Gamma(\pi)=1$, reflecting the fact that for a cylindrically symmetric system helicity is preserved in forward scattering and flipped in backward scattering, in agreement with the results contained in Chap. \ref{chap5}. 

The results of this and the previous sections suggest the following answer to Condon's question reproduced at the beginning of Sec. \ref{secc6:moa}. In the common setup of forward propagation through a solution of chiral molecules, optical activity is not only due to structural features of the individual molecules. The necessary helicity preservation is achieved through the effective cylindrical symmetry that the disordered orientations of the molecules induce onto the solution as a whole. No extra structural feature besides chirality is required of the molecule in this case. On the other hand, should scattering directions other than forward be considered, it seems that another structural feature of the molecule is necessary: Duality symmetry\footnote{The duality conditions for a dipolar scatterer can be found in Eq. (\ref{eqc3:dual}).}. The same can be said for single molecules or ordered ensembles of molecules.

\section{Designing artificial optical activity}\label{secc6:aoa}
The results of Secs. \ref{secc6:nec} and \ref{secc6:moa} can be applied to the design of structures exhibiting artificial optical activity. Such structures are a matter of current research interest. See for example \cite{Baev2007,Decker2010,Zhao2012,Ren2012}.

The requirement of breaking mirror symmetries is widely recognized in the field, but the requirement of helicity preservation does not seem to be. I speculate that this is due to the fact that natural molecular optical activity has up to now been only related to the breaking of spatial inversion symmetries, and not to helicity preservation. Two situations must be distinguished in design problems: Optical activity in forward scattering only and optical activity in non-forward scattering directions. If optical activity is only needed in forward scattering, the necessary helicity preservation can be achieved by means of geometrical symmetries of the scatterer: Discrete rotational symmetries $R_z(2\pi/n)$ with $n\ge 3$ (see Chap. \ref{chap5}). Additionally, the scatterer must break all mirror symmetries containing the propagation axis. The designs in \cite{Decker2010} and \cite{Kaschke2012} meet the two necessary conditions for optical activity in the forward direction. 

On the other hand, when optical activity in directions other than forward is required, discrete rotational symmetries do not help achieving helicity preservation. The evidence presented in \ref{secc6:disorder} suggest that helicity preservation is then only achievable by dual symmetric scatterers. Designs for non-forward directions that do not take the helicity preservation requirement into account result in rotation angles $\beta$ which depend on the input polarization angle $\alpha$, for example in \cite{Ren2012} and \cite{Papakostas2003}. Let me then assume that duality is indeed needed. A design strategy for the case of ordered arrays of inclusions like \cite{Papakostas2003,Ren2012} is, provided that the inclusions are small enough, to engineer their polarizability tensors to have a dual symmetric dipolar response according to the conditions (\ref{eqc3:dual}) and to break the appropriate mirror symmetries. For example, chiral inclusions would break them all. In Chap. \ref{chap7} I will briefly discuss two types of dual dipolar scatterers. One of them is chiral.

As a last example, the recipe for an object with zero backscattering and optical activity in forward scattering is straightforward when using results from Chap. \ref{chap5}. The object must be dual, break all the reflection symmetries across planes containing the optical axis and possess a discrete rotational $R_z(2\pi/n)$ symmetry of degree $n\ge 3$. A 2D array of chiral and dual dipolar scatterers arranged to have the discrete rotational symmetry would accomplish it.

To finalize, I would like to point out that the consideration of helicity and duality allows to treat near field situations. Enhancing the interaction between molecules and fields by means of strong near fields is a very important technique. Such strong near fields appear for example very close to resonant nanostructures. The analysis in this chapter, implicitly identifying scattering directions with plane waves, relates to the far field but its extension to the near field is straightforward. For example, if one desires to excite adsorbed molecules with a strong near field of pure handedness, the recipe is to design resonant nanostructures that have duality symmetry. The appropriate external illumination would, upon interaction with the nanostructure, excite a resonant near field with well defined helicity which would interact with the adsorbed molecules. 

\section{Faraday rotation}\label{secc6:fad}
Systems that break spatial inversion symmetries are not the only ones able to rotate linear polarization. Those that break time inversion symmetry can do it as well. The effect is similar but not identical.

When a linearly polarized electromagnetic wave propagates in a medium in the presence of an external magnetic field parallel to the propagation direction, the polarization of the wave rotates. This effect is called Faraday rotation or Faraday effect. It has applications for example in astrophysics \cite[Prob. 7.15]{Jackson1998}, and polarization dependent isolators. These isolators exploit the following effect: If the polarization is rotated by $\theta$ degrees when the wave propagates forward through the medium, it is rotated by an additional $\theta$ degrees when it propagates back after a reflection. The final rotation is $2\theta$. In the case of a solution of chiral molecules, the polarization rotation on the way back exactly cancels the polarization rotation of the way forth resulting in 0 total rotation.

It is interesting to see that this difference is a direct consequence of the properties of each system under time and spatial inversion symmetries, and the transformation properties of plane waves with well defined helicity under these two discrete transformations. The objective of this section is to show this difference in a formal way.

Consider the two different media in Fig. \ref{figc6:faraday}. The medium of Fig. \ref{figc6:faraday}-(a) is homogeneous and isotropic, and there is an external static magnetic field. The medium in Fig. \ref{figc6:faraday}-(b) is a solution of chiral molecules, also homogeneous and isotropic in effect. Here are the assumptions about the two media.

\begin{figure}[h!]
\begin{center}
\makeatletter{}\def\ASYprefix{}
\newbox\ASYbox
\newdimen\ASYdimen
\long\def\ASYbase#1#2{\leavevmode\setbox\ASYbox=\hbox{#1}\ASYdimen=\ht\ASYbox\setbox\ASYbox=\hbox{#2}\lower\ASYdimen\box\ASYbox}
\long\def\ASYaligned(#1,#2)(#3,#4)#5#6#7{\leavevmode\setbox\ASYbox=\hbox{#7}\setbox\ASYbox\hbox{\ASYdimen=\ht\ASYbox\advance\ASYdimen by\dp\ASYbox\kern#3\wd\ASYbox\raise#4\ASYdimen\box\ASYbox}\put(#1,#2){#5\wd\ASYbox 0pt\dp\ASYbox 0pt\ht\ASYbox 0pt\box\ASYbox#6}}\long\def\ASYalignT(#1,#2)(#3,#4)#5#6{\ASYaligned(#1,#2)(#3,#4){
}{
}{#6}}
\long\def\ASYalign(#1,#2)(#3,#4)#5{\ASYaligned(#1,#2)(#3,#4){}{}{#5}}
\def\ASYraw#1{
currentpoint currentpoint translate matrix currentmatrix
100 12 div -100 12 div scale
#1
setmatrix neg exch neg exch translate}
 
\makeatletter{}\setlength{\unitlength}{1pt}
\makeatletter\let\ASYencoding\f@encoding\let\ASYfamily\f@family\let\ASYseries\f@series\let\ASYshape\f@shape\makeatother{\catcode`"=12\includegraphics{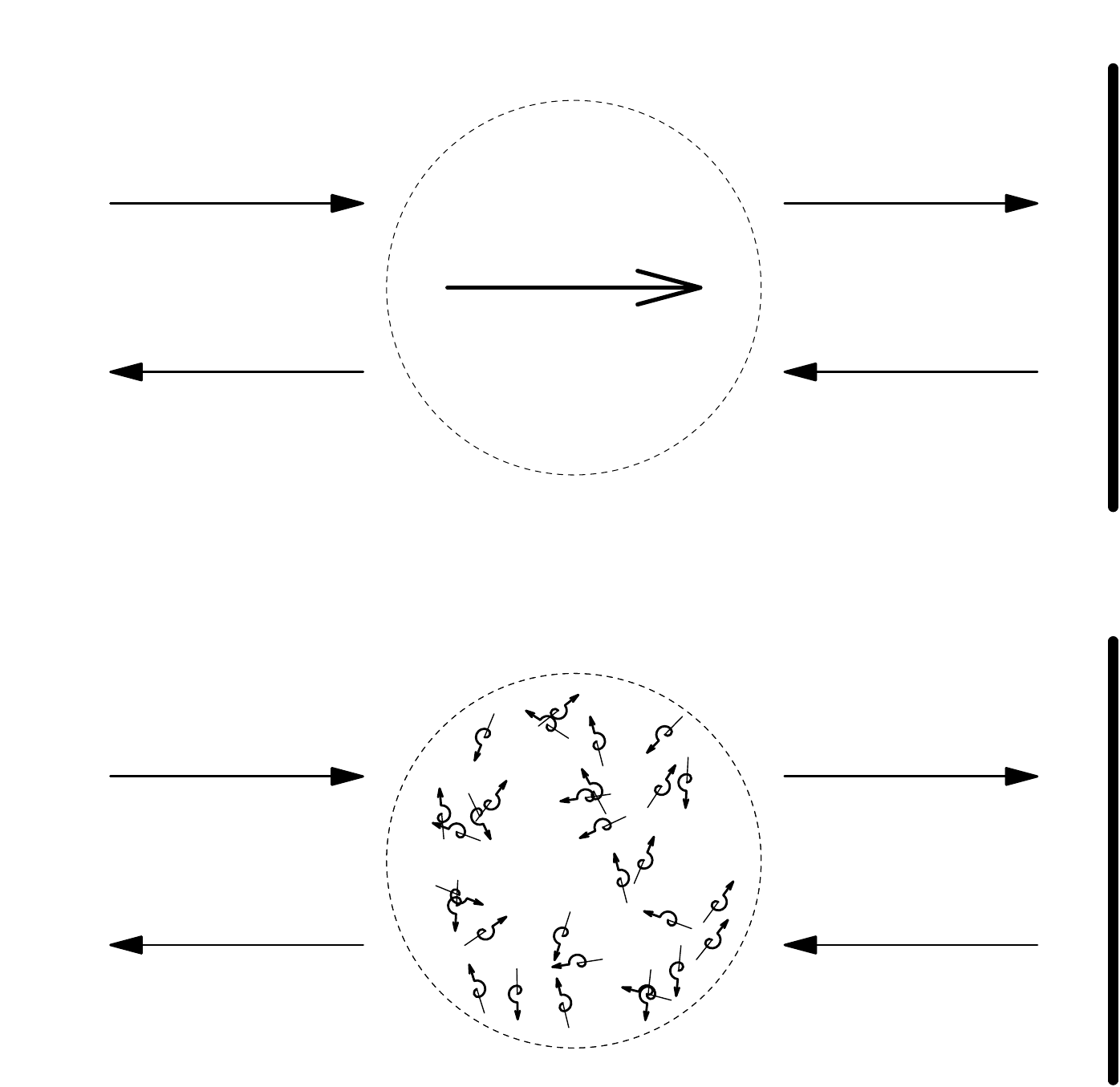}}\definecolor{ASYcolor}{gray}{0.000000}\color{ASYcolor}
\fontsize{12.000000}{14.400000}\selectfont
\usefont{\ASYencoding}{\ASYfamily}{\ASYseries}{\ASYshape}\ASYalign(-202.257486,177.197919)(-0.500000,-0.250001){{\small b) Optical activity: time inversion symmetric, spatial inversion NOT symmetric}}
\definecolor{ASYcolor}{gray}{0.000000}\color{ASYcolor}
\fontsize{12.000000}{14.400000}\selectfont
\ASYalign(-317.705583,96.021744)(-0.500000,-0.250000){$|\mathbf{p}\ +\rangle,\ |\mathbf{p}\ -\rangle$}
\definecolor{ASYcolor}{gray}{0.000000}\color{ASYcolor}
\fontsize{12.000000}{14.400000}\selectfont
\ASYalign(-363.257057,126.389393)(-0.500000,-0.500000){in}
\definecolor{ASYcolor}{gray}{0.000000}\color{ASYcolor}
\fontsize{12.000000}{14.400000}\selectfont
\ASYalign(-68.690860,96.021744)(-0.500000,-0.250000){$\alpha_b|\mathbf{p}\ +\rangle,\ \beta_b|\mathbf{p}\ -\rangle$}
\definecolor{ASYcolor}{gray}{0.000000}\color{ASYcolor}
\fontsize{12.000000}{14.400000}\selectfont
\ASYalign(-74.764389,35.286445)(-0.500000,-0.250000){$\alpha_b|\mathbf{-p}\ -\rangle,\ \beta_b|\mathbf{-p}\ +\rangle$}
\definecolor{ASYcolor}{gray}{0.000000}\color{ASYcolor}
\fontsize{12.000000}{14.400000}\selectfont
\ASYalign(-335.926173,35.286445)(-0.500000,-0.250000){$\alpha_b\beta_b|\mathbf{-p}\ -\rangle,\ \alpha_b\beta_b|\mathbf{-p}\ +\rangle$}
\definecolor{ASYcolor}{gray}{0.000000}\color{ASYcolor}
\fontsize{12.000000}{14.400000}\selectfont
\ASYalign(-366.293822,65.654095)(-0.500000,-0.500000){out}
\definecolor{ASYcolor}{gray}{0.000000}\color{ASYcolor}
\fontsize{12.000000}{14.400000}\selectfont
\ASYalign(-202.257486,177.197919)(-0.500000,-0.250001){{\small b) Optical activity: time inversion symmetric, spatial inversion NOT symmetric}}
\definecolor{ASYcolor}{gray}{0.000000}\color{ASYcolor}
\fontsize{12.000000}{14.400000}\selectfont
\ASYalign(-202.257486,383.697934)(-0.500000,-0.250001){{\small a) Faraday rotation: spatial inversion symmetric, time inversion NOT symmetric}}
\definecolor{ASYcolor}{gray}{0.000000}\color{ASYcolor}
\fontsize{12.000000}{14.400000}\selectfont
\ASYalignT(-196.234986,283.737934)(-0.500000,-1.000000){1.200000 -0.000000 -0.000000 1.200000}{$\mathbf{B}$}
\definecolor{ASYcolor}{gray}{0.000000}\color{ASYcolor}
\fontsize{12.000000}{14.400000}\selectfont
\ASYalign(-317.705583,302.521759)(-0.500000,-0.250000){$|\mathbf{p}\ +\rangle,\ |\mathbf{p}\ -\rangle$}
\definecolor{ASYcolor}{gray}{0.000000}\color{ASYcolor}
\fontsize{12.000000}{14.400000}\selectfont
\ASYalign(-363.257057,332.889408)(-0.500000,-0.500000){in}
\definecolor{ASYcolor}{gray}{0.000000}\color{ASYcolor}
\fontsize{12.000000}{14.400000}\selectfont
\ASYalign(-68.690860,302.521759)(-0.500000,-0.250000){$\alpha_a|\mathbf{p}\ +\rangle,\ \beta_a|\mathbf{p}\ -\rangle$}
\definecolor{ASYcolor}{gray}{0.000000}\color{ASYcolor}
\fontsize{12.000000}{14.400000}\selectfont
\ASYalign(-74.764389,241.786460)(-0.500000,-0.250000){$\alpha_a|\mathbf{-p}\ -\rangle,\ \beta_a|\mathbf{-p}\ +\rangle$}
\definecolor{ASYcolor}{gray}{0.000000}\color{ASYcolor}
\fontsize{12.000000}{14.400000}\selectfont
\ASYalign(-329.852643,241.786460)(-0.500000,-0.260197){$\alpha_a^2|\mathbf{-p}\ -\rangle,\ \beta_a^2|\mathbf{-p}\ +\rangle$}
\definecolor{ASYcolor}{gray}{0.000000}\color{ASYcolor}
\fontsize{12.000000}{14.400000}\selectfont
\ASYalign(-366.293822,272.154109)(-0.500000,-0.500000){out}
 
\end{center}
\caption[Faraday effect versus optical activity.]{\label{figc6:faraday} Effect of spatial and time inversion symmetry properties. System (a) has spatial inversion symmetry and breaks time inversion. System (b) has time inversion symmetry and breaks spatial inversion. Both systems are assumed to be cylindrically symmetric. The ``in'' plane waves travel through the system a first time, reflect off the mirrors (thick dark line on the right) and travel through the system a second time in the opposite sense. The coefficients relating the complex amplitudes of the ``in'' and ``out'' states can be computed using (\ref{eqc6:sa}) and (\ref{eqc6:sb}). Due to the different properties under the discrete transformations, the behavior of the two systems is different. In (a) the overall coefficients are potentially different for the two helicity states. In (b), they must be equal. Any linear polarization rotation is canceled in (b) and is doubled in (a). This is what happens in a solution of chiral molecules and a medium exhibiting Faraday rotation, respectively.
}
\end{figure}
The first medium is invariant under spatial inversion symmetry but breaks time inversion symmetry. The electromagnetic response of the medium depends on the external field $\BB$, which is even under spatial inversion but flips sign under time inversion and changes the constitutive relations. The second medium is invariant under time inversion symmetry but breaks spatial inversion symmetry.

Therefore, with respect to spatial inversion, the scattering operators of the two media (a), (b), meet:
\begin{equation}
\Pi^{-1}S_a\Pi=S_a,\ \Pi^{-1}S_b\Pi\neq S_b,
\end{equation}
which means that, for arbitrary vectors $|\eta\rangle$, $\bar{\eta}\rangle$ the symmetry enforces
\begin{equation}
	\label{eqc6:ai1}
	\langle \bar{\eta}|S_a|\eta\rangle=\langle \bar{\eta}|\Pi^{-1}S_a\Pi|\eta\rangle
\end{equation}
in system (a), and that, in general
\begin{equation}
	\label{eqc6:ai2}
	\langle \bar{\eta}|S_b|\eta\rangle\neq\langle \bar{\eta}|\Pi^{-1}S_b\Pi|\eta\rangle
\end{equation}
in system (b).

The consequences of time inversion symmetry in the scattering coefficients must be evaluated with care due to the antilinearity of the time inversion operator $Tr$. A detailed discussion can be found in \cite[Chap. 12.7.2]{Tung1985}. The bottom line is that time inversion symmetry of the system imposes that the scattering operator $S$ meets $\langle \bar{\eta}|S_a|\eta\rangle=\langle \eta|Tr^{-1}STr|\bar{\eta}\rangle$. Therefore, for the two systems under consideration, we have that
\begin{equation}
	\label{eqc6:ai3}
	\langle \bar{\eta}|S_b|\eta\rangle=\langle \eta|Tr^{-1}S_bTr|\bar{\eta}\rangle
\end{equation}
and that, in general
\begin{equation}
	\label{eqc6:ai4}
	\langle \bar{\eta}|S_a|\eta\rangle\neq\langle \eta|Tr^{-1}S_aTr|\bar{\eta}\rangle.
\end{equation}

An important clarification is in order at this point. Relations (\ref{eqc6:ai2}) and (\ref{eqc6:ai4}) do not mean that electromagnetic phenomena can break time or spatial inversion symmetries. The complete theory has both those symmetries. This is different from saying that the behavior of a given system considered in isolation breaks one of this discrete symmetries. For example, the time inversion operator $Tr$ in (\ref{eqc6:ai4}) is not meant to act on the sources that originate the $\BB$ field in Fig. \ref{figc6:faraday}-(a). Should these sources be transformed as well, the symmetry would be restored.  See \cite{Barron1972} for an equivalent discussion for a system like the one in Fig. \ref{figc6:faraday}-(b) with respect to spatial inversion. 

Let me recall the transformation properties of plane waves of well defined helicity under $\Pi$ and $Tr$ (Tab. \ref{tabc3:pwheltrans}):
\begin{equation}
	\label{eqc6:tx}
		\Pi|\pp \ \lambda\rangle=|\shortminus\pp \ \shortminus\lambda\rangle,\ Tr|\pp \ \lambda\rangle=-|\shortminus\pp \ \lambda\rangle.
\end{equation}

In both systems, the different scattering coefficients for plane waves of momenta $\pm\pp$ and helicities $\pm$ can be related pairwise using Eqs. (\ref{eqc6:ai1}), (\ref{eqc6:ai3}) and (\ref{eqc6:tx}). In the case of medium (a):
\begin{equation}
	\label{eqc6:sa}
	\begin{split}
		\alpha_a&=\langle +\ \pp|S_a|\pp +\rangle=\langle +\ \pp|\Pi^{-1}S_a\Pi|\pp \ +\rangle=\langle - \ \shortminus\pp|S_a|\shortminus\pp\ -\rangle.\\
	    \beta_a&=\langle -\ \pp|S_a|\pp -\rangle=\langle -\ \pp|\Pi^{-1}S_a\Pi|\pp \ -\rangle=\langle + \ \shortminus\pp|S_a|\shortminus\pp\ +\rangle.
	\end{split}
\end{equation}
For medium (b):
\begin{equation}
	\label{eqc6:sb}
	\begin{split}
		\alpha_b&=\langle +\ \pp|S_b|\pp +\rangle=\langle +\ \pp|Tr^{-1}S_bTr|\pp \ +\rangle=\langle + \ \shortminus\pp|S_b|\shortminus\pp\ +\rangle.\\
	    \beta_b&=\langle -\ \pp|S_b|\pp -\rangle=\langle -\ \pp|Tr^{-1}S_bTr|\pp \ -\rangle=\langle - \ \shortminus\pp|S_b|\shortminus\pp\ -\rangle.
	\end{split}
\end{equation}

In both systems, (\ref{eqc6:sa}) and (\ref{eqc6:sb}) mean that the $\alpha$ and $\beta$ coefficients are related to each other by the symmetry that is broken in each system. This allows $\alpha\neq\beta$, which causes the rotation of the linearly polarized states. 

I will model the action of the reflectors on the right of Fig. \ref{figc6:faraday} as: $|\pp\ \lambda\rangle\rightarrow |\shortminus\pp\ \shortminus\lambda\rangle$. This is consistent with the fact that reflections from a cylindrically symmetric object flip helicity (see Chap. \ref{chap5}).

Finally, note that the cylindrical symmetry of both media ensures helicity preservation on forward scattering in both forth and back transmissions.

Using all these considerations, one can calculate the overall effect of the forth (``towards the right'') and back (``towards the left'') transmissions through each medium. The doubling of the rotation angle in the time inversion breaking system is clearly seen in the coefficients acquired after the round trip: $\alpha_a^2$ and $\beta_a^2$ for the positive and negative helicity cases, respectively. The canceling of the rotation for the space inversion breaking system is also clearly seen since both helicities pick up the same $\alpha_b\beta_b$ factor.

Finally, it is worth mentioning how the properties of medium (a) fit with the necessary conditions for polarization rotation derived in Sec. \ref{secc6:nec}. The presence of the external $\BB$ field breaks all mirror reflection symmetries except the one across the plane orthogonal to $\BB$. This can be seen by recalling that a mirror reflection across a plane perpendicular to vector $\uhat$ can be written as $\Pi R_{\uhat}(\pi)$. Since $\BB$ is unchanged by parity, it will be modified by the $R_{\uhat}(\pi)$ rotation unless $\BB$ is along the direction of $\uhat$. This breaking of mirror symmetries accomplishes one of the necessary conditions from Sec. \ref{secc6:nec}. The situation with respect to the other one, helicity preservation, should in principle be as discussed throughout this chapter: Helicity preservation can be achieved by geometrical means in the forward direction and seems to require duality symmetry in non-forward directions.

\makeatletter{}\chapter[Duality in transformation optics]{Duality symmetry in transformation electromagnetics}
\label{chap7}
\epigraph{{\em Il y avait un nombre important de questions que je m'\'etais pos\'ees et, comme vous le savez, lorsqu'on se pose vraiment les questions, on donne de meilleures r\'eponses que si l'on se contente de lire les r\'eponses convenues.\\There were a significant number of questions I had asked myself and, as you know, when you really ask yourself the questions, you give better answers than when one merely reads the conventional answers.}}{Albert Messiah}
Transformation electromagnetics offers a path to the design of invisibility cloaks, perfect lenses and any other device whose action on the electromagnetic field can be casted as a spacetime coordinate transformation \cite{Leonhardt2006,Pendry2006,Leonhardt2006b}. Transformation electromagnetics\footnote{Also known as ``Transformation optics'', which is the original name used by Leonhardt and Philbin in \cite{Leonhardt2009}. Their formulation is wavelength independent and thus warrants the more general name, which is also used in the literature.} is based on the fact that Maxwell's equations in an arbitrary coordinate system or an empty region of curved spacetime are equivalent to Maxwell's equations inside a material medium in a flat spacetime background \cite{Plebanski1960}. The desired transformation specifies a spacetime metric which at its turn specifies the constitutive relations of the material. A detailed treatise in transformation electromagnetics can be found in \cite{Leonhardt2009}.

Such formidable step in the ability to manipulate electromagnetic waves comes with a correspondingly steep increase in the tunability requirements of material constitutive relations. Nature does not provide nearly enough flexibility in this aspect. We must synthesize artificial materials: Electromagnetic metamaterials \cite{Kock1956}. Transformation media are typically implemented by means of an ensemble of inclusions in an homogeneous and isotropic dielectric. These inclusions are sometimes referred to as meta atoms. The idea is to obtain the required constitutive relations from the collective response of the meta atoms. Currently, though, there is no systematic design methodology to go from the constitutive relations to the actual implementation of the metamaterial. In general, this is a highly complex task, partly because of the large number of degrees of freedom which include the electromagnetic response of the meta atoms and their three dimensional spatial arrangement. Reducing the number of degrees of freedom that have to be managed while maintaining the ability to implement general coordinate transformations is desirable.

In this chapter, I study the role of duality symmetry in transformation electromagnetic devices, in particular in their implementation by means of metamaterials. The fact that duality is an inherent symmetry of transformation electromagnetics (Sec. \ref{secc7:dto}) allows to constrain the individual response of the meta atoms without restricting the implementable transformations (Sec. \ref{secc7:dm}). Additionally, I identify the portion of a given transformation which acts equally on both helicity components and the one which has a different effect on each of them. Finally, I give two examples of families of meta atoms that can be engineered to have a helicity preserving response in the dipolar approximation.

\section{An inherent symmetry of transformation electromagnetics}\label{secc7:dto}
Let us imagine that we want to build a device that transforms the electromagnetic field in a given way. If the transformation can be written as a change of coordinates, the framework of transformation electromagnetics \cite{Leonhardt2006b,Leonhardt2009} allows us to obtain the constitutive relations that the device must have. 
Transformation electromagnetics is based on the equivalence of Maxwell's equations in two very different scenarios. The macroscopic Maxwell's equations in a general coordinate system, or a general gravitational field, have the same form as in a particular dielectric medium whose constitutive relations depend on the corresponding spacetime metric $g_{\mu\nu}$ \cite{Plebanski1960}. When written in the Riemann-Silberstein notation with the usual restriction to positive energies 
\begin{equation}
\F=\frac{1}{\sqrt{2}}\begin{bmatrix}Z_0\DD+i\BB\\-Z_0\DD+i\BB\end{bmatrix},\
\G=\frac{1}{\sqrt{2}}\begin{bmatrix}\EE+iZ_0\HH\\-\EE+iZ_0\HH\end{bmatrix}.
\end{equation}
those constitutive relations read
\begin{equation}
\label{eqc7:GF}
\F=\begin{bmatrix}A_+&0\\0&A_-\end{bmatrix}\G,
\end{equation}
with $A_\pm^{nm}=(-\sqrt{-g}g^{nm}\mp ig_{0k}\varepsilon^{nkm})/g_{00}$, where $n,m$ and $k$ run from 1 to 3, $g^{\mu\nu}$ is the inverse spacetime metric and $\varepsilon^{nkm}$ is the totally antisymmetric Levi-Civita symbol.

A crude description of the technique of transformation electromagnetics is to say that it is a recipe to find the constitutive relations that would create the optical potential that ``bends'' light in the desired way. The ``bending'' is not only spatial; it can also be spatio-temporal. In general, the coordinate change acts on spacetime and may mix space and time components. 

It is easy to show that duality symmetry is inherent to the framework. The block diagonal form of the constitutive relation (\ref{eqc7:GF}) is the necessary and sufficient condition for duality symmetry in the macroscopic equations (\ref{eqc3:dualmacro}). Duality symmetry is hence an inherent property of electromagnetism in any coordinate system and in curved space time. This is fully consistent with I. Bialynicki-Birula's realization that the two helicity components of an electromagnetic wave do not mix in a gravitational field \cite{Birula1994,Birula1996}. 

Duality symmetry (helicity preservation) is therefore also an inherent property in transformation electromagnetics. Strictly speaking, it can be seen as a necessary condition for any transformation medium.

\section[Metamaterials for transformation optics]{Metamaterials for transformation electromagnetics}\label{secc7:dm}
According to the above discussion, helicity preservation is a necessary condition for a transformation medium. Therefore, a metamaterial designed to act as a transformation medium should preserve helicity, i.e, it should be dual symmetric. In this section, I discuss the conceptually most straightforward way to build dual symmetric meta media, i.e. the use of helicity preserving meta atoms. It is also the only general way that I know of. Additionally, I give two examples of helicity preserving meta atoms. I also obtain some design insights by considering the properties of media that can affect a different action on the two helicities, and the parity transformation properties of both the meta atoms and their lattice arrangement. 

\subsection{Helicity preserving meta atoms}
The obvious way to achieve helicity preservation in an arrangement of scatterers is that each of the scatterers preserves helicity. The results of Sec. \ref{secc6:disorder} hint towards the possibility that this may be the only general way. There is also an argument for venturing that any implementable transformation can be achieved using only helicity preserving scatterers. In transformation electromagnetics, the information contained in the two helicity components of the field is inherently kept apart. Therefore, any coupling of the two helicities by the individual scatterers is an undesired effect which would need to be canceled by their collective arrangement (if such cancellation is at all possible). Some of the degrees of freedom of the arrangement would then have to be sacrificed for this purpose and they would not be available for the implementation of the desired transformation. It then seems disadvantageous to use non-helicity preserving inclusions.

From now on, I will restrict the discussion to metamaterials whose inclusions are helicity preserving. I will also assume that the inclusions are small enough so that they can be treated in the dipolar approximation. In such case, the conditions to be met by the inclusions are the duality conditions for a dipolar scatterer written in (\ref{eqc3:dual}):
\begin{equation}
	\label{eqc7:dualdipol}
\MdE=\epsilon\MmH,\ \MmE=-\frac{\MdH}{\mu},
\end{equation}
which refer to the components of the 6x6 polarizability tensor of the inclusion
\begin{equation}
\begin{bmatrix}\MdE & \MdH\\\MmE & \MmH\end{bmatrix}.
\end{equation}

Equation (\ref{eqc7:dualdipol}) is a restriction on the polarizability tensors of meta atoms in transformation electromagnetics. It reduces the number of possible dipolar tensors to those that preserve helicity. By getting rid of unwanted helicity cross-couplings at the inclusion level, the duality symmetry inherent in the transformation electromagnetics formalism is ensured when using meta atoms that meet (\ref{eqc7:dualdipol}). In a sense, it allows to concentrate the research effort on a particular class of meta atoms. I will briefly discuss two families of helicity preserving meta atoms in Sec. \ref{secc7:two}.

\subsection{Equal and distinct action on the two helicities}
A given transformation can be decomposed into a portion which acts equally on both helicity components and one which has a different action on each of them. One can then make considerations on the necessary spatial inversion properties of lattices and inclusions for metamaterials to act differently on both helicities.

Eq. (\ref{eqc7:GF}) is equivalent to:
\begin{equation}
\label{eqc7:cons}
	\begin{bmatrix}Z_0\DD\\\BB\end{bmatrix}=\begin{bmatrix}\matr{\epsilon}&\matr{\chi}\\-\matr{\chi}&\matr{\epsilon}\end{bmatrix}\begin{bmatrix}\EE\\Z_0\HH\end{bmatrix},
\end{equation}
with $\matr{\epsilon}=-\sqrt{-g}g^{nm}/g_{00}$ and $\matr{\chi}=ig_{0k}\varepsilon^{nkm}/g_{00}$. Note how $\matr{\epsilon}$ depends only on the space-space components of the metric\footnote{Because $n,m$ and $k$ run from 1 to 3 and do not take the value 0 which addresses the spacetime components.} and $\matr{\chi}$ only on the spacetime components of the metric. This separation is discussed in detail in \cite{Leonhardt2006b,Leonhardt2009}: $\matr{\epsilon}$ represents the space only part of the coordinate transformation and $\matr{\chi}$ the part that mixes space and time. For example, the transformation that results in an invisibility cloak has $\chi=0$, while that corresponding to a moving medium has a magneto-electric component $\chi\neq0$ \cite[Sec. 5]{Leonhardt2009}.

From the point of view of helicity, $\matr{\epsilon}$ and $\matr{\chi}$ also have a distinct role. In a transformation medium, the time evolution equations for the field can be written:
\begin{equation}
\label{eqc7:timeevdual}
i\partial_t \begin{bmatrix}{\matr{\epsilon}}-i{\matr{\chi}}&0\\0&{\matr{\epsilon}}+i{\matr{\chi}}\end{bmatrix}\G=\begin{bmatrix}\nabla\times&0\\0&-\nabla\times\end{bmatrix}\G.
\end{equation}

Eq. (\ref{eqc7:timeevdual}) means that $\matr{\epsilon}$ contains the part of the transformation which acts equally on both helicity components, while $\matr{\chi}$ has a different action on each helicity\footnote{To see that this is so, one can use abstract notation (see Sec. \ref{secc3:M}) to substitute the curl ($\JJ\cdot\PP$ in abstract form) by $\pm |\PP|$ for the two helicities, respectively: From Eq. (\ref{eqc3:helfrac}), $\frac{\JJ\cdot\PP}{|\PP|}|\Phi_{\pm}\rangle=\pm|\Phi_{\pm}\rangle\implies\JJ\cdot\PP|\Phi_{\pm}\rangle=\pm|\PP||\Phi_{\pm}\rangle.$ The sign difference in front of the curls in (\ref{eqc7:timeevdual}) cancels the one in $\pm |\PP|$ and the only difference left in the evolution equations of the two helicity components is the different sign in front of $\matr{\chi}$.}. From the coordinate transformation point of view, space-only transformations act equally on the two helicity components while space time mixing transformations have a different effect on each helicity.

\subsubsection{The lattice}
For the overall effective response of the metamaterial, the properties of the three dimensional arrangement need to be taken into account. For example, in a Bravais lattice with sites $\mathbf{r}(n_1,n_2,n_3)$ given by
\begin{equation}
\label{eqc7:brav}
\mathbf{r}(n_1,n_2,n_3)=n_1\mathbf{a}+n_2\mathbf{b}+n_3\mathbf{c},
\end{equation}
where $n_i$ are integers and $(\mathbf{a},\mathbf{b},\mathbf{c})$ are the lattice vectors, spatial inversion is always a symmetry of the lattice because to each point $(n_1,n_2,n_3)$ there exist its spatially inverted image at $(-n_1,-n_2,-n_3)$. The fact that a Bravais lattice has parity symmetry can be used, together with the transformation properties of $(\DD,\BB,\EE,\HH)$ under parity, to show that the lattice cannot induce non-zero values of the constitutive magneto-electric component $\matr{\chi}$. Therefore, in a Bravais lattice, the magneto-electric coupling must originate from the inclusions cross-polarizabilities $\MdH(\MmE)$. This situation is analogous to the breaking of time inversion symmetry in a magnetic crystal due not to the lattice itself, but to the alignment of the magnetic moments of the atoms in it and their transformation properties under time inversion. Figure \ref{figc7:lattice} illustrates the discussion about spatial inversion.

The case of non Bravais lattices is different since they may or may not have parity symmetry.

\subsubsection{The inclusion}
Let me now recall Eq. (\ref{eqc3:txdipoledual}), which connects the values of the fields at the location of a helicity preserving dipolar scatterer with the dipolar moments that they induce in it. Monochromatic fields are assumed:

\begin{equation}
	\label{eqc7:txdipoledual}
	\begin{split}
	&\begin{bmatrix}\mathbf{q}_+\\\mathbf{q}_-\end{bmatrix}=
	\begin{bmatrix}\MdE-i\sqrt{\frac{\epsilon}{\mu}}\MdH & 0\\0 & \MdE+i\sqrt{\frac{\epsilon}{\mu}}\MdH\end{bmatrix}\Gsix.
	\end{split}
\end{equation}
In analogy with the macroscopic case, equation (\ref{eqc7:txdipoledual}) shows that $\MdE$ has the same action on the two helicity eigenstates, while $\MdH$ acts differently because of the different sign preceding it. 

The spatial inversion properties of the inclusion are crucial to establish a priori which inclusions can and which cannot exhibit non-zero cross-polarizabilities $\MdH (\MmE)$. For example, for inclusions that are invariant under a spatial inversion (parity) operation, their cross-polarizabilities can be shown to vanish due to the spatial inversion transformation properties of the fields $(\EE,\HH)$ and the electric and magnetic dipolar moments $(\ed,\md)$. Note that this argument applies independently of the helicity preserving condition of the polarizability tensor. It is worth highlighting that duality symmetry and spatial inversion symmetry are distinct symmetries. Scatterers may have or lack either of the symmetries independently of the other one. Since parity is the only fundamental operator that flips helicity (see Tab. \ref{tabc3:gentrans}), a dual object must break spatial inversion symmetries (not necessarily parity) in order to have a different effect in the two helicities. 

\begin{figure}[h]
	\begin{center}
\subfloat{
\makeatletter{}\def\ASYprefix{}
\newbox\ASYbox
\newdimen\ASYdimen
\long\def\ASYbase#1#2{\leavevmode\setbox\ASYbox=\hbox{#1}\ASYdimen=\ht\ASYbox\setbox\ASYbox=\hbox{#2}\lower\ASYdimen\box\ASYbox}
\long\def\ASYaligned(#1,#2)(#3,#4)#5#6#7{\leavevmode\setbox\ASYbox=\hbox{#7}\setbox\ASYbox\hbox{\ASYdimen=\ht\ASYbox\advance\ASYdimen by\dp\ASYbox\kern#3\wd\ASYbox\raise#4\ASYdimen\box\ASYbox}\put(#1,#2){#5\wd\ASYbox 0pt\dp\ASYbox 0pt\ht\ASYbox 0pt\box\ASYbox#6}}\long\def\ASYalignT(#1,#2)(#3,#4)#5#6{\ASYaligned(#1,#2)(#3,#4){
}{
}{#6}}
\long\def\ASYalign(#1,#2)(#3,#4)#5{\ASYaligned(#1,#2)(#3,#4){}{}{#5}}
\def\ASYraw#1{
currentpoint currentpoint translate matrix currentmatrix
100 12 div -100 12 div scale
#1
setmatrix neg exch neg exch translate}
 
\makeatletter{}\setlength{\unitlength}{1pt}
\makeatletter\let\ASYencoding\f@encoding\let\ASYfamily\f@family\let\ASYseries\f@series\let\ASYshape\f@shape\makeatother{\catcode`"=12\includegraphics{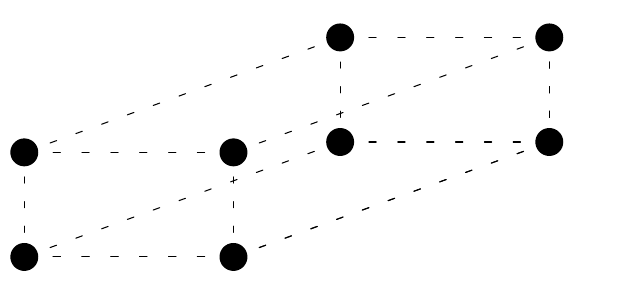}}\definecolor{ASYcolor}{gray}{0.000000}\color{ASYcolor}
\fontsize{12.000000}{14.400000}\selectfont
\usefont{\ASYencoding}{\ASYfamily}{\ASYseries}{\ASYshape}\ASYalign(-142.422182,9.701125)(-0.500000,-1.000000){$\Delta x$}
\definecolor{ASYcolor}{gray}{0.000000}\color{ASYcolor}
\fontsize{12.000000}{14.400000}\selectfont
\ASYalign(-65.306609,26.534272)(-0.318015,-0.778483){$\Delta y$}
\definecolor{ASYcolor}{gray}{0.000000}\color{ASYcolor}
\fontsize{12.000000}{14.400000}\selectfont
\ASYalign(-17.285685,61.649258)(0.000000,-0.500000){$\Delta z$}
\definecolor{ASYcolor}{gray}{0.000000}\color{ASYcolor}
\fontsize{12.000000}{14.400000}\selectfont
\ASYalign(-172.654285,79.811752)(-0.500000,-0.250000){a)}
 
}
\hspace{0.5cm}
\subfloat{
	\makeatletter{}\def\ASYprefix{}
\newbox\ASYbox
\newdimen\ASYdimen
\long\def\ASYbase#1#2{\leavevmode\setbox\ASYbox=\hbox{#1}\ASYdimen=\ht\ASYbox\setbox\ASYbox=\hbox{#2}\lower\ASYdimen\box\ASYbox}
\long\def\ASYaligned(#1,#2)(#3,#4)#5#6#7{\leavevmode\setbox\ASYbox=\hbox{#7}\setbox\ASYbox\hbox{\ASYdimen=\ht\ASYbox\advance\ASYdimen by\dp\ASYbox\kern#3\wd\ASYbox\raise#4\ASYdimen\box\ASYbox}\put(#1,#2){#5\wd\ASYbox 0pt\dp\ASYbox 0pt\ht\ASYbox 0pt\box\ASYbox#6}}\long\def\ASYalignT(#1,#2)(#3,#4)#5#6{\ASYaligned(#1,#2)(#3,#4){
}{
}{#6}}
\long\def\ASYalign(#1,#2)(#3,#4)#5{\ASYaligned(#1,#2)(#3,#4){}{}{#5}}
\def\ASYraw#1{
currentpoint currentpoint translate matrix currentmatrix
100 12 div -100 12 div scale
#1
setmatrix neg exch neg exch translate}
 
	\makeatletter{}\setlength{\unitlength}{1pt}
\makeatletter\let\ASYencoding\f@encoding\let\ASYfamily\f@family\let\ASYseries\f@series\let\ASYshape\f@shape\makeatother{\catcode`"=12\includegraphics{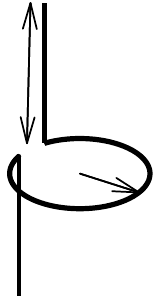}}\definecolor{ASYcolor}{gray}{0.000000}\color{ASYcolor}
\fontsize{12.000000}{14.400000}\selectfont
\usefont{\ASYencoding}{\ASYfamily}{\ASYseries}{\ASYshape}\ASYalign(-39.426677,64.816387)(-1.000000,-0.500000){{\small l}}
\definecolor{ASYcolor}{gray}{0.000000}\color{ASYcolor}
\fontsize{12.000000}{14.400000}\selectfont
\ASYalign(-19.570513,31.512062)(-0.659370,-1.000000){{\small r}}
\definecolor{ASYcolor}{gray}{0.000000}\color{ASYcolor}
\fontsize{12.000000}{14.400000}\selectfont
\ASYalign(-10.898277,76.341976)(-0.500000,-0.250000){b)}
 
}
\\
\subfloat{
\makeatletter{}\def\ASYprefix{}
\newbox\ASYbox
\newdimen\ASYdimen
\long\def\ASYbase#1#2{\leavevmode\setbox\ASYbox=\hbox{#1}\ASYdimen=\ht\ASYbox\setbox\ASYbox=\hbox{#2}\lower\ASYdimen\box\ASYbox}
\long\def\ASYaligned(#1,#2)(#3,#4)#5#6#7{\leavevmode\setbox\ASYbox=\hbox{#7}\setbox\ASYbox\hbox{\ASYdimen=\ht\ASYbox\advance\ASYdimen by\dp\ASYbox\kern#3\wd\ASYbox\raise#4\ASYdimen\box\ASYbox}\put(#1,#2){#5\wd\ASYbox 0pt\dp\ASYbox 0pt\ht\ASYbox 0pt\box\ASYbox#6}}\long\def\ASYalignT(#1,#2)(#3,#4)#5#6{\ASYaligned(#1,#2)(#3,#4){
}{
}{#6}}
\long\def\ASYalign(#1,#2)(#3,#4)#5{\ASYaligned(#1,#2)(#3,#4){}{}{#5}}
\def\ASYraw#1{
currentpoint currentpoint translate matrix currentmatrix
100 12 div -100 12 div scale
#1
setmatrix neg exch neg exch translate}
 
\makeatletter{}\setlength{\unitlength}{1pt}
\makeatletter\let\ASYencoding\f@encoding\let\ASYfamily\f@family\let\ASYseries\f@series\let\ASYshape\f@shape\makeatother{\catcode`"=12\includegraphics{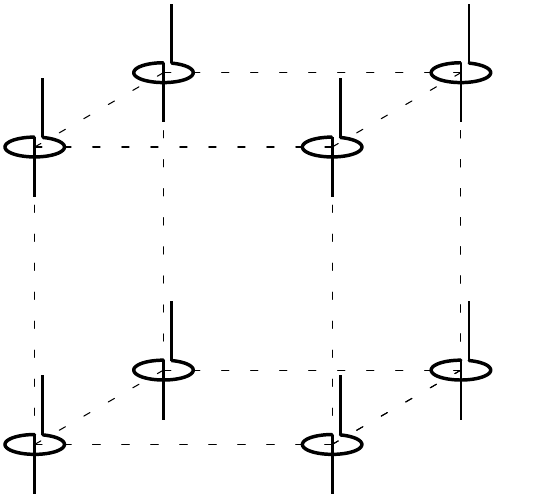}}\definecolor{ASYcolor}{gray}{0.000000}\color{ASYcolor}
\fontsize{12.000000}{14.400000}\selectfont
\usefont{\ASYencoding}{\ASYfamily}{\ASYseries}{\ASYshape}\ASYalign(-100.648280,10.981022)(-0.500000,-1.000000){$\Delta s$}
\definecolor{ASYcolor}{gray}{0.000000}\color{ASYcolor}
\fontsize{12.000000}{14.400000}\selectfont
\ASYalign(-37.242701,22.210894)(-0.211325,-1.000000){$\Delta s$}
\definecolor{ASYcolor}{gray}{0.000000}\color{ASYcolor}
\fontsize{12.000000}{14.400000}\selectfont
\ASYalign(-16.827375,79.066402)(0.000000,-0.500000){$\Delta s$}
\definecolor{ASYcolor}{gray}{0.000000}\color{ASYcolor}
\fontsize{12.000000}{14.400000}\selectfont
\ASYalign(-135.040482,132.804218)(-0.500000,-0.250000){c)}
 
}
	\end{center}
\caption[Lattice units for metamaterials.]{\label{figc7:lattice}(a,c) Lattice unit cells. (b) Single turn helix. (a) Spheres in an orthorombic lattice arrangement ($\Delta x\neq \Delta y\neq \Delta z$). The spatial inversion symmetry of this structure precludes it from exhibiting magneto-electric coupling in its effective constitutive relations. In (c), the parity symmetry of the cubic lattice is broken by the chiral inclusions (b), and a non-null magneto-electric coupling $\chi$ is allowed.}
\end{figure}
\subsection{Dual spheres and dual helices}\label{secc7:two}
I will now briefly discuss two families of helicity preserving meta atoms. They are actually two kinds of inclusions that are commonly considered for metamaterials: Dielectric spheres \cite{Wheeler2010PHDTH,Nieto2011,Soukoulis2011,Liu2012} and conducting chiral inclusions \cite{Semchenko2009,Gansel2009,Soukoulis2011}. 

In the dipolar approximation, a sphere can be engineered to meet the duality condition (\ref{eqc7:dualdipol}) by appropriately choosing its radius as a function of the material \cite{Zambrana2013b}. For a conducting helix and other conducting chiral inclusions, duality can be achieved by adjusting their geometrical dimensions. Fig. \ref{figc7:lattice} illustrates the two kinds of inclusions.

Spheres have spatial inversion symmetry. Therefore $\MdH=\MmE=0$, and it follows that for arrangements of the type (\ref{eqc7:brav}) the sphere is an inclusion that cannot produce a different effect for the two helicities. It also follows that the only condition that a sphere has to meet for it to be dual symmetric is $\MdE=\epsilon\MmH$. Consider then a small dielectric sphere with radius $r_S$ and relative electric and magnetic constants equal to $\epsilon_S$ and 1, respectively. The polarizabilities of a such a sphere when immersed in a homogeneous and isotropic medium can be derived analytically. Their expressions can be found for example in \cite[Chap. 3.4]{Wheeler2010PHDTH}. In the units I have assumed they read:

\begin{equation}
	\label{eqc7:mdemmh}
	\MdE=I\epsilon\frac{6\pi i}{k^3}a_1\ \, \ \MmH=I\frac{6\pi i}{k^3}b_1,
\end{equation}
where $I$ is the identity matrix and $\epsilon$ and $k$ are those of the host medium. The numbers $a_1$ and $b_1$ are the Mie coefficients of dipolar order\footnote{Expressions for the Mie coefficients can be found for example in \cite[Sec. 9.25]{Stratton1941}. I have reproduced them in Chap. \ref{chap3} App. \ref{appc3:fm}.}. The duality condition $\MdE=\epsilon\MmH$ is then met when $a_1=b_1$. For a given wavelength and $\epsilon_S$, the solution to the equation $a_1(r_S)=b_1(r_S)$ determines one particular radius. For that radius, the sphere is dual in the dipolar approximation. Outside the dipolar regime, if higher multipolar orders are considered, the sphere ceases to be dual. There will be some helicity change upon scattering. The idea is that for small spheres, where the non-dipolar terms are very small, the helicity change will be correspondingly small. For example, in \cite{Zambrana2013b}, the following case can be found: A sphere of 130 nm radius and a refractive index of 2.55 is dual in the dipolar regime ($a_1=b_1$) for a wavelength of 780 nm. The total relative helicity conversion due to symmetry breaking higher multipolar orders is of the order of $10^{-4}$ in helicity converted power.

Now to the other example: Conducting helices. Chiral inclusions lack spatial inversion symmetry. Consequently, non-zero electric and magnetic cross-polarizabilities $\MdH$ and $\MmE$ are allowed. This type of inclusions are inherently suitable for the implementation of transformations which mix space and time components, i.e, which have a different action for the two helicities. Helices and chiral split ring resonators (Ch-SRR) \cite{Marques2007,Marques2007b,Semchenko2009,Gansel2009} are being considered as meta atoms for operation from the microwave to the infra-red regime. Analytical expressions for their polarizability tensors have been derived under suitable approximations, \cite{Tretyakov1996,Marques2007b,Radi2013}. Using those expressions, it can be seen that helicity preservation (\ref{eqc7:dualdipol}) can be achieved around the resonant frequency of the inclusion by adjusting its dimensions. For the helix (\cite[Sec. 4.1]{Radi2013}), the key dimensional parameter is $r^2/l$, the ratio between the square of the radius of the loop and the length of the straight wire (see Fig. \ref{figc7:lattice}-(b)). For the Ch-SRR (\cite[Sec. 3]{Marques2007b}), it is the ratio between the square of the radius of the rings and the height separation between the two parallel rings composing the chiral inclusion (see Fig. 1 in \cite{Marques2007b} for a drawing of a Ch-SRR). In electromagnetic terms, the meaning of the key parameter is very similar in both the helix and the Ch-SRR cases. 

The value of the key parameter that makes the inclusion dual has a $1/\omega$ dependency. If the structure is made dual for the resonant frequency, many physically interesting phenomena occur. For example, in \cite{Semchenko2009}, it is shown that a helix meeting such condition interacts only with one of the circular polarizations, that is, is transparent for the other one. In \cite{Radi2013}, such a helix is shown to maximally interact with a given electromagnetic field, extracting the maximum possible power from it. In \cite{Marques2007b} the authors state that, under such condition, a Ch-SRR has several advantages for building negative refractive index metamaterials including wide operation bandwidth and lack of forbidden bands.

All these conditions were found in those works without consideration of the duality symmetry properties of the structure. I believe that the fact that all these interesting and apparently useful phenomena occur when the structure is dual is not a coincidence, but rather further indication that the consideration of the duality symmetry provides a useful guide for the design of meta atoms for transformation devices.

Very recently \cite{Nakata2013}, duality and helicity have been used to discuss the scattering of plane waves by metasurfaces.

\section{Outlook}\label{secc7:outlook}
Given a desired transformation, transformation electromagnetics gives the required constitutive relations. The next step is to find the lattice and inclusions that effectively result in those constitutive relations. 
Many questions arise. For example: Does this problem always have a solution? Is the solution unique? How to fabricate an inclusion with a specified polarizability tensor? 

Let me assume for a moment that there actually exists a systematic way of going from the desired constitutive relations to the polarizability tensors and lattice arrangement of the inclusions. Provided that the desired transformation of the fields can be written as a space-time coordinate change, we can use transformation electromagnetics to obtain the constitutive relations and from these, the polarizability and arrangement of the meta atoms. This situation would then be closer to a fully automated process going from the specification of the transformation to the fabrication of the device. The ability to make inclusions with specified polarizabilities also needs to be addressed. Assuming again that this step is successful, the last obstacle would then be the actual fabrication process, in particular the accuracy with which the inclusions can be fabricated and placed.

A related endeavor is the search for other families of meta atoms that can be engineered to preserve helicity.
 
\makeatletter{}\chapter{Conclusion}\label{chap8}
In my thesis, I have developed a tool for the study, understanding and engineering of the interaction of electromagnetic radiation with material systems. The tool is based on symmetries and conservation laws. At its core lies the systematic use of the electromagnetic duality symmetry and its conserved quantity, the electromagnetic helicity. Their use allows to treat the polarization degrees of freedom of the field in a straightforward way. I have applied the framework to the study of several different problems, obtaining insights and design guidelines in optical activity, zero backscattering, metamaterials for transformation optics and nanophotonics phenomena involving the electromagnetic angular momentum. This conclusion chapter contains a summary of the key contributions and the research outlook.

\section[Key contributions]{Summary of key contributions of this thesis to the field of research}
The theoretical basis and results are established in Chap. \ref{chap3}. 

Section \ref{secc3:M} shows that helicity is the natural operator for describing the non-scalar (polarization) degree of freedom of the electromagnetic field. Maxwell's equations for an infinite isotropic and homogeneous medium can be derived from a small set of assumptions by means of abstract manipulations. Looking for the equations of massless vectorial objects with positive energy automatically brings in the helicity operator in its abstract form. This, and the fact that the transformation generated by helicity, electromagnetic duality, does not affect any other property of the field except for its polarization, distinguish helicity and duality from other means of treating polarization.  Section \ref{secc3:tetmhel} explains the relationship of helicity eigenstates and the commonly used TE and TM polarization modes. Two important differences between helicity and TE/TM are that, first, the TE/TM character does not have a unique symmetry associated to it and, second, that all of the mode dependent symmetries associated with the TE/TM character affect not only the polarization but also other properties of a general field like, for example, momentum or angular momentum.

Section \ref{secc3:helpresdualsym} contains a detailed study of the conditions needed to preserve the helicity of light, that is, to achieve duality invariance. Such conditions are given for three different approximations: The microscopic equations with elementary charged particles, the macroscopic equations and the dipolar scattering approximation. I have obtained the relations between the electric and magnetic charges of hypothetical elementary particles that would separate electrodynamics into two uncoupled domains, one for each helicity. For the macroscopic and dipolar cases, I have obtained the restrictions that duality invariance imposes on the constitutive relations and the polarizability tensor of the dipolar scatterer, respectively. For these two last cases, expressions for the restrictions already existed, but had not been derived and understood considering the two ends of the conservation law: Helicity preservation and duality symmetry.  The macroscopic and dipolar cases have immediate consequences for the analysis of practical scattering problems. In particular, they allow to assert whether a given scatterer will or will not preserve the helicity of the light interacting with it. This completes the framework and gives it general applicability. Some of the contents of this chapter were published in [FCZPMT12], [FCZPT$^+$13] and [FCMT13].

In Chaps. \ref{chap4} to \ref{chap7}, I have used the framework developed in Chap. \ref{chap3} to study different problems.

In Chap. \ref{chap4}, helicity and angular momentum are used to clarify the underlying symmetry reasons for observations commonly attributed to the mechanism of ``spin to orbital angular momentum transfer'' a.k.a ``spin orbit interaction''. I show that the reason for the appearance of optical vortices in focused fields is the breaking of transverse translational symmetry by the lens. On the other hand, in the case of scattering off a cylindrically symmetric target, the vortices appear due to the breaking of duality symmetry by the scatterer. The symmetry approach avoids the separate consideration of the spin and orbital angular momentum operators. This separation is not possible for solutions of Maxwell's equations since both operators generally break the transversality of the fields they act on. The chapter also contains a study of two well known transverse operators whose sum also results in the total angular momentum operator. In particular, I have derived the transformations that they generate and have given examples of their use in light matter interactions. As expected, those transformations are not rotations. Instead, they are related to helicity and frequency dependent translations. Additionally, I have shown that there is a direct relationship between the transverse ``spin'' operators and the spatial part of the Pauli-Lubanski four vector. Most of these results are contained in the published article [FCZPMT12] or in the manuscript [FCZPMT13], which is currently under review.   

Chapter \ref{chap5} shows how the interplay of two different symmetries, discrete rotational symmetries and duality symmetry, can be used to understand and engineer the forward and backward scattering properties of an object. In particular, I have shown that a dual symmetric scatterer with a discrete rotational symmetry $R_z(2\pi/m)$ of degree $m\ge 3$ will necessarily exhibit zero backscattering, that is, no energy will be scattered backwards. Using these results, I have given a symmetry based recipe for building a two dimensional array exhibiting zero backscattering. These results were published in\footnote{In this section, I will refer to the publications listed in the publication list, located after the acknowledgments} [FC13]. The symmetry arguments relating to scattering off magnetic spheres are contained in [FCZPT$^+$13] and were extended to cylindrally symmetric objects in [ZPFCJ$^+$13].

Chapter \ref{chap6} contains a study of optical activity from the point of view of symmetries and conservation laws. I have shown that helicity preservation is a necessary condition for optical activity understood as the rotation of linear polarization between tow input/output plane waves. This necessary condition is not normally taken into account in the design of structures with artificial optical activity. This omission results in polarization rotation angles that are not independent of the input polarization state. The omission may be due to the fact that the most common setting for the observation of natural optical activity, the forward scattering direction of a solution of chiral molecules, preserves helicity ``automatically''. This preservation, shown not to happen in non-forward directions, is not related to duality symmetry. I have proved that it is due to the effective cylindrical symmetry that the random orientations of the molecules endows the solution with. The chapter contains analytical and numerical evidence strongly suggesting that in a general non-forward scattering direction, duality symmetry is a requirement for helicity preservation and it is therefore also a requirement for optical activity in those directions. Taking all these results into account I have given a symmetry based recipe for building two dimensional arrays exhibiting artificial optical activity on a general direction. Finally, I have shown how the different results that optical activity and Faraday rotation produce upon a propagation-reflection-propagation round trip are connected to the different behavior of the two systems under time and space inversion transformations. Most of these results were published in [FCVTMT13].

In Chap. \ref{chap7} I have used the duality symmetry inherent to the mathematics of transformation electromagnetics to constrain the polarizability tensors of the meta atoms used to build transformation electromagnetic devices. The chapter contains arguments that show that helicity preserving meta atoms meeting the dipolar duality conditions are the only class of meta atoms needed to build a general transformation device. Two families of dual symmetric meta atoms, one chiral and one non-chiral are briefly discussed. Additionally, the chapter provides some insight into the kinds of transformations that can be obtained by chiral and non-chiral meta atoms. These results were published in [FCMT13].

\section{Outlook}
As shown in the application chapters, the current state of the tool already allows its use in many cases of practical interest. 

I believe that other cases where it should lead to new insights are the study of the spin Hall effect of light \cite{Bliokh2008,Shitrit2011,Luo2012,Yin2013} or the action of the so called ``q-plates'' \cite{Marrucci2006}. Both cases are commonly explained as a ``spin to orbital angular momentum'' transfer. 

I also believe that the new understanding of the symmetries involved in optical activity can be used for the design of sensitive schemes for detection of chiral molecules and sorting of their chiral phase. 

Nevertheless, the tool has areas with room for improvement that deserve careful consideration. 

It is clearly interesting to attempt to increase the quantitative ability of the tool. For example, for the zero backscattering objects of Chap. \ref{chap5}, it would be important to have some information about how much energy is scattered in directions that are a small deviation away from the backward direction. In each particular system, the answer can be obtained by simulation, but it is conceivable that the manifestation of a {\em slightly} broken symmetry has general consequences. 

From the qualitative point of view, a proof of the Gell-Mann's totalitarian principle, ``everything that is not forbidden is mandatory'', maybe in the form ``everything that is not forbidden {\em by symmetry} is mandatory'', would endow the tool with more qualitative power.

With respect to extensions, a natural one would be to extend it so that multi photon states can be considered. Since multi photon states belong to products of electromagnetic Hilbert spaces, decomposing those spaces into their irreducible components and enforcing the correct permutation symmetry seems to be a plausible starting point.

An important extension would be the consideration of other elementary particles, notably the electron. The tandem helicity/duality provides a route for controlling the non-scalar degree of freedom of the electromagnetic field. A similar tandem for the electron would no doubt have an impact on the practical applications of the polarization of electrons like ``spintronics''. The fact that the electron has mass is a crucial difference. Helicity is a Lorentz invariant for massless fields but is not Lorentz invariant for massive particles: It changes with Lorentz boosts. From the point of view of group theory, the free electromagnetic field may be seen as being composed of two separate entities of different helicity. This is not true for the electron and will certainly impact the extension of the framework to treat its polarization. On the other hand, the equations that govern the propagation of spacetime curvature in vacuum have been shown to be symmetric under a duality transformation \cite{Maartens1998}. This is consistent with the fact that the gravitational field is massless, and should allow a smoother extension of the theory of helicity and duality that I have developed to the gravitational field.

Finally, I believe that one of the interesting applications of symmetries and conservation laws, in particular duality, is the quest for a systematic design strategy for transformation electromagnetic devices. As briefly mentioned in Sec. \ref{secc7:outlook}, it would be desirable to find a systematic way to, given the constitutive relations specified by the desired transformation, obtain the polarizability tensors and three dimensional arrangement of the meta medium exhibiting such effective response.
 
\makeatletter{}

\newpage
\thispagestyle{empty}
\hspace{-6cm}
\vspace{-9cm}
\noindent\includegraphics{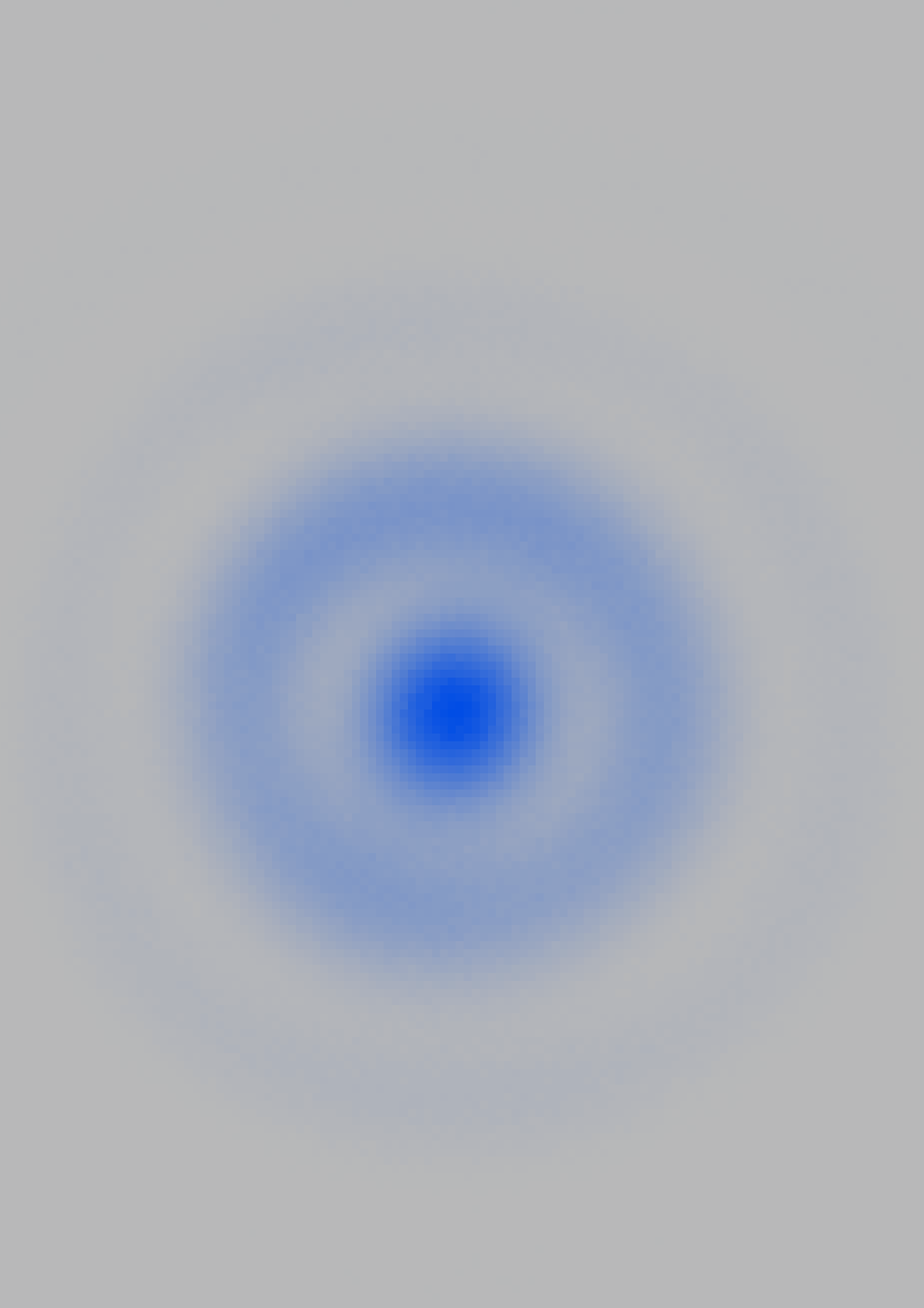}
\end{document}